%% file: main.tex
\documentclass[12pt]{article}
\usepackage{a4wide,graphicx,amsmath,amssymb,hyperref}
\usepackage[numbers,sort&compress]{natbib}
\usepackage{hypernat}
\renewcommand{\vec}[1]{\boldsymbol{#1}}

\begin{document}
  \titlepage
  \begin{flushright}
    IPPP/08/95 \\
    DCPT/08/190 \\
    Cavendish-HEP-08/16 \\
    11th June 2009 \\
  \end{flushright}
  
  \vspace*{0.5cm}
  
  \begin{center}
    {\Large \bf Parton distributions for the LHC} \\
    \vspace*{1cm}
    \textsc{A.D. Martin$^a$, W.J. Stirling$^b$, R.S. Thorne$^c$ and G. Watt$^{c}$} \\
    \vspace*{0.5cm}
    $^a$ Institute for Particle Physics Phenomenology, University of Durham, DH1 3LE, UK \\
    $^b$ Cavendish Laboratory, University of Cambridge, CB3 0HE, UK\\
    $^c$ Department of Physics and Astronomy, University College London, WC1E 6BT, UK
  \end{center}
  
  \vspace*{0.5cm}
  
  \begin{abstract}
    We present updated leading-order, next-to-leading order and next-to-next-to-leading order parton distribution functions (``MSTW 2008'') determined from global analysis of hard-scattering data within the standard framework of leading-twist fixed-order collinear factorisation in the $\overline{\rm MS}$ scheme.  These parton distributions supersede the previously available ``MRST'' sets and should be used for the first LHC data-taking and for the associated theoretical calculations.  New data sets fitted include CCFR/NuTeV dimuon cross sections, which constrain the strange quark and antiquark distributions, and Tevatron Run II data on inclusive jet production, the lepton charge asymmetry from $W$ decays and the $Z$ rapidity distribution.  Uncertainties are propagated from the experimental errors on the fitted data points using a new dynamic procedure for each eigenvector of the covariance matrix.  We discuss the major changes compared to previous MRST fits, briefly compare to parton distributions obtained by other fitting groups, and give predictions for the $W$ and $Z$ total cross sections at the Tevatron and LHC.
  \end{abstract}
  
  \newpage
  \tableofcontents
  
  \include{introduction}
  \include{surveyexperimental}
  \include{overviewtheoretical}
  \include{treatmentofheavyflavours}
  \include{globalpartonanalyses}
  \include{erroranalysis}
  \include{inclusivestructurefunctions}
  \include{dimuondata}
  \include{heavyflavourdata}
  \include{lowenergydrellyan}
  \include{wandzattevatron}
  \include{jetdata}
  \include{longitudinal}
  \include{comparison}
  \include{predictions}
  \include{conclusions}
  \include{acknowledgements}

\input{references}
\end{document}

%% file: introduction.tex
\section{Introduction} \label{sec:introduction}
In deep-inelastic scattering (DIS), and ``hard'' proton--proton (or proton--antiproton) high-energy collisions, the scattering proceeds via the partonic constituents of the hadron.  To predict the rates of the various processes a set of universal parton distribution functions (PDFs) is required.  These distributions are best determined by global fits to all the available DIS and related hard-scattering data.  The fits can be performed at leading-order (LO), next-to-leading order (NLO) or at next-to-next-to-leading order (NNLO) in the strong coupling $\alpha_S$.  Over the last couple of years there has been a considerable improvement in the precision, and in the kinematic range, of the experimental measurements for many of these processes, as well as new types of data becoming available.  In addition, there have been valuable theoretical developments, which increase the reliability of the global analyses.  It is therefore timely, particularly in view of the forthcoming experiments at the Large Hadron Collider (LHC) at CERN, to perform new global analyses which incorporate all of these improvements.

The year 2008 marked the twentieth anniversary of the publication of the first MRS analysis of parton distributions~\cite{Martin:1987vw}, which was, indeed, the first NLO global analysis.  Following this initial MRS analysis there have been numerous updates over the years, necessitated by both the regular appearance of new data sets and by theoretical developments~\cite{Martin:1988nk,Harriman:1990hi,Kwiecinski:1990ru,Martin:1992as,Martin:1992zi,Martin:1994kn,Martin:1995ws,Glover:1996ae,Martin:1996as,Martin:1996eva,Martin:1998sq,Martin:1999ww,Martin:2001es,Martin:2002dr,Martin:2002aw,Martin:2003sk,Martin:2004ir,Martin:2004dh,Martin:2006qz,Martin:2007bv}.  In the modern era, the MRST98 (NLO) sets~\cite{Martin:1998sq} were the first of our sets to take full advantage of a large amount of new HERA structure function data, and the first to incorporate heavy quarks in a consistent and rigorous way in the default PDF set (using a general-mass variable flavour number scheme approach; see Section~\ref{sec:heavyflavour}).  The uncertainty in the PDFs was explored by producing a modest number of additional sets with different gluon distributions and different values of the strong coupling.  The following year, the NLO set was updated (MRST99), a LO set was produced, and new sets corresponding to a different treatment of higher-twist contributions and the DIS scheme were presented~\cite{Martin:1999ww}.  The year 2001 saw a major update~\cite{Martin:2001es}, with a NNLO (MRST 2001) set, based on an approximation of the NNLO splitting functions~\cite{vanNeerven:1999ca,vanNeerven:2000uj}, produced for the first time~\cite{Martin:2002dr}, alongside new NLO and LO sets.  The grid interpolation was also improved to allow for faster and more accurate access to the PDFs in the public interface code.  The following year the Hessian approach (see Section~\ref{sec:erroranalysis}) was used to produce a ``parton distributions with errors'' package (MRST 2001 E) comprising a central NLO set and 30 extremum sets~\cite{Martin:2002aw}.  The central NLO and NNLO sets were updated slightly in the same year (MRST 2002)~\cite{Martin:2002aw}, using an improved approximation to the NNLO splitting functions~\cite{vanNeerven:2000wp}.  In 2003, fits were performed in which the $x$ and $Q^2$ range of DIS structure function data was restricted to ensure stability with respect to cuts on the data, and corresponding NLO and NNLO ``conservative'' variants of the MRST 2002 sets were derived (MRST 2003 C)~\cite{Martin:2003sk}.  The next major milestone was in 2004, with a substantial update of the NLO and NNLO sets (MRST 2004)~\cite{Martin:2004ir}, the latter using the full NNLO splitting functions~\cite{Moch:2004pa,Vogt:2004mw} for the first time and both incorporating a ``physical'' parameterisation of the gluon distribution in order to better describe the high-$E_T$ Tevatron jet data.  A NLO set incorporating ${\cal O}(\alpha)$ QED corrections in the DGLAP evolution equations was also produced for the first time (MRST 2004 QED)~\cite{Martin:2004dh}, together with fixed-flavour-number LO and NLO variants~\cite{Martin:2006qz}.  Finally, in 2006 a NNLO set ``with errors'' was produced for the first time (MRST 2006 NNLO)~\cite{Martin:2007bv}, using a new general-mass variable flavour number scheme and with broader grid coverage in $x$ and $Q^2$ than in previous sets.

In this paper we present the new MSTW 2008 PDFs at LO, NLO and NNLO.  These sets are a major update to the currently available MRST 2001 LO~\cite{Martin:2002dr}, MRST 2004 NLO~\cite{Martin:2004ir} and MRST 2006 NNLO~\cite{Martin:2007bv} PDFs.
\begin{figure}
  \centering
  \includegraphics[width=0.82\textwidth]{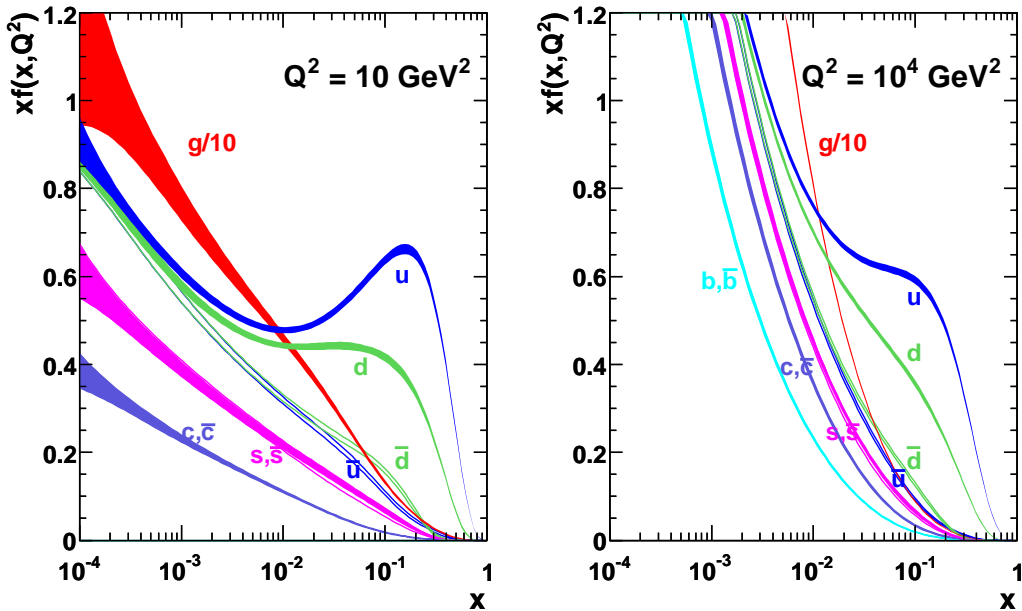}
  \caption{MSTW 2008 NLO PDFs at $Q^2 = 10$ GeV$^2$ and $Q^2 = 10^4$ GeV$^2$.}
  \label{fig:nloallpdfs}
\end{figure}
The ``end products'' of the present paper are grids and interpolation code for the PDFs, which can be found at Ref.~\cite{mstwpdf}.  An example is given in Fig.~\ref{fig:nloallpdfs}, which shows the NLO PDFs at scales of $Q^2 = 10$ GeV$^2$ and $Q^2 = 10^4$ GeV$^2$, including the associated one-sigma (68\%) confidence level (C.L.) uncertainty bands.

The contents of this paper are as follows.  The new experimental information is summarised in Section~\ref{sec:surveyexperimental}.  An overview of the theoretical framework is presented in Section~\ref{sec:overviewtheoretical} and the treatment of heavy flavours is explained in Section~\ref{sec:heavyflavour}.  In Section~\ref{sec:globalfit} we present the results of the global fits and in Section~\ref{sec:erroranalysis} we explain the improvements made in the error propagation of the experimental data to the PDF uncertainties, and their consequences.  Then we present a more detailed discussion of the description of different data sets included in the global fit: inclusive DIS structure functions (Section~\ref{sec:inclusivedata}), dimuon cross sections from neutrino--nucleon scattering (Section~\ref{sec:dimuon}), heavy flavour DIS structure functions (Section~\ref{sec:heavyflavourdata}), low-energy Drell--Yan production (Section~\ref{sec:lowmassDY}), $W$ and $Z$ production at the Tevatron (Section~\ref{sec:wztevatron}), and inclusive jet production at the Tevatron and at HERA (Section~\ref{sec:jetdata}).  In Section~\ref{sec:longitudinal} we discuss the low-$x$ gluon and the description of the longitudinal structure function, in Section~\ref{sec:comparison} we compare our PDFs with other recent sets, and in Section~\ref{sec:totalpredictions} we present predictions for $W$ and $Z$ total cross sections at the Tevatron and LHC.  Finally, we conclude in Section~\ref{sec:conclusions}.  Throughout the text we will highlight the numerous refinements and improvements made to the previous MRST analyses.

%% file: surveyexperimental.tex
\section{Survey of experimental developments} \label{sec:surveyexperimental}
Since the most recent MRST analyses~\cite{Martin:2002dr,Martin:2004ir,Martin:2007bv} a large number of new data sets suitable for inclusion in the global fit have become available, or are included for the first time. Some of these are entirely new types of data, while others supersede existing sets, either improving the precision, extending the kinematic range, or both.  Here, we list the new data that we include in the global fit, together with an indication of the parton distributions that they mainly constrain.

\begin{enumerate}
  \renewcommand{\labelenumi}{(\roman{enumi})}

\item Compared to the analysis in Ref.~\cite{Martin:2004ir} there is no new large $ep \to eX$ deep-inelastic $F_2$ structure function data set.  However, note that we now fit to the measured reduced cross section values
\begin{equation} \label{eq:sigred}
  \tilde \sigma(x,Q^2) = F_2(x,Q^2) -\frac{y^2}{1+(1-y)^2}F_L(x,Q^2),
\end{equation}
where $y=Q^2/(xs)$ and $\sqrt{s}$ is the $ep$ centre-of-mass energy.  In fact, this was already done in Refs.~\cite{Martin:2006qv,Martin:2007bv}.  Since $y\ll 1$ in most of the kinematic range, $\tilde \sigma(x,Q^2)$ is effectively the same as $F_2(x,Q^2)$.  However, at HERA, for the lowest $x$-values at a given $Q^2$, the value of $y$ can become as large as $0.7$--$0.8$, and the effect of $F_L(x,Q^2)$ becomes apparent~\cite{Adloff:1996yz,Gogitidze:2001jz}.  We now include a small, but important, number of data points from H1 in Ref.~\cite{Lobodzinska:2003yd}.  The effect of $F_L(x,Q^2)$ is seen in the data as a flattening of the growth of $\tilde \sigma(x,Q^2)$ as $x$ decreases to very small values (for fixed $Q^2$), leading eventually to a turnover.  Hence, for precise analysis of the HERA structure function data it is particularly important to fit any theoretical prediction to the measured $\tilde \sigma(x,Q^2)$, rather than to model-dependent extracted values of $F_2(x,Q^2)$.  In previous analyses we have used the next best approach to fitting to $\tilde \sigma(x,Q^2)$ directly by making the correction using our own prediction for $F_L(x,Q^2)$, maintaining self-consistency.  We now also include data on $F_L(x,Q^2)$ at high $x$ from fixed-target experiments~\cite{Benvenuti:1989rh,Arneodo:1996qe,Whitlow:1990gk}, which provide a weak constraint on the gluon for $x \sim 0.1$.

\item The structure functions $F_2$ and $xF_3$ for the charged-current deep-inelastic processes $e^-p \to \nu_e X$ and $e^+p \to \bar{\nu}_e X$ have been measured at HERA~\cite{Adloff:2003uh,Chekanov:2003vw}.  At large $x$ these processes are dominated by the $u$ and $d$ valence quark distributions respectively, although the event rate is low in this domain.  In principle, these data are valuable as they constrain exactly the same partonic combinations as the data which exist for the neutrino processes $\bar{\nu}_\mu N \to \mu^+ X$ and $\nu_\mu N \to \mu^- X$, but without the problems of having to allow for the effects of the heavy nuclear target, $N$, but the precision of published data is still relatively low.  We include the data on $e^+p \to \bar{\nu}_e X$ since these are more precise, and constrain the less well-known down valence distribution.

\item We include new neutrino structure function data from NuTeV (with an iron target)~\cite{Tzanov:2005kr} and CHORUS (with a lead target)~\cite{Onengut:2005kv}, though consistency requires the use of only data with $x<0.5$, as discussed later in Section~\ref{sec:neutrinoinclusive}.  These provide information on the flavour decomposition of quarks for $0.01 \lesssim x \lesssim 0.5$.

\item Data also from the neutrino DIS experiments NuTeV and CCFR on dimuon production give a direct~\cite{Goncharov:2001qe} constraint on the strange quark content of the proton for $0.01 \lesssim x \lesssim 0.2$, replacing the previous direct, but model-dependent, constraint on the strange distribution~\cite{Bazarko:1994tt}.  Furthermore, the separation into neutrino and antineutrino cross sections allows a separation into quark and antiquark contributions and therefore a determination of the strangeness asymmetry.

\item There are now improved measurements of the structure functions, $F_2^{c\bar{c}}$ and $F_2^{b\bar{b}}$~\cite{Adloff:1996xq,Adloff:2001zj,Aktas:2005iw,Aktas:2004az,Breitweg:1999ad,Chekanov:2003rb,Chekanov:2007ch}, for heavy quark production at HERA.  We include all of the published data on $F_2^{c\bar{c}}$ in the analysis.  Since these HERA data are driven by the small-$x$ gluon distribution, via the $g \to {H\overline{H}}$ transition, they probe the gluon for $x \sim 0.001$.  However, they also provide useful information on the mass of the charm quark and are a test of our procedure for including heavy flavours.  Since the data on $F_2^{b\bar{b}}$ are far less numerous and of lower precision, they do not provide any further constraints on the gluon and so we simply compare to these.

\item There are Tevatron Run II data on the lepton charge asymmetry from $W$ decays~\cite{Acosta:2005ud,Abazov:2007pm,Abazov:2008qv}.  These are more precise than the Run I data~\cite{Abe:1998rv} and provide constraints on the down quark distribution for $x \gtrsim 0.05$, and also to a smaller extent on the up and down sea quarks for $x \approx 0.1$.

\item For the first time there are data from CDF~\cite{Han:2008} and D0~\cite{Abazov:2007jy} on the $Z$ ($\to\ell^+\ell^-$) rapidity distribution which constrain the quarks for $x \gtrsim 0.05$ with a different weighting than DIS data and at a significantly higher $Q^2$ scale.

\item There now exist data for inclusive jet production from Run II at the Tevatron from CDF~\cite{Abulencia:2007ez,Aaltonen:2008eq} and D{\O}~\cite{Abazov:2008hu}.  These data exist in different rapidity bins, are more precise than the previous Run I data and go out to larger jet $p_T$ values.  They provide constraints on the gluon (and quark) distributions in the domain $0.01 \lesssim x \lesssim 0.5$.

\item We include data from HERA on inclusive jet production in DIS~\cite{Chekanov:2002be,Chekanov:2006xr,Aktas:2007pb}.  (We do not include the photoproduction data because of potential sensitivity to photon PDFs.)  These data constrain the gluon for $0.01 \lesssim x \lesssim 0.1$.

\end{enumerate}

The total data that we use in the new global analyses are listed in Table \ref{tab:chisquared} of Section~\ref{sec:globalfit}, together with the individual $\chi^2$ values for each data set for the LO, NLO and NNLO fits.  A rough indication of the particular parton distributions that the various data constrain is also given in Table 1.
\begin{table}
  \begin{center}
    \begin{tabular}{llll}
      \hline
      \hline
      Process & Subprocess & Partons & $x$ range \\ \hline
      $\ell^\pm\,\{p,n\}\to\ell^\pm\,X$ & $\gamma^*q\to q$ & $q,\bar{q},g$ & $x\gtrsim 0.01$ \\
      $\ell^\pm\,n/p\to\ell^\pm\,X$ & $\gamma^*\,d/u\to d/u$ & $d/u$ & $x\gtrsim 0.01$ \\
      $pp\to \mu^+\mu^-\,X$ & $u\bar{u},d\bar{d}\to\gamma^*$ & $\bar{q}$ & $0.015\lesssim x\lesssim 0.35$ \\
      $pn/pp\to \mu^+\mu^-\,X$ & $(u\bar{d})/(u\bar{u})\to \gamma^*$ & $\bar{d}/\bar{u}$ & $0.015\lesssim x\lesssim 0.35$ \\
      $\nu (\bar{\nu})\,N \to \mu^-(\mu^+)\,X$ & $W^*q\to q^\prime$ & $q,\bar{q}$ & $0.01 \lesssim x \lesssim 0.5$ \\
      $\nu\,N \to \mu^-\mu^+\,X$ & $W^*s\to c$ & $s$ & $0.01\lesssim x\lesssim 0.2$ \\
      $\bar{\nu}\,N \to \mu^+\mu^-\,X$ & $W^*\bar{s}\to\bar{c}$ & $\bar{s}$ & $0.01\lesssim x\lesssim 0.2$ \\\hline
      $e^\pm\,p \to e^\pm\,X$ & $\gamma^*q\to q$ & $g,q,\bar{q}$ & $0.0001\lesssim x\lesssim 0.1$ \\
      $e^+\,p \to \bar{\nu}\,X$ & $W^+\,\{d,s\}\to \{u,c\}$ & $d,s$ & $x\gtrsim 0.01$ \\
      $e^\pm p\to e^\pm\,c\bar{c}\,X$ & $\gamma^*c\to c$, $\gamma^* g\to c\bar{c}$ & $c$, $g$ & $0.0001\lesssim x\lesssim 0.01$ \\
      $e^\pm p\to\text{jet}+X$ & $\gamma^*g\to q\bar{q}$ & $g$ & $0.01\lesssim x\lesssim 0.1$ \\ \hline
      $p\bar{p}\to \text{jet}+X$ & $gg,qg,qq\to 2j$ & $g,q$ & $0.01\lesssim x\lesssim 0.5$ \\
      $p\bar{p}\to (W^\pm\to\ell^{\pm}\nu)\,X$ & $ud\to W,\bar{u}\bar{d}\to W$ & $u,d,\bar{u},\bar{d}$ & $x\gtrsim 0.05$ \\
      $p\bar{p}\to (Z\to\ell^+\ell^-)\,X$ & $uu,dd\to Z$ & $d$ & $x\gtrsim 0.05$
      \\
      \hline
      \hline
    \end{tabular}
  \end{center}
  \caption{The main processes included in the current global PDF analysis ordered in three groups: fixed-target experiments, HERA and the Tevatron.  For each process we give an indication of their dominant partonic subprocesses, the primary partons which are probed and the approximate range of $x$ constrained by the data.}
\end{table}

%% file: overviewtheoretical.tex
\section{Overview of theoretical framework} \label{sec:overviewtheoretical}

In this section we first give a brief overview of the standard theoretical formalism used, and then present a summary of the theoretical improvements and changes in methodology in the global analysis.  A more detailed discussion of the various items is given later in separate sections.

We work within the standard framework of leading-twist fixed-order collinear factorisation in the $\overline{\rm MS}$ scheme, where structure functions in DIS, $F_i(x,Q^2)$, can be written as a convolution of coefficient functions, $C_{i,a}$, with PDFs of flavour $a$ in a hadron of type $A$, $f_{a/A}(x,Q^2)$, i.e.
\begin{equation}
  F_i(x,Q^2) = \sum_{a=q,g}\;C_{i,a}\otimes f_{a/A}(x,Q^2).
\end{equation}
Similarly, in hadron--hadron collisions, hadronic cross sections can be written as process-dependent partonic cross sections convoluted with the same universal PDFs, i.e.
\begin{equation}
  \sigma_{AB} = \sum_{a,b=q,g}\;\hat{\sigma}_{ab}\otimes f_{a/A}(x_1,Q^2)\otimes f_{b/B}(x_2,Q^2).
\end{equation}

The scale dependence of the PDFs is given by the DGLAP evolution equation in terms of the calculable splitting functions, $P_{aa^\prime}$, i.e.
\begin{equation} \label{eq:dglap}
  \frac{\partial f_{a/A}}{\partial \ln Q^2} = \sum_{a^\prime=q,g} P_{aa^\prime}\otimes f_{a^\prime/A}.
\end{equation}
The DIS coefficient functions, $C_{i,a}$, the partonic cross sections, $\hat{\sigma}_{ab}$, and the splitting functions, $P_{aa^\prime}$, can each be expanded as perturbative series in the running strong coupling, $\alpha_S(Q^2)$.  The strong coupling satisfies the renormalisation group equation, which up to NNLO reads
\begin{equation} \label{eq:rge}
  \frac{{\rm d}}{{\rm d}\ln Q^2}\left(\frac{\alpha_S}{4\pi}\right) = -\beta_0\left(\frac{\alpha_S}{4\pi}\right)^2-\beta_1\left(\frac{\alpha_S}{4\pi}\right)^3-\beta_2\left(\frac{\alpha_S}{4\pi}\right)^4-\ldots.
\end{equation}

The input for the evolution equations, \eqref{eq:dglap} and \eqref{eq:rge}, $f_{a/A}(x,Q_0^2)$ and $\alpha_S(Q_0^2)$, at a reference input scale, taken to be $Q_0^2 = 1$ GeV$^2$, must be determined from a global analysis of data.  In the present study we use a slightly extended form, compared to previous MRST fits, of the parameterisation of the parton distributions at the input scale $Q_0^2=1 ~{\rm GeV}^2$:
\begin{align}
  xu_v(x,Q_0^2) &= A_u\,x^{\eta_1} (1-x)^{\eta_2} (1 + \epsilon_u\,\sqrt{x} + \gamma_u\,x),  \label{eq:uv}\\
  xd_v(x,Q_0^2) &= A_d\,x^{\eta_3} (1-x)^{\eta_4} (1 + \epsilon_d\,\sqrt{x} + \gamma_d\,x), \label{eq:dv} \\
  xS(x,Q_0^2) &= A_S\,x^{\delta_S} (1-x)^{\eta_S} (1 + \epsilon_S\,\sqrt{x} + \gamma_S\,x),  \label{eq:S} \\
  x\Delta(x,Q_0^2) &= A_\Delta\,x^{\eta_\Delta} (1-x)^{\eta_S+2} (1 + \gamma_\Delta\,x + \delta_\Delta\,x^2),\\
  xg(x,Q_0^2) &= A_g\,x^{\delta_g} (1-x)^{\eta_g} (1 + \epsilon_g\,\sqrt{x} + \gamma_g\,x) + A_{g^\prime}\,x^{\delta_{g^\prime}} (1-x)^{\eta_{g^\prime}}, \label{eq:inputxg}\\
  x(s+\bar{s})(x,Q_0^2) & = A_{+}\,x^{\delta_S}\,(1-x)^{\eta_{+}} (1 + \epsilon_S\,\sqrt{x} + \gamma_S\,x),\\
  x(s-\bar{s})(x,Q_0^2) & = A_{-}\,x^{\delta_{-}} (1-x)^{\eta_{-}} (1-x/x_0), \label{eq:sv}
\end{align}
where $\Delta\equiv \bar{d}-\bar{u}$, $q_v\equiv q-\bar{q}$, and where the light quark sea contribution is defined as
\begin{equation} \label{eq:Ssum}
  S \equiv 2(\bar{u}+\bar{d})+s+\bar{s}.
\end{equation}
The input PDFs listed in Eqs.~\eqref{eq:uv}--\eqref{eq:sv} are subject to three constraints from number sum rules:
\begin{equation} \label{eq:numbersumrule}
  \int_0^1\!{\rm d}x\;u_v(x,Q_0^2) = 2,\qquad\int_0^1\!{\rm d}x\;d_v(x,Q_0^2) = 1,\qquad\int_0^1\!{\rm d}x\;s_v(x,Q_0^2) = 0,
\end{equation}
together with the momentum sum rule:
\begin{equation}
  \int_0^1\!{\rm d}x\;x\left[u_v(x,Q_0^2)+d_v(x,Q_0^2)+S(x,Q_0^2)+g(x,Q_0^2)\right] = 1.
\end{equation}
We use these four constraints to determine $A_g$, $A_u$, $A_d$ and $x_0$ in terms of the other parameters.  There are therefore potentially $34-4 = 30$ free PDF parameters in the fit, including $\alpha_S$.  The values of the parameters obtained in the LO, NLO and NNLO fits are given in Table~\ref{tab:parameters} in Section~\ref{sec:globalfit} below.  (In practice, we fix $\delta_- = 0.2$ in \eqref{eq:sv} due to extreme correlation with $A_{-}$ and $\eta_-$.)  For the LO fit, where there is no tendency for the input gluon distribution to go negative at small $x$, the second term of the parameterisation \eqref{eq:inputxg} is omitted.

The major changes to the analysis framework, compared to previous MRST analyses, are listed below.

\begin{enumerate}
  \renewcommand{\labelenumi}{(\roman{enumi})}

\item We produce and present PDF sets at LO, NLO and NNLO in $\alpha_S$.  The last of these uses the splitting functions calculated in Refs.~\cite{Moch:2004pa,Vogt:2004mw} and the (massless) coefficient functions for structure functions calculated in Refs.~\cite{vanNeerven:1991nn,Zijlstra:1991qc,Zijlstra:1992kj,Zijlstra:1992qd,Moch:2004xu,Vermaseren:2005qc} together with the massive coefficient functions described in Section~\ref{sec:heavyflavour}.  We do not include here a set of modified LO PDFs for use in LO Monte Carlo generators of the form described in Refs.~\cite{Sherstnev:2007nd,Sherstnev:2008dm}.  This will be the topic of a further study.  We produce eigenvector sets describing the uncertainty due to experimental errors in all cases.  This is the first time this has been done at LO.  We have significantly modified the method of determining the size of the uncertainties, no longer using a simple $\Delta \chi^2 = 50$ above the global minimum to determine the $90\%$ confidence level uncertainty for the PDFs.  We now have a {\it dynamic} determination of the tolerance for each eigenvector direction, which gives qualitatively similar results to previously.  That is, we now have no {\it a priori} fixed value for the tolerance.  This method is described in detail in Section~\ref{sec:erroranalysis}.  We also include, for the first time, the uncertainties in the PDFs due to the uncertainties in the normalisations of the individual data sets included in the global fit.  We present the grids for the eigenvector PDFs as both one-sigma ($68\%$) and $90\%$ confidence level uncertainties, rather than just the latter.

\item We now use the simpler and more conventional definition of $\alpha_S$ defined in Ref.~\cite{Vogt:2004ns}, i.e.~we solve \eqref{eq:rge} truncated at the appropriate order, starting from an input value of $\alpha_S(Q_0^2)$.  This is particularly important at NNLO where the coupling becomes discontinuous at the heavy flavour matching points.  Further discussion of the difference between the MRST and MSTW definitions of $\alpha_S$ will be given elsewhere.  The value of $\alpha_S(Q_0^2)$ is now one of the fit parameters and replaces the $\Lambda_{\rm QCD}$ parameter used in MRST fits.  Since there is more than one definition of $\Lambda_{\rm QCD}$ in common use, replacing it by $\alpha_S(Q_0^2)$ reduces the potential scope for misuse.

\item The new definition of the coupling is then identical to that used in the \textsc{pegasus}~\cite{Vogt:2004ns} (which uses moment space evolution) and \textsc{hoppet}~\cite{Salam:2008qg} (which uses $x$ space evolution) programs, therefore we can now test our evolved PDFs by benchmarking against those two programs, which were shown to display excellent agreement with each other~\cite{Giele:2002hx,Dittmar:2005ed}.  We use our best-fit input parameterisations at $Q_0^2 = 1$ GeV$^2$ in each case, which are more complicated than those in the toy PDFs used for the benchmarks in Refs.~\cite{Giele:2002hx,Dittmar:2005ed}, and perform trial evolutions at LO, NLO and NNLO up to $Q^2 = 10^4$ GeV$^2$, finding agreement with both \textsc{pegasus} and \textsc{hoppet} to an accuracy of ${\cal O}(0.1\%)$ or less in most regions, with discrepancies a little larger only in regions where the PDFs are very small or at extremely low $x$ where extrapolations are used.  (We see larger discrepancies with \textsc{pegasus} at $x\sim 0.9$, where the default choice of the Mellin-inversion contour is inaccurate when using our input parameterisation.)  We intend to return to more comprehensive comparisons in the future, but for the moment are satisfied that any discrepancies between evolution codes are orders of magnitude lower than the uncertainties on the PDFs.

\item For structure function calculations, both neutral and charged current, we use an improved general-mass variable flavour number scheme (GM-VFNS) both at NLO and more particularly at NNLO.  This is based on the procedure defined in Ref.~\cite{Thorne:2006qt}, though there are some slight modifications described in detail in Section~\ref{sec:heavyflavour}.  The new procedure is much easier to generalise to higher orders.  At NLO there is simply a change in the details, but at NNLO there is a major correction to the transition across the heavy flavour matching points, first implemented in Ref.~\cite{Martin:2007bv}, which demonstrates that NNLO sets before 2006 should be considered out-of-date.  No other available NNLO sets which include heavy flavours currently treat the transition across the matching points in such a complete manner.  A full GM-VFNS is not currently available for Drell--Yan production of virtual photons, or $W$ and $Z$ bosons, or for inclusive jet production.  For the low-mass Drell--Yan data in our fit, heavy flavours contribute $\ll 1\%$ of the total, so the inaccuracy invoked by the approximation of using the zero-mass scheme is negligible.  All other processes are at scales such that the charm mass $m_c$ is effectively very small, and the zero-mass scheme is a very good approximation.  The approximation should still be reliable for processes induced by bottom quarks, and in this case the relative contribution is small.

\item Vector boson production data can now be described in a fully exclusive way at NNLO.  We use the \textsc{fewz} code~\cite{Melnikov:2006kv}, and compare with the results obtained using the \textsc{resbos} code~\cite{Balazs:1997xd} which includes NLO+NNLL $p_T^W$-resummation effects.  This allows, for example, a detailed description of the $W$ asymmetry data accounting for the $W$ width ($\Gamma_W$) and lepton decay effects.

\item We now implement \textsc{fastnlo}~\cite{Kluge:2006xs}, based on \textsc{nlojet++}~\cite{Nagy:2001fj,Nagy:2003tz}, which allows the inclusion of the NLO hard cross section corrections to both the Tevatron and HERA jet data in the fitting routine.  This improvement replaces the $K$-factors and pseudo-gluon data that were previously necessary to speed up the fitting procedure.

\item We provide~\cite{mstwpdf} a new form of the grids that list the parton distributions over an extended $Q^2$ range and which are more dense in $x$.  The extended public grids allows an improved description around the heavy quark thresholds, where at NNLO discontinuities in parton distributions appear.  (We also make improvements to the internal grids used in our evolution code, so that the heavy quark thresholds lie exactly on the grid points, which was not the case in the MRST analyses.)  Moreover, the grids now contain more parton flavours: $s\ne\bar{s}$ at all orders due to the (mild) evidence from the dimuon data, and additionally $c\ne\bar{c}$ and $b\ne\bar{b}$ at NNLO, due to automatic generation of a small $\mathcal{O}(\alpha_S^2)$ quark--antiquark difference during the NNLO evolution, so at this order the $H$ distribution is (very) slightly different from $\overline{H}$~\cite{Catani:2004nc}.  (This small asymmetry at NNLO was omitted in previous MRST analyses.)
  
\item Two features of the parameterisation are worth emphasising.  First, as since Ref.~\cite{Martin:2001es}, we allow the gluon to have a very general form at low $x$, where it is by far the dominant parton distribution. We will discuss the consequences of this in Section~\ref{sec:inputparamunc}.  Second, we parameterise $s \pm \bar{s}$ as functions of $x$, rather than assuming, as previously,
\begin{equation}
  s=\bar{s}=\frac{\kappa}{2}(\bar{u}+\bar{d})
\end{equation}
at the input scale, where $\kappa$ was a constant fixed by the neutrino dimuon production ($\kappa \sim 0.4$--$0.5$); see Section~\ref{sec:dimuon}.  Implicit in our parameterisation is the assumption that the strange sea will have approximately the same shape as the up and down sea at small $x$.  This extra parametric freedom of the gluon and strange quark distributions means that all partons are less constrained, particularly at low $x$.  This, in turn, leads to a more realistic estimate of the uncertainties on all parton distributions.

\item We fix the heavy quark masses at $m_c = 1.40$ GeV and $m_b = 4.75$ GeV, changed from the MRST default values of $m_c = 1.43$ GeV and $m_b = 4.30$ GeV.  Our value of $m_b=4.75$ GeV is close to the calculated $\overline{\rm MS}$ mass transformed to the pole mass value, using the three-loop relation between the pole and $\overline{\rm MS}$ masses, of $m_b = 4.800$ GeV~\cite{Kuhn:2007vp,Kuhn:2007tn}, while our value of $m_c = 1.40$ GeV is smaller than the calculated pole mass value of $m_c = 1.666$ GeV~\cite{Kuhn:2007vp,Kuhn:2007tn}.  If allowed to go free in the global fits, the best-fit values are $m_c = 1.39$ GeV at NLO and $m_c = 1.27$ GeV at NNLO.  In a subsequent paper we will present a more detailed discussion of the sensitivity of different data sets to the charm quark mass and present the best-fit values including a determination of the uncertainty.  We allow a maximum of five flavours in the evolution and do not include top quarks.

\end{enumerate}

%% file: treatmentofheavyflavours.tex
\section{Treatment of heavy flavours} \label{sec:heavyflavour}

The correct treatment of heavy flavours in an analysis of parton distributions is essential for precision measurements at hadron colliders.  For example, the cross section for $W$ production at the LHC depends crucially on precise knowledge of the charm quark distribution.  Moreover, a correct method of fitting the heavy flavour contribution to structure functions is important because of the knock-on effect it can have on all parton distributions.  However, it has become clear in recent years that it is a delicate issue to obtain a proper treatment of heavy flavours.  There are various choices that can be made, and also many ways in which subtle mistakes can occur.  Both the choices and the mistakes can lead to changes in parton distributions which may be similar to, or even greater than, the quoted uncertainties --- though the mistakes usually lead to the more dramatic changes.  Hence, we will here provide a full description of our procedure, along with a comparison to alternatives and some illustrations of pitfalls which must be avoided.

\begin{figure}[ht]
  \centering
  \begin{minipage}{0.3\textwidth}
    (a)\\
    \includegraphics[width=\textwidth,clip]{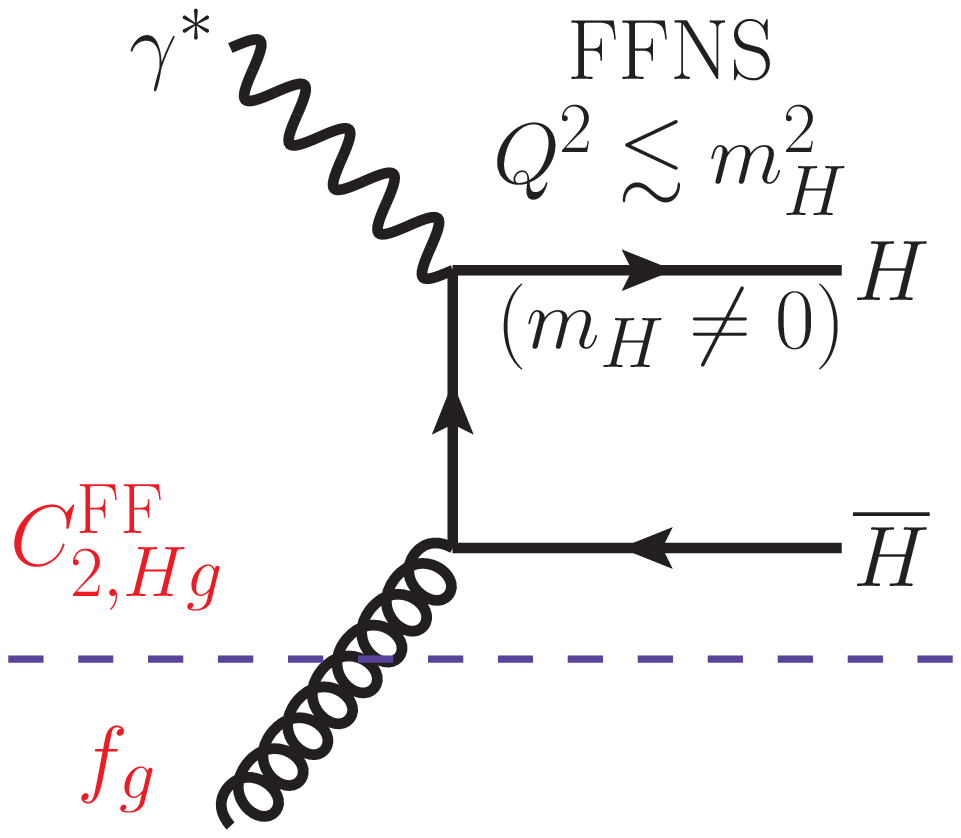}%
  \end{minipage}\hspace{0.1\textwidth}
  \begin{minipage}{0.3\textwidth}
    (b)\\
    \includegraphics[width=\textwidth,clip]{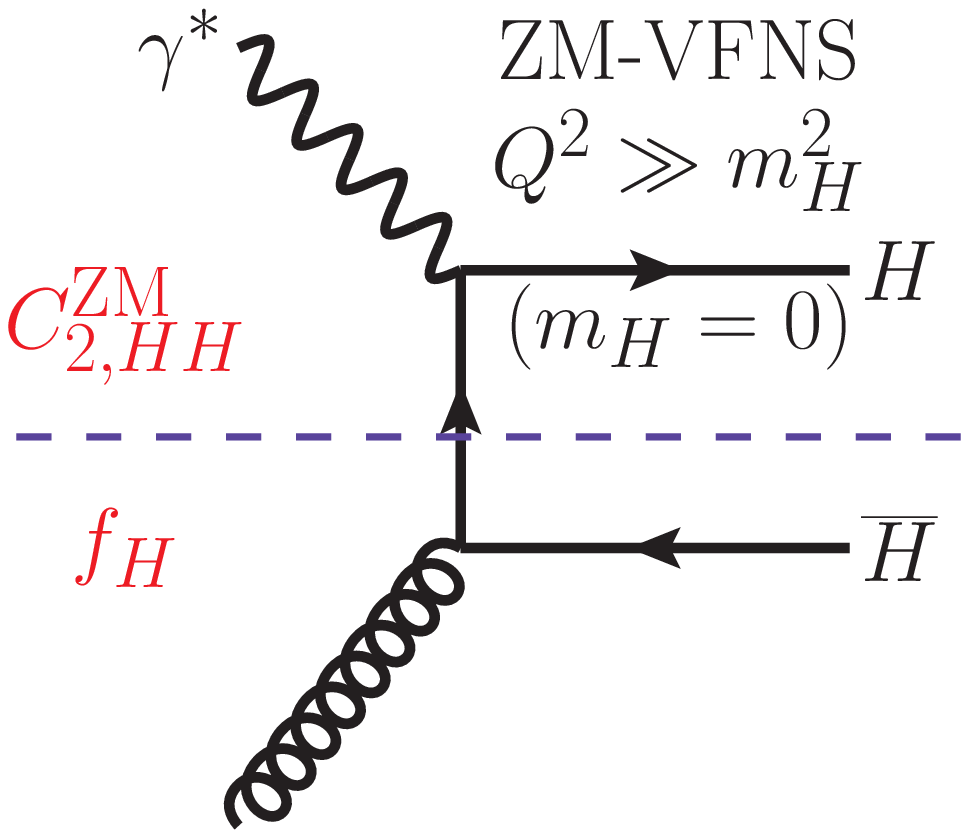}
  \end{minipage}
  \caption{Diagrams contributing at leading-order to the heavy flavour structure function, $F_2^H$, in (a) the fixed flavour number scheme (FFNS) valid for $Q^2\lesssim m_H^2$, and (b) the zero-mass variable flavour number scheme (ZM-VFNS) valid for $Q^2\gg m_H^2$.}
  \label{fig:f2heavydiagrams}
\end{figure}

First we describe the two distinct regimes for describing heavy quarks where the pictures are relatively simple.  These are the so-called fixed flavour number scheme (FFNS) and zero-mass variable flavour number scheme (ZM-VFNS); see Fig.~\ref{fig:f2heavydiagrams}.

\subsection{Fixed flavour number scheme}
First, there is the region where the hard scale of the process is similar to, or smaller than, the quark mass\footnote{Throughout this section we use $H$ to denote a heavy quark: $H=c,~b$ or $t$.}, i.e.~$Q^2\lesssim m_H^2$.  In this case it is most natural to describe the massive quarks as final-state particles, and not as partons within the proton.  This requirement defines the FFNS, where only light quarks are partons, and the number of flavours is fixed.  We label the number of quark flavours appearing as parton distributions by $n_f$.  Note, however, that there are multiple instances of FFNSs with different numbers of {\it active} flavours $n_f$.  The number $n_f$ is normally equal to 3, where up, down and strange are the light quarks, but we can treat charm as a light quark while bottom remains heavy at high scales, i.e.~$n_f=4$, and it may also be equal to 5 if only top quarks are treated as heavy.  In each example of a FFNS the structure functions are given by\footnote{To simplify the discussion, we will use the convention that the factorisation scale and renormalisation scale are both equal to $Q^2$.  It is also the choice we make in the analysis.  Alternative choices are possible, and cause no problems in principle, but can add considerable technical complications.}
\begin{equation} \label{ffns}
  F_i(x,Q^2)=\sum_k C^{{\rm FF}, n_f}_{i,k}(Q^2/m_H^2)\otimes f^{n_f}_k(Q^2).
\end{equation}
This approach contains all the $m_H$-dependent contributions, and as it is conceptually simple it is frequently used in analyses of structure functions.  Even in this case one must be careful to be self-consistent in defining all quantities, i.e.~parton distributions, coefficient functions and coupling constant in the same renormalisation schemes (which is often not done).  The mistake made by not doing so can lead to errors in the gluon distribution of order the size of the uncertainty~\cite{Martin:2006qz}.

Despite its conceptual simplicity, the FFNS has some potential problems.  It does not sum $\alpha_S^m \ln^l( Q^2/m_H^2)$  ($l\leq m$) terms in the perturbative expansion.  Thus the accuracy of the fixed-order expansion becomes increasingly uncertain as $Q^2$ increases above $m_H^2$.  For example, $\alpha_S(M_W^2)\ln (M_W^2/m_c^2)$ is approximately equal to unity, and so there is no guarantee an expansion in this variable will converge.  As well as this problem of principle, there are additional practical issues.  Since calculations including full mass dependence are complicated, there are only a few cross sections known even to NLO in $\alpha_S$ within this framework, so the resulting parton distributions are not universally useful.  In addition, even for neutral-current structure functions, the FFNS coefficient functions are known only up to NLO~\cite{Laenen:1992zk,Harris:1995tu}, and are not calculated at NNLO --- that is, the $\alpha_S^3$ coefficient\footnote{We add a subscript $H$ to distinguish the $g \to H$ coefficient function $C_{2,Hg}$ from the usual coefficient function $C_{2,g}$ describing the $g \to q$ transition of light quarks.}, $C^{{\rm FF},n_f,(3)}_{2,Hg}$, for $F_2$ is unknown, so one cannot determine parton distributions at NNLO in this scheme.

\subsection{Zero-mass variable flavour number scheme}

All of the problems of the FFNS are solved in the so-called zero-mass variable flavour number scheme (ZM-VFNS).  Here, the heavy quarks evolve according to the splitting functions for massless quarks and the resummation of the large logarithms in $Q^2/m_H^2$ is achieved by the introduction of heavy-flavour parton distributions and the solution of the evolution equations.  It assumes that at high scales, $Q^2\gg m_H^2$, the massive quarks behave like massless partons, and the coefficient functions are simply those in the massless limit, e.g.~for structure functions
\begin{equation} \label{ZMVFNS}
  F_i(x,Q^2) = \sum_k C^{{\rm ZM}, n_f}_{i,j}\otimes f^{n_f}_j(Q^2),
\end{equation}
where $n_f-3$ is the number of active heavy quarks, with masses above some transition point for turning on the heavy flavour distribution, typically at a scale similar to $m_H^2$.  This is technically simpler than the FFNS, and many more cross sections are known in this scheme.  The nomenclature of ``zero-mass'' is a little misleading because some mass dependence is included in the boundary conditions for evolution.  The parton distributions in different quark number regimes are related to each other perturbatively, i.e.
\begin{equation} \label{transition}
  f^{n+1}_j(Q^2)= \sum_k A_{jk}(Q^2/m_H^2)\otimes f^{n}_k(Q^2),
\end{equation}
where the perturbative matrix elements $A_{jk}(Q^2/m_H^2)$~\cite{Buza:1996wv} containing $\ln(Q^2/m_H^2)$ terms are known to NNLO, i.e.~${\cal O}(\alpha_S^2)$.\footnote{When testing our NNLO evolution code against \textsc{pegasus}~\cite{Vogt:2004ns} we traced the source of a discrepancy in our implementation of $A_{Hg}$ to a typo in Eq.(B.3) of the preprint version of Ref.~\cite{Buza:1996wv} (fixed in the journal version).  Specifically, the term $C_A\,T_f\,(16z+16z^2)\,\left[{\rm Li}_2(-z)+\ln z\ln(1+z)\right]$ has $16$ rather than $16z$ in the preprint version.}  They relate $f^{n}_i(Q^2)$ and $f^{n+1}_i(Q^2)$, guaranteeing the correct evolution for both regimes. At NLO in the $\overline{\rm MS}$ scheme they constrain the heavy quarks to begin their evolution from a zero value at $Q^2=m_H^2$, and the other partons to be continuous at this choice of transition point, hence making it the natural choice.

\subsection{General-mass variable flavour number schemes}
  
The ZM-VFNS has many advantages.  However, it has the failing that it simply ignores ${\cal O}(m_H^2/Q^2)$ corrections to the coefficient functions, and hence it is inaccurate in the region where $Q^2$ is not so much greater than $m_H^2$.  The nomenclature {\it scheme} may be thought of as misleading in a similar way that {\it zero-mass} might.  Scheme usually implies an alternative choice in ordering the expansion, or a particular separation of contributions between coefficient functions and parton distributions, i.e.~the inherent ambiguity in a perturbative QCD calculation allows a choice, the effects of which become increasingly smaller as higher orders are included.  The ZM-VFNS misses out ${\cal O}(m_H^2/Q^2)$ contributions completely, and there is a permanent error of this order.  This clearly already happens at LO and NLO.  The error induced by fitting to HERA structure function data using a ZM-VFNS was shown to be up to $\sim 6\%$ in the small-$x$ light-quark distributions by CTEQ in Ref.~\cite{Tung:2006tb}, resulting in a systematic error of $\sim 8\%$ in predictions for vector boson production at the LHC.  This NLO result seems to provide ample evidence for the use of a general-mass variable flavour number scheme (GM-VFNS) to provide default parton distributions, and indeed this is the approach we have adopted in previous global analyses~\cite{Martin:1998sq,Martin:1999ww,Martin:2001es,Martin:2004ir,Martin:2007bv}.  A definition of a GM-VFNS was first proposed in Ref.~\cite{Aivazis:1993pi} --- the ACOT scheme, and the MRST group have been using the alternative TR scheme~\cite{Thorne:1997ga} as the default since MRST98~\cite{Martin:1998sq}.  The CTEQ group have since also adopted this convention~\cite{Tung:2006tb,Nadolsky:2008zw}.

However, the inherent problems with the ZM-VFNS are thrown into particularly sharp relief at NNLO.  At ${\cal O}(\alpha_S^2)$ the $A_{jk}(Q^2/m_H^2)$ are no longer zero at $Q^2=m_H^2$, and lead to discontinuities in the parton distributions.  This is similar to the discontinuity at $\mu^2=m_H^2$ in $\alpha_S(\mu^2)$ at NNLO.  For the coupling constant this discontinuity is rather small.  However, the corresponding discontinuities in parton distributions can be significant, about $10\%$ for the gluon, but it turns out that $c(x,m_c^2)$ is quite considerably negative at small $x$, see e.g.~Fig.~3 of Ref.~\cite{Martin:2007bv}.  As it happens, the effect of the NNLO massless coefficient function is to make this worse rather than better at $x \sim 0.0001-0.01$, and the charm contribution to $F_2(x,Q^2)$ is negative at the transition point to such a degree that there is an ${\cal O}(10\%)$ discontinuity in the total structure function at $Q^2=m_c^2$, as illustrated in Fig.~1 of Ref.~\cite{Thorne:2006qt}.

Hence, for a precise analysis of structure functions, and other data, one must use a GM-VFNS which smoothly connects the two well-defined limits of $Q^2\leq m_H^2$ and $Q^2\gg m_H^2$.  This is already the case at lower orders, but becomes imperative at NNLO.  However, despite the above reasoning, a GM-VFNS is not always used as default even today.  This is undoubtedly often because of the extra complication compared to the simple ZM-VFNS.  However, part of the reason may be because the definition is not unique.  The reasons for this and the consequences will now be discussed, along with a detailed outline of the prescription used in our analysis.  A less detailed, but more introductory and comparative discussion of the schemes used by MRST/MSTW and CTEQ can be found in Ref.~\cite{Thorne:2008xf}.

A GM-VFNS can be defined by demanding equivalence of the $n_f=n$ (FFNS) and $n_f=n+1$-flavour (GM-VFNS) descriptions above the transition point for the new parton distributions (they are by definition identical below this point), at all orders, i.e.
\begin{eqnarray}
  F_i(x,Q^2)=\sum_k C^{{\rm FF},n}_{i,k}(Q^2/m_H^2)\otimes f^{n}_k(Q^2) &=& \sum_j C^{{\rm VF},n+1}_{i,j}(Q^2/m_H^2)\otimes f^{n+1}_j(Q^2)\\ \nonumber
  &\equiv & \sum_{j,k} C^{{\rm VF},n+1}_{i,j}(Q^2/m_H^2)\otimes A_{jk}(Q^2/m_H^2)\otimes f^{n}_k(Q^2). \label{GMVFNS}
\end{eqnarray}
The description where the number of active partons is taken to be $n_f=n$ must be identical to that when it increases, i.e.~$n_f=n+1$.  Hence, the GM-VFNS coefficient functions satisfy\footnote{It is implicit that the coupling constant is a function of $n$ flavours on the left-hand side and of $n+1$ flavours on the right-hand side.}
\begin{equation} \label{VFNSdef}
  C^{{\rm FF},n}_{i,k}(Q^2/m_H^2) = \sum_j C^{{\rm VF},n+1}_{i,j}(Q^2/m_H^2)\otimes A_{jk}(Q^2/m_H^2),
\end{equation}
which, for example, at ${\cal O}(\alpha_S)$ gives for $F_2(x,Q^2)$:
\begin{equation} \label{LOVFNSdef}
  C^{{\rm FF},n,(1)}_{2,g}(Q^2/m_H^2) = C^{{\rm VF},n+1,(0)}_{2, HH}(Q^2/m_H^2)\otimes P^{(0)}_{qg}\ln(Q^2/m_H^2) + C^{{\rm VF},n+1,(1)}_{2,g}(Q^2/m_H^2).
\end{equation}
The GM-VFNS coefficient functions, $C^{{\rm VF},n_f}_{i,j}$, are constrained to tend to the massless limits for $Q^2\gg m_H^2$ and the $A_{jk}(Q^2/m_H^2)$ are such that this happens self-consistently.  However, the $C^{{\rm VF},n_f,(m)}_{i,j}(Q^2/m_H^2)$ are only uniquely defined in this massless limit $Q^2/m_H^2 \to \infty$.  For finite $Q^2/m_H^2$ one can swap ${\cal O}(m_H^2/Q^2)$ terms between $C^{{\rm VF},n+1,(0)}_{2, HH}(Q^2/m_H^2)$ and $C^{{\rm VF},n+1,(1)}_{2,g}(Q^2/m_H^2)$ while maintaining the exact definition in \eqref{LOVFNSdef}.  It is clear that this general feature applies to all relationships in \eqref{VFNSdef}.  Although the equivalence \eqref{VFNSdef} was first pointed out in general in Ref.~\cite{Buza:1996wv}, and \eqref{LOVFNSdef} is effectively used in defining the original ACOT scheme, the freedom to swap ${\cal O}(m_H^2/Q^2)$ terms without violating the definition of a GM-VFNS was first noticed in Ref.~\cite{Thorne:1997ga} and put to use to define the TR scheme, as described below.  This freedom to redistribute ${\cal O}(m_H^2/Q^2)$ terms can be classified as a change in {\it scheme} since it leads to an ambiguity in the result at a fixed order, but the ambiguity becomes higher order if the order of the calculation increases, much like the renormalisation and factorisation scheme (and scale) ambiguities.  Moreover, it is a change of scheme which does not change the definition of the parton distributions, only the coefficient functions.  This is perhaps a surprising result, which occurs because there is a redundancy in \eqref{VFNSdef}, there being one more coefficient function above the transition point than below, i.e.~that for the heavy quark.

The original ACOT prescription~\cite{Aivazis:1993pi} calculated the coefficient functions for single heavy quark scattering from a virtual photon exactly.  This might seem the most natural definition.  However, it assumes that immediately above the transition point a {\it single} heavy quark or antiquark can exist in isolation.  Hence, each coefficient function violates the real physical threshold $W^2>4m_H^2$, since only {\it one} heavy quark is produced in the final state, rather than a quark--antiquark {\it pair}.  Moreover, this definition requires the calculation of mass-dependent coefficient functions, which becomes progressively more difficult at higher orders.  As mentioned above, in the TR scheme~\cite{Thorne:1997ga} the ambiguity in the definition of $C^{{\rm VF},n_f,(0)}_{2, HH}(Q^2/m_H^2)$ was recognised and exploited for the first time.  To be precise, the constraint that $(\partial F^{H}_2/\partial \ln Q^2)$ was continuous at the transition point (in the gluon sector) was applied to define the heavy-quark coefficient functions.  This imposed the correct threshold dependence on all coefficient functions and improved the smoothness at $Q^2=m_H^2$, and did not involve the explicit calculation of new mass-dependent diagrams.  However, it did involve complications, since it required the convolution of the formal inverse of splitting functions with coefficient functions, which itself becomes technically difficult at higher orders.

Since these early definitions there have been various modifications, including a precise definition of an ACOT-like scheme up to NNLO by Chuvakin, Smith and van Neerven~\cite{Chuvakin:1999nx}.\footnote{Indeed in Ref.~\cite{Chuvakin:1999nx} it is pointed out that at NNLO a further complication appears for the first time.  There are $\ln^3(Q^2/m_H^2)$ divergences at ${\cal O}(\alpha_S^2)$ coming from gluon splitting into heavy quark--antiquark pairs.  These divergences arise from heavy quark emission diagrams which cancel with opposite ones originating from virtual heavy quark loops in the ``light quark'' coefficient functions.  This cancellation is achieved in a physically meaningful manner by imposing a cut on the softness of the heavy quark final state in the former process, i.e.~the unobservable soft process cancels with the virtual corrections in the ``light quark'' cross section, and a well-behaved remainder enters the ``heavy quark'' cross section.  In practice, after cancellation both contributions are very small (both are quark- rather than gluon-initiated) at NNLO and we currently include the total in the ``light quark'' sector.  In the ${\cal O}(\alpha_S^2)$ FFNS they are usually combined instead in the ``heavy quark'' contribution.}  A major simplification was achieved when the flexibility in the choice of heavy-quark coefficient functions was used to define the ACOT($\chi$) prescription~\cite{Tung:2001mv,Kretzer:2003it}, which in the language used in this paper (and in Ref.~\cite{Thorne:1997ga}) would be defined by
\begin{equation} \label{acotchi}
  C^{{\rm VF},n_f,(0)}_{2, HH}(z,Q^2/m_H^2) = e_H^2\,z(1+4m_H^2/Q^2)\;\delta\left(z-\frac{Q^2}{Q^2+4m_H^2}\right).
\end{equation}
This gives the LO definition
\begin{equation} 
  F^{H,(0)}_2(x,Q^2)=e_H^2\,(x/x_{\rm max})(H+\overline{H})(x/x_{\rm max}, Q^2),
\end{equation}
where $x_{\rm max}= Q^2/(Q^2+4m_H^2)$. It automatically reduces to the massless limit $C^{{\rm ZM},(0)}_{2, HH}(z)= e_H^2\,z\;\delta(1-z)$ for $Q^2/m_H^2 \to \infty$, and also imposes the true physical threshold 
\begin{equation}
  W^2 =Q^2(1-x)/x \ge 4m_H^2.
\end{equation}
This choice of the LO heavy-flavour coefficient function has been adopted in our current prescription, which we denote the ${\rm TR}^\prime$ scheme, described in detail in Ref.~\cite{Thorne:2006qt}.  For the GM-VFNS to remain simple (and physical) at all orders, $m$, it is necessary to choose
\begin{equation} 
  C^{{\rm VF},n_f,(m)}_{2, HH}(z,Q^2/m_H^2)= C^{{\rm ZM},n_f,(m)}_{2, HH}(z/x_{\rm max}),
\end{equation}
which is the implicit ACOT($\chi$) definition, and is our choice.  It removes one of the sources of ambiguity in defining a GM-VFNS. However, there are others. 

One major issue in a complete definition of the GM-VFNS, is that of the ordering of the perturbative expansion.  This ambiguity comes about because the ordering in $\alpha_S$ for $F_2^H(x,Q^2)$ is different for the number of active flavours $n_f=n$ and $n_f=n+1$ regions:
\begin{equation}
  \begin{array}{ccc}
    & n-{\rm flavour}& n+1-{\rm flavour} \\
    & & \\
    {\rm LO} &  \frac{\alpha_S}{4\pi} C^{{\rm FF},n,(1)}_{2, Hg}\otimes g^{n} & C^{{\rm VF},n+1,(0)}_{2, HH}\otimes (H+\overline{H})\\\nonumber
    {\rm NLO} &  \left(\frac{\alpha_S}{4\pi}\right)^2 \left(C^{{\rm FF},n,(2)}_{2, Hg}\otimes g^{n}+C^{{\rm FF},n,(2)}_{2, Hq}\otimes \Sigma^{n}\right) & \frac{\alpha_S}{4\pi}\left(C^{{\rm VF},n+1,(1)}_{2, HH}\otimes (H+\overline{H}) + C^{{\rm VF},n+1,(1)}_{2, Hg}\otimes g^{n+1}\right)\\ 
    {\rm NNLO} & \left(\frac{\alpha_S}{4\pi}\right)^3 \sum_j C^{{\rm FF},n,(3)}_{2, Hj}\otimes f_j^{n} & \left(\frac{\alpha_S}{4\pi}\right)^2\sum_j C^{{\rm VF},n+1,(2)}_{2, Hj}\otimes f^{n+1}_j
  \end{array}
\end{equation}
with obvious generalisation to even higher orders.  This means that switching directly from an $n$ flavours fixed order to the $n+1$ fixed order leads to a discontinuity in $F_2^H(x,Q^2)$.  As with the discontinuities in the ZM-VFNS already discussed this is not just a problem in principle --- the discontinuity is comparable to the errors on data, particularly at small $x$.

Hence, any definition of a GM-VFNS must make some decision how to deal with this, and the ACOT-type schemes have always made a different choice to that for the TR-type schemes used in our analyses.  The ACOT-type schemes simply define the same order of $\alpha_S$ both below and above the transition point.  For example at NLO the definition is
\begin{equation} \label{ACOTNLO}
  F_2^H(x,Q^2) ~=~ \frac{\alpha_S}{4\pi} C^{{\rm FF},n,(1)}_{2, Hg}\otimes g^{n} ~~\to ~~ \frac{\alpha_S}{4\pi} \left(C^{{\rm VF},n+1,(1)}_{2, HH}\otimes (H+\overline{H}) + C^{{\rm VF},n+1,(1)}_{2, Hg}\otimes g^{n+1}\right).
\end{equation}
This clearly maintains continuity in the structure function across the transition point.  However, it only contains information on LO heavy flavour evolution below $Q^2=m_H^2$, since $C^{{\rm FF},n,(1)}_{2, Hg}$ only contains information on the LO splitting function, but the heavy quarks evolve using NLO splitting functions above $Q^2=m_H^2$ --- a big change at small $x$.  The TR scheme, defined in Ref.~\cite{Thorne:1997ga}, and all subsequent variations used in our analyses, try to maintain the correct ordering in each region as closely as possible.  For example at LO we have the definition
\begin{eqnarray} \label{trlo}
  F_2^H(x,Q^2) &= & \frac{\alpha_S(Q^2)}{4\pi} C^{{\rm FF},n,(1)}_{2, Hg}(Q^2/m_H^2)\otimes g^{n}(Q^2) \\ \nonumber 
  &\to & \frac{\alpha_S(m_H^2)}{4\pi} C^{{\rm FF},n,(1)}_{2, Hg}(1)\otimes g^{n}(m_H^2)+ C^{{\rm VF},n+1,(0)}_{2, HH}(Q^2/m_H^2)\otimes (H+\overline{H})(Q^2),
\end{eqnarray}
i.e.~we freeze the ${\cal O}(\alpha_S)$ term when going upwards through $Q^2=m_H^2$.  This generalises to higher orders by freezing the term with the highest power of $\alpha_S$ in the definition for $Q^2< m_H^2$ when moving upwards above $m_H^2$.  Hence, the definition of the ordering is consistent within each region, except for the addition of a constant term (which does not affect evolution) above $Q^2=m_H^2$, which becomes progressively less important at higher $Q^2$, and whose power of $\alpha_S$ increases as the order of the perturbative expansion increases.

This definition of the ordering means that in order to define our VFNS at NNLO~\cite{Thorne:2006qt} we need to use the ${\cal O}(\alpha_S^3)$ heavy-flavour coefficient functions for $Q^2 \leq m_H^2$ (and this contribution will be frozen for $Q^2>m_H^2$).  As mentioned above, these coefficient functions are not yet calculated.  However, as explained in Ref.~\cite{Thorne:2006qt}, we can model this contribution using the known leading threshold logarithms~\cite{Laenen:1998kp} and leading $\ln(1/x)$ terms derived from the $k_T$-dependent impact factors~\cite{Catani:1990eg}.  This results in a significant contribution at small $Q^2$ and $x$ with some model dependence.  However, variation in the free parameters does not lead to a large change, as discussed in Section~\ref{sec:heavyflavourdata}.\footnote{It should be stressed that this model is only valid for the region $Q^2 \leq m_H^2$, and would not be useful for a full NNLO FFNS since it contains no information on the large $Q^2/m_H^2$ limits of the coefficient functions.  A more general approximation to the ${\cal O}(\alpha_S^3)$ coefficient functions could be attempted, but full details would require first the calculation of the ${\cal O}(\alpha_S^3)$ matrix element $A_{Hg}$.  This more tractable project is being investigated at present~\cite{Bierenbaum:2008tm}.  An approximation to the $\alpha_S^3$ coefficient functions using logarithmically-enhanced terms near threshold and exact scale-dependent terms has recently been proposed in Ref.~\cite{Alekhin:2008hc}.}  Up to this small model dependence we have a full NNLO GM-VFNS with automatic continuity of structure functions across heavy flavour transition points.\footnote{There are actually ${\cal O}(\alpha_S^3)$ discontinuities due to terms such as $C^{{\rm VF},n+1,(1)}_{2, HH}\otimes (H+\overline{H})$ and $C^{{\rm VF},n+1,(1)}_{2, Hg}\otimes g^{n+1}$, i.e.~${\cal O}(\alpha_S)$ coefficient functions convoluted with ${\cal O}(\alpha_S^2)$ discontinuities in partons.  These would be cancelled at NNNLO by discontinuities in ${\cal O}(\alpha_S^3)$ coefficient functions.  In practice the imposition of the correct threshold behaviour in all coefficient functions minimises these effects and they are very small.}  This is certainly the most complete treatment of heavy-flavour effects currently used in any NNLO analysis.

\subsubsection*{Scheme dependence}
 
Although all definitions of the GM-VFNS become very similar at very high $Q^2$, the difference in choice can be phenomenologically important.  For example, our definition effectively includes exactly one higher order than ACOT-type schemes for $Q^2<m_H^2$, and the value of this contribution at $Q^2=m_H^2$ is carried to higher $Q^2$.  Since at small $x$ and near threshold the higher orders in $\alpha_S$ are accompanied by large corrections, this leads to large differences below the transition point, which are still important a little way above the transition point. This is shown for $F_2^{c\bar{c}}(x,Q^2)$ in Fig.~2 of Ref.~\cite{Thorne:2006qt} where the two choices are shown for the same parton distributions.  More clearly, this difference in the definition of the ordering is the main difference in the NLO predictions from MRST and CTEQ in the comparison to H1 data on $F_2^{b\bar{b}}(x,Q^2)$~\cite{Aktas:2005iw}, shown in Fig.~4 of the same paper~\cite{Thorne:2006qt}.

The inclusion of the complete GM-VFNS in a global fit at NNLO first appeared in Ref.~\cite{Martin:2007bv}, and led to some important changes compared to our previous NNLO analyses, which had a much more approximate inclusion of heavy flavours (which was explained clearly in the Appendix of~\cite{Martin:2002dr}).  A consequence of including the positive ${\cal O}(\alpha_S^3)$ coefficient functions at low $Q^2$ is that the NNLO $F_2^{c\bar{c}}(x,Q^2)$ automatically starts from a higher value at low $Q^2$.  However, at high $Q^2$, the structure function is dominated by $(c+\bar c)(x,Q^2)$.  This has started evolving from a significantly negative value at $Q^2=m_c^2$.  The parton distributions in an NNLO fit readjust so that the light flavours evolve similarly to those at NLO, in order to fit the data.  Since the heavy flavour quarks evolve at the same rate as light quarks, but at NNLO start from a negative starting value, they remain lower than at NLO for higher $Q^2$.  Hence, there is a general trend: $F_2^{c\bar{c}}(x,Q^2)$ is flatter in $Q^2$ at NNLO than at NLO, as shown in Fig.~4 of Ref.~\cite{Martin:2007bv}.  It is also flatter than in our earlier (approximate) NNLO analyses.  We found that this had an important effect on the gluon distribution when we updated our NNLO analysis.  As seen in Fig.~5 of Ref.~\cite{Martin:2007bv}, it led to a larger gluon for $x \sim 0.0001$--$0.01$, as well as a larger value of $\alpha_S(M_Z^2)$, both compensating for the naturally flatter evolution, and consequently leading to more evolution of the light quark sea.  Both the gluon and the light quark sea were up to $6$--$7\%$ greater than in the 2004 set~\cite{Martin:2004ir} for $Q^2=10^4$ GeV$^2$, the increase maximising at $x= 0.0001$--$0.001$.  As a result there was a $6\%$ increase in the predictions for $\sigma_W$ and $\sigma_Z$ at the LHC.  This surprisingly large change is a \emph{correction} rather than a reflection of the uncertainty due to the freedom in choosing heavy flavour schemes.  The treatment of heavy flavour at NNLO is the same in the sets presented in this paper as for the MRST 2006 set, and as we will see, there are no further changes in predictions of the same size as this increase in going from the 2004 to the 2006 analyses.  This demonstrates that the MRST 2004 NNLO distributions should now be considered to be obsolete.

Our 2006 NNLO parton update~\cite{Martin:2007bv} was made because this was the first time the heavy flavour prescription had been treated precisely at NNLO and also because there was previously no MRST NNLO set with uncertainties. The data used in the analysis were very similar to the 2004 set, and since a consistent GM-VFNS was already used at NLO, and a set with uncertainties already existed, no new corresponding release of a NLO set was made along with the 2006 NNLO set.  With the benefit of hindsight, it is interesting to check the effect on the distributions due to the change in the prescription for the GM-VFNS at NLO without complicating the issue by also changing many other things in the analysis.  To this end we have obtained an unpublished MRST 2006 NLO set, which is fit to exactly the same data as the MRST 2006 NNLO set.\footnote{We do not intend to officially release the MRST 2006 NLO set, since it is superseded by the present MSTW 2008 NLO analysis, but it is nevertheless available on request.}

\begin{figure}
  (a)\hspace{0.5\textwidth}(b)\\
  \includegraphics[width=0.5\textwidth]{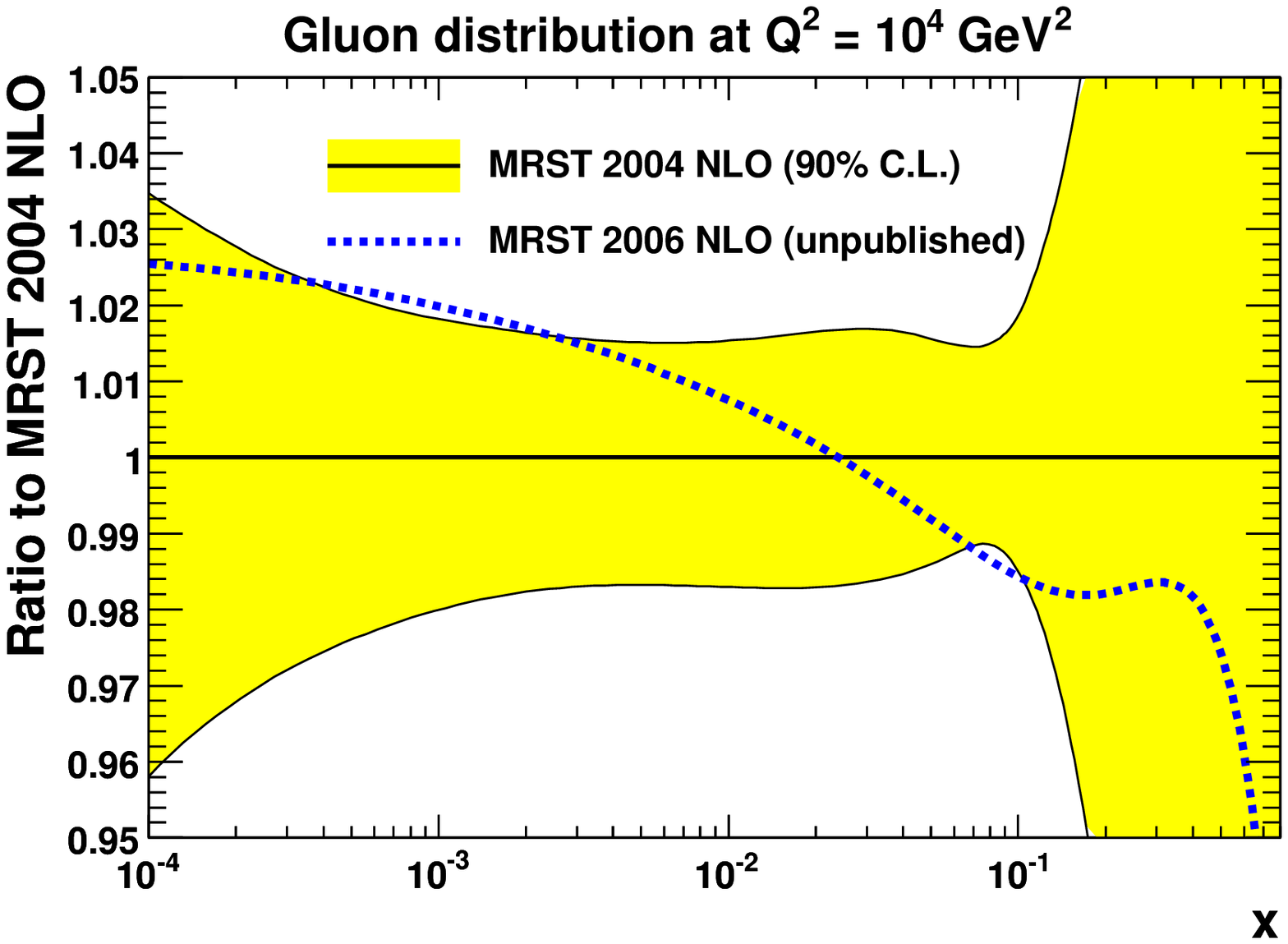}%
  \includegraphics[width=0.5\textwidth]{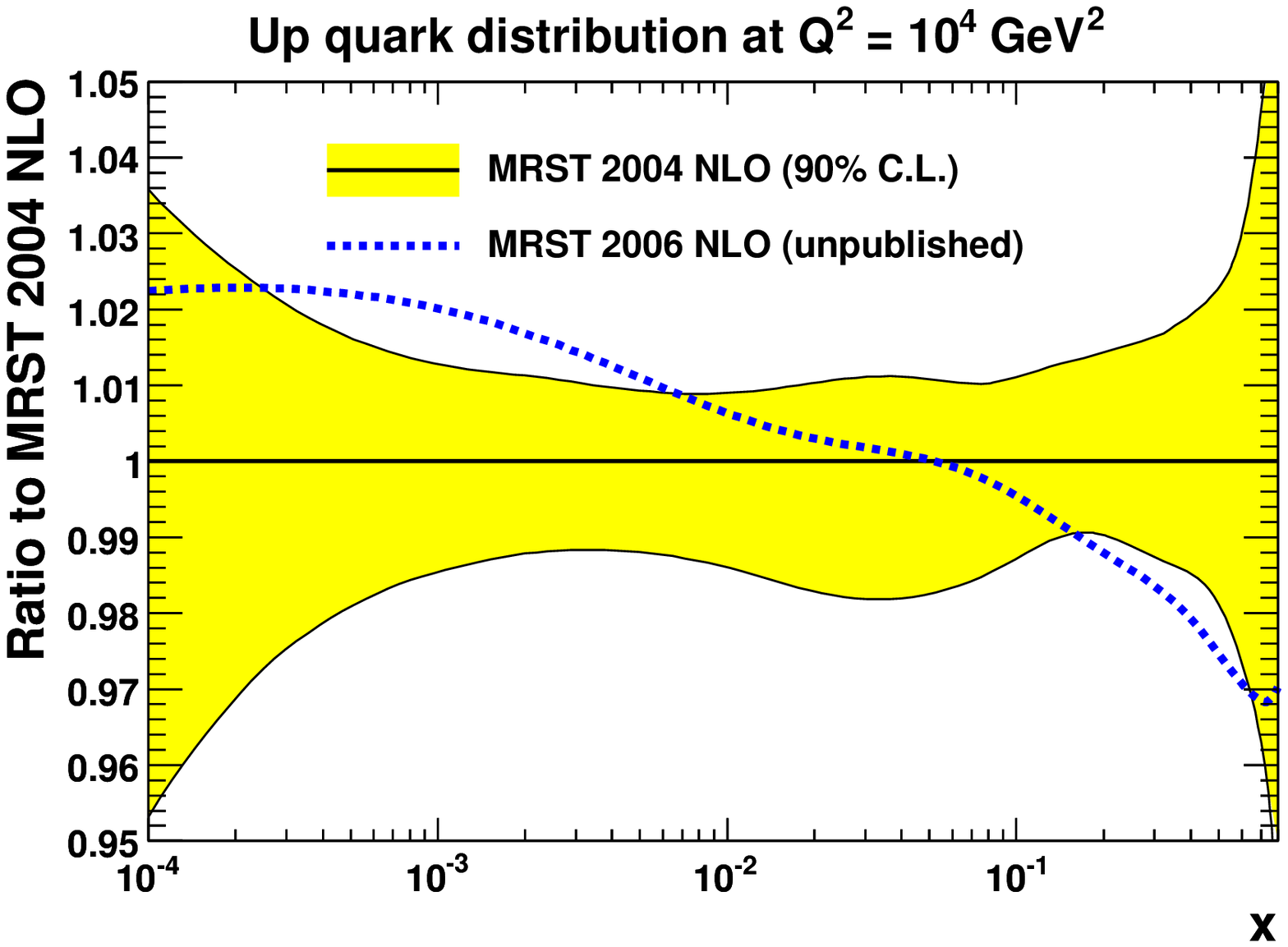}
  \caption{A comparison of the unpublished MRST 2006 NLO parton distributions to the MRST 2004 NLO distributions at $Q^2 = 10^4$ GeV$^2$. In order to illustrate the significance of the size of the differences, the fractional uncertainty on the MRST 2001 distributions is used for the 2004 distributions. The corresponding comparison at NNLO can be seen in Fig.~5 of Ref.~\cite{Martin:2007bv}.}
  \label{fig:06to04}
\end{figure}
The comparison of the up quark and gluon distributions for the MRST 2006 NLO set and the MRST 2004 NLO set, i.e.~the comparable plot to  Fig.~5 of Ref.~\cite{Martin:2007bv} for NNLO, is shown in Fig.~\ref{fig:06to04}.  As can be seen it leads to the same trend for the parton distributions as at NNLO, i.e.~an increase in the small-$x$ gluon and light quarks, but the effect is much smaller ---  a maximum of a $2\%$ change.  Also, the value of the coupling constant increases by $0.001$ from the 2004 value of $\alpha_S(M_Z^2)=0.120$.  Again, this is similar to, but smaller than, the change at NNLO.  Hence, we can conclude that the change in our choice of the heavy-flavour coefficient function alone leads to changes in the distributions of up to $2\%$, and since the change is simply a freedom we have in making a definition, this is a theoretical uncertainty on the parton distributions, much like the frequently invoked scale uncertainty.  Like the latter, it should decrease as we go to higher orders.  The ambiguity simultaneously moves to higher order, but it is difficult to check this explicitly since our main reason for making our change in the choice of heavy-quark coefficient functions was the difficulty of applying the original procedure in Ref.~\cite{Thorne:1997ga} at NNLO.  Certainly an absolute maximum of $2\%$ of the 6--7\% change, in the predictions for $\sigma_W$ and $\sigma_Z$ at the LHC in going from the 2004 to the 2006 NNLO parton sets, is due to true ambiguities, and the remaining $5\%$ is due to the correction of the flaws in the previous approach.

To close this section, we note that the MRST 2006 NLO distributions lead to a nearly $3\%$ increase in the predictions for $\sigma_W$ and $\sigma_Z$ at the LHC compared to MRST 2004 NLO, though there is very little change at the Tevatron, where the typical values of $x$ probed are nearly an order of magnitude higher.  As with the distributions themselves, this variation is a similar size to the quoted uncertainty in the cross sections, and again this is a genuine theoretical uncertainty.  Our most up-to-date predictions for LHC and Tevatron $W$ and $Z$ total cross sections, and comparison to the various previous sets, will be made in Section \ref{sec:totalpredictions}.

\subsubsection*{Longitudinal structure function}

One also has to be careful in defining the GM-VFNS for the longitudinal structure function.  In one sense this is not so important since the contribution from the longitudinal structure function to the total reduced cross section measured in DIS experiments is small, and the errors on direct measurements of $F_L$ are comparatively large.  However, the importance is increased by the larger ambiguity inherent in the definition of $F_L$ as compared to $F_2$.  

This large ambiguity occurs because if one calculates the coefficient function for a single massive quark scattering off a virtual photon, there is an explicit zeroth-order contribution
\begin{equation}
  C^{{\rm VF},n_f,(0)}_{L,HH} (z,Q^2/m_H^2) = e_H^2\,z\frac{4m_H^2}{Q^2}\;
  \delta\left(z-\frac{Q^2}{Q^2+m_H^2}\right). 
\end{equation}
This disappears at high $Q^2$ and the correct zero-mass limit is reached.
 
Such a zeroth-order coefficient function is implicit in the original ACOT definition of a GM-VFNS, and leads to a peculiar behaviour of $F_L^{c\bar{c}}$ just above $Q^2=m^2_c$.  It is convoluted with the heavy flavour distribution, which for $Q^2$ just above $m_c^2$ is small in magnitude.  However, the coefficient function is large near $m_c^2$, while the unsubtracted (i.e.~FFNS) gluon and singlet-quark coefficient functions are suppressed by a factor of $v^3$, where $v$ is the velocity of the heavy quark in the centre-of-mass frame, and are very small for low $Q^2$.  This means that this zeroth-order heavy-flavour contribution dominates just above $Q^2=m_c^2$, despite the fact that in a FFNS, where the $c\bar{c}$ pair has to be created, as it must in reality, the contribution is absent.

The contribution from $C^{{\rm VF},n_f,(0)}_{L,cc} \otimes (c + \bar c)$ turns on rapidly just above $m_c^2$, dominating other contributions, then dies away as $m^2_H/Q^2$ becomes small.  This leads to a distinct bump in $F^{c\bar{c}}_L(x,Q^2)$ for $Q^2$ just above $m_c^2$, as pointed out in Ref.~\cite{Thorne:1997ga}.  In principle this cancels between orders in a properly defined GM-VFNS, as this contribution implicitly appears in the subtraction terms for the gluon and singlet-quark coefficient functions with opposite sign to its explicit contribution.  However, the cancellation is imperfect at finite order, and even the partially cancelled contribution dominates at NLO.  If this coefficient function is implemented it leads to peculiar behaviour for $Q^2$ slightly above $m_c^2$.  At NNLO, where heavy-flavour distributions begin at $Q^2=m_c^2$ with negative values, the ``bump'' in $F^{c\bar{c}}_L(x,Q^2)$ is negative, as illustrated in Fig.~18 of Ref.~\cite{Chuvakin:1999nx}, highlighting the unphysical nature of this contribution.

Hence, as in Ref.~\cite{Thorne:1997ga} we choose to ignore the explicit single heavy-quark--photon scattering results.  We define the longitudinal sector in what seems to us to be the most physical generalisation of the definition for $F^H_2(x,Q^2)$, as explained in Ref.~\cite{Thorne:2006qt}.  The heavy-quark coefficient functions are simply those for the light quarks, with the upper limit of integration moved from $1$ to $x_{\rm max}=Q^2/(Q^2+4m_H^2)$.  Thus the physical threshold of $W^2\geq 4m_H^2$ is contained in all terms, and there are no spurious zeroth-order terms.  These could only make a contribution if one works in the framework of single heavy quark scattering in the region of low $Q^2$ where the parton model for the heavy quark is least appropriate.  The definition of the SACOT-type scheme~\cite{Kramer:2000hn}, and particularly the SACOT($\chi$) scheme used in the global fits~\cite{Tung:2006tb}, also avoids this undesirable zeroth-order coefficient function.  We will discuss our predictions for $F_L(x,Q^2)$ in Section~\ref{sec:longitudinal}.

\subsubsection*{Charged-current structure functions}

The extension of the GM-VFNS to charged currents is most important for the heavy-flavour contribution to neutrino structure function data from CCFR, NuTeV and CHORUS, where the dominant process is $\nu + s \to \ell^- + c$ (with small Cabibbo mixing) and the charge conjugate process.  As such, this is particularly important for an analysis of the dimuon production data from CCFR and NuTeV, discussed in more detail in Section~\ref{sec:dimuon}.  There are also charged-current data from HERA, but this is less precise (though it does not need nuclear corrections, unlike the neutrino data), and is at sufficiently high $Q^2$ that the heavy quarks are effectively massless, though the GM-VFNS is applied in this case also.

The general procedure for the GM-VFNS for charged-current deep-inelastic scattering works on the same principles as for neutral currents --- one can now produce a {\it single} charm quark from a strange quark so the threshold is now at $x_{\rm max}=Q^2/(Q^2+m_c^2)$.  However, as explained in Ref.~\cite{Thorne:2006qt}, there is a complication because the massive FFNS coefficient functions are not known at ${\cal O} (\alpha_S^2)$ (only asymptotic limits~\cite{Buza:1997mg} have been calculated).  These coefficient functions are needed in our GM-VFNS at low $Q^2$ at NLO, and at all $Q^2$ at NNLO --- though in the latter case the definition of the GM-VFNS means that the $\ln(Q^2/m_c^2)$ terms are subtracted, and the ${\cal O}(m_c^2/Q^2)$ terms die away at high $Q^2$, so the GM-VFNS coefficient functions tend to the precisely known massless limits for large $Q^2/m_c^2$.

The initial proposal to deal with this, outlined in Ref.~\cite{Thorne:2006qt}, was to assume that the mass-dependence in the ${\cal O} (\alpha_S^2)$ coefficient functions is the same as for the neutral-current functions, but with the threshold in $4m_c^2$ replaced by a threshold in $m_c^2$. It was noted that this meant that the coefficient functions at least satisfy the threshold requirements, and tend smoothly to the correct massless limits, so were very likely to be an improvement on the ZM-VFNS.  However, in the course of the analyses performed in the latest global fit we have noticed various complications. One consideration is that the neutrino cross sections are given by expressions of the form
\begin{equation}
  \frac{{\rm d}^2\sigma^{\rm CC}}{{\rm d}x\,{\rm d}y} ~~\propto~~ \left(1-y +\frac{y^2}{2}-\frac{M_Nxy}{2E_{\nu}}\right) F_2^{\rm CC}(x,Q^2) -\frac{y^2}{2}F_L^{\rm CC}(x,Q^2) \pm y\left(1-\frac{y}{2}\right)xF_3^{\rm CC}(x,Q^2),
\end{equation}
where  $y \sim 0.3-0.8$.  Unlike the general case for neutral currents, where the $F_2(x,Q^2)$ term dominates, all three terms are important for charged currents and in the heavy-flavour contribution there are significant cancellations between them, so all need to be treated very carefully.

\begin{figure}
  \begin{center}
    \includegraphics[height=5cm]{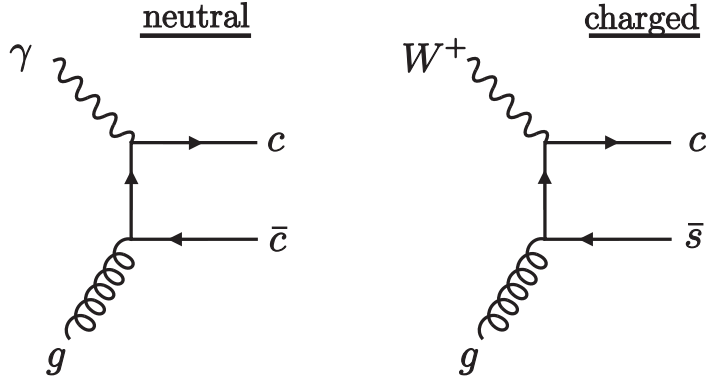}
    \caption{The gluon coefficient function at ${\cal O} (\alpha_S)$ in NC and CC processes.}
    \label{fig:NCCC}
  \end{center}
\end{figure}

Let us first discuss $F_2^{\rm CC}(x,Q^2)$.  At ${\cal O} (\alpha_S^2)$ this is dominated by the gluon contribution.  The simple prescription, suggested in Ref.~\cite{Thorne:2006qt}, gives large corrections to $F^{\rm CC}_2(x,Q^2)$ at low $Q^2$  because the lower threshold compared to the neutral-current case leads to a longer convolution length.  However, let us consider the comparison of the gluon coefficient functions at ${\cal O} (\alpha_S)$, represented by the two diagrams in Fig.~\ref{fig:NCCC}.

The neutral-current coefficient function is infrared finite and positive.  On the other hand the charged-current coefficient function diverges due to the collinear emission of a {\it light} (i.e.~strange) quark.  After subtraction of this divergence, via the usual factorisation theorem in the $\overline{\rm MS}$ scheme, there is an approximate factor of $(1+2\ln(1-z))$ in the coefficient function at low $Q^2$.  It is negative for high $z$ even in the FFNS.  At higher $Q^2$, the finite parts of the neutral-current and charged-current GM-VFNS coefficient functions, after the subtraction of the  $\ln(Q^2/m_c^2)$ terms, converge to the same massless expression, but from qualitatively different forms at lower $Q^2$.

At ${\cal O} (\alpha_S^2)$, the neutral-current FFNS coefficient function is positive with threshold $\log$ enhancements.  On the other hand, obtaining the approximate charged-current expression by just a change in threshold limits leads to a large contribution, which, bearing in mind the above discussion, is unlikely to be accurate.  At ${\cal O} (\alpha_S^2)$, the charged-current diagram will still have an extra emission of a strange quark and a corresponding collinear subtraction.  So we choose to obtain the charged-current coefficient function by both the change in threshold kinematics and by introducing a $(1+2(m_c^2/Q^2)\ln(1-z))$ factor.  This still leads to the correct large $Q^2/m_c^2$ limit, but is very likely to be a more accurate representation of the low $Q^2$ region.  The same procedure is applied to the similar ${\cal O}(\alpha_S^2)$ singlet-quark contribution.

We now consider $xF_3^{\rm CC}(x,Q^2)$ at ${\cal O} (\alpha_S^2)$.  For massless quarks this contribution is zero for initial gluons and singlet quarks, but the coefficient function $C^{{\rm CC},n_f}_{3,g}(x,m_c^2,Q^2)$ is non-zero for finite $Q^2/m_c^2$.  However, it must vanish as both $Q^2/m_c^2 \to \infty$ and $W^2/m_c^2 \to \infty$.  Hence, our model for $C^{{\rm CC},n_f,(2)}_{3,g}(x,m_c^2,Q^2)$ is weighted by a factor $m_c^2/\hat W^2 = m_c^2z/((1-z)Q^2)$, as is the singlet-quark contribution.  It is important that a suppression of this type is implemented. Otherwise the contribution is potentially anomalously important at low $Q^2$ and $x$.  

Finally, there is a complication in the ordering for the longitudinal charged-current heavy flavour production.  In the massless limit the lowest-order contribution, simplified by neglecting the Cabibbo-suppressed $d$-quark contribution, is
\begin{equation}
  F_{L}^{{\rm CC},c}(x,Q^2) = 2\,\alpha_S (C_{L,g}^{\rm CC}(x) \otimes g(x,Q^2) +
  C_{L,q}^{\rm CC}(x) \otimes s(x,Q^2)).
\end{equation}
However, for a massive quark there is a zeroth-order contribution
\begin{equation}
  F_{L}^{{\rm CC},c}(x,Q^2) = 2\,\frac{m_c^2}{m_c^2+Q^2}\,\xi s(\xi,Q^2),
  \label{loccfllq}
\end{equation}
where $\xi = x(1+m_c^2/Q^2)$.  Note that this is unlike the neutral-current case for $F_L(x,Q^2)$, where there was also a zeroth-order contribution.  Here it is due to a real physical process, i.e.~$W^++s \to c$, rather than one which only makes sense in the limit where the charm quark is most definitely behaving like a massless parton.  Hence, for the charged-current case, the zero-order contribution must be included.  This means that there is a difference in orders below and above the transition point, i.e.~the FFNS begins at zeroth order whereas the ZM-VFNS begins at first order  --- opposite to the case for the neutral-current $F_2(x,Q^2)$. Again a choice in ordering must be made.  We choose to obtain the correct limits in both regimes along with maintaining continuity.  That is, we use \eqref{loccfllq} to define the LO contribution for $Q^2<m_c^2$, whereas for $Q^2>m_c^2$ the LO contribution is defined by
\begin{equation} \label{loccfl}
  F_{L}^{{\rm CC},c}(x,Q^2) = 2\,\frac{m_c^2}{m_c^2+Q^2}\,\xi s(\xi,Q^2)+2\left(1-\frac{m_c^2}{Q^2}\right)\alpha_S (C_{L,g}(x) \otimes g(x,Q^2) + C_{L,q}(x) \otimes s(x,Q^2)).
\end{equation}
At high $Q^2$ the first term naturally dies away leading to the normal massless limit.  We can easily generalise the prescription to higher orders by including the next term in the $\alpha_S$ expansion on both sides of the transition point, with the $(1-m_c^2/Q^2)$ factor always multiplying the highest-order term in the region above the transition point.

In principle, we could also make some use of the ${\cal O}(\alpha_S^3)$ charged-current coefficient functions in the same way as we do for neutral currents.  However, the region where they make a significant impact is overwhelmingly in the small $x$ and $Q^2$ regime accessed only by HERA neutral-current measurements.  The amount of modelling required for these terms in charged-current processes is large.  Since they are unlikely to have much effect at all, we simply omit them.

With the chosen prescription for charged-current heavy flavour described above, we find that the description of the data, and the resulting parton distributions, are rather stable in going from LO $\to$ NLO $\to$ NNLO.  This will be discussed further in later sections.  

\subsection{Intrinsic heavy flavour}

Throughout the above discussion we have made the assumption that all heavy quark ($c,b$) flavour is perturbatively generated from the gluon and lighter flavours.  Indeed, this has always been the assumption in our analyses.  It is justified up to corrections of ${\cal O}(\Lambda_{QCD}^2/m_H^2)$.  This potential power-suppressed correction is known as intrinsic heavy flavour.  Although it is parametrically small it may be relatively enhanced at high values of $x$~\cite{Brodsky:1980pb}, and a plausible measure of the contribution is that at low scales the integrated number density to find intrinsic charm is $\leq 1\%$.  There is little within the current global fit which can tightly constrain this possible intrinsic flavour contribution.  The possibility of different types of intrinsic charm contribution has been studied recently by CTEQ~\cite{Pumplin:2007wg}.  They conclude that a global fit can allow up to a factor $\sim3$ times more high-$x$ enhanced intrinsic charm than the traditional value of $1\%$, and that a sea-like contribution can carry a momentum fraction up to $2\%$ of that of the proton (the integrated number density being infinite in this case). Indeed, for the sea-like distribution, they find a preference for a non-zero value, but we believe that this merely compensates for the terms which are systematically absent in the ACOT($\chi$) scheme at low $Q^2$ compared to our ${\rm TR}^\prime$ scheme or to the FFNS.  There seems to be no theoretical impetus to consider a sea-like intrinsic heavy flavour to be anything other than tiny.  We will consider the possibilities for a high-$x$ enhanced contribution in Section~\ref{sec:heavyflavourdata}.  However, in order to do so, we discuss briefly how it fits into the GM-VFNS framework.

Intrinsic flavour and GM-VFNS definitions were discussed in Ref.~\cite{Thorne:1998xv}.  We briefly recall them here.  Allowing an intrinsic heavy quark distribution actually removes the redundancy in the definition of the coefficient functions in the GM-VFNS, and two different definitions of a GM-VFNS will no longer be identical if formally summed to all orders, though they will only differ by contributions depending on the intrinsic flavour.  The reason is as follows.  Consider using identical parton distributions, including the intrinsic heavy quarks, in two different flavour schemes.  The heavy-quark coefficient functions at each order are different by ${\cal O}(m_H^2/Q^2)$.  This difference has been constructed to disappear at all orders when combining the parton distributions other than the intrinsic heavy quarks, but will  persist for the intrinsic contribution.  The intrinsic heavy-flavour distributions are of ${\cal O}(\Lambda_{QCD}^2/m_H^2)$, and when combined with the difference in coefficient functions the mass-dependence cancels leading to a difference in structure functions of ${\cal O}(\Lambda_{QCD}^2/Q^2)$.  It has been shown~\cite{Collins:1998rz} that for a given GM-VFNS the calculation of the structure functions is limited in accuracy to ${\cal O}(\Lambda_{QCD}^2/Q^2)$.  Hence, when including intrinsic charm, the scheme ambiguity is of the same order as the best possible accuracy one can obtain in leading-twist QCD, which is admittedly better than that obtained from ignoring the intrinsic heavy flavour (if it exists) as $Q^2$ increases above $m_H^2$.  It is intuitively obvious that better accuracy will be obtained from defining a GM-VFNS where all coefficient functions respect threshold kinematics, e.g.~the ACOT($\chi$) and ${\rm TR}^\prime$ schemes, than from some older schemes which do not, and which violate physical thresholds when combined with intrinsic heavy-flavour contributions.

%% file: globalpartonanalyses.tex
\section{Global parton analyses} \label{sec:globalfit}

In this section we present a summary of three global PDF analyses performed using the theoretical formalism outlined in Section \ref{sec:overviewtheoretical}, together with the treatment of heavy flavours given in Section \ref{sec:heavyflavour}.  We perform fits at LO, NLO and NNLO starting from input parton distributions parameterised at $Q_0^2=1$ GeV$^2$ in the form shown in Eqs.~(\ref{eq:uv})--(\ref{eq:sv}).

\subsection{Choice of data sets} \label{sec:choicedata}

The data sets included in the analyses are listed in Table~\ref{tab:chisquared}, together with the $\chi^2$ values, defined in Section \ref{sec:chisquared}, corresponding to each individual data set for each of the three fits, and the number of individual data points fitted for each data set.  The data sets in Table \ref{tab:chisquared} are ordered according to the type of process.  First we have the fixed-target data, which are subdivided into $\ell^\pm N$ structure functions, low-mass Drell--Yan (DY) cross sections, and $\nu N$ structure functions and dimuon cross sections.  Then we list data collected at HERA, and finally data collected at the Tevatron.  More detailed discussion of the description of the individual data sets will be given later in Sections \ref{sec:inclusivedata}--\ref{sec:jetdata}.
\begin{table}
  \centering
{\footnotesize
  \begin{tabular}{l|c|c|c}
    \hline \hline
    Data set & LO & NLO & NNLO \\ \hline
    BCDMS $\mu p$ $F_2$~\cite{Benvenuti:1989rh} & 165 / 153 & 182 / 163 & 170 / 163 \\ 
    BCDMS $\mu d$ $F_2$~\cite{Benvenuti:1989fm} & 162 / 142 & 190 / 151 & 188 / 151 \\ 
    NMC $\mu p$ $F_2$~\cite{Arneodo:1996qe} & 137 / 115 & 121 / 123 & 115 / 123 \\ 
    NMC $\mu d$ $F_2$~\cite{Arneodo:1996qe} & 120 / 115 & 102 / 123 & 93 / 123 \\ 
    NMC $\mu n/\mu p$~\cite{Arneodo:1996kd} & 131 / 137 & 130 / 148 & 135 / 148 \\ 
    E665 $\mu p$ $F_2$~\cite{Adams:1996gu} & 59 / 53 & 57 / 53 & 63 / 53 \\ 
    E665 $\mu d$ $F_2$~\cite{Adams:1996gu} & 49 / 53 & 53 / 53 & 63 / 53 \\ 
    SLAC $ep$ $F_2$~\cite{Whitlow:1991uw,Whitlow:1990dr} & 24 / 18 & 30 / 37 & 31 / 37 \\ 
    SLAC $ed$ $F_2$~\cite{Whitlow:1991uw,Whitlow:1990dr} & 12 / 18 & 30 / 38 & 26 / 38 \\ 
    NMC/BCDMS/SLAC $F_L$~\cite{Arneodo:1996qe,Benvenuti:1989rh,Whitlow:1990gk} & 28 / 24 & 38 / 31 & 32 / 31 \\ \hline 
    E866/NuSea $pp$ DY~\cite{Webb:2003bj} & 239 / 184 & 228 / 184 & 237 / 184 \\ 
    E866/NuSea $pd/pp$ DY~\cite{Towell:2001nh} & 14 / 15 & 14 / 15 & 14 / 15 \\ \hline 
    NuTeV $\nu N$ $F_2$~\cite{Tzanov:2005kr} & 49 / 49 & 49 / 53 & 46 / 53 \\ 
    CHORUS $\nu N$ $F_2$~\cite{Onengut:2005kv} & 21 / 37 & 26 / 42 & 29 / 42 \\ 
    NuTeV $\nu N$ $xF_3$~\cite{Tzanov:2005kr} & 62 / 45 & 40 / 45 & 34 / 45 \\ 
    CHORUS $\nu N$ $xF_3$~\cite{Onengut:2005kv} & 44 / 33 & 31 / 33 & 26 / 33 \\ 
    CCFR $\nu N\to \mu\mu X$~\cite{Goncharov:2001qe} & 63 / 86 & 66 / 86 & 69 / 86 \\ 
    NuTeV $\nu N\to \mu\mu X$~\cite{Goncharov:2001qe} & 44 / 40 & 39 / 40 & 45 / 40 \\ \hline 
    H1 MB 99 $e^+p$ NC~\cite{Lobodzinska:2003yd} & 9 / 8 & 9 / 8 & 7 / 8 \\ 
    H1 MB 97 $e^+p$ NC~\cite{Adloff:2000qk} & 46 / 64 & 42 / 64 & 51 / 64 \\ 
    H1 low $Q^2$ 96--97 $e^+p$ NC~\cite{Adloff:2000qk} & 54 / 80 & 44 / 80 & 45 / 80 \\ 
    H1 high $Q^2$ 98--99 $e^-p$ NC~\cite{Adloff:2000qj} & 134 / 126 & 122 / 126 & 124 / 126 \\ 
    H1 high $Q^2$ 99--00 $e^+p$ NC~\cite{Adloff:2003uh} & 153 / 147 & 131 / 147 & 133 / 147 \\ 
    ZEUS SVX 95 $e^+p$ NC~\cite{Breitweg:1998dz} & 35 / 30 & 35 / 30 & 35 / 30 \\ 
    ZEUS 96--97 $e^+p$ NC~\cite{Chekanov:2001qu} & 118 / 144 & 86 / 144 & 86 / 144 \\ 
    ZEUS 98--99 $e^-p$ NC~\cite{Chekanov:2002ej} & 61 / 92 & 54 / 92 & 54 / 92 \\ 
    ZEUS 99--00 $e^+p$ NC~\cite{Chekanov:2003yv} & 75 / 90 & 63 / 90 & 65 / 90 \\ 
    H1 99--00 $e^+p$ CC~\cite{Adloff:2003uh} & 28 / 28 & 29 / 28 & 29 / 28 \\ 
    ZEUS 99--00 $e^+p$ CC~\cite{Chekanov:2003vw} & 36 / 30 & 38 / 30 & 37 / 30 \\ 
    H1/ZEUS $ep$ $F_2^{\rm charm}$~\cite{Adloff:1996xq,Adloff:2001zj,Aktas:2005iw,Aktas:2004az,Breitweg:1999ad,Chekanov:2003rb,Chekanov:2007ch} & 110 / 83 & 107 / 83 & 95 / 83 \\ 
    H1 99--00 $e^+p$ incl.~jets~\cite{Aktas:2007pb} & 109 / 24 & 19 / 24 & --- \\ 
    ZEUS 96--97 $e^+p$ incl.~jets~\cite{Chekanov:2002be} & 88 / 30 & 30 / 30 & ---  \\ 
    ZEUS 98--00 $e^\pm p$ incl.~jets~\cite{Chekanov:2006xr} & 102 / 30 & 17 / 30 & --- \\ \hline 
    D{\O} II $p\bar{p}$ incl.~jets~\cite{Abazov:2008hu} & 193 / 110 & 114 / 110 & 123 / 110 \\ 
    CDF II $p\bar{p}$ incl.~jets~\cite{Abulencia:2007ez} & 143 / 76 & 56 / 76 & 54 / 76 \\ 
    CDF II $W\to \ell\nu$ asym.~\cite{Acosta:2005ud} & 50 / 22 & 29 / 22 & 30 / 22 \\ 
    D{\O} II $W\to \ell\nu$ asym.~\cite{Abazov:2007pm} & 23 / 10 & 25 / 10 & 25 / 10 \\ 
    D{\O} II $Z$ rap.~\cite{Abazov:2007jy} & 25 / 28 & 19 / 28 & 17 / 28 \\ 
    CDF II $Z$ rap.~\cite{Han:2008} & 52 / 29 & 49 / 29 & 50 / 29 \\ \hline 
    All data sets & \textbf{3066 / 2598} & \textbf{2543 / 2699} & \textbf{2480 / 2615} \\
    \hline \hline
  \end{tabular}
}
\caption{The values of $\chi^2 / N_{\rm pts.}$ for the data sets included in the global fits.  For the NuTeV $\nu N\to \mu\mu X$ data, the effective number of degrees of freedom is quoted instead of $N_{\rm pts.}$ since smearing effects mean that nearby data points are highly correlated~\cite{Goncharov:2001qe}.  The details of corrections to data, kinematic cuts applied and definitions of $\chi^2$ are contained in the text.}
\label{tab:chisquared}
\end{table}

For the DIS data we make cuts on the exchanged boson virtuality of $Q^2\ge 2$ GeV$^2$ and on the squared boson--nucleon centre-of-mass energy, $W^2=Q^2(1/x-1)+m_N^2$, of $W^2\ge 15$ GeV$^2$ at NLO and NNLO, while $W^2\ge 20$ GeV$^2$ at LO due to the absence of higher-order large-$x$ terms.  The cut on $W^2$ is necessary to exclude data for which potentially large higher-twist contributions are present, such as target mass corrections~\cite{Schienbein:2007gr}.  For the $\nu(\bar{\nu})N$ $xF_3$ data we only include data with $W^2\ge25$ GeV$^2$ for all three fits.  The higher-twist contributions for $F_2^{\nu(\bar{\nu})N}(x,Q^2)$ do not contribute to the Adler sum rule, i.e.~$\int_0^1\!{\rm d}x\;F^{\rm HT}_2(x,Q^2)=0$, which means they must be well behaved as $x\to 0$. However, there is no such restriction on the higher-twist corrections for $xF_3$, and renormalon calculations imply that they are large~\cite{Dokshitzer:1995qm,Dasgupta:1996hh}.  The CHORUS data extend into the region of expected large higher-twist corrections, so the above cut is necessary to avoid contamination (which is qualitatively similar to that predicted in Refs.~\cite{Dokshitzer:1995qm,Dasgupta:1996hh}).

As in previous MRST fits, data on the deuteron structure function $F_2^d$, and the ratio $F_2^d/F_2^p$~\cite{Arneodo:1996kd}, are corrected for nuclear shadowing effects~\cite{Badelek:1994qg}.  No such corrections are applied to the E866/NuSea $pd/pp$ Drell--Yan ratio~\cite{Towell:2001nh}, though we expect such corrections to be small compared to the experimental errors, and it is not clear how such corrections would be applied.  For other nuclear target data we apply an improved nuclear correction discussed in detail in Section \ref{sec:neutrinoinclusive}.  We apply a correction of $-3.4\%$ to the luminosity of the published H1 MB 97 data~\cite{Adloff:2000qk} following a luminosity reanalysis~\cite{VargasTrevino:2007zz}.

The Tevatron high-$p_T$ jet data are included in the fit at each order. The full NNLO corrections are not known in this case, but the approximation based on threshold corrections~\cite{Kidonakis:2000gi} are included in the \textsc{fastnlo} package~\cite{Kluge:2006xs}.  There is no guarantee that these give a very good approximation to the full NNLO corrections, but in this case the NLO corrections themselves are of the same order as the systematic uncertainties on the data.  The threshold corrections are the only realistic source of large NNLO corrections, so the fact that they provide a correction which is smooth in $p_T$ and small compared to NLO (and compared to the systematic uncertainties on the data) strongly implies that the full NNLO corrections would lead to very little change in the PDFs.  Since these jet data are the only good direct constraint on the high-$x$ gluon we choose to include them in the NNLO fit judging that the impact of leaving them out would be more detrimental than any inaccuracies in including them without knowing the full NNLO hard cross section.  The HERA jet data, for which the NNLO coefficient functions are also unknown at present, are omitted from our NNLO fit.  This is because they do not provide as important a constraint as the Tevatron jet data, and also because the NLO corrections are much larger than for Tevatron jet data (due to the lower scales) and there is not even some partial information on the NNLO result available.  However, we compare the HERA jet data to results calculated with the NNLO PDFs using the NLO coefficient functions in Section \ref{sec:jetdata}, finding surprisingly good agreement.

\subsection{\texorpdfstring{$\chi^2$}{Chi-squared} definition} \label{sec:chisquared}

The global goodness-of-fit measure is defined as $\chi_{\rm global}^2 = \sum_n \chi^2_n$, where, in all previous MRST fits,
\begin{equation} \label{eq:chisqmrst}
  \chi_n^2(\{a\},\mathcal{N}_n) \;=\; \sum_{i=1}^{N_{\rm pts.}} \frac{\left(D_{n,i}-T_{n,i}(\{a\})/\mathcal{N}_n\right)^2}{(\sigma_{n,i}^{\rm uncorr.})^2 + \sum_k (\sigma_{n,k,i}^{\rm corr.})^2} \;+\; \chi_{\mathcal{N}_n}^2.
\end{equation}
Here, $n$ labels a particular data set, or a combination of data sets, with a common (fitted) normalisation $\mathcal{N}_n$, $i$ labels the number of individual data points in that data set, and $k$ labels the individual correlated systematic errors for a particular data set.  The individual data points $D_{n,i}$ have uncorrelated (statistical and systematic) errors $\sigma_{n,i}^{\rm uncorr.}$ and correlated systematic errors $\sigma_{n,k,i}^{\rm corr.}$.  The theory prediction $T_{n,i}(\{a\})$ depends on the input PDF parameters $\{a\}$.  The MRST fits added all uncorrelated and correlated experimental errors in quadrature in the $\chi^2$ definition \eqref{eq:chisqmrst}, except for Tevatron jet production.  Here, the full correlated error information was used but the procedure was more complicated; see Section \ref{sec:jetdata}.  The second term in \eqref{eq:chisqmrst} is a possible penalty term to account for deviations of the fitted data set normalisations $\mathcal{N}_n$ from the nominal values of 1.  (In practice, the MRST fits took $\chi_{\mathcal{N}_n}^2=0$, although constraints were applied as explained below.)  We will now discuss improvements to the treatment of data set normalisations made in the present analysis, then we will discuss improvements in the treatment of correlated systematic errors.

\subsubsection*{Treatment of data set normalisations}

In the MRST analyses, the normalisations of several data sets were floated and then fixed, checking that these normalisations lay inside the quoted one-sigma error range.  For example, the H1 data sets from different running periods were all taken to have a common normalisation of exactly 1, while the ZEUS data sets from different running periods were all taken to have a common normalisation of 0.98.  In general, HERA data from different runs are allowed to have different normalisations, therefore we now split the H1 and ZEUS data into different running periods, and allow a different normalisation for each of these running periods.\footnote{We take the normalisations of the H1 and ZEUS data on $F_2^{c\bar{c}}$ to be fixed at 1, although here the experimental errors are anyway much larger than for the total reduced cross sections, and the $F_2^{c\bar{c}}$ data do not provide a strong constraint on the PDFs.}  The normalisations of the data sets are now all allowed to go free at the same time as the PDF parameters, including in the determination of the PDF uncertainties discussed in Section \ref{sec:erroranalysis}.

Note that the (fitted) normalisation parameters $\mathcal{N}_n$ are included in \eqref{eq:chisqmrst} in the correct manner to avoid bias in the result, i.e.~if the central value of the data is rescaled by a factor $\mathcal{N}_n$ then so are the uncertainties.  If fitting a single data set with no normalisation constraint on the theory we would automatically obtain $\mathcal{N}_n=1$.  However, the fit does allow data sets to have different relative normalisations, and additionally the sum rules on the PDFs do influence the normalisation to some extent.

The MRST analyses did not include any penalty term in the $\chi^2$ definition for data set normalisations $\mathcal{N}_n$ differing from the nominal values of 1.  The usual choice for the penalty term for the data set normalisations is
\begin{equation} \label{eq:gausnormpenalty}
  \chi_{\mathcal{N}_n}^2 = \left(\frac{1-\mathcal{N}_n}{\sigma_n^{\mathcal{N}}}\right)^2,
\end{equation}
where $\sigma_n^{\mathcal{N}}$ is the one-sigma normalisation error for data set $n$.  Alternatively, it has been proposed that normalisation uncertainties should behave according to a log-normal distribution, where the penalty term is~\cite{Blobel:2003wa}
\begin{equation} \label{eq:lognormpenalty}
  \chi_{\mathcal{N}_n}^2 = \ln\left(\mathcal{N}_n\right)\,\left(3+\frac{\ln\left(\mathcal{N}_n\right)}{\ln\left(1+\left(\sigma_n^{\mathcal{N}}\right)^2\right)}\right),
\end{equation}
which reduces to the usual quadratic term \eqref{eq:gausnormpenalty} for small $\sigma_n^{\mathcal{N}}$ and for small deviations of $\mathcal{N}_n$ from 1.  We find that, using \eqref{eq:gausnormpenalty}, the best-fit data set normalisations tend to stray outside their nominal one-sigma range, with all the largest shifts being in the downwards direction.  It has long been known that both LO and NLO fits would prefer to have more than $100\%$ momentum for the PDFs if allowed (e.g.~at NLO in Ref.~\cite{Vogt:1999ik}, the extra momentum in the fit regions for the conservative NLO PDFs in Ref.~\cite{Martin:2003sk}, and the large momentum violation at LO seen in Ref.~\cite{Sherstnev:2007nd}) so imposing momentum conservation on PDFs conversely leads to a preference for lower data set normalisations.  However, this is a diminishing effect as we increase the order of the QCD calculation, implying that it is an artifact of an incomplete theory at lower orders, and as such is an undesirable systematic effect. Additionally, it has been claimed that normalisation uncertainties are expected to behave more like a box shape than the usual Gaussian behaviour~\cite{Devenish:2004pb}, and indeed, the term in \eqref{eq:lognormpenalty} does alter the shape of the penalty in the downwards direction in this type of manner, albeit to a very small degree for errors of only a few percent.  Hence, taking into account the systematic effect at low perturbative orders and that normalisations are very unlikely to be perfectly Gaussian, we use a more severe \emph{quartic} penalty term for the normalisations, i.e.
\begin{equation} \label{eq:normpenalty}
  \chi_{\mathcal{N}_n}^2 = \left(\frac{1-\mathcal{N}_n}{\sigma_n^{\mathcal{N}}}\right)^4.
\end{equation}
This binds the normalisations more strongly to the range $\mathcal{N}_n\in[1-\sigma_n^{\mathcal{N}},1+\sigma_n^{\mathcal{N}}]$, but in practice it is far more the case that it stops them from floating too low.  If more than one data set has a common normalisation, the penalty term $ \chi_{\mathcal{N}_n}^2$ is divided amongst those data sets according to the number of data points.

The resulting (re)normalisation factors $\mathcal{N}_n$ for each fit are listed in Table \ref{tab:normalisations}.
\begin{table}
  \centering
{\footnotesize
  \begin{tabular}{l|c|c|c|c}
    \hline \hline
    Data set & $\sigma_n^{\mathcal{N}}$ & LO & NLO & NNLO \\ \hline
    BCDMS $\mu p$ $F_2$~\cite{Benvenuti:1989rh} & 3\% & 0.9667 & 0.9644 & 0.9678 \\ 
    BCDMS $\mu d$ $F_2$~\cite{Benvenuti:1989fm} & 3\% & 0.9667 & 0.9644 & 0.9678 \\ 
    NMC $\mu p$ $F_2$~\cite{Arneodo:1996qe} & 2\% & 1.0083 & 0.9982 & 0.9999 \\ 
    NMC $\mu d$ $F_2$~\cite{Arneodo:1996qe} & 2\% & 1.0083 & 0.9982 & 0.9999 \\ 
    NMC $\mu n/\mu p$~\cite{Arneodo:1996kd} & --- & 1 & 1 & 1 \\ 
    E665 $\mu p$ $F_2$~\cite{Adams:1996gu} & 1.85\% & 1.0146 & 1.0052 & 1.0024 \\ 
    E665 $\mu d$ $F_2$~\cite{Adams:1996gu} & 1.85\% & 1.0146 & 1.0052 & 1.0024 \\ 
    SLAC $ep$ $F_2$~\cite{Whitlow:1991uw,Whitlow:1990dr} & 1.9\% & 1.0227 & 1.0125 & 1.0078 \\ 
    SLAC $ed$ $F_2$~\cite{Whitlow:1991uw,Whitlow:1990dr} & 1.9\% & 1.0227 & 1.0125 & 1.0078 \\ 
    NMC/BCDMS/SLAC $F_L$~\cite{Arneodo:1996qe,Benvenuti:1989rh,Whitlow:1990gk} & --- & 1 & 1 & 1 \\ \hline 
    E866/NuSea $pp$ DY~\cite{Webb:2003bj} & 6.5\% & 1.0629 & 1.0086 & 1.0868 \\ 
    E866/NuSea $pd/pp$ DY~\cite{Towell:2001nh} & --- & 1 & 1 & 1 \\ \hline 
    NuTeV $\nu N$ $F_2$~\cite{Tzanov:2005kr} & 2.1\% & 0.9987 & 0.9997 & 0.9992 \\ 
    CHORUS $\nu N$ $F_2$~\cite{Onengut:2005kv} & 2.1\% & 0.9987 & 0.9997 & 0.9992 \\ 
    NuTeV $\nu N$ $xF_3$~\cite{Tzanov:2005kr} & 2.1\% & 0.9987 & 0.9997 & 0.9992 \\ 
    CHORUS $\nu N$ $xF_3$~\cite{Onengut:2005kv} & 2.1\% & 0.9987 & 0.9997 & 0.9992 \\ 
    CCFR $\nu N\to \mu\mu X$~\cite{Goncharov:2001qe} & 2.1\% & 0.9987 & 0.9997 & 0.9992 \\ 
    NuTeV $\nu N\to \mu\mu X$~\cite{Goncharov:2001qe} & 2.1\% & 0.9987 & 0.9997 &  0.9992 \\ \hline 
    H1 MB 99 $e^+p$ NC~\cite{Lobodzinska:2003yd} & 1.3\% & 0.9861 & 1.0098 & 1.0090 \\ 
    H1 MB 97 $e^+p$ NC~\cite{Adloff:2000qk} & 1.5\% & 0.9863 & 0.9921 & 0.9953 \\ 
    H1 low $Q^2$ 96--97 $e^+p$ NC~\cite{Adloff:2000qk} & 1.7\% & 1.0029 & 1.0095 & 1.0172 \\ 
    H1 high $Q^2$ 98--99 $e^-p$ NC~\cite{Adloff:2000qj} & 1.8\% & 0.9782 & 0.9851 & 0.9860 \\ 
    H1 high $Q^2$ 99--00 $e^+p$ NC~\cite{Adloff:2003uh} & 1.5\% & 0.9762 & 0.9834 & 0.9842 \\ 
    ZEUS SVX 95 $e^+p$ NC~\cite{Breitweg:1998dz} & 1.5\% & 0.9944 & 0.9948 & 1.0004 \\ 
    ZEUS 96--97 $e^+p$ NC~\cite{Chekanov:2001qu} & 2\% & 0.9735 & 0.9811 & 0.9871 \\ 
    ZEUS 98--99 $e^-p$ NC~\cite{Chekanov:2002ej} & 1.8\% & 0.9771 & 0.9855 & 0.9862 \\ 
    ZEUS 99--00 $e^+p$ NC~\cite{Chekanov:2003yv} & 2.5\% & 0.9656 & 0.9761 & 0.9762 \\ 
    H1 99--00 $e^+p$ CC~\cite{Adloff:2003uh} & 1.5\% & 0.9762 & 0.9834 & 0.9842 \\ 
    ZEUS 99--00 $e^+p$ CC~\cite{Chekanov:2003vw} & 2.5\% & 0.9656 & 0.9761 & 0.9762 \\ 
    H1/ZEUS $ep$ $F_2^{\rm charm}$~\cite{Adloff:1996xq,Adloff:2001zj,Aktas:2005iw,Aktas:2004az,Breitweg:1999ad,Chekanov:2003rb,Chekanov:2007ch} & --- & 1 & 1 & 1 \\ 
    H1 99--00 $e^+p$ incl.~jets~\cite{Aktas:2007pb} & 1.5\% & 0.9762 & 0.9834 & --- \\ 
    ZEUS 96--97 $e^+p$ incl.~jets~\cite{Chekanov:2002be} & 2\% & 0.9735 & 0.9811 & --- \\ 
    ZEUS 98--00 $e^\pm p$ incl.~jets~\cite{Chekanov:2006xr} & 2.5\% & 0.9656 & 0.9761 & --- \\ \hline 
    D{\O} II $p\bar{p}$ incl.~jets~\cite{Abazov:2008hu} & 6.1\% & 0.9353 & 1.0596 & 1.0759 \\ 
    CDF II $p\bar{p}$ incl.~jets~\cite{Abulencia:2007ez} & 5.8\% & 0.8779 & 0.9646 & 0.9900 \\ 
    CDF II $W\to \ell\nu$ asym.~\cite{Acosta:2005ud} & --- & 1 & 1 & 1 \\ 
    D{\O} II $W\to \ell\nu$ asym.~\cite{Abazov:2007pm} & --- & 1 & 1 & 1 \\ 
    D{\O} II $Z$ rap.~\cite{Abazov:2007jy} &  --- & 1 & 1 & 1 \\ 
    CDF II $Z$ rap.~\cite{Han:2008} & 5.8\% & 0.8779 & 0.9646 & 0.9900 \\ 
    \hline \hline
  \end{tabular}
}
\caption{The fitted normalisations $\mathcal{N}_n$ of the data sets included in the global fit, together with the one-sigma normalisation errors, $\sigma_n^{\mathcal{N}}$, for each data set $n$.}
\label{tab:normalisations}
\end{table}

\subsubsection*{Correlated systematic errors}

The insensitivity of both the central values of the PDFs and their uncertainties to the complete inclusion of correlation information for the DIS data was confirmed in the benchmark PDF fits in the proceedings of the HERA--LHC workshop~\cite{Dittmar:2005ed} and an extensive discussion of the effect of correlations for HERA data was presented in the appendix of Ref.~\cite{Martin:2001es}.  To summarise the latter, the main effect was by far the systematic shifts due to changes in normalisation which we now treat in full (indeed a strong hint of the necessity for the normalisation correction~\cite{VargasTrevino:2007zz} of the H1 MB 97 data~\cite{Adloff:2000qk} was seen).  Beyond this, the introduction of full correlations led to rather small changes and those were, in general, precisely such as to flatten the evolution of $F_2(x,Q^2)$ allowing the gluon to reduce.  As previous comments on normalisations and momentum conservation suggest, this is itself likely correlated to shortcomings of the theory at low orders.  Hence, we maintain that the lack of the correlations has little effect on our results, and continue to make this simplification.  This is particularly the case since we note that for the preliminary averaged HERA cross section data~\cite{averagedata}, sources of correlated uncertainty are often dramatically reduced (sometimes by a factor of 3--4), and in the averaging between H1 and ZEUS the data points frequently move relative to each other in a rather different manner than a data set does relative to theory in a fit.  We intend to include full correlations for the averaged HERA data, when they are published and will be better understood, but at this stage they will also be very small indeed.  We also note that in some cases a textbook treatment of correlated uncertainties can lead to peculiar results.  Indeed in Ref.~\cite{wukidis06} it was seen that the high-$y$ turnover in reduced cross section data due to the influence of $F_L(x,Q^2)$ could be eliminated completely in a fit by letting the correlated systematic uncertainty due to the photoproduction background move by about two-sigma.  This seems unlikely (though not impossible), and it is certainly the case that the distribution of uncertainties in this background are far from Gaussian, so the conventional treatment may give surprising results. It is definitely the case that the absolute $\chi^2$ values for the best fit do change depending on how the correlations are treated, but given that our new tolerance determination, described in detail in Section \ref{sec:tolerance}, relies on changes in $\chi^2$ relative to the best fit, we are confident that we are even less sensitive to such details than in previous studies, e.g.~the aforementioned~\cite{Dittmar:2005ed,Martin:2002aw}.

Nevertheless, for selected newly-added data sets we do include the full correlated error information.  Instead of \eqref{eq:chisqmrst}, the $\chi^2$ is given by~\cite{Pumplin:2002vw}
\begin{equation} \label{eq:chisqcorr}
  \chi_n^2(\{a\},\mathcal{N}_n) \;=\; \sum_{i=1}^{N_{\rm pts.}} \left(\frac{\hat{D}_{n,i}-T_{n,i}(\{a\})/\mathcal{N}_n}{\sigma_{n,i}^{\rm uncorr.}}\right)^2 \;+\; \sum_{k=1}^{N_{\rm corr.}}r_{n,k}^2 \;+\; \chi_{\mathcal{N}_n}^2,
\end{equation}
where $\hat{D}_{n,i} \equiv D_{n,i} - \sum_{k=1}^{N_{\rm corr.}}r_{n,k}\,\sigma_{n,k,i}^{\rm corr.}$ are the data points allowed to shift by the systematic errors in order to give the best fit.  Minimising $\chi_n^2$ with respect to $r_{n,k}$ gives the analytic result that~\cite{Pumplin:2002vw}
\begin{equation} \label{eq:rksolution}
  r_{n,k}(\{a\},\mathcal{N}_n) = \sum_{k^\prime=1}^{N_{\rm corr.}} (A^{-1})_{kk^\prime} B_{k^\prime}(\{a\},\mathcal{N}_n),
\end{equation}
where
\begin{equation}
  A_{kk^\prime} = \delta_{kk^\prime} + \sum_{i=1}^{N_{\rm pts.}}\frac{\sigma_{n,k,i}^{\rm corr.}\;\sigma_{n,k^\prime,i}^{\rm corr.}}{\sigma_{n,i}^{\rm uncorr.}}, \quad
  B_{k}(\{a\},\mathcal{N}_n) = \sum_{i=1}^{N_{\rm pts.}}\frac{\sigma_{n,k,i}^{\rm corr.}\left(D_{n,i}-T_{n,i}(\{a\})/\mathcal{N}_n\right)}{(\sigma_{n,i}^{\rm uncorr.})^2}.
\end{equation}
Therefore, the optimal shifts of the data points by the systematic errors are solved for analytically, while the input PDF parameters $\{a\}$, together with the data normalisations $\mathcal{N}_n$, must be determined by numerical minimisation of $\chi_{\rm global}^2$ in the usual way.

In practice, we use \eqref{eq:chisqcorr} only for the jet-energy-scale uncertainties in the ZEUS inclusive jet data~\cite{Chekanov:2002be,Chekanov:2006xr}, the 4 contributions to the correlated uncertainty in the H1 inclusive jet data~\cite{Aktas:2007pb}, the 16 sources of correlated uncertainty in the CDF Run II inclusive jet data~\cite{Abulencia:2007ez}, the 22 sources of correlated uncertainty in the D{\O} Run II inclusive jet data~\cite{Abazov:2008hu}, and the 6 sources of correlated uncertainty in the CDF Run II $Z$ rapidity distribution~\cite{Han:2008}.  In many other cases the uncertainties (other than normalisation) are presented without any information on correlations between systematic uncertainties, or are assumed to be uncorrelated. For the remainder of the data sets, the systematic errors (other than the normalisation error) are simply treated as being uncorrelated and added in quadrature with the statistical uncertainties, i.e.~we use \eqref{eq:chisqmrst} with the penalty term \eqref{eq:normpenalty}.

\subsubsection*{Minimisation of \texorpdfstring{$\chi^2$}{chi-squared} and calculation of Hessian matrix}

To determine the best fit at NLO and NNLO we need to minimise the $\chi_{\rm global}^2$ with respect to 28 free input PDF parameters, together with $\alpha_S(Q_0^2)$, 3 parameters ($r_1$, $r_2$, $r_3$) associated with nuclear corrections (see Section \ref{sec:inclusivedata}), and 17 different data set normalisations, giving a total of 49 free parameters.  This is a difficult task and is unlikely to be possible with the widely-used \textsc{minuit} package~\cite{Minuit}, where the practical maximum number of free parameters is around 15.  Indeed, the CTEQ group have found it necessary to extend \textsc{minuit} to use an improved iterative method in the numerical calculation of the Hessian matrix~\cite{Pumplin:2000vx}.  Other fitting groups, such as Alekhin and H1/ZEUS, use far fewer PDF parameters, and so presumably do not suffer from the same problems when using the standard \textsc{minuit} package.  As in previous MRS/MRST analyses~\cite{Martin:1988nk,Harriman:1990hi,Kwiecinski:1990ru,Martin:1992as,Martin:1992zi,Martin:1994kn,Martin:1995ws,Glover:1996ae,Martin:1996as,Martin:1996eva,Martin:1998sq,Martin:1999ww,Martin:2001es,Martin:2002dr,Martin:2002aw,Martin:2003sk,Martin:2004ir,Martin:2004dh,Martin:2006qz,Martin:2007bv}, we instead use the Levenberg--Marquardt method~\cite{Levenberg:1944,Marquardt:1963} described in Numerical Recipes~\cite{NumericalRecipes}, which combines the advantages of the inverse-Hessian method and the steepest descent method for minimisation.  The method requires knowledge of the gradient and Hessian matrix of the $\chi_{\rm global}^2$, which we provide partially as analytic expressions in our fitting code, only using numerical finite-difference computations for the derivatives of the theory predictions with respect to the fitted parameters.  A linearisation approximation is made in the calculation of the Hessian matrix to avoid the presence of potentially destabilising second-derivative terms~\cite{NumericalRecipes}.  For example, the contribution to the global Hessian matrix $H_{lm}=\sum_{n} H_{lm}^n$ from a data set $n$ with $\chi^2_n$ defined by \eqref{eq:chisqmrst} is
\begin{equation}
  H_{lm}^{n} \equiv \frac{1}{2}\,\frac{\partial{\chi^2_n}}{\partial a_l\,\partial a_m} \;=\; \sum_{i=1}^{N_{\rm pts.}}\frac{1}{(\sigma_{n,i}^{\rm uncorr.})^2 + \sum_k (\sigma_{n,k,i}^{\rm corr.})^2}\;\frac{\partial T_{n,i}(\{a\})/\mathcal{N}_n}{\partial a_l}\;\frac{\partial T_{n,i}(\{a\})/\mathcal{N}_n}{\partial a_m},
\end{equation}
while in the more complicated case of treating correlated systematic errors \eqref{eq:chisqcorr}, the corresponding expression is
\begin{equation}
  H_{lm}^{n} \;=\; \sum_{i=1}^{N_{\rm pts.}}\frac{1}{(\sigma_{n,i}^{\rm uncorr.})^2}\;\frac{\partial \hat{T}_{n,i}}{\partial a_l}\;\frac{\partial \hat{T}_{n,i}}{\partial a_m}\;+\;\sum_{k=1}^{N_{\rm corr.}}\;\frac{\partial r_{n,k}}{\partial a_l}\;\frac{\partial r_{n,k}}{\partial a_m},
\end{equation}
where $\hat{T}_{n,i}(\{a\},\mathcal{N}_n)\equiv T_{n,i}(\{a\})/\mathcal{N}_n+\sum_{k=1}^{N_{\rm corr.}}r_{n,k}(\{a\},\mathcal{N}_n)\,\sigma_{n,k,i}^{\rm corr.}$.  Similar expressions are used for the elements of the Hessian matrix corresponding to data set normalisations $\mathcal{N}_n$.

\subsection{Discussion of fit results}

For each of the three fits, the optimal values of the input PDF parameters of Eqs.~(\ref{eq:uv})--(\ref{eq:sv}) and of the QCD coupling $\alpha_S$ are given in Table \ref{tab:parameters}.
\begin{table}
  \centering
{\footnotesize
  \begin{tabular}{l||rl|rl|rl}
    \hline \hline
    Parameter & \multicolumn{2}{c|}{LO} & \multicolumn{2}{c|}{NLO} & \multicolumn{2}{c}{NNLO} \\\hline
    $\alpha_S(Q_0^2)$ & $0.68183$ & & $0.49128$ & & $0.45077$ & \\
    $\alpha_S(M_Z^2)$ & $0.13939$ & & $0.12018$ & & $0.11707$ & \\ \hline
    $A_u$ & $1.4335$ & & $0.25871$ & & $0.22250$ & \\
    $\eta_1$ & $0.45232$ & $^{+0.022}_{-0.018}$ & $0.29065$ & $^{+0.019}_{-0.013}$ & $0.27871$ & $^{+0.018}_{-0.014}$ \\
    $\eta_2$ & $3.0409$ & $^{+0.079}_{-0.067}$ & $3.2432$ & $^{+0.062}_{-0.039}$ & $3.3627$ & $^{+0.061}_{-0.044}$ \\
    $\epsilon_u$ & $-2.3737$ & $^{+0.54}_{-0.48}$ & $4.0603$ & $^{+1.6}_{-2.3}$ & $4.4343$ & $^{+2.4}_{-2.7}$ \\
    $\gamma_u$ & $8.9924$ & & $30.687$ & & $38.599$ & \\ \hline
    $A_d$ & $5.0903$ & & $12.288$ & & $17.938$ & \\
    $\eta_3$ & $0.71978$ & $^{+0.057}_{-0.082}$ & $0.96809$ & $^{+0.11}_{-0.11}$ & $1.0839$ & $^{+0.12}_{-0.11}$ \\
    $\eta_4-\eta_2$ & $2.0835$ & $^{+0.32}_{-0.45}$ & $2.7003$ & $^{+0.50}_{-0.52}$ & $2.7865$ & $^{+0.50}_{-0.44}$ \\
    $\epsilon_d$ & $-4.3654$ & $^{+0.28}_{-0.22}$ & $-3.8911$ & $^{+0.31}_{-0.29}$ & $-3.6387$ & $^{+0.27}_{-0.28}$ \\
    $\gamma_d$ & $7.4730$ & & $6.0542$ & & $5.2577$ & \\ \hline
    $A_S$ & $0.59964$ & $^{+0.036}_{-0.030}$ & $0.31620$ & $^{+0.030}_{-0.021}$ & $0.64942$ & $^{+0.047}_{-0.041}$ \\
    $\delta_S$ & $-0.16276$ & & $-0.21515$ & & $-0.11912$ & \\
    $\eta_S$ & $8.8801$ & $^{+0.33}_{-0.33}$ & $9.2726$ & $^{+0.23}_{-0.33}$ & $9.4189$ & $^{+0.25}_{-0.33}$ \\
    $\epsilon_S$ & $-2.9012$ & $^{+0.33}_{-0.37}$ & $-2.6022$ & $^{+0.71}_{-0.96}$ & $-2.6287$ & $^{+0.49}_{-0.51}$ \\
    $\gamma_S$ & $16.865$ & & $30.785$ & & $18.065$ & \\ \hline
    $\int_0^1\!{\rm d}x\;\Delta(x,Q_0^2)$ & $0.091031$ & $^{+0.012}_{-0.009}$ & $0.087673$ & $^{+0.013}_{-0.011}$ & $0.078167$ & $^{+0.012}_{-0.0091}$ \\
    $A_\Delta$ & $8.9413$ & & $8.1084$ & & $16.244$ & \\
    $\eta_\Delta$ & $1.8760$ & $^{+0.24}_{-0.30}$ & $1.8691$ & $^{+0.23}_{-0.32}$ & $2.0741$ & $^{+0.18}_{-0.35}$ \\
    $\gamma_\Delta$ & $8.4703$ & $^{+2.0}_{-0.3}$ & $13.609$ & $^{+1.1}_{-0.6}$ & $6.7640$ & $^{+0.77}_{-0.41}$ \\
    $\delta_\Delta$ & $-36.507$ & & $-59.289$ & & $-36.090$ & \\ \hline
    $A_g$ & $0.0012216$ & & $1.0805$ & & $3.4055$ & \\
    $\delta_g$ & $-0.83657$ & $^{+0.15}_{-0.14}$ & $-0.42848$ & $^{+0.066}_{-0.057}$ & $-0.12178$ & $^{+0.23}_{-0.16}$ \\
    $\eta_g$ & $2.3882$ & $^{+0.51}_{-0.50}$ & $3.0225$ & $^{+0.43}_{-0.36}$ & $2.9278$ & $^{+0.68}_{-0.41}$ \\
    $\epsilon_g$ & $-38.997$ & $^{+36}_{-35}$ & $-2.2922$ & & $-2.3210$ & \\
    $\gamma_g$ & $1445.5$ & $^{+880}_{-750}$ & $3.4894$ & & $1.9233$ & \\
    $A_{g^\prime}$ & --- & & $-1.1168$ & & $-1.6189$ & \\
    $\delta_{g^\prime}$ & --- & & $-0.42776$ & $^{+0.053}_{-0.047}$ & $-0.23999$ & $^{+0.14}_{-0.10}$ \\
    $\eta_{g^\prime}$ & --- & & $32.869$ & $^{+6.5}_{-5.9}$ & $24.792$ & $^{+6.5}_{-5.2}$ \\ \hline
    $A_+$ & $0.10302$ & $^{+0.029}_{-0.017}$ & $0.047915$ & $^{+0.0095}_{-0.0076}$ & $0.10455$ & $^{+0.019}_{-0.016}$ \\
    $\eta_+$ & $13.242$ & $^{+2.9}_{-1.4}$ & $9.7466$ & $^{+1.0}_{-0.8}$ & $9.8689$ & $^{+1.0}_{-0.6}$ \\
    $A_-$ & $-0.011523$ & $^{+0.009}_{-0.018}$ & $-0.011629$ & $^{+0.009}_{-0.023}$ & $-0.0093692$ & $^{+0.006}_{-0.024}$ \\
    $\eta_-$ & $10.285$ & $^{+16}_{-6}$ & $11.261$ & $^{+22}_{-6}$ & $9.5783$ & $^{+26}_{-5}$ \\
    $x_0$ & $0.017414$ & & $0.016050$ & & $0.018556$ & \\ \hline
    $r_1$ & $-0.39484$ & & $-0.57631$ & & $-0.80834$ & \\
    $r_2$ & $-1.0719$ & & $0.81878$ & & $1.2669$ & \\
    $r_3$ & $-0.28973$ & & $-0.083208$ & & $0.15098$ & \\
    \hline \hline
  \end{tabular}
}
  \caption{The optimal values of $\alpha_S$ and the input PDF parameters at $Q_0^2 = 1$ GeV$^2$ determined from the global analysis.  The one-sigma errors are calculated using \eqref{eq:Fp} and \eqref{eq:Fm} using the 68\% C.L.~tolerance discussed in Section~\ref{sec:erroranalysis}, and are shown only for the 20 parameters allowed to go free when determining the eigenvector PDF sets.  The parameters $A_u$, $A_d$, $A_g$ and $x_0$ are determined from sum rules and are not fitted parameters.  Similarly, $A_\Delta$ is determined from $\int_0^1\!{\rm d}x\;\Delta(x,Q_0^2)$.  The three parameters $r_i$, defined in \eqref{eq:nucmod}, are associated with the nuclear corrections to the neutrino data; see Section \ref{sec:neutrinoinclusive}.  The parameter values are given to five significant figures solely for accuracy in the case of reproduction of the PDFs.}
  \label{tab:parameters}
\end{table}

Notice that the LO fit gives a significantly larger $\alpha_S(M_Z^2)$ than the world-average value of $0.1176 \pm 0.0020$~\cite{Amsler:2008zz}, and that the parametrisation of the input gluon is simpler since the fit never shows any tendency for the gluon to go negative, or even to turn over at small $x$.  Moreover, the LO fit gives a much worse description of the data than that obtained at NLO and NNLO.  It fails for HERA structure functions because the evolution is too slow on account of missing higher-order small-$x$ enhanced terms in the quark--gluon splitting function.  It fails for some fixed-target data due to the absence of higher-order large-$x$ enhanced terms in the coefficient functions.  For the HERA jets it is missing large NLO corrections to the cross sections.  For Tevatron jets we find that the LO evolution cannot give, in detail, the correct shape to fit the data, contrary to previous findings for the less precise Run I data which were fit very well~\cite{Martin:2002dr}.  It is also missing large NLO corrections to the cross section for the Drell--Yan data, and as before we have applied a $K$-factor (previously fixed to 1.3), but have improved this to $(1+\alpha_S(M^2)\,C_F\,\pi/2)$, which improves the fit quality considerably.  The deficiencies due to higher-order terms cannot be completely mimicked by the very large coupling $\alpha_S$ (although this works quite well at high $x$) and a large small-$x$ gluon, even when the normalisations of many data sets choose to lie rather low.  So, in conclusion, the LO fit is unsatisfactory.  On the other hand, if required, it is possible to obtain a modified LO PDF set which gives a much improved description of the data by relaxing the momentum conservation constraint and using the NLO coupling $\alpha_S$~\cite{Sherstnev:2007nd}, and this can be even further improved by a modification of the scale of the coupling~\cite{Sherstnev:2008dm}.

The quality of the fit at NLO is much better, with no obviously badly fit data sets (except to some extent the D{\O} $W\to\ell\nu$ charge asymmetry, which will be discussed later).  Again the normalisations tend to lie a little low, though there are systematic differences between data sets, e.g.~the SLAC DIS data are normalised about $1.5\%$ higher than NMC, which is $3\%$ higher than BCDMS, and in general H1 sets are $1\%$ or so higher in normalisation than ZEUS.  These trends have long been known.  We also see that the fitted normalisation of the D{\O} Run II jet data is about $8\%$ higher than that for CDF Run II data (in the latter case the normalisation of the inclusive jet data is tied to the normalisation of the $Z$ rapidity distribution data), but the luminosity uncertainties in each are compatible with this.  The fit quality at NNLO is almost identical to that at NLO, and is rather similar for most data sets.  However, it is striking that the normalisations are clustered about unity much more symmetrically in this case, with those for the E866/NuSea Drell--Yan cross sections and the D{\O} Run II jet data more than one-sigma above unity.  At NLO and NNLO the values of $\alpha_S(M_Z^2)$ of $0.1202$ and $0.1171$ are both consistent with the world average value of $0.1176 \pm 0.0020$~\cite{Amsler:2008zz}, though the NLO value is just a little high, suggesting a partial compensation for missing higher orders.  A detailed study of the uncertainty in $\alpha_S$ awaits a further publication, but previous estimates of an uncertainty on our determination of $\alpha_S(M_Z^2)$ of $\pm 0.002$ from data and $\pm 0.003$ from theory are unlikely to change very significantly.

The errors that are shown on the input parameters in Table \ref{tab:parameters} are calculated with \eqref{eq:Fp} and \eqref{eq:Fm} using the 68\% C.L.~tolerance discussed in Section~\ref{sec:erroranalysis}, and are given only for the 20 parameters allowed to go free when determining the eigenvector PDF sets.  It is important to note that although they give some idea of PDF uncertainties they are only quantitatively meaningful when the information on the correlations between parameters is included, and this is most easily presented and utilised in terms of orthogonal eigenvectors.  The eigenvector PDF sets are defined and discussed in Section \ref{sec:hessian}.  Indeed, in Section \ref{sec:erroranalysis} we describe, in detail, the procedure that we use to determine the uncertainty on the parton distributions and on the predictions of cross sections.  We then turn to the more detailed description of the various data sets that are obtained in the global fits in Sections \ref{sec:inclusivedata}--\ref{sec:jetdata}.

%% file: erroranalysis.tex
\section{Error propagation from experimental data points} \label{sec:erroranalysis}

Uncertainties in global PDF analyses can be divided into two general categories, which may loosely be called ``theoretical'' and ``experimental''.  The \emph{theoretical} errors include those associated with the choice of the form of the input parameterisation, the neglected higher-order QCD (including enhanced $\ln(1/x)$ or $\ln(1-x)$ terms) and electroweak corrections, parton recombination and other higher-twist corrections, the choice of data sets and kinematic cuts, the choice of nuclear corrections for the neutrino-initiated data and the treatment of heavy flavours.  These uncertainties are often difficult to quantify {\it a priori} until a better calculation or prescription becomes available, but see the attempt made in Ref.~\cite{Martin:2003sk}.

On the other hand, in principle there are well-defined procedures for propagating \emph{experimental} uncertainties on the fitted data points through to the PDF uncertainties.  Three main methods have been used to do this.
\begin{enumerate}
\item The \emph{Lagrange multiplier} method~\cite{Pumplin:2000vx,Stump:2001gu}, which does not rely on linear error propagation, but requires the ability to perform a global fit.
\item The \emph{Hessian} method~\cite{Pumplin:2001ct}, which is based on linear error propagation and involves the production of eigenvector PDF sets suitable for convenient use by the end user.
\item The use of \emph{Monte Carlo sampling}~\cite{Giele:1998gw,Giele:2001mr}, which has recently been used in conjunction with neural networks to determine NLO PDFs from a DIS-only fit~\cite{Ball:2008by}.
\end{enumerate}
The first two methods were originally used by CTEQ~\cite{Pumplin:2000vx,Pumplin:2001ct,Stump:2001gu} and then by MRST~\cite{Martin:2002aw}.  In principle, the Lagrange multiplier method is superior to the Hessian approach, but it suffers from the enormous practical disadvantage that a series of new global fits has to be done every time one considers a new quantity.  Fortunately, it turns out that for those quantities that have been considered by both methods, the uncertainties have been found to be comparable.  We therefore concentrate here on the most commonly used Hessian method.  We will not discuss the use of Monte Carlo sampling in this paper.

\subsection{Review of Hessian method in global PDF analyses} \label{sec:hessian}
The basic procedure is discussed in detail in Refs.~\cite{Pumplin:2001ct,Martin:2002aw}.  Here, we briefly review the important points.  The traditional propagation of experimental uncertainties assumes that the global goodness-of-fit quantity, $\chi^2_{\rm global}$, is quadratic about the global minimum, which has parameters $\{a_1^0,\ldots,a_n^0\}$.  In this case we can write
\begin{equation} \label{eq:hessian}
  \Delta\chi^2_{\rm global} \equiv \chi^2_{\rm global} - \chi_{\rm min}^2 = \sum_{i,j=1}^n H_{ij}(a_i-a_i^0)(a_j-a_j^0),
\end{equation}
where the Hessian matrix $H$ has components
\begin{equation}
  H_{ij} = \left.\frac{1}{2}\frac{\partial^2\,\chi^2_{\rm global}}{\partial a_i\partial a_j}\right|_{\rm min}.
\end{equation}
The uncertainty on a quantity $F(\{a_i\})$ is then obtained from linear error propagation:
\begin{equation} \label{eq:heserror}
  \Delta F = T \sqrt{\sum_{i,j=1}^n\frac{\partial F}{\partial a_i}C_{ij}\frac{\partial F}{\partial a_j}},
\end{equation}
where $C\equiv H^{-1}$ is the covariance matrix, also known as the error matrix, and $T = (\Delta\chi^2_{\rm global})^{1/2}$ is the tolerance for the required confidence interval.  This formula \eqref{eq:heserror} has the disadvantage that PDF uncertainties are not readily calculable for general observables, since the derivative of the observable $F$ with respect to each parameter $a_i$ is needed.

It is convenient to diagonalise the covariance (or Hessian) matrix~\cite{Pumplin:2001ct,Martin:2002aw}, and work in terms of the eigenvectors and eigenvalues.  Since the covariance matrix is symmetric it has a set of {\it orthonormal} eigenvectors $\vec{v}_k$ defined by
\begin{equation} \label{eq:eigeq}
  \sum_{j=1}^n C_{ij} v_{jk} = \lambda_k v_{ik},
\end{equation}
where $\lambda_k$ is the $k$th eigenvalue and $v_{ik}$ is the $i$th component of the $k$th orthonormal eigenvector ($k = 1,\ldots,n$).  The parameter displacements from the global minimum can be expanded in a basis of rescaled eigenvectors $e_{ik}\equiv \sqrt{\lambda_k}v_{ik}$, that is,
\begin{equation} \label{eq:component}
  a_i - a_i^0 = \sum_{k=1}^n e_{ik} z_k.
\end{equation}
Then it can be shown, using the orthonormality of $\vec{v}_k$, that \eqref{eq:hessian} reduces to
\begin{equation} \label{eq:hessiandiag}
  \chi^2_{\rm global} = \chi^2_{\rm min} + \sum_{k=1}^n z_k^2,
\end{equation}
that is, $\sum_{k=1}^n z_k^2\le T^2$ is the interior of a hypersphere of radius $T$.  Pairs of eigenvector PDF sets $S_k^\pm$ can then be produced to span this hypersphere, at a fixed value of $\alpha_S$, with parameters given by
\begin{equation} \label{eq:eigenstept}
  a_i(S_k^\pm) = a_i^0 \pm t\,e_{ik},
\end{equation}
with $t$ adjusted to give the desired $T = (\Delta\chi^2_{\rm global})^{1/2}$.  In the quadratic approximation, $t=T$.  For the larger eigenvalues $\lambda_k$, where there are significant deviations from the ideal quadratic behaviour, $t$ is adjusted iteratively to give the desired value of $T$.  Then uncertainties on a quantity $F$, which may be an individual PDF at particular values of $x$ and $Q^2$, or a derived quantity such as a cross section, can be calculated with\footnote{It can be shown that \eqref{eq:heserror} reduces to \eqref{eq:symmunc} in the quadratic approximation ($t=T$)~\cite{Pumplin:2001ct,Martin:2002aw}, but we treat \eqref{eq:symmunc} as the fundamental definition in the departure of this ideal limit.  In this paper, we will generally use \eqref{eq:Fp} and \eqref{eq:Fm} to calculate asymmetric PDF uncertainties.}
\begin{equation} \label{eq:symmunc}
  \Delta F = \frac{1}{2}\sqrt{\sum_{k=1}^n \left[F(S_k^+)-F(S_k^-)\right]^2},
\end{equation}
or asymmetric errors can be calculated with
\begin{align}
  (\Delta F)_+ &= \sqrt{\sum_{k=1}^n \left\{{\rm max}\left[\;F(S_k^+)-F(S_0),\;F(S_k^-)-F(S_0),\;0\right]\right\}^2}, \label{eq:Fp} \\
  (\Delta F)_- &= \sqrt{\sum_{k=1}^n \left\{{\rm max}\left[\;F(S_0)-F(S_k^+),\;F(S_0)-F(S_k^-),\;0\right]\right\}^2}, \label{eq:Fm}
\end{align}
where $S_0$ is the central PDF set.  Correlations between two quantities can also be calculated; see, for example, Ref.~\cite{Nadolsky:2008zw}.  Defining a correlation cosine between two quantities $F$ and $G$,
\begin{equation}
  \cos\phi_{FG} = \frac{1}{4\,\Delta F\,\Delta G}\,\sum_{k=1}^n\,\left[F(S_k^+)-F(S_k^-)\right]\,\left[G(S_k^+)-G(S_k^-)\right],
\end{equation}
where the uncertainties $\Delta F$ and $\Delta G$ are calculated using \eqref{eq:symmunc}, then values of $\cos\phi_{FG}\approx 1$ mean that $F$ and $G$ are correlated, values of $\approx -1$ mean that they are anticorrelated, while values of $\approx 0$ mean that they are uncorrelated.  A tolerance ellipse in the $F$--$G$ plane can then be defined by the two parametric equations:
\begin{align}
  F &= F(S_0) + \Delta F\,\cos\theta, \label{eq:ellipseF} \\
  G &= G(S_0) + \Delta G\,\cos(\theta+\phi_{FG}), \label{eq:ellipseG}
\end{align}
where $\theta\in [0,2\pi]$.  We will show examples in Section~\ref{sec:totalpredictions}.  Note that using e.g.~90\% C.L.~PDFs results in a probability less than 90\% for the 2-D tolerance ellipse~\cite{Nadolsky:2008zw,Amsler:2008zz,NumericalRecipes}.

To determine the ``best-fit'' parameters we allow all the input PDF parameters of Eqs.~\eqref{eq:uv}--\eqref{eq:sv} to vary.  However, when investigating in detail the small departures from the global minimum we notice a certain amount of redundancy in parameters.  A striking example is for the NLO and NNLO parameterisations of the gluon distribution \eqref{eq:inputxg}.  Small changes in the values of three of the parameters can be compensated almost exactly by changes in the remaining four.  This high degree of correlation between parameters means that very small changes in $\chi^2$ will be obtained.  However, at some point the compensation starts to fail significantly and the $\chi^2$ increases dramatically. Hence, the redundancy leads to a severe breaking of the quadratic behaviour in $\Delta\chi^2$, and some very flat directions in the eigenvector space (that is, very large eigenvalues of the covariance matrix) and cubic, quartic etc.~terms dominate.  During the process of diagonalisation this bad behaviour feeds through into the whole set of eigenvectors to some extent.  Therefore, in order that the Hessian method works at all well we have to lessen the redundancy in the input parameters.  In order to do this we simply fix some of the parameters at their best-fit values, so that the Hessian matrix only depends on a subset of parameters that are sufficiently independent that the quadratic approximation is reasonable.  We finish up with 20 reasonably well-behaved eigenvectors in total, i.e.~those coming from the combinations of the 20 parameters that are assigned errors in Table \ref{tab:parameters}.  However, we emphasise that the other parameters are fixed at their best-fit values, rather than simply set to zero.  The problem of a certain amount of redundancy in parameters is a general feature of the full global fits obtained by CTEQ and MRST which have sufficient parametric freedom to account for all features of the data.

\begin{figure}
  \centering
  \includegraphics[width=0.8\textwidth]{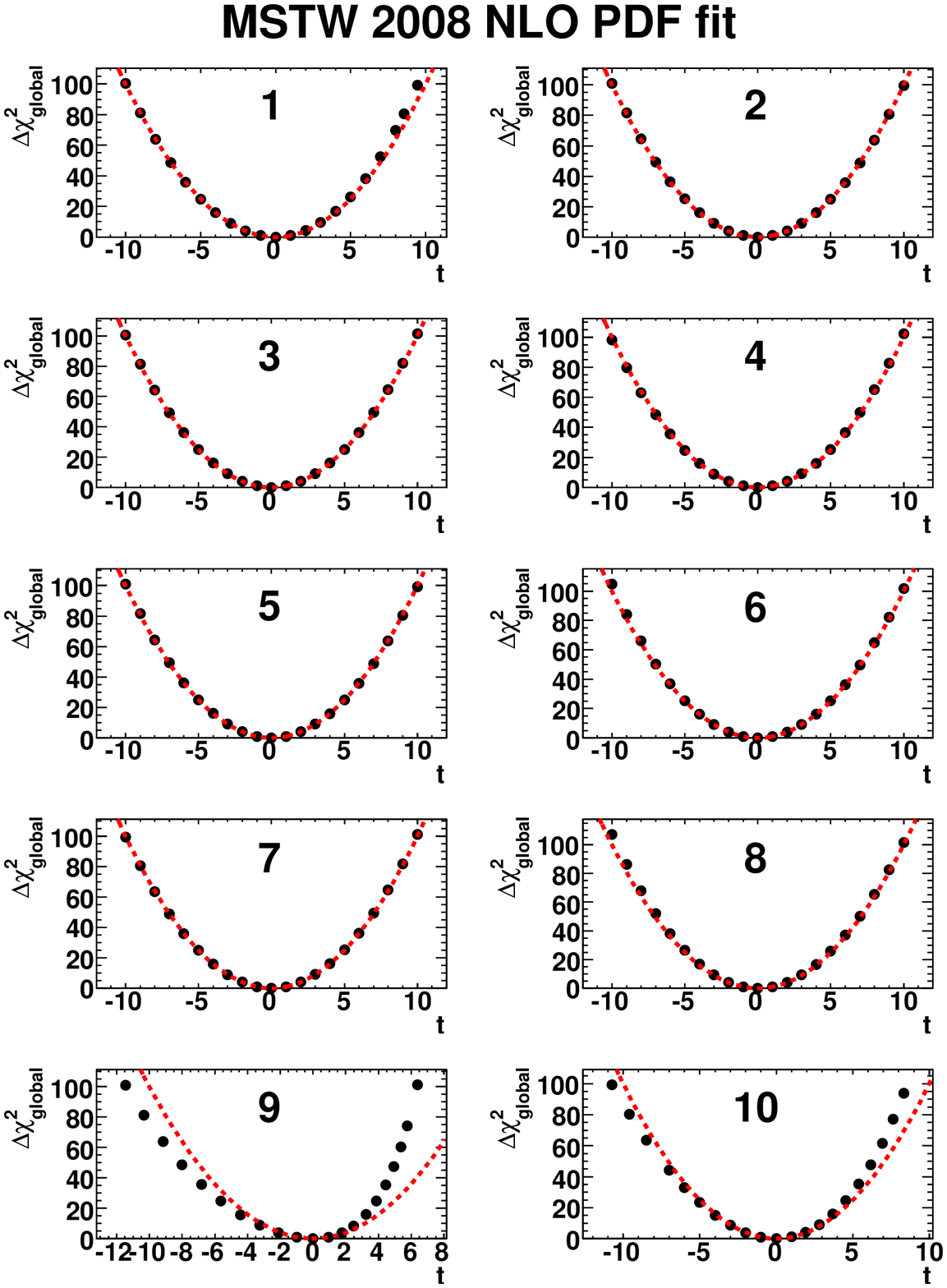}
  \caption{The points ($\bullet$) show $\Delta\chi^2_{\rm global}$ as a function of the distance along each eigenvector direction, $t$, defined in \eqref{eq:eigenstept}, for eigenvectors numbered 1--10 corresponding to the 10 smallest eigenvalues.  The dashed curve is the ideal case, $\Delta\chi^2_{\rm global} = t^2$.}
  \label{fig:globalchisq1}
\end{figure}
\begin{figure}
  \centering
  \includegraphics[width=0.8\textwidth]{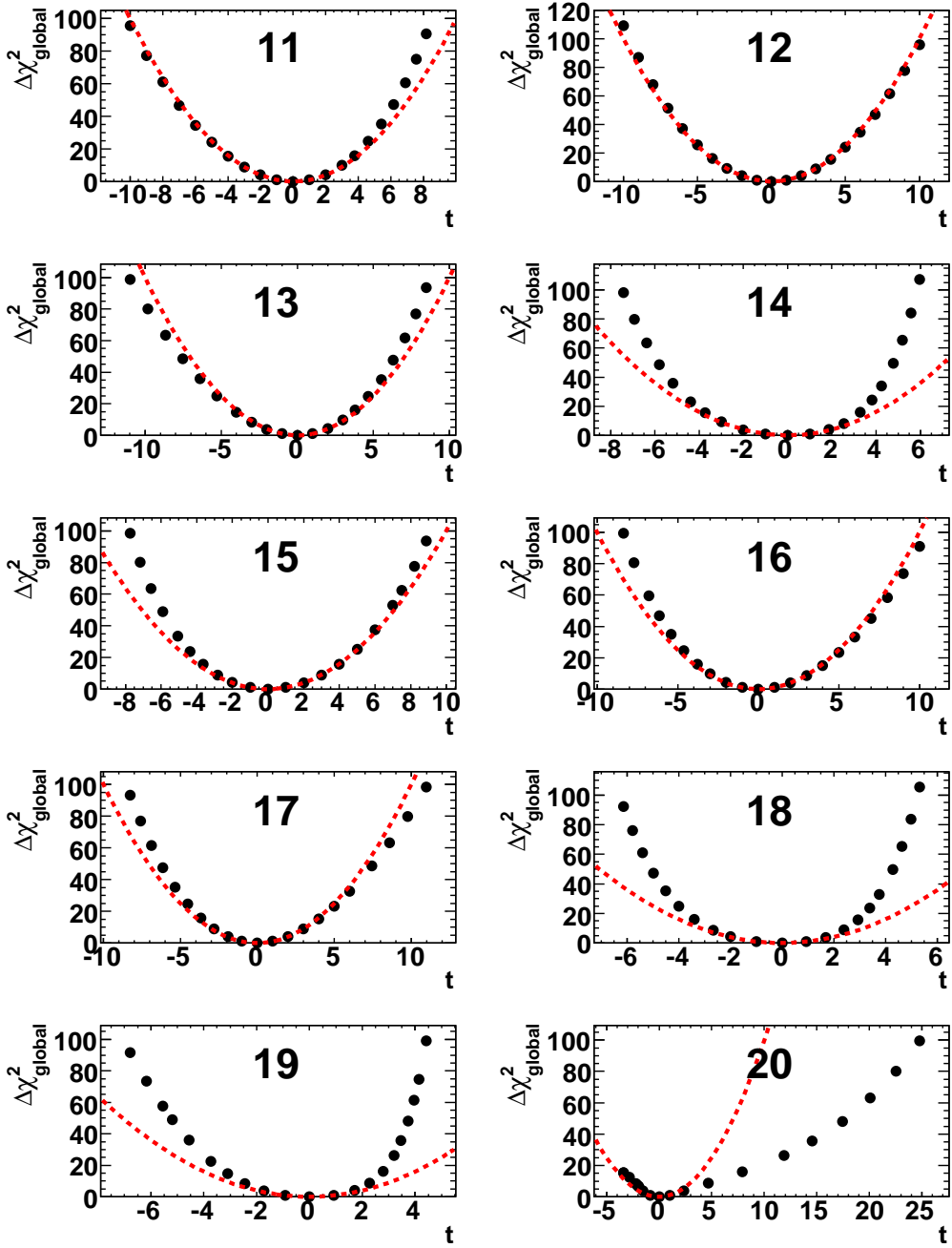}
  \caption{The points ($\bullet$) show $\Delta\chi^2_{\rm global}$ as a function of the distance along each eigenvector direction, $t$, defined in \eqref{eq:eigenstept}, for eigenvectors numbered 11--20 corresponding to the 10 largest eigenvalues.  The dashed curve is the ideal case, $\Delta\chi^2_{\rm global} = t^2$.}
  \label{fig:globalchisq2}
\end{figure}
The behaviour of $\Delta\chi^2_{\rm global}$ for the 20 eigenvectors in the current NLO fit is shown in Figs.~\ref{fig:globalchisq1} and \ref{fig:globalchisq2}.  For the lowest 13 or so eigenvalues the quadratic approximation is very good.  Above this we see some departure from quadratic behaviour, though there is still good agreement with the ideal case for relatively low $t$.  The peculiar behaviour for eigenvector 20 is related to the fact that this is mainly concerned with the high-$x$ part of $s_v=s-\bar{s}$ at input, and this will be discussed in Section \ref{sec:inputparamunc}.

\subsection{Dynamic determination of tolerance} \label{sec:tolerance}

Ideally, with the standard ``parameter-fitting'' criterion, we would expect the errors to be given by the choice of tolerance $T = (\Delta\chi^2_{\rm global})^{1/2}=1$ for the $68\%$ (one-sigma) confidence level (C.L.) limit\footnote{To be precise, rather than using exactly 68\% we use ${\rm erf}(1/\sqrt{2})=0.682689$.}, or $T^2=2.71$ for the $90\%$ C.L.~limit~\cite{NumericalRecipes,Amsler:2008zz}.  This is appropriate if fitting consistent data sets with ideal Gaussian errors to a well-defined theory.  However, in practice, there are some inconsistencies between the independent fitted data sets, and unknown experimental and theoretical uncertainties, so the parameter-fitting criterion is not appropriate for global PDF analyses.  One possible approach is to attempt to only include completely consistent data sets.  This was tried in Ref.~\cite{Giele:2001mr}, but it was found that the ZEUS and NMC data were inconsistent with H1, BCDMS and E665, so only the latter three data sets were used in their fit.  Clearly, this does not seem like a practical approach to follow.

So how can we choose the value of the tolerance $T$ so as to obtain more reliable errors?  Instead, we can appeal to the much weaker ``hypothesis-testing'' criterion, where the eigenvector PDF sets are treated as alternative hypotheses.  Very roughly, a fit is judged to be ``good'' if each data set $n$, consisting of $N$ data points, has $\chi_n^2\simeq N\pm\sqrt{2N}$~\cite{Collins:2001es}.  More precisely, ranges of $\chi_n^2$ corresponding to a 90\% C.L.~limit, for example, can be calculated, then the value of the tolerance $T = (\Delta\chi^2_{\rm global})^{1/2}$ can be chosen to ensure that each data set is described within its 90\% C.L.~limit.  Assuming that $\chi_n^2$ follows the $\chi^2$-distribution with $N$ degrees of freedom, which has a probability density function
\begin{equation} \label{eq:chi2pdf}
  P_N(\chi^2) = \frac{(\chi^2)^{N/2-1}\,\mathrm{e}^{-\chi^2/2}}{2^{N/2}\,\Gamma(N/2)},
\end{equation}
then the 90th percentile, $\xi_{90}$, is obtained by solving
\begin{equation}
  \int_0^{\xi_{90}}\!{\rm d}\chi^2\;P_N(\chi^2) = 0.90.
\end{equation}
Similarly for the 68th percentile, $\xi_{68}$, and for the most probable value, $\xi_{50}\simeq N$.
\begin{figure}
  \centering
  \includegraphics[width=0.6\textwidth]{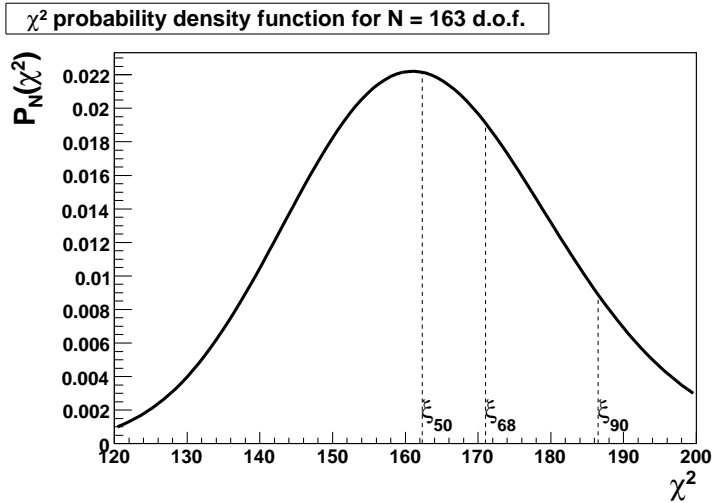}
  \caption{Probability density function of the $\chi^2$-distribution with 163 degrees of freedom.}
  \label{fig:plotchisqdistribution}
\end{figure}
In Fig.~\ref{fig:plotchisqdistribution} we show \eqref{eq:chi2pdf} for $N=163$ corresponding to a typical data set included in the global fit (BCDMS $\mu p$ $F_2$).  We have $\xi_{50} = 162.3$, $\xi_{68}-\xi_{50}=8.7$ and $\xi_{90}-\xi_{50} = 24.2$, cf.~$\sqrt{2N} = 18.1$.

We now describe the new procedure we adopt in determining the tolerance for each eigenvector direction: it follows closely, but significantly extends, similar studies carried out earlier by the CTEQ group~\cite{Stump:2001gu,Pumplin:2001ct,Pumplin:2002vw}.  Since the new procedure is quite complicated, we will break the explanation down into steps, supplemented by example plots from the current NLO analysis.
\begin{figure}
  \centering
  \includegraphics[width=0.8\textwidth]{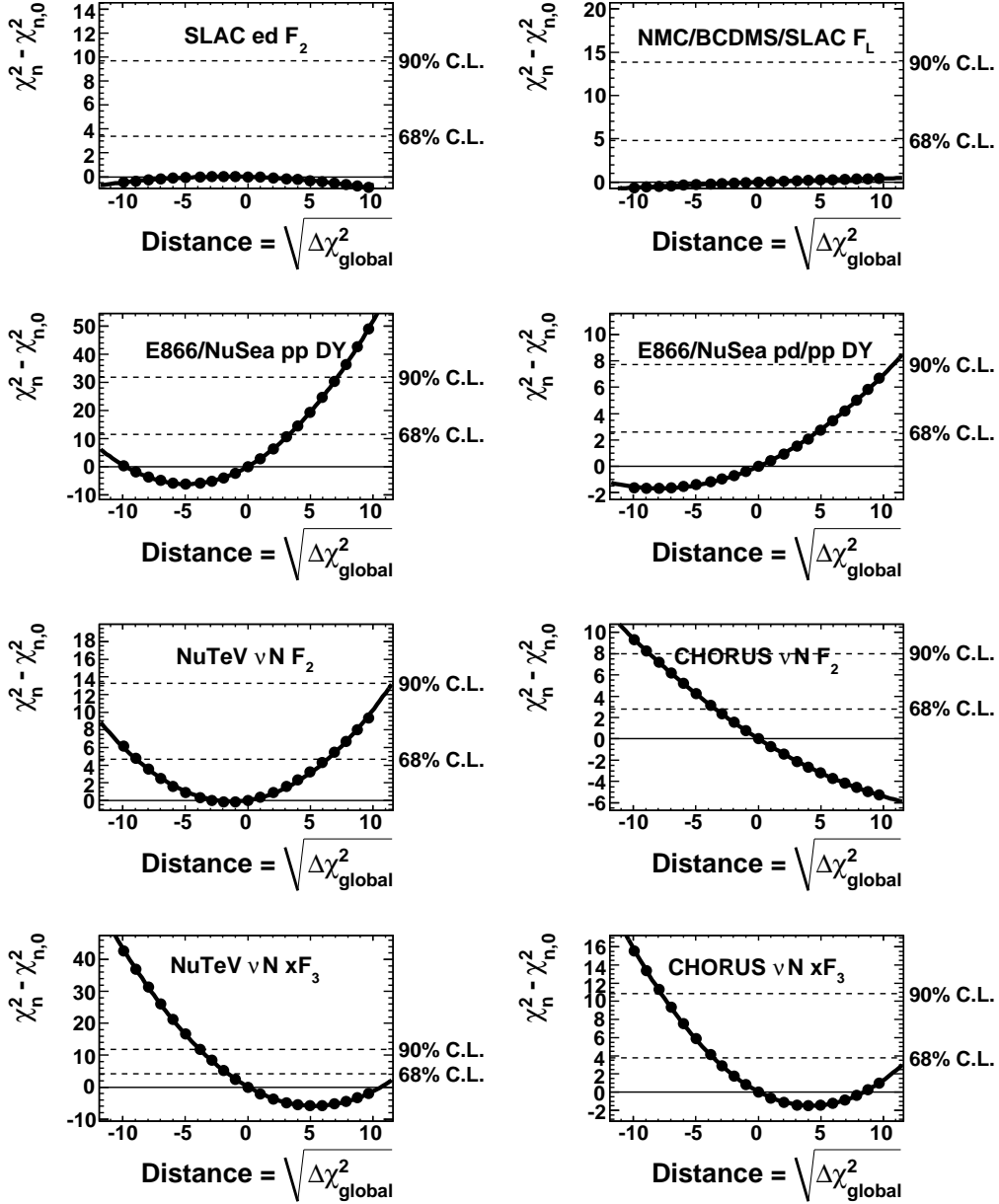}
  \caption{Change in $\chi_n^2$ for a subset of data sets $n$, compared to the values at the global minimum ($\chi_{n,0}^2$), when moving along eigenvector number 13 in units of $(\Delta\chi_{\rm global}^2)^{1/2}$.  The 90\% and 68\% C.L.~regions determined according to \eqref{eq:90percentCL} are indicated.}
  \label{fig:scanchisq13}
\end{figure}
\begin{enumerate}
  \renewcommand{\labelenumi}{(\roman{enumi})}
\item First we plot the change in $\chi^2_n$ for each data set, as the difference from the value at the \emph{global} minimum, $\chi^2_{n,0}$, when moving along a particular eigenvector direction in units of $(\Delta\chi^2_{\rm global})^{1/2}$.  An example for a subset of the most constraining data sets for eigenvector number 13 is shown in Fig.~\ref{fig:scanchisq13}.  The points ($\bullet$) are generated for fixed values of the distance, $(\Delta\chi^2_{\rm global})^{1/2}$, between 0 and 10 in each ``$+$'' or ``$-$'' eigenvector direction.  Note that we adjust $t$ in \eqref{eq:eigenstept} until the desired $(\Delta\chi^2_{\rm global})^{1/2}$ is obtained: these two quantities are equal only in the strict quadratic approximation.  These points ($\bullet$) are then fitted to a quadratic\footnote{The one exception is for the very asymmetric $\chi^2_n$ profile for the NuTeV dimuon data plotted against the distance along eigenvector number 20, associated with the $\eta^-$ parameter controlling the high-$x$ behaviour of $s_v=s-\bar{s}$, which we instead fit to a quartic function.} function shown by the solid lines.  Note from Fig.~\ref{fig:scanchisq13} that the minimum value of $\chi^2_n$ for a particular data set does not in general occur at the global minimum, indicating some tension between data sets included in the global fit, for example, between the E866/NuSea $pp$ Drell--Yan data and the NuTeV and CHORUS $\nu N$ $xF_3$ data.
\item We define the 90\% C.L.~region for each data set by the condition that
  \begin{equation} \label{eq:90percentCL}
    \chi_n^2 < \left(\frac{\chi_{n,0}^2}{\xi_{50}}\right)\xi_{90},
  \end{equation}
and similarly for the 68\% C.L.~region.  Note that we define the $90\%$ C.L.~region \eqref{eq:90percentCL} for $\chi_n^2$ by renormalising $\xi_{90}$ by a factor $\chi_{n,0}^2/\xi_{50}$~\cite{Stump:2001gu}.  This procedure is necessary to take account of the fact that the value of $\chi_{n,0}^2$ at the global minimum may be quite far from the most probable value of $\xi_{50}$ of the particular data set $n$.  For the example of the BCDMS $\mu p$ $F_2$ data, the value of the $\chi_n^2$ in our NLO fit is 182.2, which would lie outside the 68\% C.L.~region shown in Fig.~\ref{fig:plotchisqdistribution} if the absolute values of $\chi_n^2$ were used instead.  After applying the rescaling factor of $\chi_{n,0}^2/\xi_{50}=1.12$, the 90\% C.L.~region is given by $\chi_n^2-\chi_{n,0}^2<27.2$ and the 68\% C.L.~region by $\chi_n^2-\chi_{n,0}^2<9.8$.  The 90\% and 68\% C.L.~regions determined in this way are shown as the horizontal dashed lines in Fig.~\ref{fig:scanchisq13}.
\item For each eigenvector, we then record the value of the distance, $(\Delta\chi^2_{\rm global})^{1/2}$, for which the $\chi_n^2$ for each data set is minimised, together with the 90\% and 68\% C.L.~limits defined by the intercepts of the quadratic curves with the horizontal dashed lines in Fig.~\ref{fig:scanchisq13}.  These values are shown for eigenvectors 9 and 13 in Fig.~\ref{fig:evectors}, where the points ($\bullet$) indicate the values of the distance, $(\Delta\chi^2_{\rm global})^{1/2}$, for which $\chi_n^2$ is minimised, while the inner error bars extend across the 68\% C.L.~region and the outer error bars extend across the 90\% C.L.~region defined by \eqref{eq:90percentCL}.
\begin{figure}
  (a)\\
  \begin{minipage}{\textwidth}
    \centering
    \includegraphics[width=0.76\textwidth]{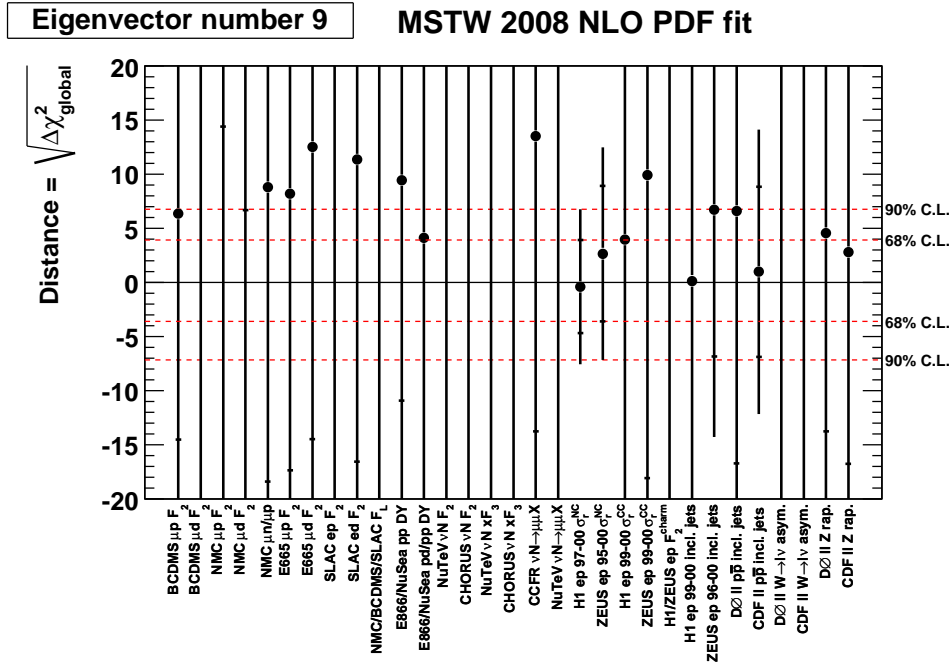}
  \end{minipage}
  (b)\\
  \begin{minipage}{\textwidth}
    \centering
    \includegraphics[width=0.76\textwidth]{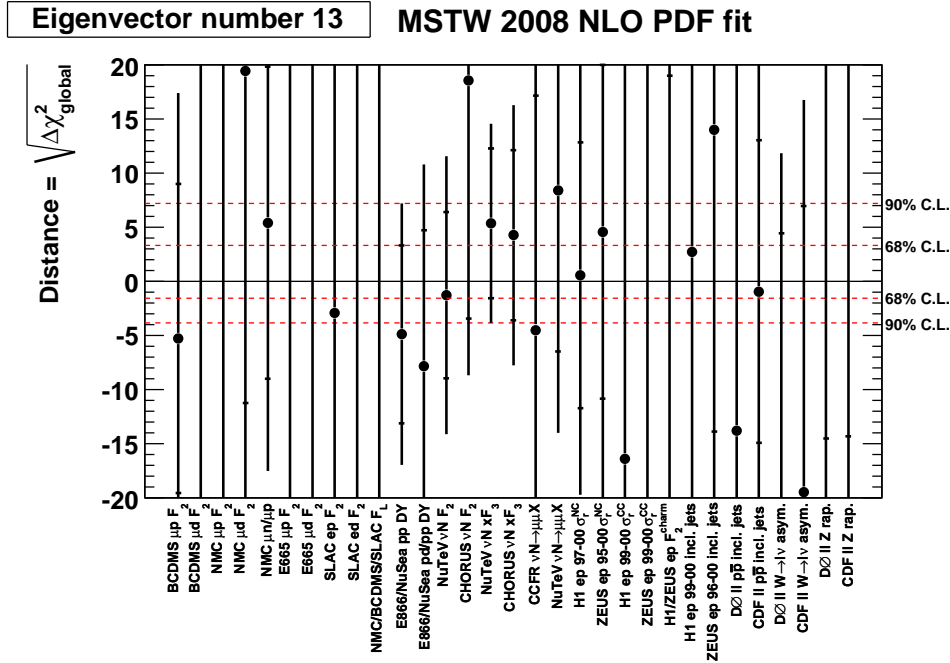}
  \end{minipage}
  \caption{Ranges of $(\Delta\chi^2_{\rm global})^{1/2}$ along (a) eigenvector 9 and (b) eigenvector 13 for which data sets are satisfied within their $90\%$ C.L.~limit (outer error bars) or $68\%$ C.L.~limit (inner error bars).  The points ($\bullet$) indicate the minimum with respect to each particular data set.  The tolerance, indicated by the horizontal dashed lines, is chosen to ensure that all data sets are described within their $68\%$ or $90\%$ C.L.~limits defined by \eqref{eq:90percentCL}.}
  \label{fig:evectors}
\end{figure}
\item For each of the eigenvectors, we choose the values of the tolerance $T=(\Delta\chi^2_{\rm global})^{1/2}$, indicated by the horizontal dashed lines in Fig.~\ref{fig:evectors}, so that all data sets are described within their 90\% or 68\% C.L.~regions.  For eigenvector 9 we see that the tolerance in the ``$+$'' direction is fixed by the H1 NC data and in the ``$-$'' direction by the ZEUS NC data, while for eigenvector 13 we see that the tolerance in the ``$+$'' direction is fixed by the E866/NuSea $pp$ Drell--Yan data and in the ``$-$'' direction by the NuTeV $\nu N$ $xF_3$ data.
\end{enumerate}

\begin{figure}[t]
  \includegraphics[width=\textwidth]{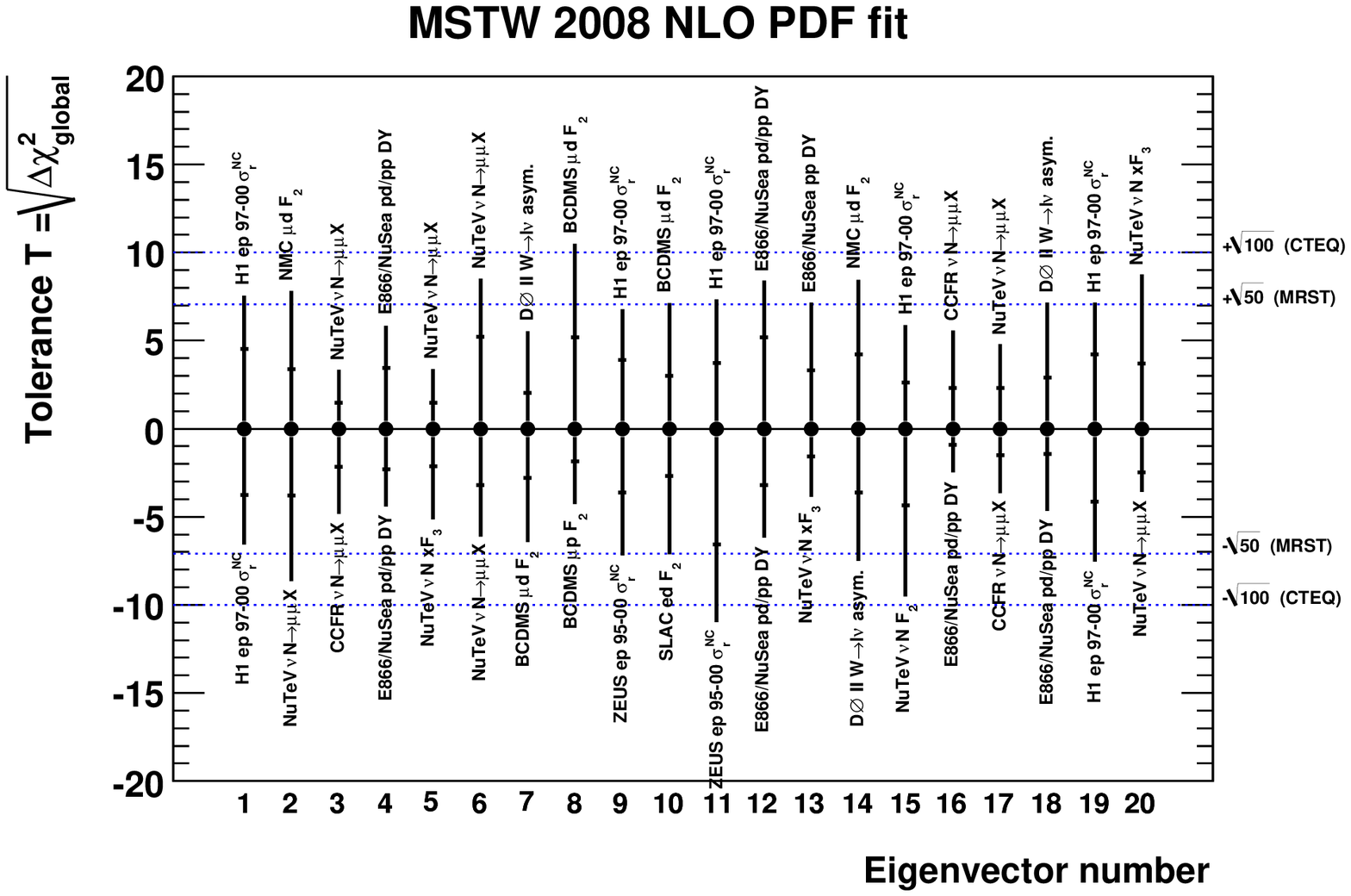}
  \caption{Tolerance for each eigenvector direction determined dynamically from the criteria that each data set must be described within its $90\%$ C.L.~(Eq.~\eqref{eq:90percentCL}) (outer error bars) or $68\%$ C.L.~limit (inner error bars).  The labels give the name of the data set which sets the $90\%$ C.L.~tolerance for each eigenvector direction.}
  \label{fig:nlotolerance}
\end{figure}
We summarise the values of the tolerance obtained by this procedure in Fig.~\ref{fig:nlotolerance} for each of the 20 eigenvectors of the NLO fit.  The inner error bars indicate the tolerance for a $68\%$ C.L.~limit while the outer error bars indicate the tolerance for a $90\%$ C.L.~limit.  The labels placed at the end of the error bars indicate the name of the data set which fixes the value of the tolerance for a $90\%$ C.L.~limit.  Note that the results shown for eigenvectors 9 and 13 correspond to the values that we extracted from the plots of Fig.~\ref{fig:evectors}.

The procedure just outlined is an extension of that previously employed by CTEQ and MRST.  These previous analyses used a fixed value of the tolerance $T = (\Delta\chi_{\rm global}^2)^{1/2}$ in producing eigenvector PDF sets, namely $T=\sqrt{100}$ for CTEQ and $T=\sqrt{50}$ for MRST.  Analogous plots to those shown in Fig.~\ref{fig:evectors} were originally presented by CTEQ in Appendix B.4 of Ref.~\cite{Pumplin:2002vw}.  Inspection of Fig.~\ref{fig:nlotolerance} shows that our new dynamic tolerance required to ensure that all data sets are described within their $90\%$ C.L.~limits are almost all in the region $T\sim \sqrt{50}$, which is close to the MRST value and suggests that the CTEQ tolerance ($T=\sqrt{100}$) is too large.\footnote{Note, however, that recent CTEQ analyses~\cite{Tung:2006tb,Nadolsky:2008zw} apply unspecified weight factors to the data from certain experiments in their $\chi_{\rm global}^2$ definition, which in practice achieves a similar effect to our dynamic tolerance method.}  However, an even smaller tolerance is obtained for some eigenvectors, in particular those associated with strange quarks and the $\bar{d}-\bar{u}$ difference, both of which are constrained by a small number of data sets generally each with a relatively small number of data points.

\subsection{Uncertainties on input PDFs} \label{sec:partonerrors}

We use the values of the \emph{dynamic} tolerance shown in Fig.~\ref{fig:nlotolerance} to generate the PDF eigenvector sets according to \eqref{eq:eigenstept}, which can be written as
\begin{equation} \label{eq:eigensteptdynamic}
  a_i(S_k^\pm) = a_i^0 \pm t_k^\pm\,e_{ik},
\end{equation}
where $t_k^\pm$ is adjusted to give $T_k^\pm$, with $T_k^\pm$ the values shown in Fig.~\ref{fig:nlotolerance}.  We provide two different sets for each fit corresponding to either a 90\% or 68\% C.L.~limit.  Note that the ratio of the PDF uncertainties calculated using these two sets is not simply an overall factor of $\sqrt{2.71} = 1.64$, as it would be if choosing the tolerance according to the usual parameter-fitting criterion.  Even in the simplest case, where the data set fixing the tolerance is the same for the 90\% and 68\% C.L.~limits, and assuming linear error propagation, then the ratio of the $T_k^\pm$ values would be $(\xi_{90}-\xi_{50})/(\xi_{68}-\xi_{50})$, which is a function of the number of data points $N$ in the data set which fixes the tolerance, and takes a value around $1.7$ for typical $N\sim 10$--$1000$.

\begin{figure}
  \centering
  \includegraphics[width=0.85\textwidth]{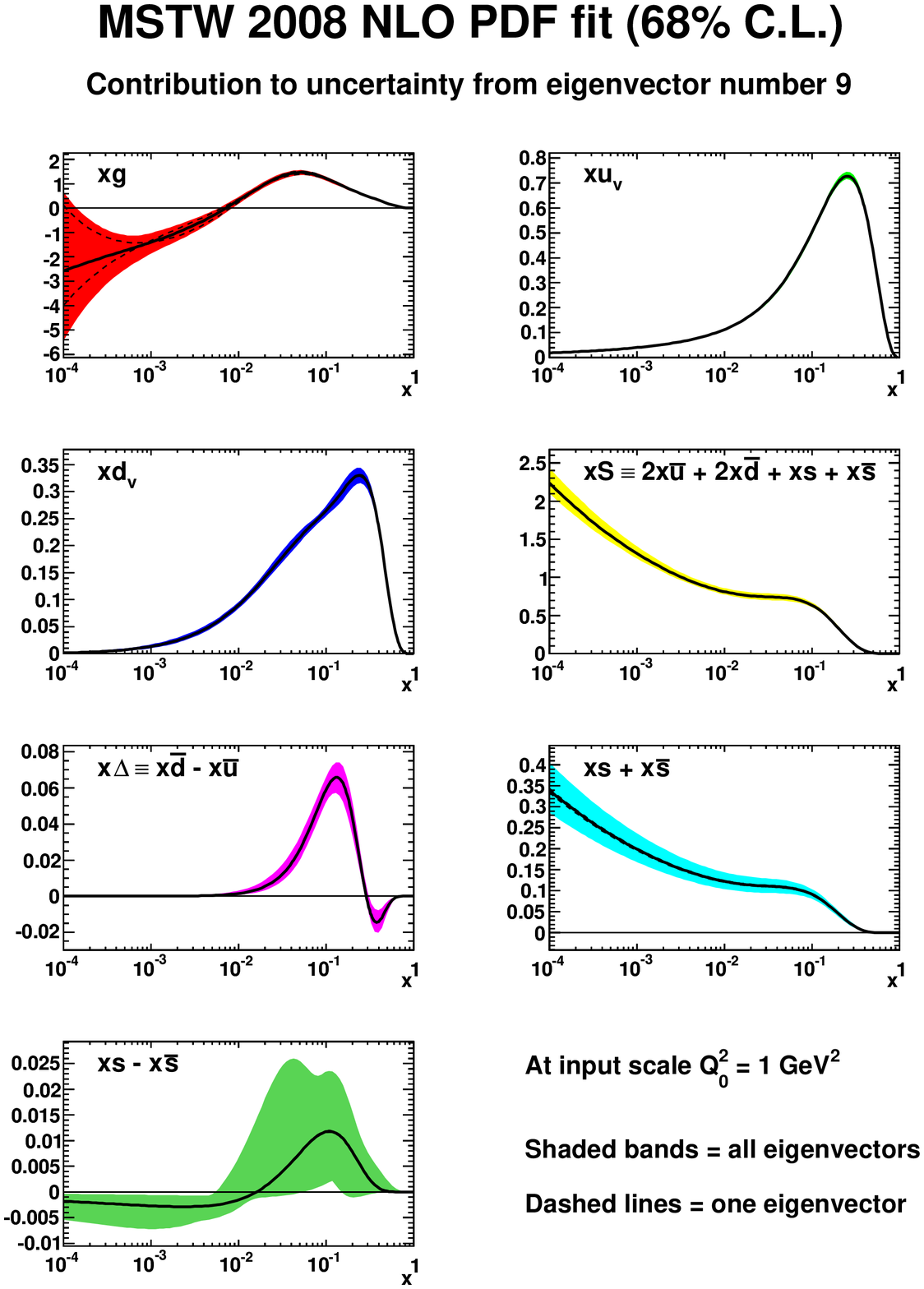}
  \caption{The total uncertainty on the input PDFs (shaded bands), given by \eqref{eq:Fp} and \eqref{eq:Fm}, and the contribution from $k=9$ in these equations (dashed lines).  For this eigenvector, the dashed lines are hidden under the central solid lines with the exception of $xg$ and $xs+x\bar{s}$.}
  \label{fig:absoluteinputuncertainty9}
\end{figure}
The total one-sigma uncertainty on the input parton distributions at $Q_0^2 = 1$ GeV$^2$, calculated using \eqref{eq:Fp} and \eqref{eq:Fm}, is shown as the shaded bands in Fig.~\ref{fig:absoluteinputuncertainty9}.  Note that the somewhat peculiar shape of the uncertainty on $s-\bar{s}$ is clearly due to the limited freedom in the input parameterisation: only two free parameters controlling the normalisation and the power of $(1-x)$.

\begin{figure}
  \centering
  \includegraphics[width=0.85\textwidth]{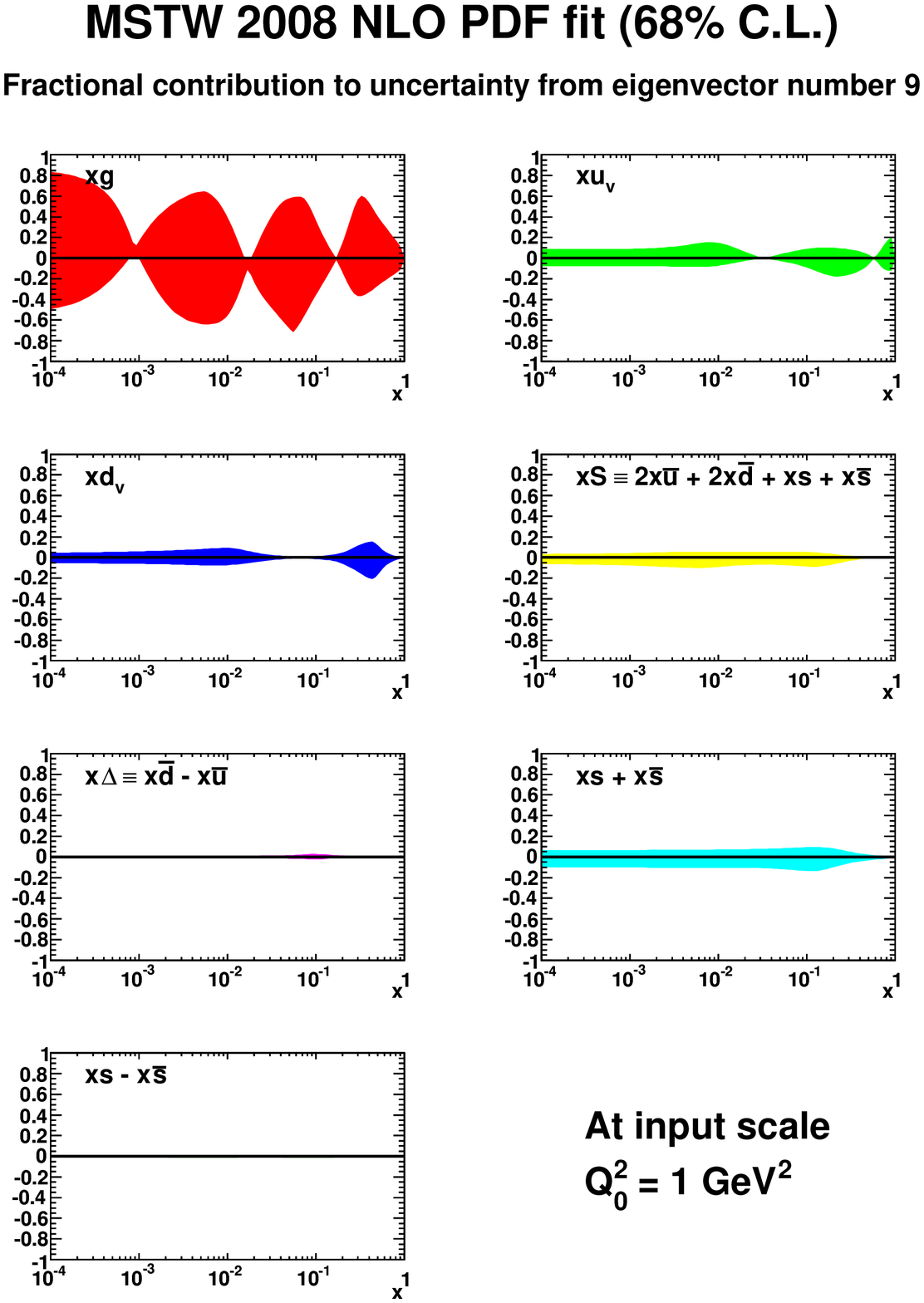}
  \caption{The fractional contribution to the uncertainty on the input PDFs from eigenvector 9, defined by the contributions to \eqref{eq:Fp} and \eqref{eq:Fm} from $k=9$, divided by the total uncertainties.}
  \label{fig:fractionalinputuncertainty9}
\end{figure}

\begin{figure}
  \centering
  \includegraphics[width=0.85\textwidth]{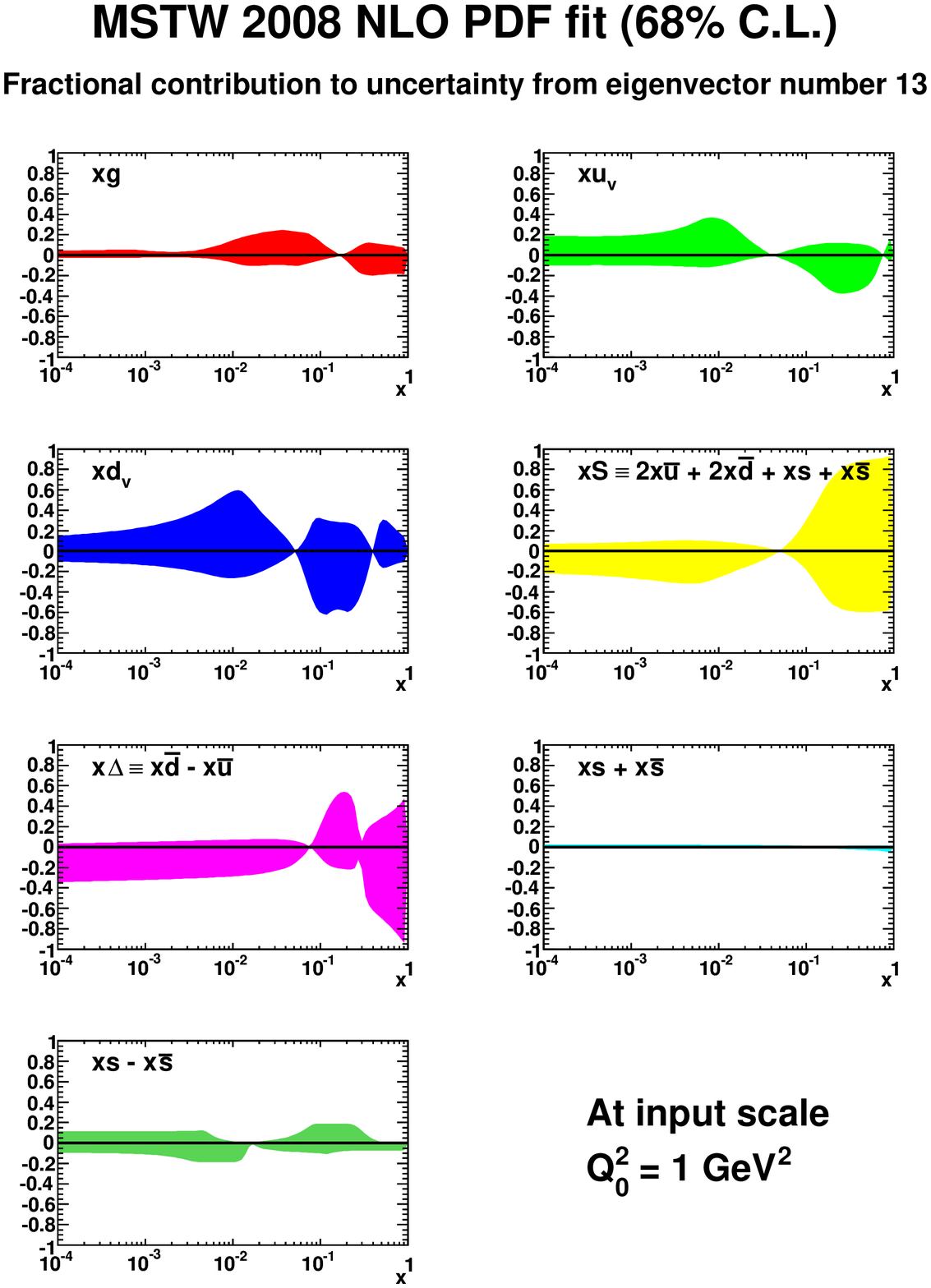}
  \caption{The fractional contribution to the uncertainty on the input PDFs from to eigenvector 13, defined by the contributions to \eqref{eq:Fp} and \eqref{eq:Fm} from $k=13$, divided by the total uncertainties.}
  \label{fig:fractionalinputuncertainty13}
\end{figure}

Recall that each eigenvector $\vec{v}_k$ is a linear combination of the input PDF parameter displacements $a_i-a_i^0$ and that the corresponding eigenvector PDF sets $S_k^\pm$ are responsible for a certain uncertainty on the parton distributions.  Sometimes an eigenvector corresponds to one particular input distribution, but often a single eigenvector contributes to the uncertainty on a variety of the input distributions.  The dashed lines in Fig.~\ref{fig:absoluteinputuncertainty9} show the contribution to the uncertainty from a single eigenvector (in this case, number 9), i.e.~only the $k=9$ term in \eqref{eq:Fp} and \eqref{eq:Fm}.  Clearly, eigenvector 9 mainly contributes to the uncertainty on the input gluon distribution.  This can be made clearer by plotting the fractional contribution to the uncertainty on the input PDFs from a single eigenvector, i.e.~the contributions to \eqref{eq:Fp} and \eqref{eq:Fm} from a single value of $k$ divided by the total uncertainties.  The fractional contribution to the uncertainty from eigenvector number 9 is shown in Fig.~\ref{fig:fractionalinputuncertainty9} and that from eigenvector number 13 in Fig.~\ref{fig:fractionalinputuncertainty13}.

\begin{table}[t]
  \centering
  \begin{tabular}{c|c|c|c|c|c|c|c}
    \hline \hline
    Eigenvector & $g$ & $u_v$ & $d_v$ & $S({\rm ea})$ & $\bar{d}-\bar{u}$ & $s+\bar{s}$ & $s-\bar{s}$ \\ \hline
     1 & 5 2 1 & --- & --- & --- & --- & --- & --- \\
     2 & 1 0 0 &3 3 3& --- & 1 1 1 & --- & 2 3 3 & --- \\
     3 & 0 1 0 &1 2 2& --- & --- & --- & 4 4 5 & --- \\
     4 & --- & 0 1 1 & --- & 0 1 1 & 1 5 {\bf 6} & 2 2 3 & --- \\
     5 & 1 1 1 & 1 1 2 & --- & 2 2 2 & 0 0 1 & 1 1 2 & 4 4 4 \\
     6 & 3 4 3 & 3 3 5 & --- & {\bf 7 7 7} & 0 2 2 & 4 {\bf 6 6} & {\bf 9 9 9} \\
     7 & 1 1 1 & --- & 4 4 {\bf 6} & 1 1 1 & 0 0 1 & --- & 1 0 0 \\
     8 & 1 2 1 & 4 4 {\bf 6} & 1 2 2 & --- & 0 2 2 & --- & --- \\
     9 & {\bf 8 7 6} & 1 1 2 & 1 1 2 & 1 1 1 & --- & 1 1 1 & --- \\
    10 & 2 1 1 & 0 1 3 & 2 5 {\bf 6} & 2 2 1 & 1 2 2 & --- & 1 1 1 \\
    11 & {\bf 8 6 8} & {\bf 7 6 7} & {\bf 8 8 8} & {\bf 6 6} 5 & 4 4 2 & 4 5 5 & --- \\
    12 & 2 4 4 & 1 1 4 & 2 2 2 & 1 1 2 & {\bf 10 9 8 } & 1 1 1 & --- \\
    13 & 1 2 2 & 4 4 4 & {\bf 6 6 6} & 3 2 {\bf 9} & 3 3 {\bf 9} & --- & 2 2 2 \\
    14 & 2 2 1 & 1 1 4 & {\bf 9 9 9} & 1 1 1 & 2 4 3 & 0 1 0 & --- \\
    15 & {\bf 7} 4 3 & 5 4 5 & 3 4 2 & {\bf 7 7} 5 & 3 3 2 & 2 4 2 & --- \\
    16 & 0 1 1 & 2 1 1 & 2 2 2 & 2 2 3 & 0 1 {\bf 7} & 4 4 {\bf 6} & 1 1 1 \\
    17 & 1 1 1 & 1 1 0 & 1 2 2 & 2 0 2 & 1 1 5 & {\bf 7 6 8} & 1 1 1 \\
    18 & 1 2 2 & {\bf 8 9} 4 & 2 2 5 & 1 2 1 & 0 5 5 & 1 0 1 & 1 1 1 \\
    19 & {\bf 6 8 9} & 4 5 4 & 1 1 1 & 2 3 3 & 1 2 2 & 2 2 3 & 0 0 1 \\
    20 & 1 1 0 & 2 4 2 & 2 4 4 & 1 2 6 & 0 2 4 & 1 2 4 & {\bf 9 10 10} \\
    \hline \hline
  \end{tabular}
\caption{The three numbers in each entry are the fractional contributions to the uncertainties ($\Delta F_k/\Delta F$ for the $F=g,u_v,\ldots$ input distributions, where $\Delta F_k$ is the contribution to \eqref{eq:Fp} and \eqref{eq:Fm} from a single eigenvector $k$) in the small $x$ ($x<0.01$), medium $x$ ($0.01<x<0.1$) and large $x$ ($x>0.1$) regions, respectively, arising from eigenvector $k$ in the NLO global fit, using 68\% C.L.~errors.  Each number has been multiplied by ten; for example, 4 denotes 0.4.  Major contributions are shown in bold type.  For a precise value of $x$, and symmetric errors, the sum of the squares of each column should be 100.  However, the entries shown are the maximum fraction in each interval of $x$, so often do not satisfy this condition.}
\label{tab:fractions}
\end{table}
A general summary of all the plots of the type shown in Figs.~\ref{fig:fractionalinputuncertainty9} and \ref{fig:fractionalinputuncertainty13} is presented in Table \ref{tab:fractions}.  The bold entries show which eigenvectors $k$ are mainly responsible for the uncertainty on the various input distributions.  The three numbers in each entry correspond to (ten times) the fractional contribution to the uncertainties ($\Delta F_k/\Delta F$ for the $F=g,u_v,\ldots$ input distributions, where $\Delta F_k$ is the contribution to \eqref{eq:Fp} and \eqref{eq:Fm} from a single eigenvector $k$) in the small $x$ ($x<0.01$), medium $x$ ($0.01<x<0.1$) and large $x$ ($x>0.1$) regions, respectively, for the NLO global fit with 68\% C.L.~uncertainties.  We see that each input distribution is sensitive to only a small number of eigenvectors.  An indication of which data sets are particularly important for which flavour of parton distribution can be obtained by correlating the results of Fig.~\ref{fig:nlotolerance} with the table.  For example, from Table~\ref{tab:fractions} we see that the gluon distribution is particularly sensitive to eigenvectors 9, 11 and 19.  The tolerance for each of these eigenvectors is fixed by the H1 and ZEUS NC data, apart from for the 68\% C.L.~tolerance for eigenvector 19 in the ``$-$'' direction, where it is fixed by the CDF Run II high-$p_T$ jet data.

In fact, the entries in Table \ref{tab:fractions} can be used in conjunction with the eigenvector plots, such as those used as examples in Fig.~\ref{fig:evectors}, to pinpoint which particular data sets constrain the various parton distributions.  For example, the entries in Table \ref{tab:fractions} show that eigenvector 13 gives the biggest contribution to the high-$x$ sea distribution uncertainty and is also very significant for the $\bar{d}-\bar{u}$ distribution at high $x$.  Inspection of the plot of eigenvector 13 versus the data, Fig.~\ref{fig:evectors}(b), shows that the tightest constraints on the eigenvector 13 come from the data sets collected in the following experiments: E866/NuSea $pp$ and $pd/pp$ Drell--Yan, NuTeV and CHORUS $\nu N$ $F_2$ and NuTeV and CHORUS $\nu N$ $xF_3$.  The former are the obvious direct constraint on the sea quarks, while the others influence the decomposition into the well-determined valence quarks and sea quarks.  Eigenvector 13 also contains information on the $d_v$ uncertainty, which is determined by the neutrino DIS data which weight down quarks more highly than neutral-current DIS on proton targets.  Similarly, eigenvector 9 only influences the gluon uncertainty to any great degree, so it is no surprise that it is constrained by the H1 and ZEUS NC data (and CDF Run II high-$p_T$ jet data); see Fig.~\ref{fig:evectors}(a).  Surprisingly, the summary plot in Fig.~\ref{fig:nlotolerance} shows that the Tevatron jet data are never the main constraint on any eigenvector, apart from for the 68\% C.L.~tolerance for eigenvector 19 in the ``$-$'' direction (not shown here).  This does not necessarily mean that the Tevatron jet data provide \emph{no} constraint, however, and this will be discussed in detail in Section \ref{sec:jetdata}.

A complete set of plots related to the error analysis and PDF uncertainties discussed in Sections \ref{sec:tolerance} and \ref{sec:partonerrors}, similar to the examples shown in Figs.~\ref{fig:globalchisq1}--\ref{fig:fractionalinputuncertainty13}, is available at Ref.~\cite{mstwpdf}.  These supplementary plots should prove valuable for the end users of the PDFs, in investigating the sensitivity of different observables to different eigenvector directions, and in finding the corresponding data sets which constrain those particular eigenvector directions.

\subsection{Effect of free data set normalisations on PDF uncertainties}

As already noted, one of our significant improvements in the uncertainty determination in the present analysis is the inclusion of the data normalisation errors while determining the PDF uncertainties.  In previous MRST (and CTEQ) fits these have been held fixed during the error propagation.  To account for free data set normalisations, we first calculate the $39\times 39$ covariance matrix with an additional 19 free parameters\footnote{To be precise, there are 16 free parameters associated with data set normalisations together with the 3 parameters $(r_1,r_2,r_3)$ associated with nuclear corrections \eqref{eq:nucmod}.  We do not allow the normalisation of the neutrino data to go free because it is very highly correlated with $r_1$.} corresponding to data set normalisations, in addition to the usual 20 free input PDF parameters.  In principle the inclusion of free data set normalisations in PDF uncertainties can be done in two different ways.

The simplest way is to diagonalise the $39\times 39$ covariance matrix to obtain 39 eigenvectors and to determine the tolerance values dynamically as described above.  However, this method involves a large number of eigenvectors, many of which actually consist largely of the variation of data set normalisations.  It is clearly more physically intuitive to use eigenvectors corresponding only to input PDF parameters.

An alternative method, which has this advantage and which we use in the present analysis, is to remove the rows and columns from the $39\times 39$ covariance matrix corresponding to data set normalisations, leaving only a $20\times 20$ covariance matrix corresponding to input PDF parameters, which contains, however, information on free data set normalisations from the inversion of the Hessian matrix~\cite{NumericalRecipes}.  Then the $20\times 20$ covariance matrix can be diagonalised to give 20 eigenvectors, and the tolerance values and eigenvector PDF sets produced as described above, with the modification that every time the PDF parameters are changed, when moving along an eigenvector direction, the $\chi^2_{\rm global}$ should be minimised with respect to the data set normalisations.\footnote{Similarly, the $\chi^2_{\rm global}$ should be minimised with respect to the shifts in correlated systematic errors, $r_k$ in Eq.~\eqref{eq:chisqcorr}, each time the PDF parameters are changed, but this is done automatically by using Eq.~\eqref{eq:rksolution}.}

\begin{figure}
  \centering
  \includegraphics[width=0.5\textwidth]{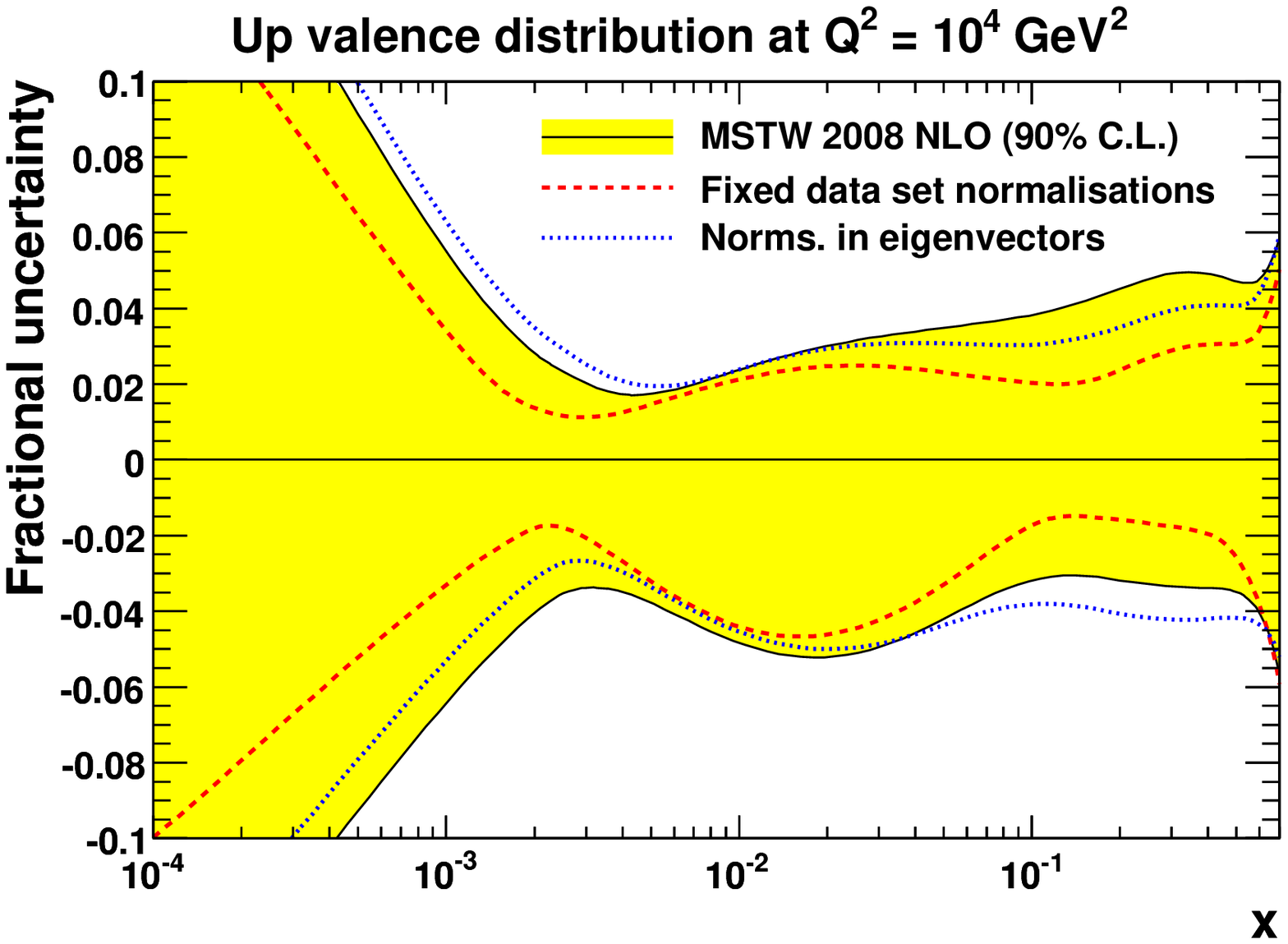}%
  \includegraphics[width=0.5\textwidth]{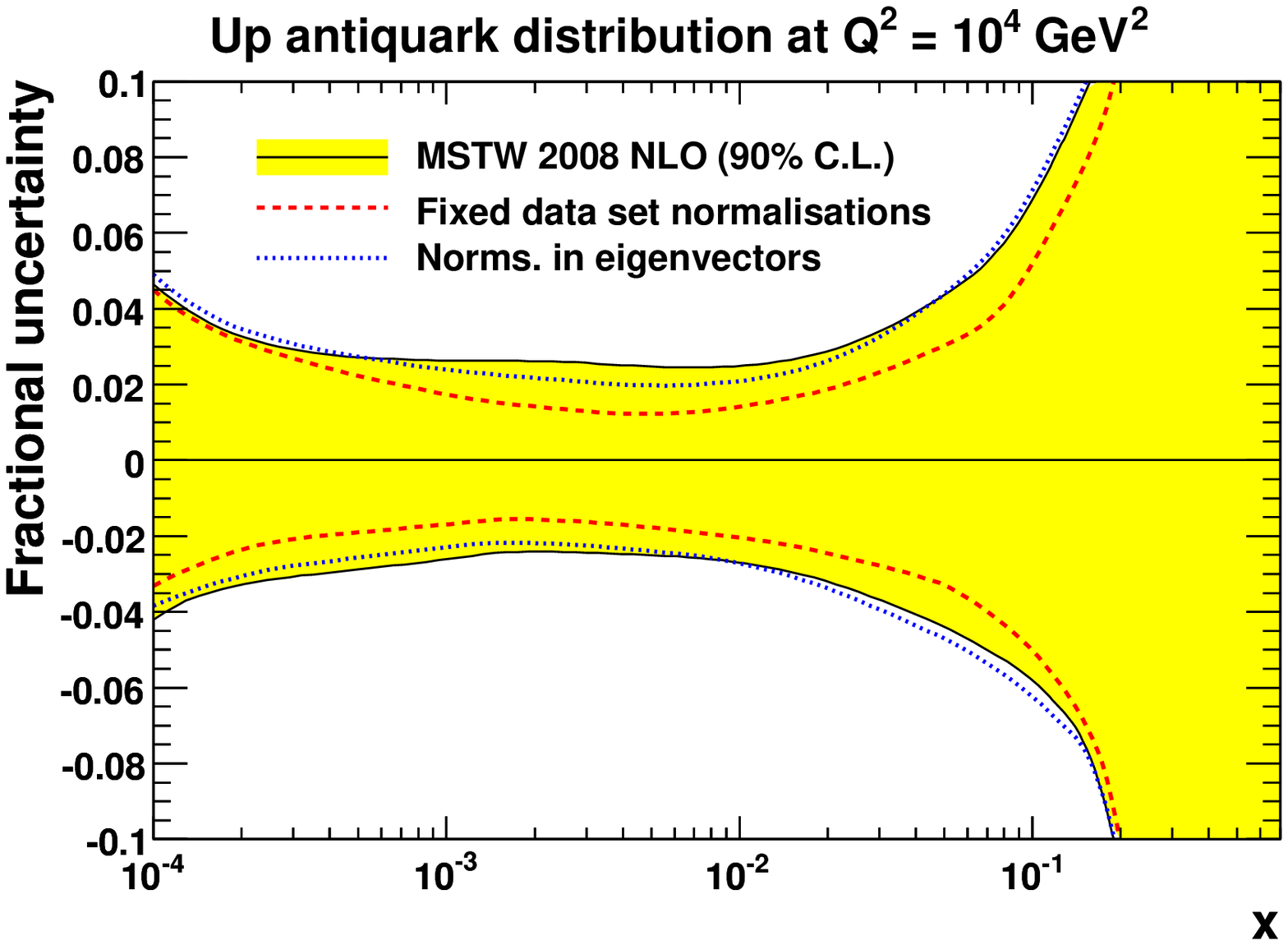}\\
  \includegraphics[width=0.5\textwidth]{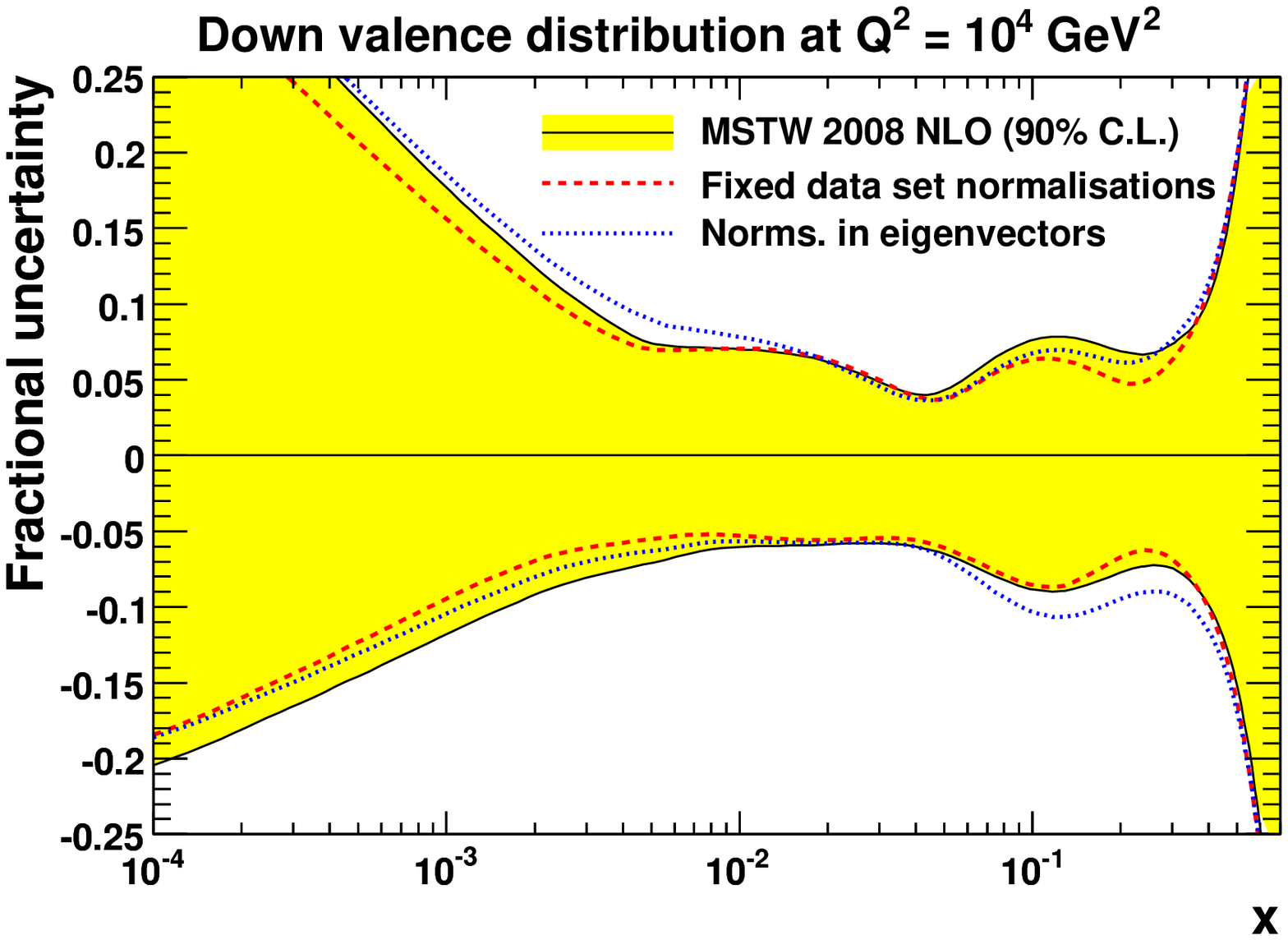}%
  \includegraphics[width=0.5\textwidth]{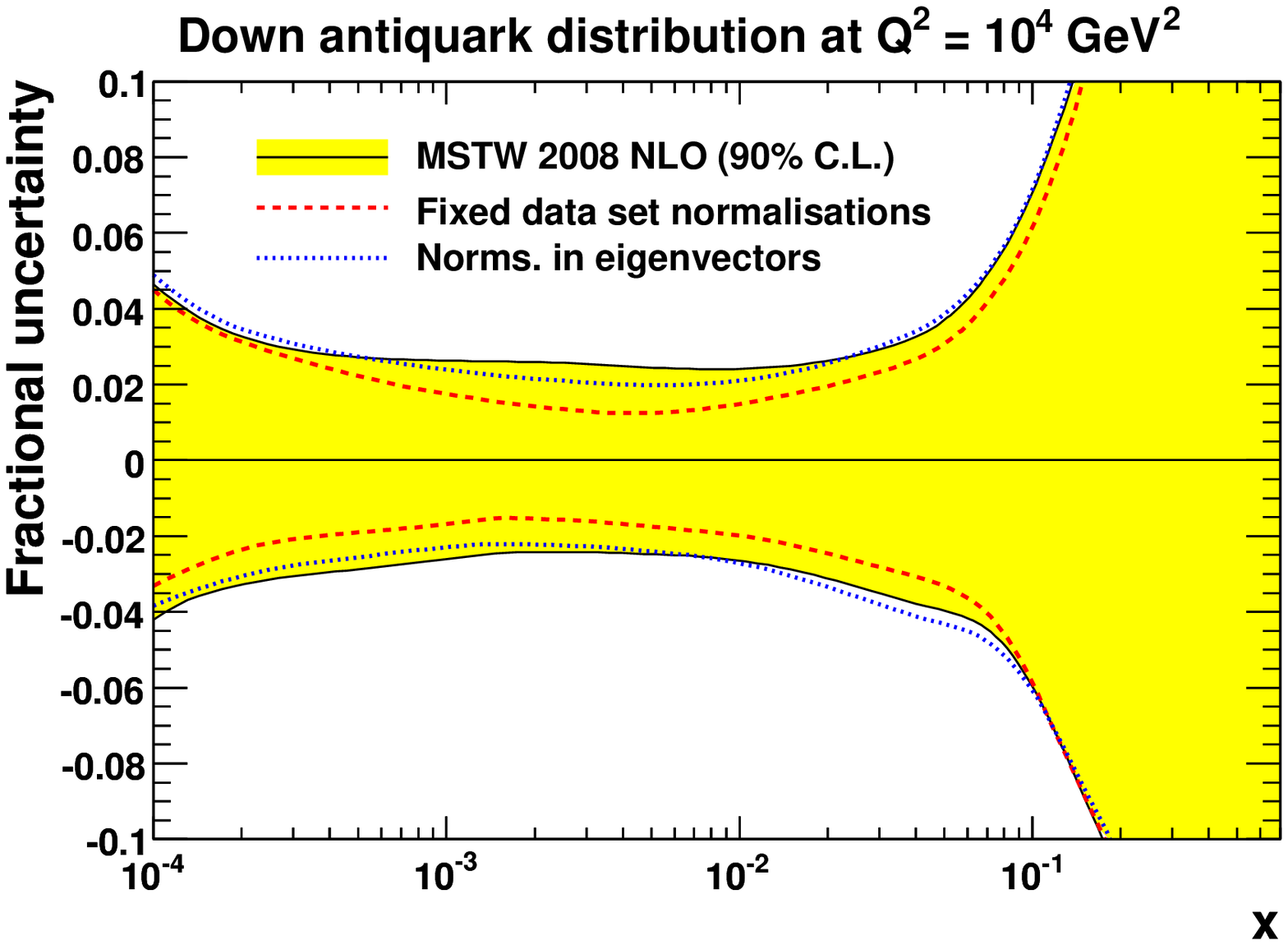}\\
  \includegraphics[width=0.5\textwidth]{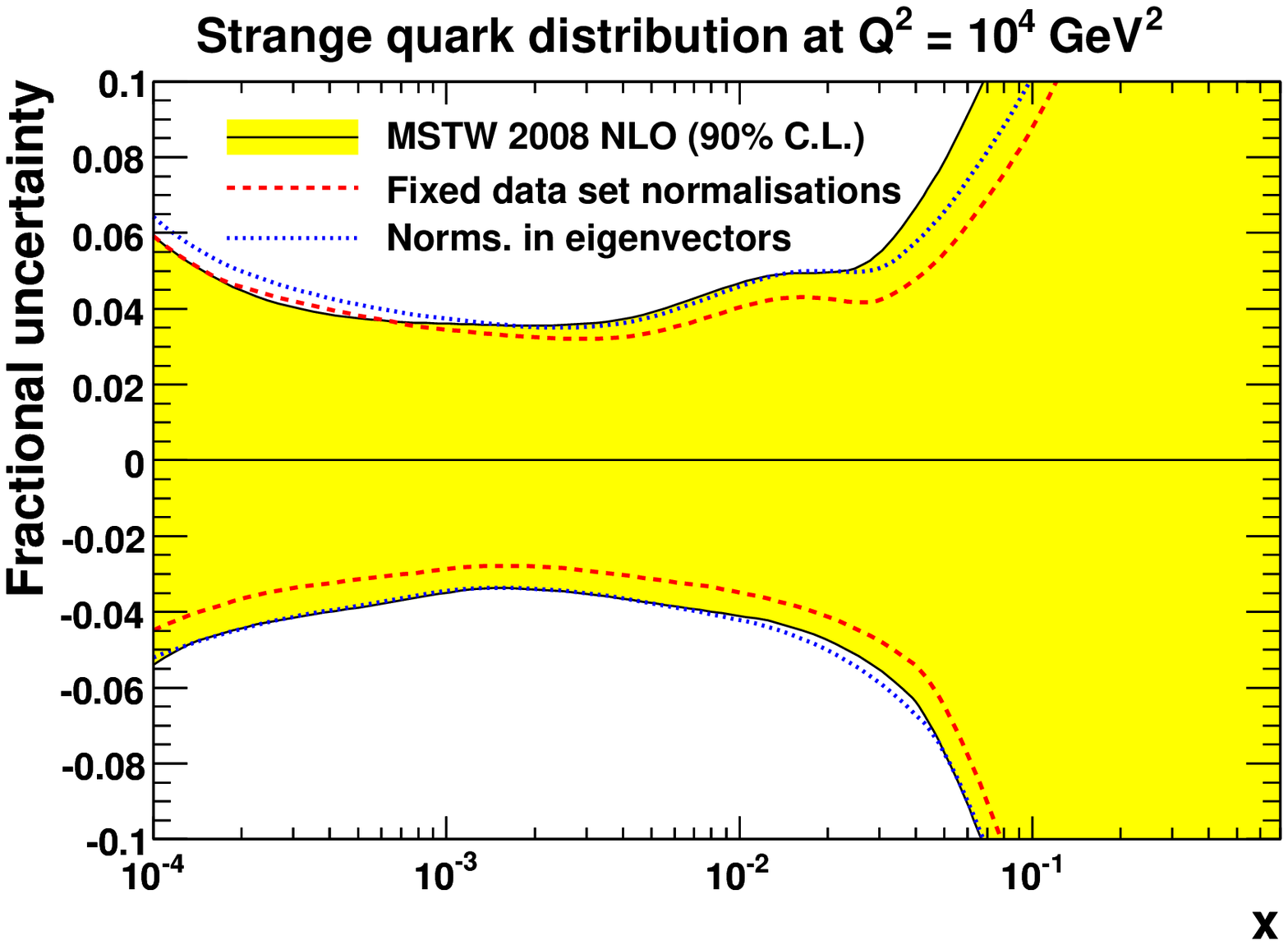}%
  \includegraphics[width=0.5\textwidth]{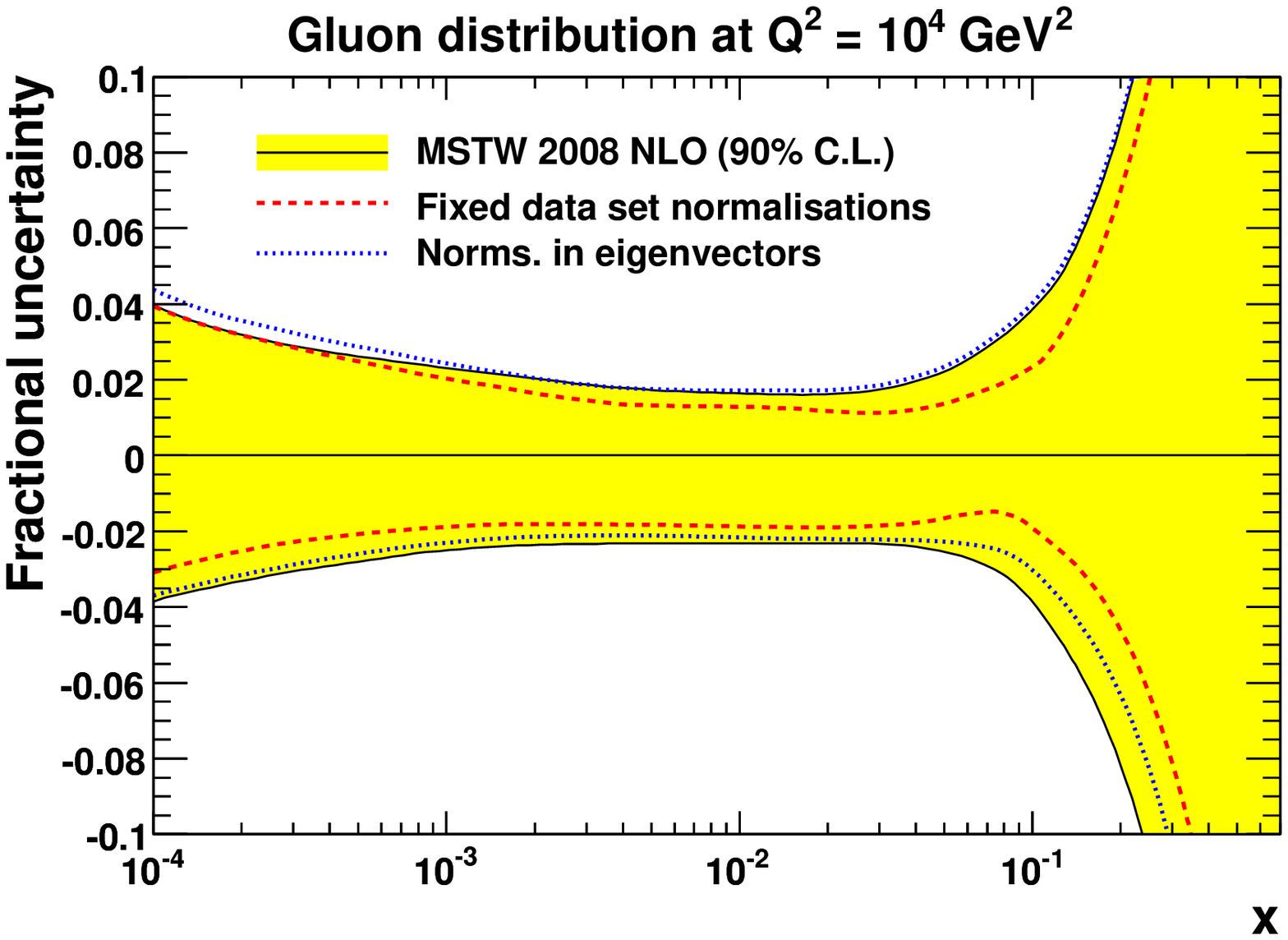}\\
  \caption{The fractional uncertainty of the MSTW 2008 NLO PDFs at $Q^2 = 10^4$ GeV$^2$ using 20 eigenvectors and allowing for free data set normalisations.  We also show the fractional uncertainty obtained using 20 eigenvectors with fixed data set normalisations (as in the MRST analyses) and with 39 eigenvectors including 19 parameters associated with data set normalisations in additional to the 20 input PDF parameters.}
  \label{fig:fractionalNLO}
\end{figure}

In the limit of quadratic behaviour for the $\chi_{\rm global}^2$ function and equal tolerance for each eigenvector we would obtain the same results for PDF uncertainties in each case.  In Fig.~\ref{fig:fractionalNLO} we show the effect of allowing data set normalisations to go free according to these two methods, compared to the case where data set normalisations are fixed as in the MRST analyses.  The new procedure leads to an increase in uncertainties, but in most places this increase is very small.  The exception is those parton distributions where the uncertainty is smallest, e.g.~the up valence quark for $x \gtrsim 0.05$ and the light sea for $x$ between $0.001$ and $0.01$.  The absolute increase in uncertainty is $1\%$ or less, but in these cases it is a significant contribution to the total.  We see that the approach using data set normalisations in eigenvectors gives very similar results, demonstrating that despite our quartic penalty term for normalisations there is little breakdown of the quadratic approximation.  Indeed, the uncertainty on normalisations determined from the fit and applying the tolerance approach is usually rather similar to the quoted experimental uncertainty.  Hence, we do not actually encounter large penalties from varying normalisations, as we rarely vary by much more than the experimental one-sigma uncertainty.

\subsection{Input parameterisation and uncertainties} \label{sec:inputparamunc}

There is a strong relationship between the input PDF parameterisation and the uncertainties which will be obtained.  Too restrictive a parameterisation can limit the calculated uncertainty in an artificial manner.  The parameterisation for the input PDFs in our analysis were presented in Section \ref{sec:overviewtheoretical}, specifically in Eqs.~\eqref{eq:uv}--\eqref{eq:sv}.  The 28 free PDF parameters listed there allow a large degree of flexibility, but are sufficiently few that only a few turning points are possible for each distribution, presumably the case in reality for quantities closely related to physical variables.  Indeed, as outlined in Section \ref{sec:hessian}, many of the 28 free PDF parameters turn out to be very strongly (anti)correlated, and only 20 parameters remain after removing all combinations where this is the case, whose uncertainties are quoted in Table \ref{tab:parameters}.  This is 5 more parameters than in the MRST error sets~\cite{Martin:2002aw,Martin:2007bv}.  Here we briefly discuss the truly free parameters for each parton flavour, i.e.~those parameters allowed to go free when calculating the covariance matrix used for error propagation, and outline the resulting effect on the uncertainty.

\subsubsection*{Valence quarks}

For the up and down valence quark distributions we use the standard ``MRS-type'' parameterisation,
\begin{equation} \label{eq:MRStype}
  xv(x,Q_0^2)= A_v\,x^{\delta_v}\,(1-x)^{\eta_v}\,(1+{\epsilon_v}\,x^{0.5} +\gamma_v\,x),
\end{equation}
where the normalisation parameter $A_v$ is determined from the number sum rules \eqref{eq:numbersumrule}.  In both cases the 3 parameters left free in the determination of the eigenvectors are $\delta_v$, $\eta_v$ and $\epsilon_v$.  These correspond to three reasonably independent regions of $x$.  At high $x$ the uncertainty is determined very largely by $\eta_v$.  In detail,
\begin{eqnarray}
  v\pm \Delta v &\sim& A_v\,(1+\epsilon_v+\gamma_v)\,(1-x)^{\eta_v \mp \Delta{\eta_v}} \nonumber \\
  &\sim& v\,(1-x)^{\mp \Delta{\eta_v}} \\
  &\sim& v\,[1 \mp \Delta{\eta_v}\,\ln (1-x)], \nonumber
\end{eqnarray}
where the last line applies if $|\Delta{\eta_v}\,\ln (1-x)|$ is small enough for the linear expansion of the exponential to be valid.  Hence, the parameterisation allows the uncertainty to become very large as $x \to 1$, as required by the lack of data in this region.  It is not as flexible as some of the competing parameterisations, e.g.~the CTEQ6M PDFs~\cite{Pumplin:2002vw}.  However, it does limit the possibility of unusual shapes at very high $x$, or arguably undesirable features, e.g.~the gluon becoming the hardest distribution at very high $x$ ($x>0.9$), as found in the CTEQ6.1M set.  At small $x$ the uncertainty is determined very largely by $\delta_v$.  Again in more detail,
\begin{eqnarray}
  x(v\pm \Delta v) &\sim& A_v\,x^{\delta_v \mp \Delta {\delta_v}} \nonumber \\
  &\sim& xv\,x^{\mp \Delta {\delta_v}} \\
  &\sim& xv\,[1 \pm \Delta {\delta_v} \ln(1/x)], \nonumber
\end{eqnarray}
where again the last line applies if $|\Delta {\delta_v}\,\ln (1/x)|$ is small enough for the linear expansion of the exponential to be valid.  Similarly to the case at high $x$, the uncertainty can become very large at small $x$, but once the behaviour has been determined by the data at fairly low $x$, fixing the one parameter, even lower $x$ values are simply an extrapolation.  The third free parameter, $\epsilon_v$, controls the variation of the input distribution at intermediate values of $x$ ($\gamma_v$ is more strongly correlated to $\eta_v$).  In practice the uncertainties at very high $x$, and even more so at very small $x$, are also constrained by the sum rules.  In the region where valence PDFs are constrained by data, i.e.~$x=0.01$--$0.75$, about $75\%$ of the total number of quarks are found, so the number found in the extrapolation regions is relatively small and very well determined.

\begin{figure}
  (a)\hspace{0.5\textwidth}(b)\\
  \includegraphics[width=0.5\textwidth]{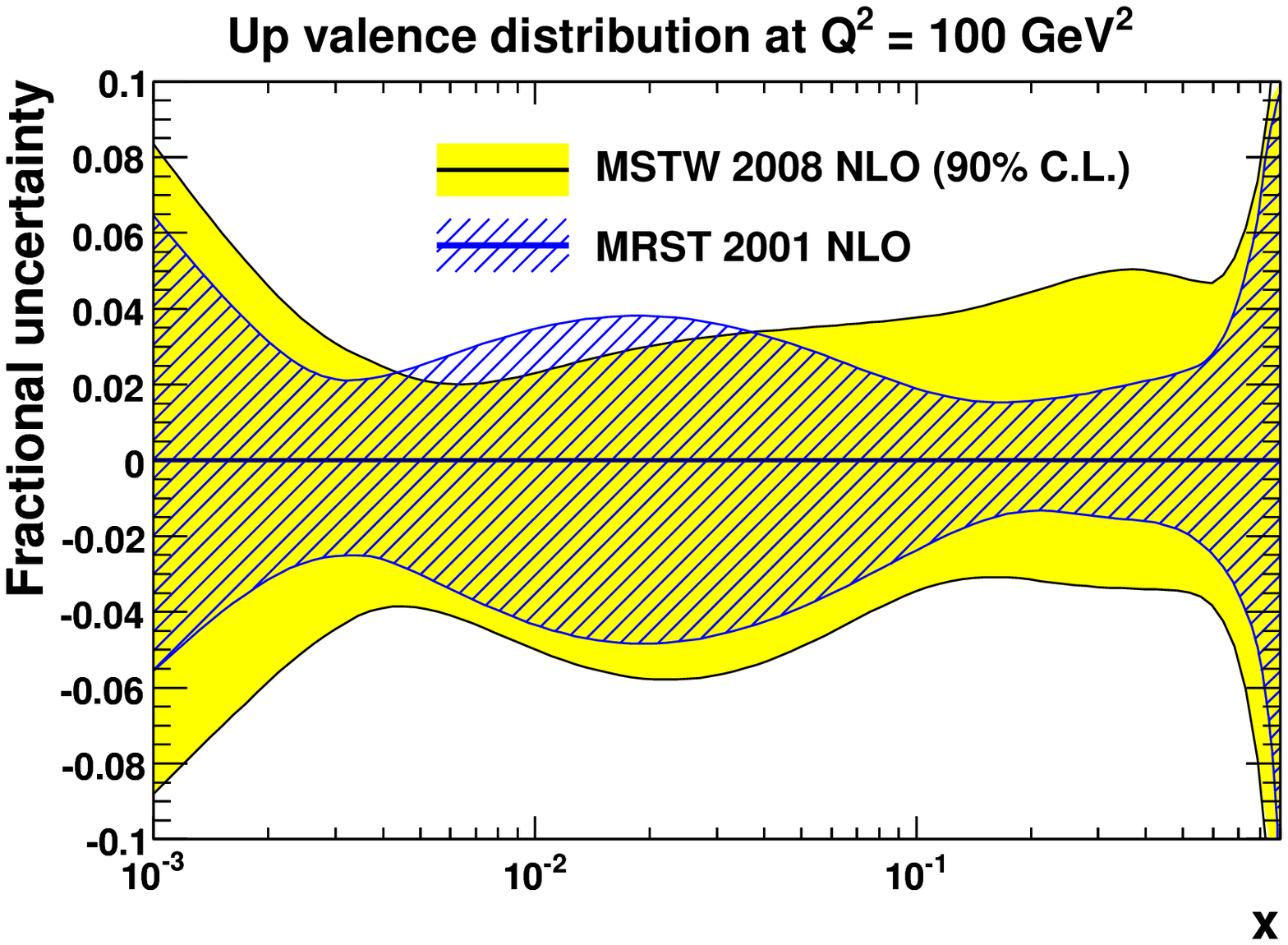}%
  \includegraphics[width=0.5\textwidth]{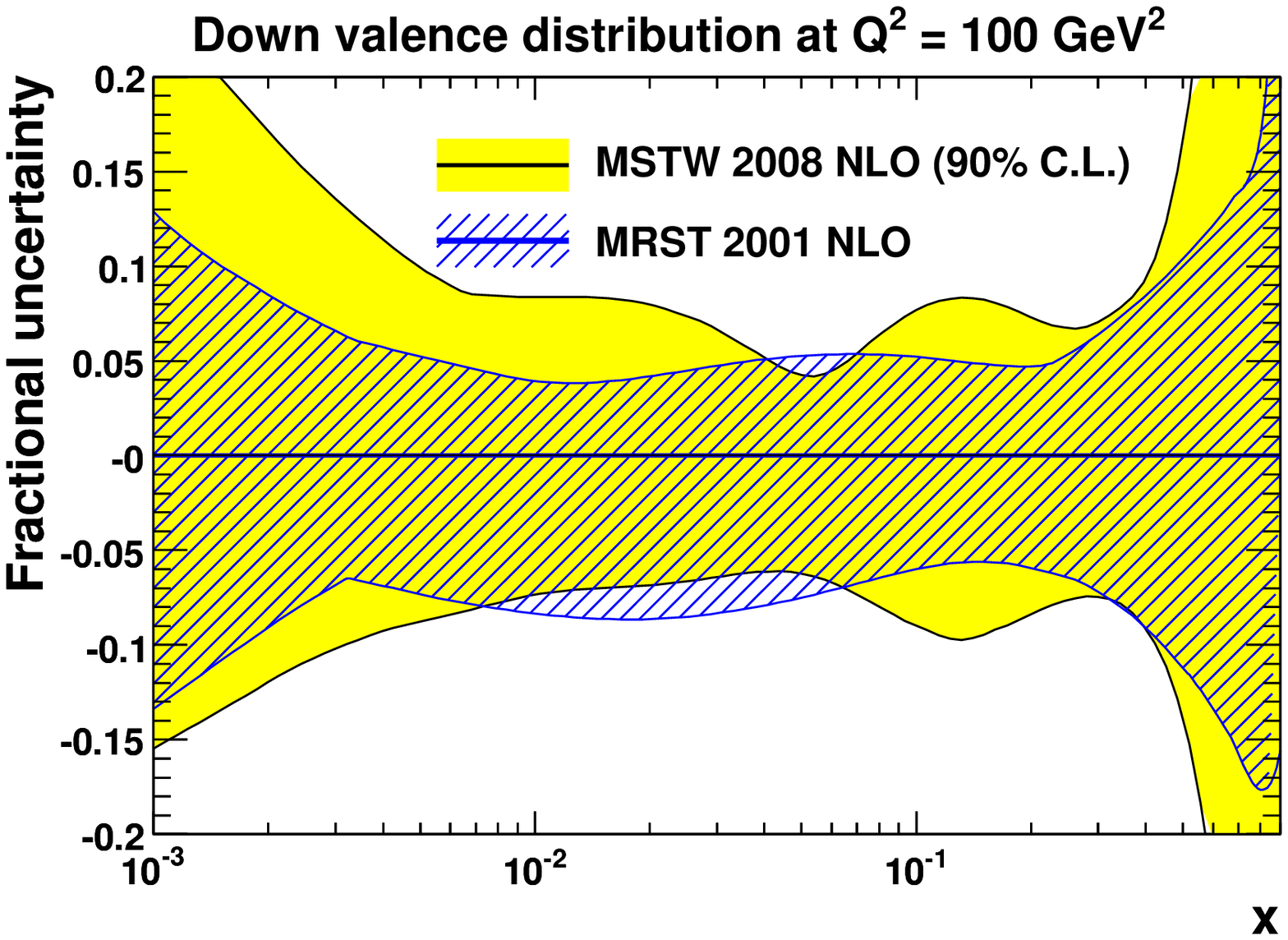}
  \caption{The fractional uncertainty on the (a) up valence quark and (b) down valence quark distributions at $Q^2=100$ GeV$^2$ from the MSTW 2008 NLO fit compared to that from the MRST 2001 NLO fit~\cite{Martin:2002aw}.}
  \label{fig:vquncertainty}
\end{figure}
The fractional uncertainty for the up valence distribution at NLO is shown in Fig.~\ref{fig:vquncertainty}(a) along with that from the MRST 2001 analysis~\cite{Martin:2002aw}.  In this case the parameters left free when determining the eigenvectors are the same and the uncertainties are larger mainly because of the fact that data set normalisations were fixed in the MRST analysis; see Fig.~\ref{fig:fractionalNLO}.  The fractional uncertainty on $u_v$ increases rapidly for very high $x$ and also for $x\lesssim0.003$.  In the latter case the data constraint for $x\lesssim0.01$ at lower $Q^2$ has moved to a smaller $x$ due to evolution.  The uncertainty for the down valence distribution is shown in Fig.~\ref{fig:vquncertainty}(b).  The free parameters in the MRST 2001 analysis were $\eta_d$, $\epsilon_d$ and $\gamma_d$.  Hence, the MRST 2001 input parameterisation allowed less flexibility in the $d_v$ uncertainty at small $x$ and from the sum rule this fed in to the unconstrained high-$x$ region also.  This is clearly illustrated in Fig.~\ref{fig:vquncertainty}(b).

\subsubsection*{Gluon}

The gluon distribution is the most complicated case for PDF uncertainties and parameterisations.  The input gluon in the original MRST fits~\cite{Martin:1998sq,Martin:1999ww} had the same form of parameterisation as Eq.~\eqref{eq:MRStype}, with $A_g$ determined by the momentum sum rule.  However, with the advent of improved HERA data included in the MRST 2001 fit~\cite{Martin:2001es}, it was noticed that if the evolution is started from $Q_0^2=1$ GeV$^2$ the input gluon preferred to be negative at very small $x$ and an additional term $A_{g^\prime}\,x^{\delta_{g^\prime}}\,(1-x)^{\eta_{g^\prime}}$ was added to facilitate this.  To this date all competing parameterisations, apart from the very recent NNPDF1.0 set~\cite{Ball:2008by}\footnote{Despite the NNPDF1.0 input parameterisation formally behaving as $xg\sim x^{-0.2}\,{\rm NN}_g(x)$, the neural network parameterisation ${\rm NN}_g(x)$ should be sufficiently flexible that there is no dependence on the factor $x^{-0.2}$~\cite{Ball:2008by}.}, have a gluon which at small $x$ behaves like $xg(x,Q_0^2) \sim x^{\delta_g}$, i.e.~is controlled by a single power.  This means that
\begin{equation} \label{eq:simpgluon}
  g \pm \Delta {g} \sim g\,[1 \pm \Delta {\delta_g} \ln (1/x)],
\end{equation}
i.e.~the uncertainty grows linearly with $\ln(1/x)$ and there is no scope for a rapidly expanding uncertainty as data constraints run out.  This is much more of an issue for the gluon than for valence quarks, as the momentum sum rule offers a far less direct constraint than the number sum rules as $x \to 0$.  However, there is another complication to consider, namely
\begin{equation} \label{eq:simpgluonunc}
  \Delta g(x,Q_0^2) \sim g(x,Q_0^2)\,\Delta {\delta_g} \ln (1/x),
\end{equation}
and so as $g(x,Q_0^2)$ becomes smaller then so does $\Delta {g}(x,Q_0^2)$.  If $g(x,Q_0^2)$ is very small, then the absolute input uncertainty for the gluon is very small, and at higher $Q^2$ the uncertainty is therefore determined entirely by evolution from higher-$x$, i.e.~by the region where the gluon distribution is better determined.  Most PDF fitting groups find that $xg(x,Q^2)$ is indeed small at low $Q^2$ and small $x$.  In this region the MRST (since 2001) and MSTW gluon distributions have the form,
\begin{equation}
  xg(x,Q_0^2) = xg_1(x,Q_0^2) +  xg_2(x,Q_0^2) \sim A_g\,x^{\delta_{g}} + A_{g^\prime}\,x^{\delta_{g^\prime}},
\end{equation}
which is more flexible than a single power.  Not only does it allow the gluon to become negative at very small $x$, but it is also particularly important for the uncertainty,
\begin{equation}
  \Delta{g}(x,Q_0^2) \sim \pm g_1(x,Q_0^2)\,\Delta{\delta_{g}} \ln (1/x) \pm g_2(x,Q_0^2)\,\Delta{\delta_{g^\prime}} \ln (1/x),
\end{equation}
where $g_1$ and $g_2$ represent the two independent terms in the gluon parameterisation.  The interplay between the two terms allows for a large uncertainty at $x \lesssim 10^{-4}$ where the data constraint, from the $Q^2$ dependence of $F_2(x,Q^2)$ at HERA, diminishes rapidly.

\begin{figure}
  \centering
  \includegraphics[width=0.7\textwidth]{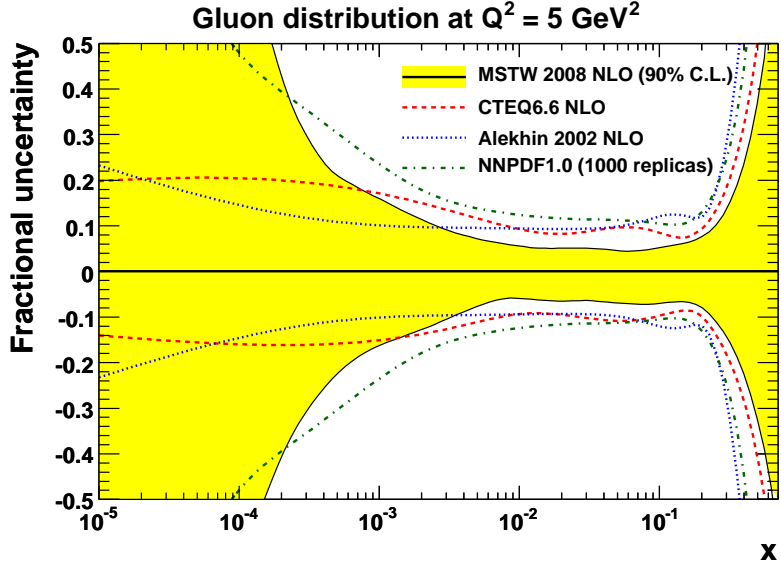}
  \caption{A comparison of the fractional uncertainty for the present MSTW, CTEQ6.6~\cite{Nadolsky:2008zw}, Alekhin~\cite{Alekhin:2002fv} and NNPDF1.0~\cite{Ball:2008by} NLO gluon distributions at $Q^2=5$ GeV$^2$.  All uncertainty bands represent a 90\% C.L.~limit.}
  \label{fig:gluonerrors}
\end{figure}
A comparison of the fractional uncertainty for the present MSTW, CTEQ6.6~\cite{Nadolsky:2008zw}, Alekhin~\cite{Alekhin:2002fv} and NNPDF1.0~\cite{Ball:2008by} NLO gluon distributions is shown in Fig.~\ref{fig:gluonerrors}.  The fractional uncertainty for MSTW blows up very quickly at small $x$, whereas that for the Alekhin fit gets bigger relatively slowly (the same would be qualitatively true of the H1~\cite{Adloff:2000qk} and ZEUS~\cite{Chekanov:2005nn} fits), while that for CTEQ saturates, or can even decrease slightly, completely contrary to what is expected from the degree of constraint from data.  These results are related to the input forms and values of $Q_0^2$, and can be understood fairly easily.  MRST/MSTW parameterise the starting distributions at $Q_0^2=1$ GeV$^2$, and since we allow both positive and negative small-$x$ contributions the uncertainty can be very large.  We believe that this represents the true uncertainty at low $x$.  The Alekhin, H1 and ZEUS gluon distributions are input at a higher scale.  They behave like $x^{-\lambda}$ at small $x$, so the uncertainty is due to the uncertainty in one parameter, as shown in \eqref{eq:simpgluon}.  Recent CTEQ gluon distributions are input at $Q_0^2=1.69$ GeV$^2$, and behave like $x^{\lambda}$ at small $x$ where $\lambda$ is large and positive, i.e.~the input gluon is valence-like.\footnote{A study of the input gluon parameterisation by CTEQ was made in Appendix D of Ref.~\cite{Pumplin:2002vw}, but it did not address the impact of the input parameterisation on the uncertainties.}  This requires fine tuning: when evolving backwards from a steep gluon at higher scales a valence-like gluon only exists for a very narrow range of $Q^2$ (if at all).  In this case the small-$x$ input gluon distribution is tiny, so there is a very small absolute error, as seen from \eqref{eq:simpgluonunc}.  At higher $Q^2$ all the uncertainty is due to evolution and driven by the  higher $x$ and more well-determined gluon distribution.  Hence the very small $x$ gluon is no more uncertain than at $x=0.001$--$0.01$.\footnote{The same features exist for the ``dynamical parton distributions'' of Ref.~\cite{Gluck:2007ck}.  However, in that case the starting scale is $Q_0^2=0.5$ GeV$^2$, and the PDFs do not fit data as well as the ``standard'' PDFs which are obtained from evolution at a higher starting scale.}  This does not seem to be realistic to us.  The NNPDF1.0~\cite{Ball:2008by} uncertainty shown in Fig.~\ref{fig:gluonerrors}, which is obtained using a very different approach for the input parameterisation and error propagation, will be discussed in Section \ref{sec:comparison}.

\begin{figure}
  (a)\hspace{0.5\textwidth}(b)\\
  \includegraphics[width=0.5\textwidth]{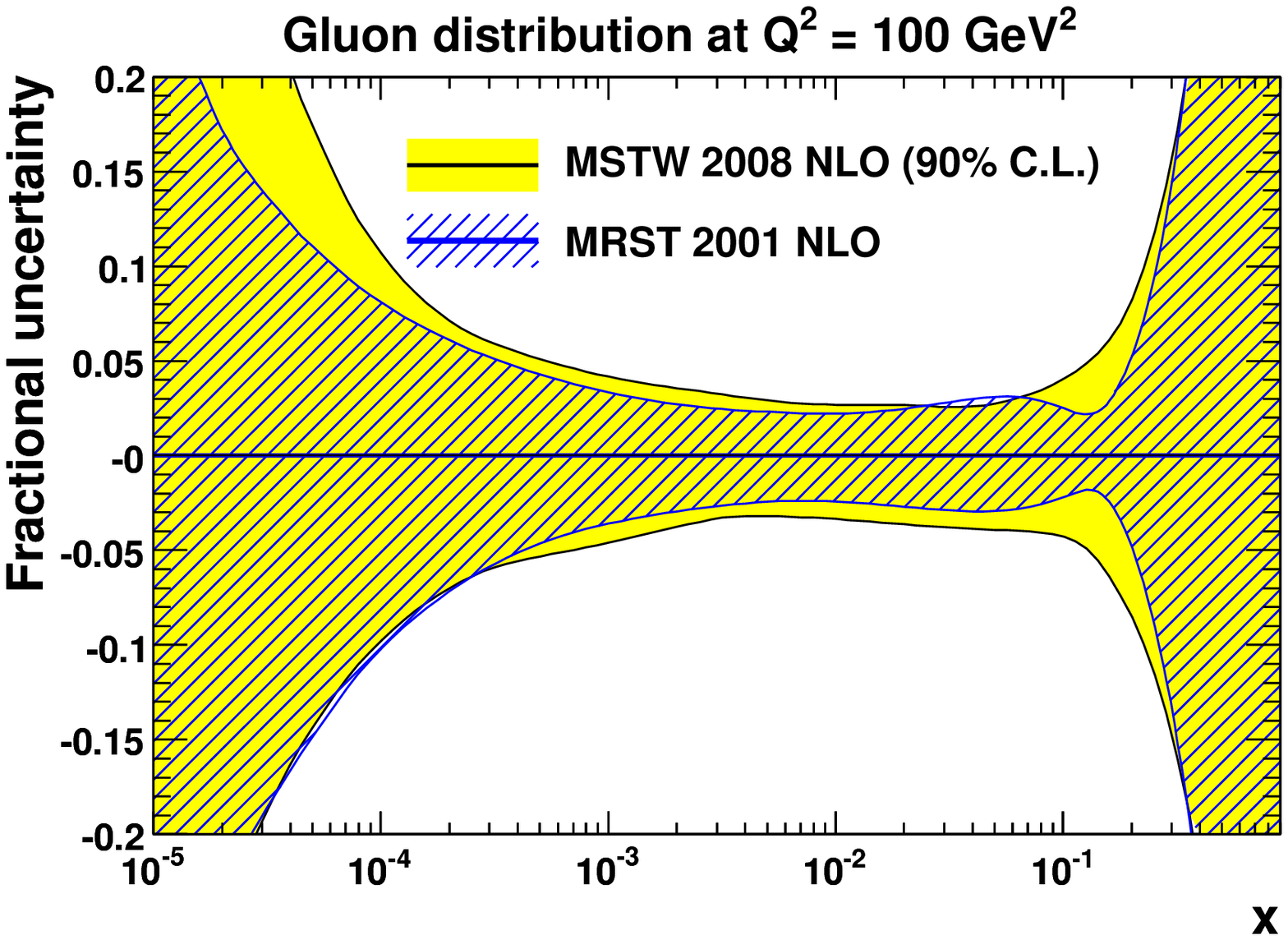}%
  \includegraphics[width=0.5\textwidth]{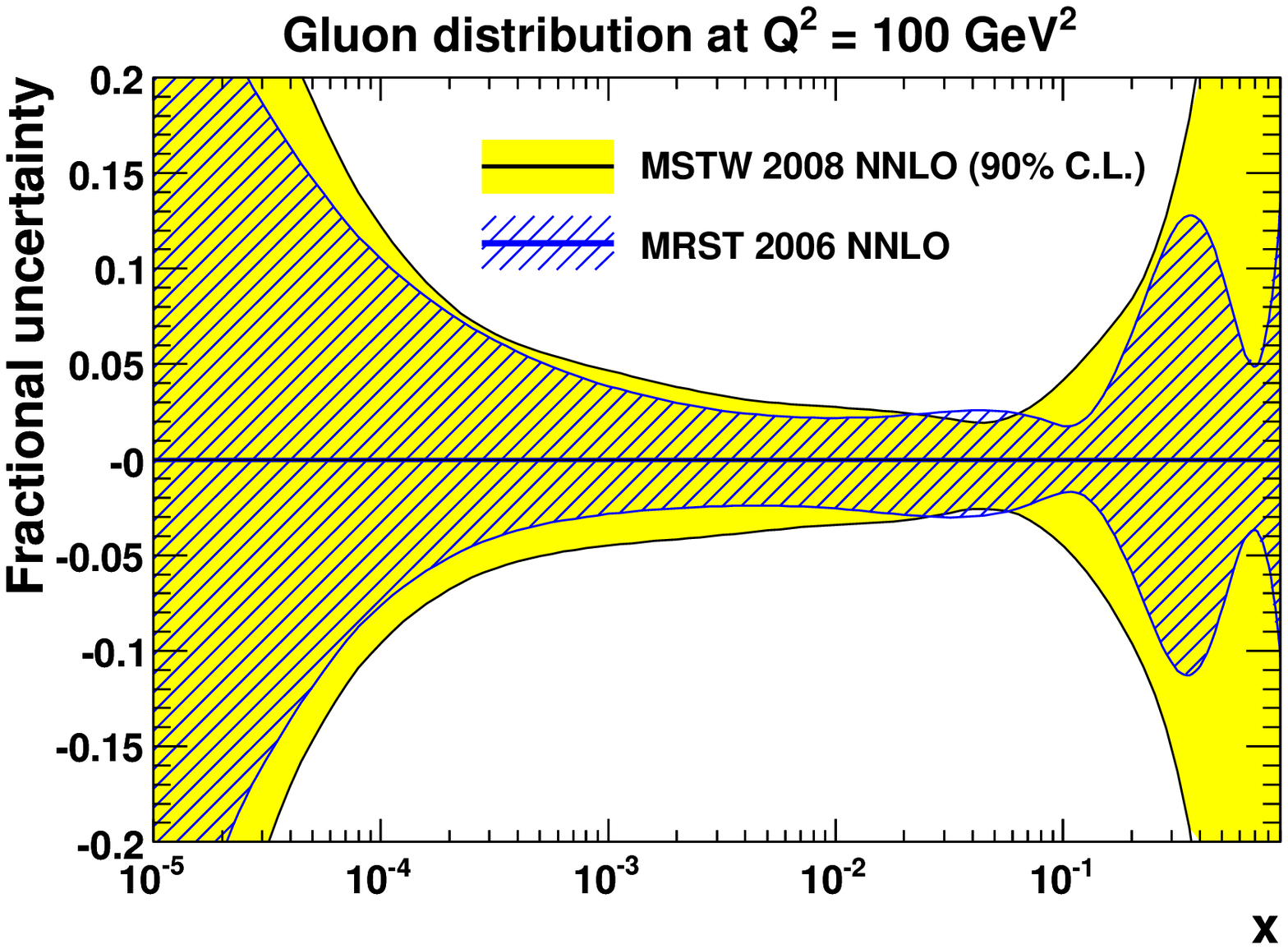}
  \caption{The fractional uncertainty on the gluon distribution at $Q^2=100$ GeV$^2$ (a) from the MSTW 2008 NLO fit compared to that from the MRST 2001 NLO fit~\cite{Martin:2002aw} and (b) from the MSTW 2008 NNLO fit compared to that from the MRST 2006 NNLO fit~\cite{Martin:2007bv}.}
  \label{fig:gluuncertainty}
\end{figure}
The above discussion of the input gluon parameterisation applies to all NLO and NNLO MRST/MSTW sets since 2001.  However, in the MSTW 2008 analysis we have one more free parameter in the uncertainty determination, i.e.~as well as $\delta_g$, $\delta_{g^\prime}$ and $\eta_g$ we also have $\eta_{g^\prime}$.  The latter contributes to the uncertainty at intermediate $x$ since its central value is much greater than that of $\eta_g$.  The comparison of the uncertainty in the gluon distribution to that from the MRST 2001 and 2006 analyses is shown in Fig.~\ref{fig:gluuncertainty}.  As well as increased uncertainty at small $x$ one sees that the ``neck'' at $x \approx 0.13$ has disappeared in the new analysis.

Note that the fractional uncertainty for the gluon distribution from the MRST 2006 NNLO set, shown in Fig.~\ref{fig:gluuncertainty}(b), has an additional narrowing or ``neck'' at $x=0.7$.  This has a simple explanation.  The MRST 2004~\cite{Martin:2004ir} and 2006~\cite{Martin:2007bv} PDF sets defined the input $\overline{\rm MS}$ gluon parameterisation via a transformation from the DIS to the $\overline{\rm MS}$ scheme, where the input gluon distribution in the DIS scheme was given by the standard parameterisation \eqref{eq:inputxg}, giving an enhancement of the $\overline{\rm MS}$ gluon at high $x$ originating from the high-$x$ quarks.  Hence, in practice, at high $x$, i.e.~above $x \approx 0.4$--$0.5$, the MRST 2006 gluon distribution was dominated by the quark distributions via the ${\rm DIS}\to\overline{\rm MS}$ transformation.  This meant that for $x\gtrsim 0.6$ the relative uncertainty on the gluon effectively became that on the quarks, which is much smaller, but does start diverging again at even higher $x$, as can clearly be seen in Fig.~\ref{fig:gluuncertainty}(b).  As we will see in Section \ref{sec:jetdata}, the high-$x$ enhancement in the gluon distribution is disfavoured by Tevatron Run II jet data, and so the present MSTW 2008 fits revert to the standard parameterisation \eqref{eq:inputxg} for the input gluon distribution directly in the $\overline{\rm MS}$ scheme.  Hence the uncertainty simply expands as $x \to 1$, as for the other PDFs.

The LO gluon distribution has no tendency to turn over at small $x$, so the second term in \eqref{eq:inputxg} is redundant.  In this case we use the standard ``MRS-type'' parameterisation \eqref{eq:MRStype}, but with all 4 parameters ($\delta_g$, $\eta_g$, $\epsilon_g$, $\gamma_g$) free in determining the eigenvectors.

\subsubsection*{Sea asymmetry, \texorpdfstring{$\bar{d}-\bar{u}$}{dbar - ubar}}
\begin{figure}
  \centering
  \includegraphics[width=0.5\textwidth]{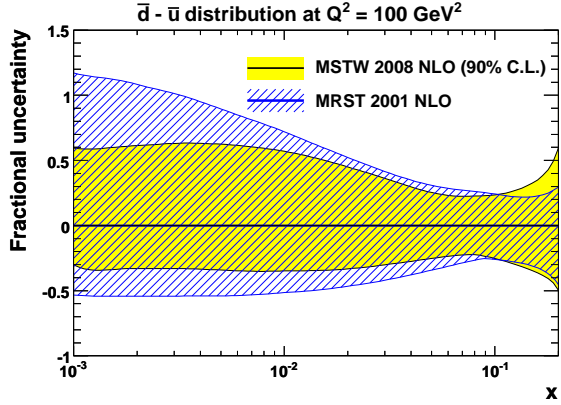}
  \caption{The fractional uncertainty on the $\bar{d}-\bar{u}$ distribution at $Q^2=100$ GeV$^2$ from the MSTW 2008 NLO fit compared to that from the MRST 2001 NLO fit~\cite{Martin:2002aw}.}
  \label{fig:Deltauncertainty}
\end{figure}
The sea asymmetry $\bar{d}-\bar{u}$ is parameterised at input as
\begin{equation}
  x(\bar{d}-\bar{u})(x,Q_0^2) = A_{(\bar{d}-\bar{u})}\,x^{\eta_{(\bar{d}-\bar{u})}}\,(1-x)^{\eta_S+2}(1+\gamma_{(\bar{d}-\bar{u})}\,x+\delta_{(\bar{d}-\bar{u})}\,x^2),
\end{equation}
and the parameters contributing to the eigenvectors are the same as in the MRST 2001 fit.  The distribution is constrained by data for $0.01 \lesssim x \lesssim 0.2$.  At small $x$ the fact that we choose a simple power means that once the data for $0.01\lesssim x\lesssim 0.1$ cause the distribution to head to zero there is no scope for it to do otherwise at smaller $x$.  We see no reason to suppose that, as a nonsinglet quantity, it should not go to zero as $x$ goes to zero, due to considerations from Regge theory.  At high $x$ we choose the $(1-x)$ power to be $\eta_S+2$, since  $\bar{d}(x,Q^2)-\bar{u}(x,Q^2)$ is becoming very small, and we attempt to constrain $\bar{u},\bar{d}\geq 0$.  Applying this constraint on the power of $(1-x)$ we then need the $x^2$ term instead of the more usual $\sqrt{x}$ term to give sufficient flexibility at large $x$.  The 3 eigenvector parameters control the normalisation, high $x$ and low $x$.  In fact, in some PDF sets one does then get negative central values of the $\bar{d}(x,Q^2)$ at the highest $x$, but at a level very small compared to the uncertainties.  The uncertainties for $\bar{d}-\bar{u}$ in the MSTW 2008 PDF set are slightly smaller than for MRST 2001, as seen in Fig.~\ref{fig:Deltauncertainty}, but this is due to the inclusion of new data and also smaller values of the tolerance.

\subsubsection*{Light sea quarks}
\begin{figure}
  (a)\hspace{0.5\textwidth}(b)\\
  \includegraphics[width=0.5\textwidth]{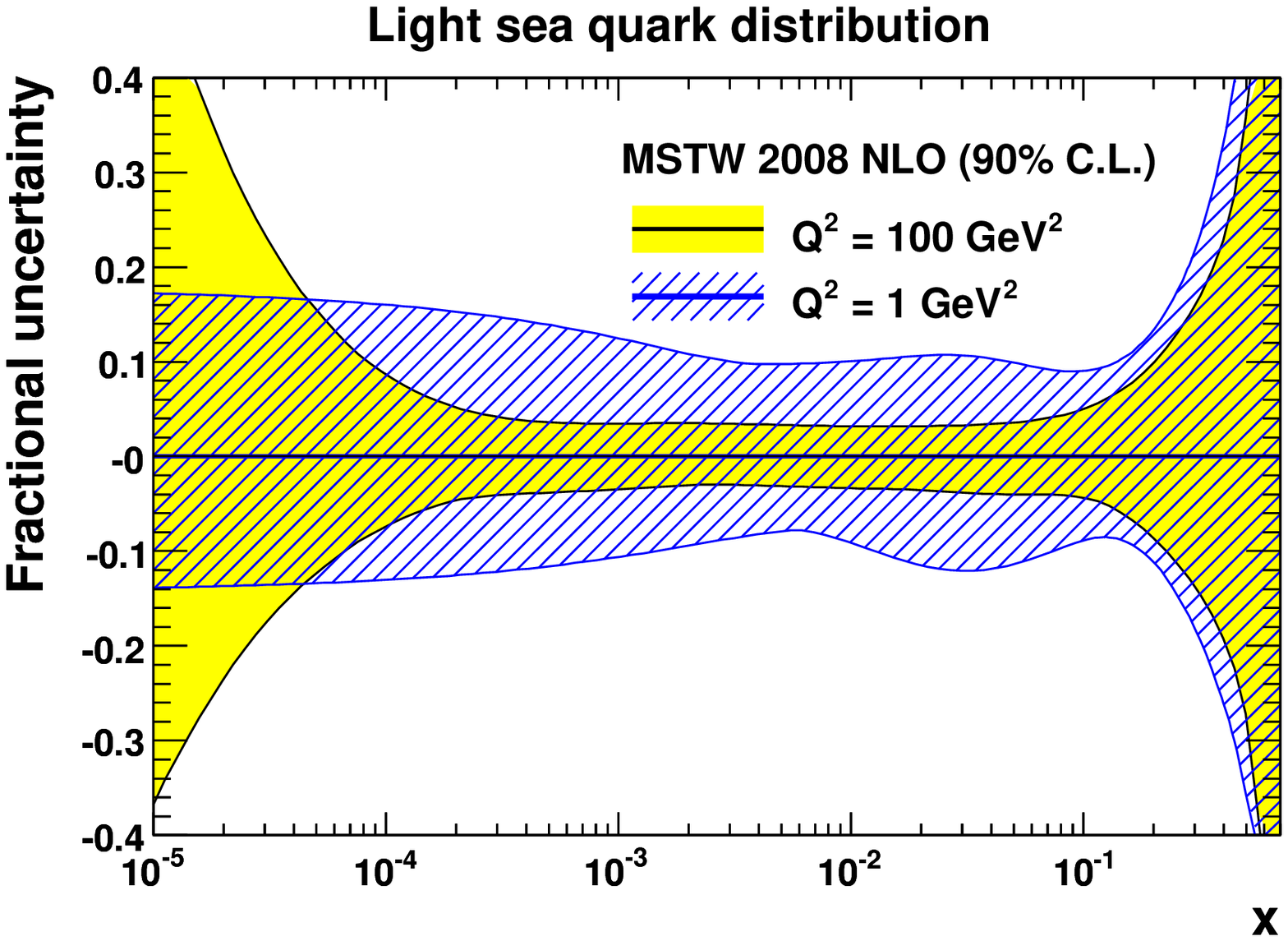}%
  \includegraphics[width=0.5\textwidth]{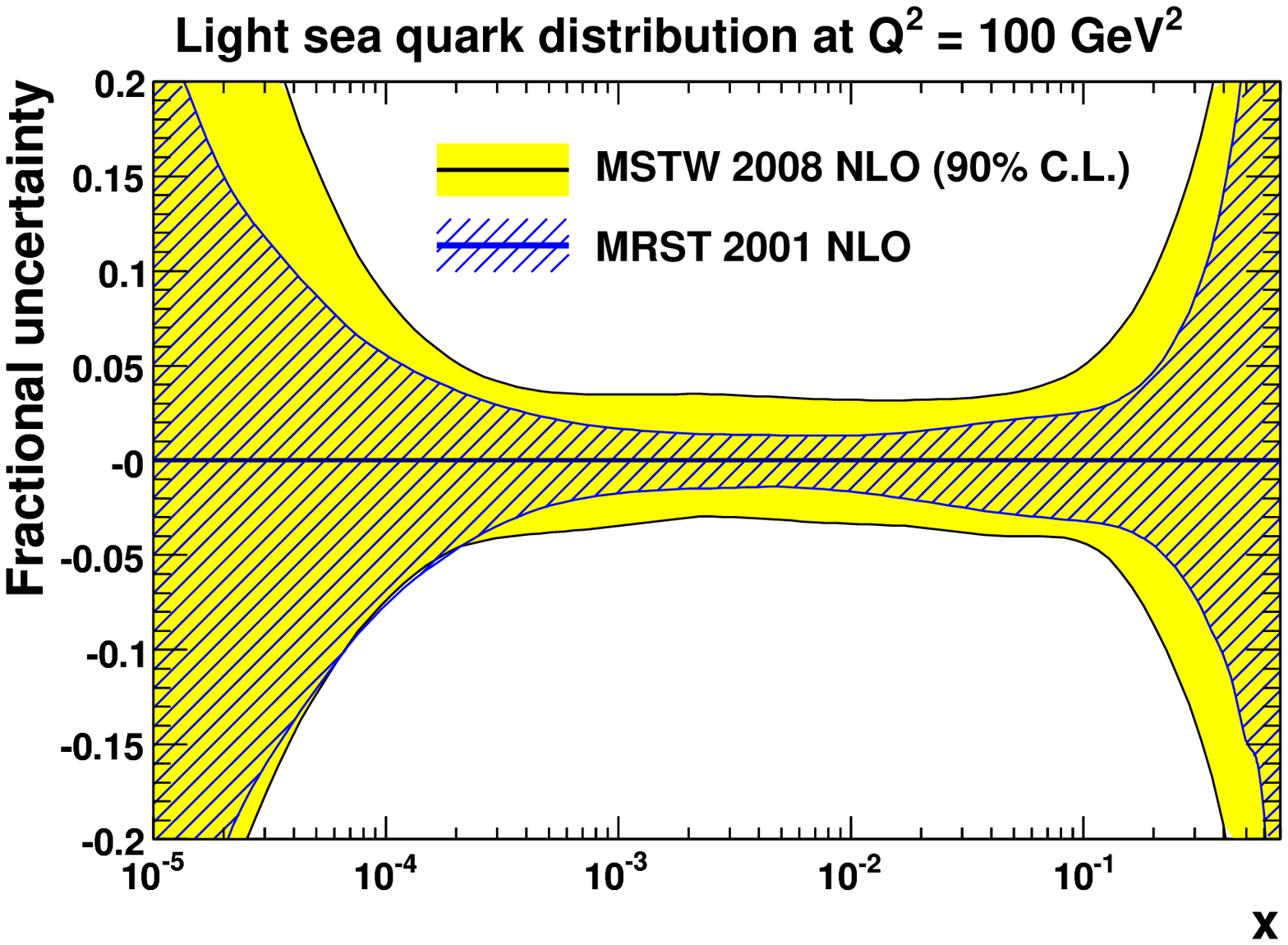}
  \caption{The fractional uncertainty on the light sea quark distribution $S\equiv 2\bar{u}+2\bar{d}+s+\bar{s}$ at $Q^2=100$ GeV$^2$ from the MSTW 2008 NLO fit compared to (a) the input sea quark distribution at $Q^2=1$ GeV$^2$ and (b) the sea quark distribution from the MRST 2001 NLO fit~\cite{Martin:2002aw}.}
  \label{fig:seauncertainty}
\end{figure}
The sea distribution $S\equiv 2\bar{u}+2\bar{d}+s+\bar{s}$ is parameterised by the standard ``MRS-type'' form,
\begin{equation}
  xS(x,Q_0^2) = A_S\,x^{\delta_S} (1-x)^{\eta_S} (1 + \epsilon_S\,\sqrt{x} + \gamma_S\,x),
\end{equation}
In this case there is no sum rule constraint, so in principle there are 5 free parameters.  Again, only 3 can contribute to the eigenvectors due to very large correlations.  We would expect these to be $\delta_S$, $\eta_S$ and $\epsilon_S$, but in practice $\delta_S$ has too large correlations with the gluon parameters, so we take $A_S$ instead.  Hence, in this case, at small $x$ the uncertainty at input is due to $A_S$, and the uncertainty $\Delta{S}(x,Q_0^2)$ is simply proportional to $S(x,Q_0^2)$.  However, at higher $Q^2$ the uncertainty is controlled by the gluon evolution which leads very quickly to a much larger uncertainty at very small $x$.  This feature is shown in Fig.~\ref{fig:seauncertainty}(a).  Perhaps we underestimate the sea uncertainty at small $Q^2$ and very small $x$ with this choice of free parameters.  However, the uncertainty does follow the data constraint, i.e.~$F_2(x,Q^2)$ constrains the quark distributions down to $x=10^{-5}$ quite precisely but only for a narrow range of small $Q^2$.  In Fig.~\ref{fig:seauncertainty}(b) we compare the sea distribution at $Q^2 = 100$ GeV$^2$ to that from the MRST 2001 analysis.  The uncertainties are larger in the new analysis mainly due to the extra freedom in the strange quark parameterisation, which feeds into the uncertainties for $\bar{u}$ and $\bar{d}$, and also due to allowing data set normalisations to go free; see Fig.~\ref{fig:fractionalNLO}.

\subsubsection*{Strange quarks}

As mentioned before we now parameterise the strange quarks separately rather than assume that
\begin{equation} \label{eq:olds+sbar}
  s(x,Q_0^2) + \bar s(x,Q_0^2) = \kappa\left[\bar{u}(x,Q_0^2)+\bar{d}(x,Q_0^2)\right],
\end{equation}
with $\kappa \approx 0.4$--$0.5$.  We could use a completely free parameterisation, but the data only exist for $x\gtrsim 0.01$ so this would lead to an enormous uncertainty for $x\lesssim 0.01$.  This is essentially what is done for the CTEQ6.6 distributions~\cite{Nadolsky:2008zw}, but it is questionable whether this is realistic.  As a rough constraint, for example, Regge trajectory considerations suggest that all flavour distributions have the same power as $x \to 0$, and we certainly think that it is sensible to impose this.  However, we apply more detailed constraints from perturbative QCD reasoning.  The strange quark has some non-insignificant mass, and we assume that it is this which qualitatively leads to the suppression compared to up and down sea quarks.  This assumption gives us a handle on the type of parameterisation to use.

When charm and bottom quarks begin their evolution they evolve like massless quarks, but always lag behind.  This leads to some suppression at all $x$ for finite $Q^2$.  But at small $x$ this suppression is roughly just a normalisation suppression, i.e.~evolution makes the small-$x$ powers $\delta_i$ the same.  Indeed, other than the overall normalisation suppression, further effects are mainly at high $x$.  This reasoning leads to us choosing only the normalisation $A_{+}$ and the high-$x$ power $\eta_{+}$ as free parameters for $s(x,Q_0^2)+\bar s(x,Q_0^2)$.  Indeed, effectively we choose a parameterisation of the form
\begin{eqnarray}
  xs(x,Q_0^2)+x\bar{s}(x,Q_0^2) &=& A_{+}\,x^{\delta_S}\,(1-x)^{\eta_{+}} (1 + \epsilon_S\,\sqrt{x} + \gamma_S\,x) \nonumber \\&\equiv& {\tilde A}\,(1-x)^{{\tilde\eta}}\,xS(x,Q_0^2),
\end{eqnarray}
where ${\tilde A}\equiv A_+/A_S$, ${\tilde\eta}\equiv \eta_+-\eta_S$ and the small-$x$ power is fixed to $\delta_S$.  Introducing distinct $\epsilon$ and $\gamma$ parameters do not improve the fit quality and lead to instability in eigenvectors, so these are also fixed to the same values as the total sea.  Therefore, apart from the normalisation $A_{+}$ and the high-$x$ power $\eta_{+}$, all other parameters in the $s+\bar{s}$ parameterisation are fixed to be the same as the total sea $S$.

\begin{figure}
  (a)\hspace{0.5\textwidth}(b)\\
  \includegraphics[width=0.5\textwidth]{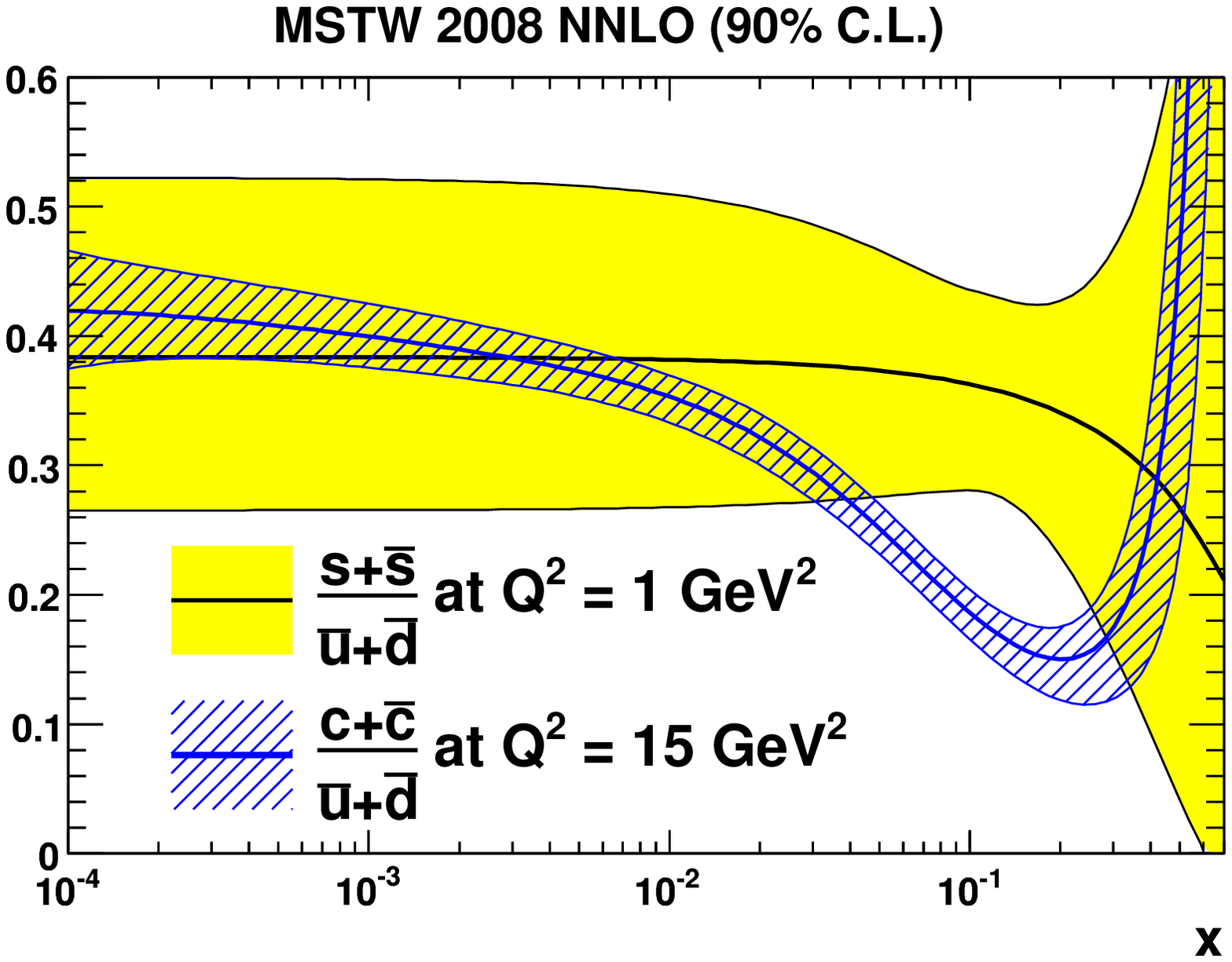}%
  \includegraphics[width=0.5\textwidth]{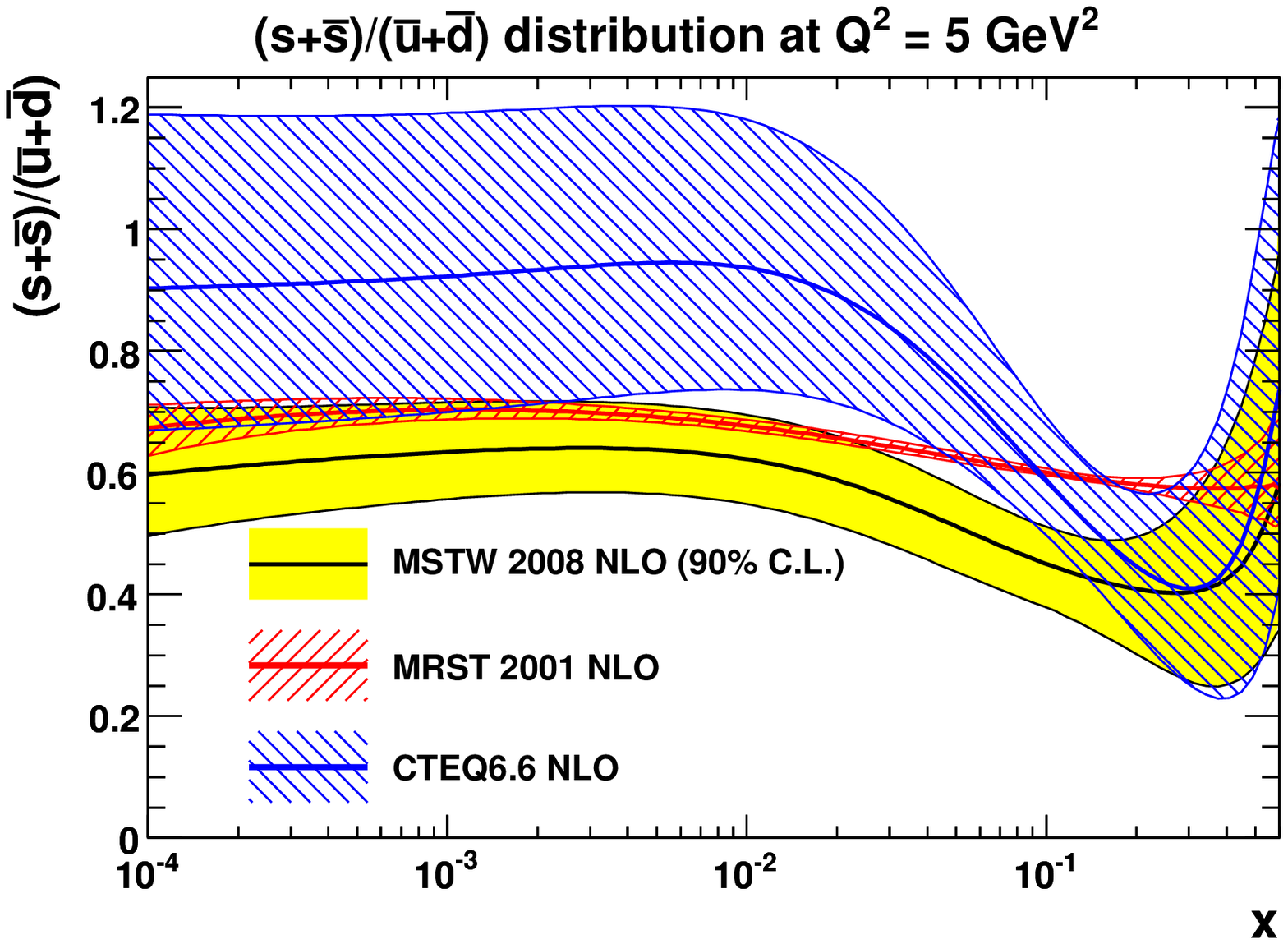}
  \caption{(a) $s+\bar{s}$ at $Q^2=1$ GeV$^2$ compared to $c+\bar{c}$ at $Q^2=15$ GeV$^2$, both as a fraction of $\bar{u}+\bar{d}$.  (b) $(s+\bar{s})/(\bar{u}+\bar{d})$ at $Q^2 = 5$ GeV$^2$ from the current NLO analysis compared to the same quantity from the MRST 2001~\cite{Martin:2002aw} and CTEQ6.6~\cite{Nadolsky:2008zw} analyses.}
  \label{fig:massratio}
\end{figure}

The results of the fit will be discussed in much more detail in Section \ref{sec:dimuon}, but at input there is about a $35\%$ normalisation suppression of the strange distribution compared to the average of $\bar{u}$ and $\bar{d}$ at $Q_0^2$ and some additional high-$x$ suppression.  Hence the suppression at the borderline nonperturbative scale $Q_0^2=1$ GeV$^2$ is now $\sim 0.3$ which is the value used in hadronisation models (i.e.~the probability to generate $s\bar{s}$ pairs compared to $u\bar{u}$ or $d\bar{d}$ pairs).  This is shown in Fig.~\ref{fig:massratio}(a) along with the suppression of $c+\bar{c}$ at $Q^2=15$ GeV$^2$, i.e.~evolved through $\sim 7$--$8$ times the mass scale.  This charm suppression is very similar to that for $s+\bar{s}$ at $Q^2=1$ GeV$^2$, implying a suppression due to evolution from slightly more than $0.1$ GeV$^2$, similar to $m_s^2$.  We would not of course expect exact correspondence, especially since the strange evolution is in a nonperturbative regime; however, the comparison is surprisingly good except that $c+\bar{c}$ is more suppressed at $x\sim 0.1$ (which is the implication for $s+\bar{s}$ from recent HERMES data on $K^{\pm}$ production~\cite{Airapetian:2008qf}).  This gives us confidence that the difference between the total sea and the strange distribution is mainly due to mass suppression, and that our limited parameterisation is well motivated.  In Fig.~\ref{fig:massratio}(b) we compare the ratio $(s+\bar{s})/(\bar{u}+\bar{d})$ at $Q^2 = 5$ GeV$^2$ obtained from the present analysis to the same quantity from the MRST 2001~\cite{Martin:2002aw} and CTEQ6.6~\cite{Nadolsky:2008zw} analyses.  The MRST 2001 ratio is slightly larger and has much smaller uncertainties, since the input was taken in the fixed form \eqref{eq:olds+sbar} with $\kappa=0.5$.  The CTEQ6.6 ratio in Fig.~\ref{fig:massratio} is much larger for $x\lesssim 0.01$, and has larger uncertainties, due to the more flexible parameterisation in a region where there are no constraining data, but is surprisingly also larger in the region $0.01\lesssim x \lesssim 0.2$ where the dimuon data should provide a constraint.

The strange asymmetry is not at all well-constrained and the parameterisation,
\begin{equation}
  xs_{v}(x,Q_0^2)\equiv xs(x,Q_0^2) - x\bar{s}(x,Q_0^2) = A_-\,x^{\delta_-}\,(1-x)^{\eta_-}\,(1-x/x_0),
\end{equation}
is the simplest we can think of with the correct features.  The number sum rule of zero strangeness \eqref{eq:numbersumrule} determines $x_0$, and hence it is a function of the other parameters.  In practice the data are not sufficient to constrain $A_-$ and $\delta_-$ independently, so we fix the latter to $\delta_-=0.2$.  Eigenvector 20 (at NLO and NNLO) consists mainly of $\eta_-$, hence its peculiar nature shown in Fig.~\ref{fig:globalchisq2}.  There is little constraint from data, but the allowed values of the parameter are limited.

\subsubsection*{Summary}
In summary, we parameterise with simple forms which do not allow bumps or shoulders in general.  Some distributions are restricted by theory assumptions, i.e.~the distributions stay extremely small ($\bar{d}-\bar{u}$) or behave like mass-suppressed quarks (strange).  We believe that we obtain a good representation of the possible uncertainty in regions where there is a data constraint, since there is an enormous amount of constraining data.  We concede that we may underestimate the uncertainty in regions of extrapolation, e.g.~the small-$x$ valence quarks, $\bar{d}-\bar{u}$ and strange distributions.  However, for some of these the uncertainty is essentially unlimited without making \emph{some} theory assumptions.

\subsection{Fit to a reduced dataset} \label{sec:reduced}

As part of the first proceedings of the HERA--LHC workshop~\cite{Dittmar:2005ed}, a ``benchmark'' PDF fit was proposed by S.~Alekhin consisting of a limited number of DIS data sets with fairly conservative cuts of $Q^2\ge9$ GeV$^2$ and $W^2\ge15$ GeV$^2$.  The original aim was to provide a test of different fitting codes starting from a common input and theoretical assumptions and fitting to the same data, analogous to the benchmark tables for PDF evolution provided by G.~Salam and A.~Vogt~\cite{Giele:2002hx,Dittmar:2005ed}.  The benchmark fit was also carried out by R.~Thorne~\cite{Dittmar:2005ed} using the MRST fitting code, albeit with some differences in the data sets used and the treatment of correlated systematic errors, but good agreement was found with the benchmark PDFs of S.~Alekhin.  The benchmark PDFs were also compared to those from the MRST 2001 NLO global fit, and significant discrepancies were found, well outside the error bands, suggesting some inconsistency of the data sets included in the global fit compared to those in the benchmark fit, and an inadequacy of the method for error propagation~\cite{Dittmar:2005ed}.  However, the benchmark PDF fit of Alekhin/Thorne used a more restrictive parameterisation for the input gluon and sea quark distributions than that used in the global fits.  Also, the one-sigma PDF uncertainties for the benchmark fit were defined by $\Delta\chi^2 = 1$, which is clearly inadequate.  The benchmark fit was repeated using the MSTW 2008 analysis framework as part of a contribution to the second HERA--LHC proceedings~\cite{Dittmar:2009ii}.  Although the discrepancy between the benchmark and global PDFs was reduced compared to the previous study~\cite{Dittmar:2005ed}, there were still sizeable discrepancies in the gluon at low-$x$ and in the down valence distribution.  It was realised that some of these discrepancies were due to different values of $\alpha_S$ used and due to the unnecessary assumption that $\bar{d} = \bar{u}$ in the benchmark fit.  It is therefore interesting to repeat the benchmark fit here, but fixing $\alpha_S$ to the value in the MSTW 2008 global fit and allowing for $\bar{d}\ne\bar{u}$.  In this way, we can investigate potential data set inconsistency and the (in)adequacy of the error propagation by fitting to a reduced number of data sets, but without any bias due to unnecessarily different input assumptions compared to the global fit.  An analogous investigation has recently been carried out by the NNPDF group in Section~5.5 of Ref.~\cite{Ball:2008by}.

We use the same data sets as specified in the MSTW version of the HERA--LHC benchmark fit~\cite{Dittmar:2009ii}, that is, we fit BCDMS data on $F_2^p$~\cite{Benvenuti:1989rh}, NMC data on $F_2^p$~\cite{Arneodo:1996qe}, NMC data on $F_2^d/F_2^p$~\cite{Arneodo:1996kd}, H1 96--97 $e^+p$ NC data on $\tilde{\sigma}$~\cite{Adloff:2000qk} and ZEUS 96--97 $e^+p$ NC data on $\tilde{\sigma}$~\cite{Chekanov:2001qu}, with cuts of $Q^2\ge9$ GeV$^2$ and $W^2\ge15$ GeV$^2$.  The heavy flavour treatment is the same as in the global fit, that is, using the GM-VFNS with $m_c = 1.4$ GeV and $m_b = 4.75$ GeV.  The strong coupling is fixed at the value obtained in the NLO global fit.  The input PDF parameterisation is taken to be the same as in the global fit, with the following exceptions.  None of the data sets included constrain strangeness, and indeed, negative strange quark distributions are obtained if allowed to go free.  Therefore, we fix the input $s_v=s-\bar{s}$ and $(s+\bar{s})/(\bar{u}+\bar{d})$ to be the same as in the global fit.

\begin{table}
  \centering
  \begin{tabular}{l||c|c}
    \hline \hline
    Data set & Reduced fit & Global fit \\ \hline
    BCDMS $\mu p$ $F_2$~\cite{Benvenuti:1989rh} & 125 / 157 & 177 / 157 \\
    NMC $\mu p$ $F_2$~\cite{Arneodo:1996qe} & 72 / 67 & 75 / 67 \\
    NMC $\mu n/\mu p$~\cite{Arneodo:1996kd} & 59 / 73 & 65 / 73 \\
    H1 96--97 $e^+p$ NC~\cite{Adloff:2000qk} & 46 / 80 & 47 / 86 \\
    ZEUS 96--97 $e^+p$~\cite{Chekanov:2001qu} & 155 / 206 & 163 / 206 \\ \hline
    All data sets & \textbf{458 / 589} & \textbf{526 / 589} \\
    \hline \hline
  \end{tabular}
\caption{$\chi^2 / N_{\rm pts.}$ for the data sets included in the reduced fit compared to the corresponding values in the global fit.}
\label{tab:reducedchisq}
\end{table}
In Table \ref{tab:reducedchisq} we compare the $\chi^2$ values for the reduced fit to the $\chi^2$ values for the same data points in the global fit.  The deterioration in the quality of the fit (total $\Delta\chi^2 = 68 \gg 1$) with the inclusion of more data sets is evident, especially for the BCDMS $F_2^p$ data.  Both the reduced fit and the global fit give an acceptable description of all data sets included in the reduced fit, and so any error propagation based on $\Delta\chi^2 = 1$ is clearly inadequate.

\begin{figure}[t]
  \includegraphics[width=\textwidth]{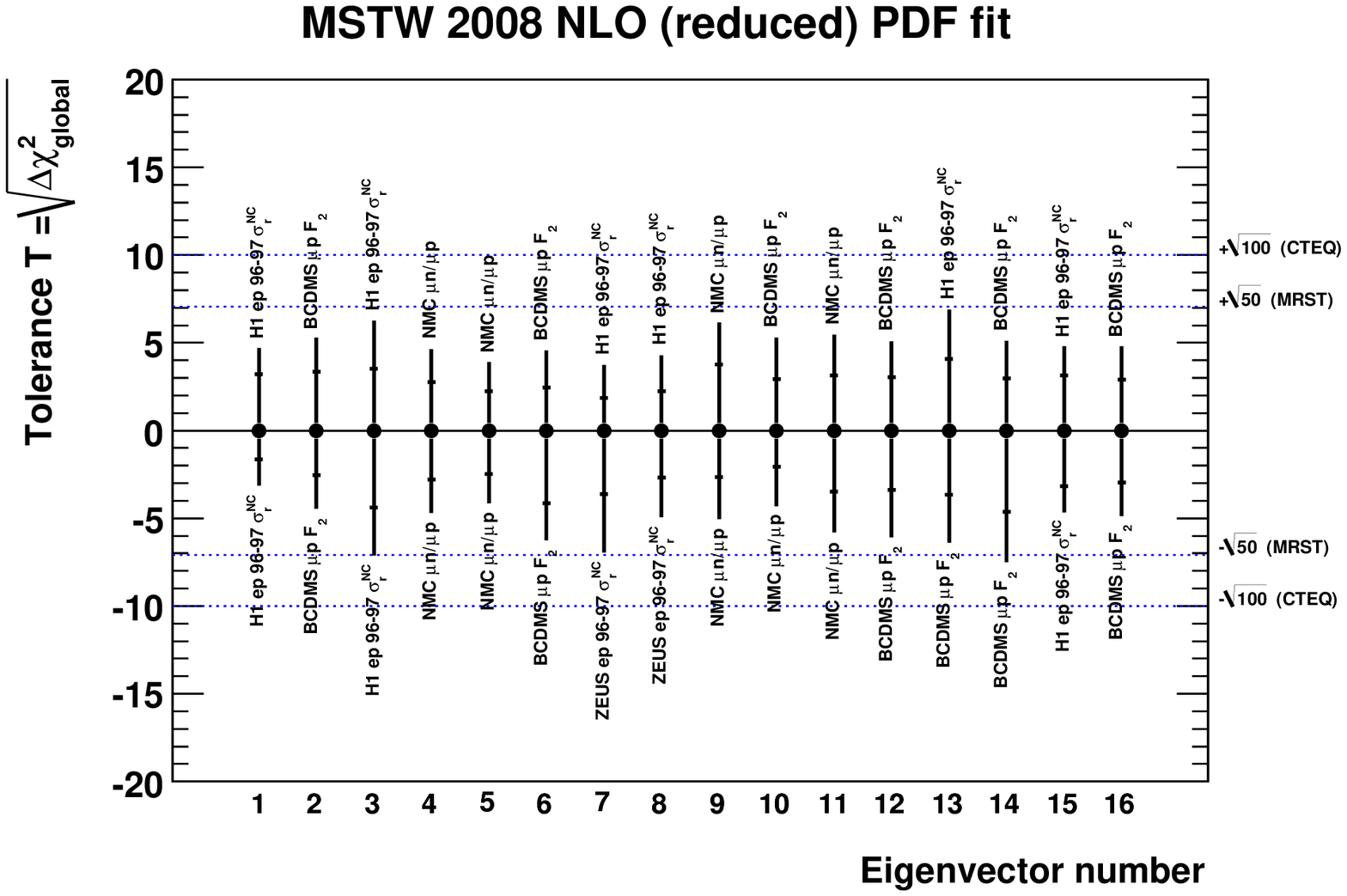}
  \caption{Tolerance for each eigenvector in the reduced fit.}
  \label{fig:reducedtolerance}
\end{figure}
Application of the dynamic tolerance procedure described above to the reduced fit gives values shown in Fig.~\ref{fig:reducedtolerance}.  Typical values for the tolerance are $T\sim 3$ for 68\% C.L.~uncertainties and $T\sim 5$ for 90\% C.L.~uncertainties, that is, slightly smaller than the typical values in the global fit, and with less variation between different eigenvector directions.  We note that the departure from ideal quadratic behaviour ($T=t$) for higher eigenvector numbers is more severe for the reduced fit than the global fit, since the PDF parameters are much less constrained in the reduced fit and hence there are more relatively ``flat'' directions in eigenvector space.

\begin{figure}
  (a)\hspace{0.5\textwidth}(b)\\
  \includegraphics[width=0.5\textwidth]{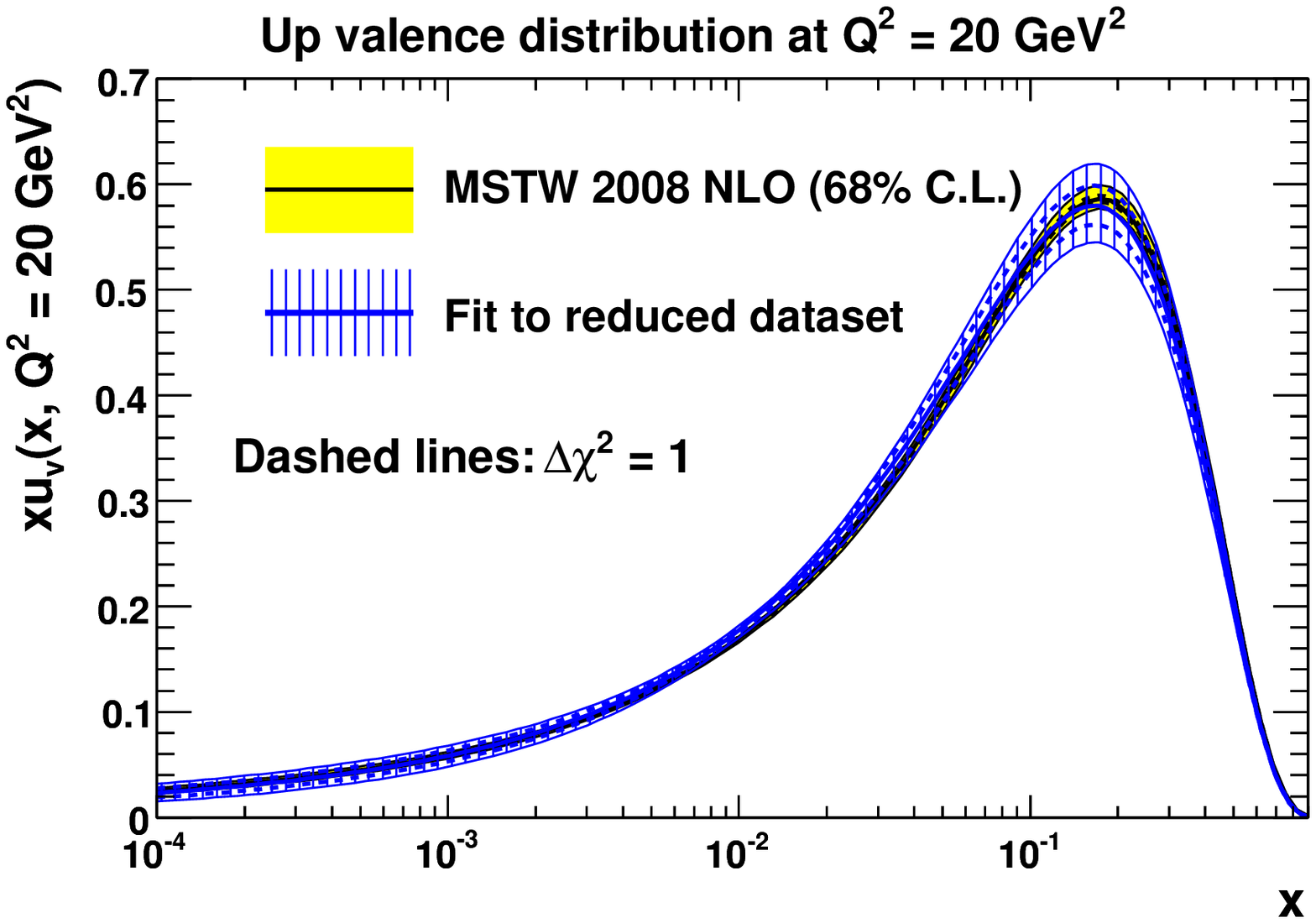}%
  \includegraphics[width=0.5\textwidth]{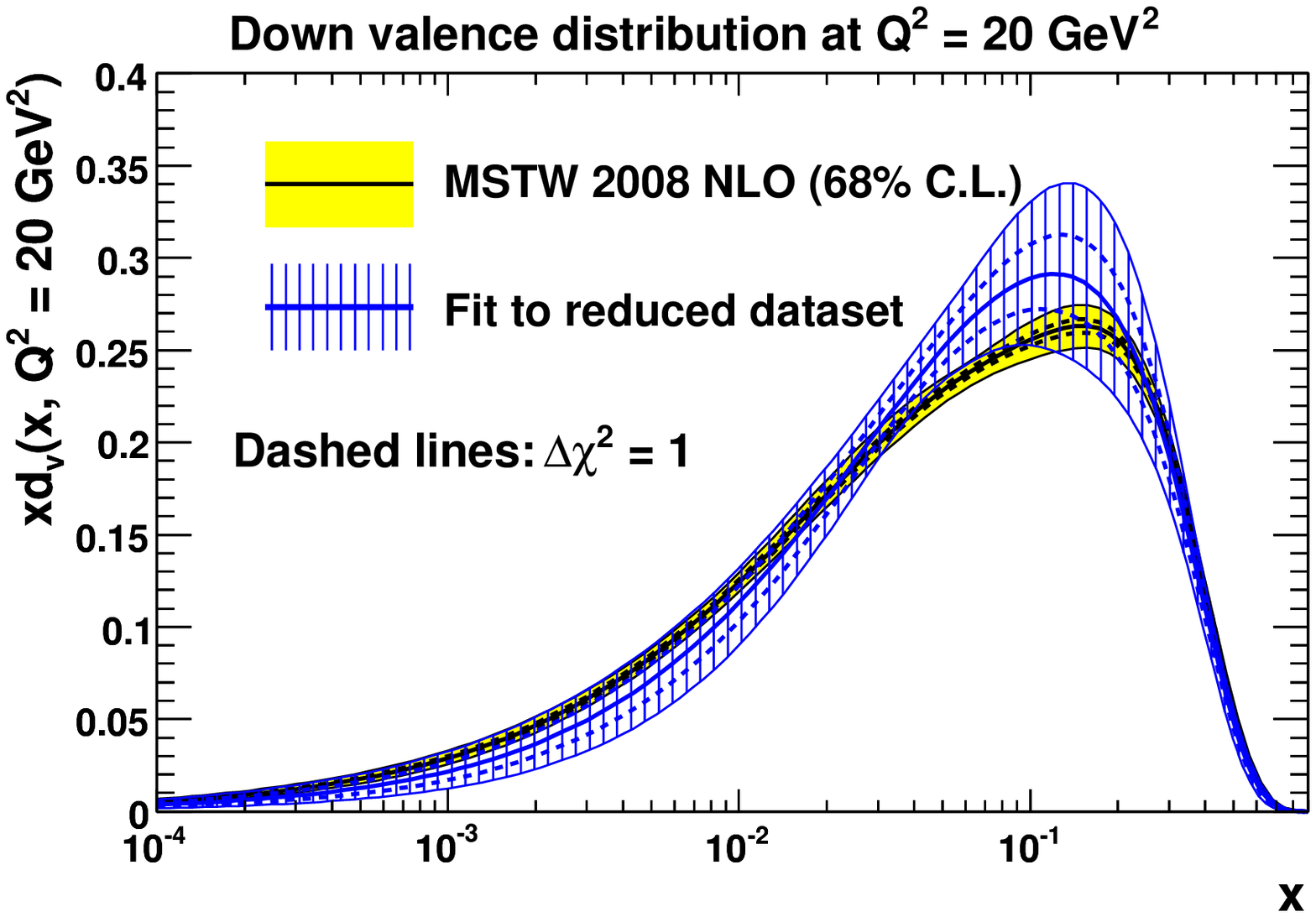}\\
  \includegraphics[width=0.5\textwidth]{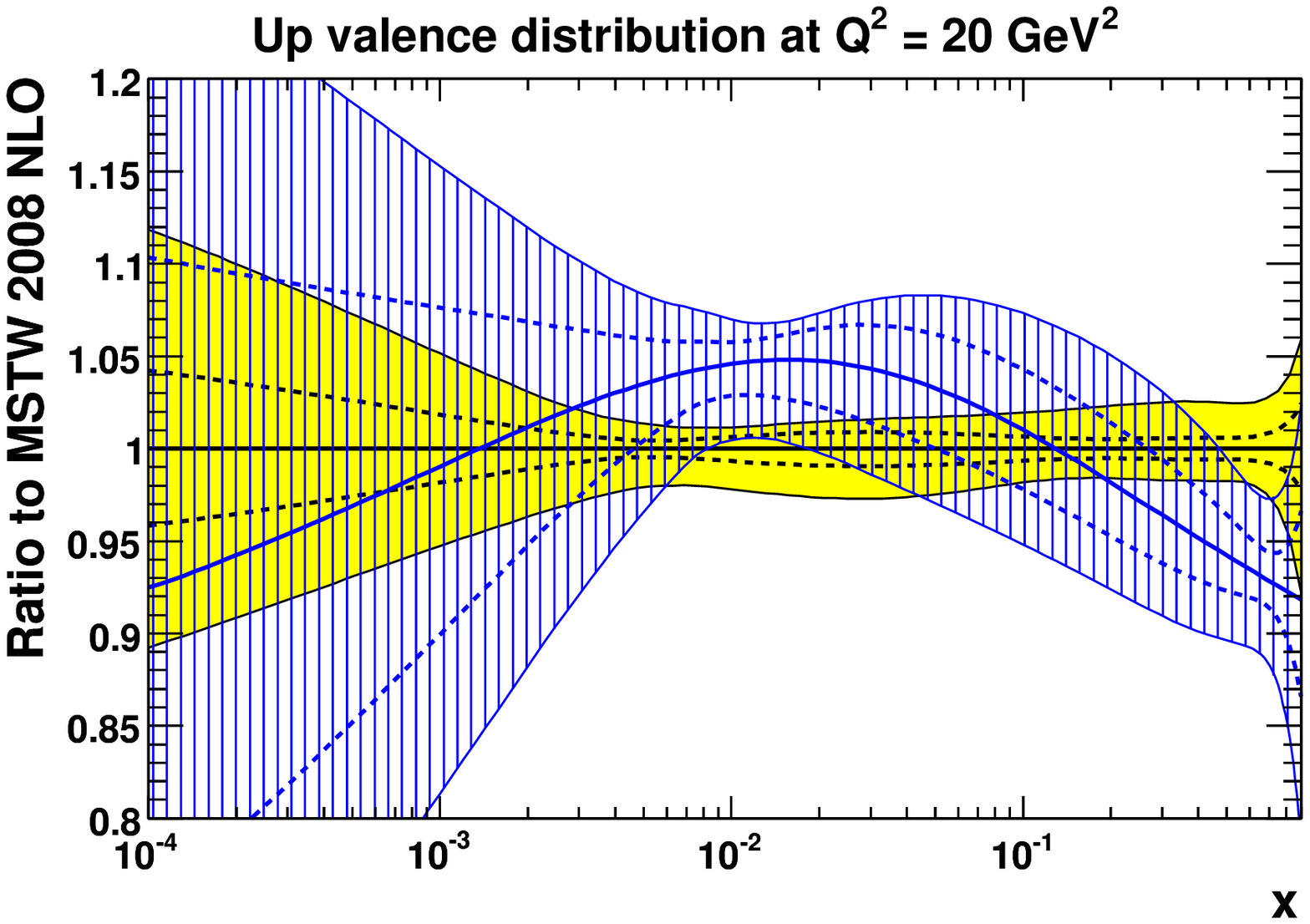}%
  \includegraphics[width=0.5\textwidth]{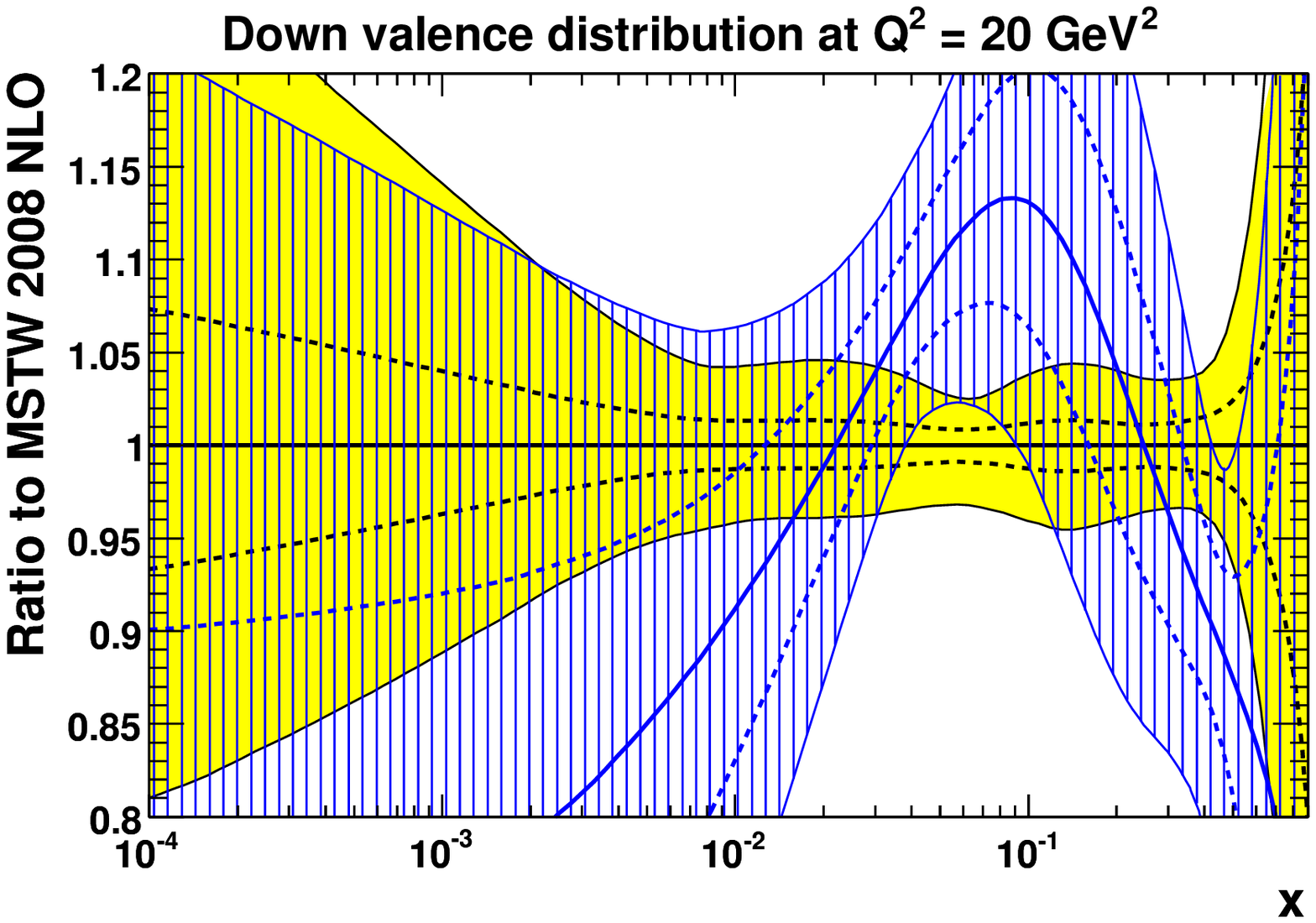}
  \caption{Comparison of the (a) up and (b) down valence quark distributions from the global and reduced NLO fits.}
  \label{fig:reducedvalenceplots}
\end{figure}
In Fig.~\ref{fig:reducedvalenceplots} we compare the $u_v$ and $d_v$ distributions at $Q^2 = 20$ GeV$^2$ from the reduced and global fits.  The dashed lines indicate the uncertainty bands obtained with $T=1$, while the outer uncertainty bands are obtained using the dynamic tolerance shown in Fig.~\ref{fig:reducedtolerance}.  For the valence quarks one can see that there are differences in detail, particularly for $d_v$.  Using uncertainties determined by $\Delta \chi^2=1$ there are still significant discrepancies in places, e.g.~$d_v$ at $x=0.05$, but using the dynamic tolerance the two fits are consistent, though the global fit leads to much smaller uncertainties, as expected.
\begin{figure}
  (a)\hspace{0.5\textwidth}(b)\\
  \includegraphics[width=0.5\textwidth]{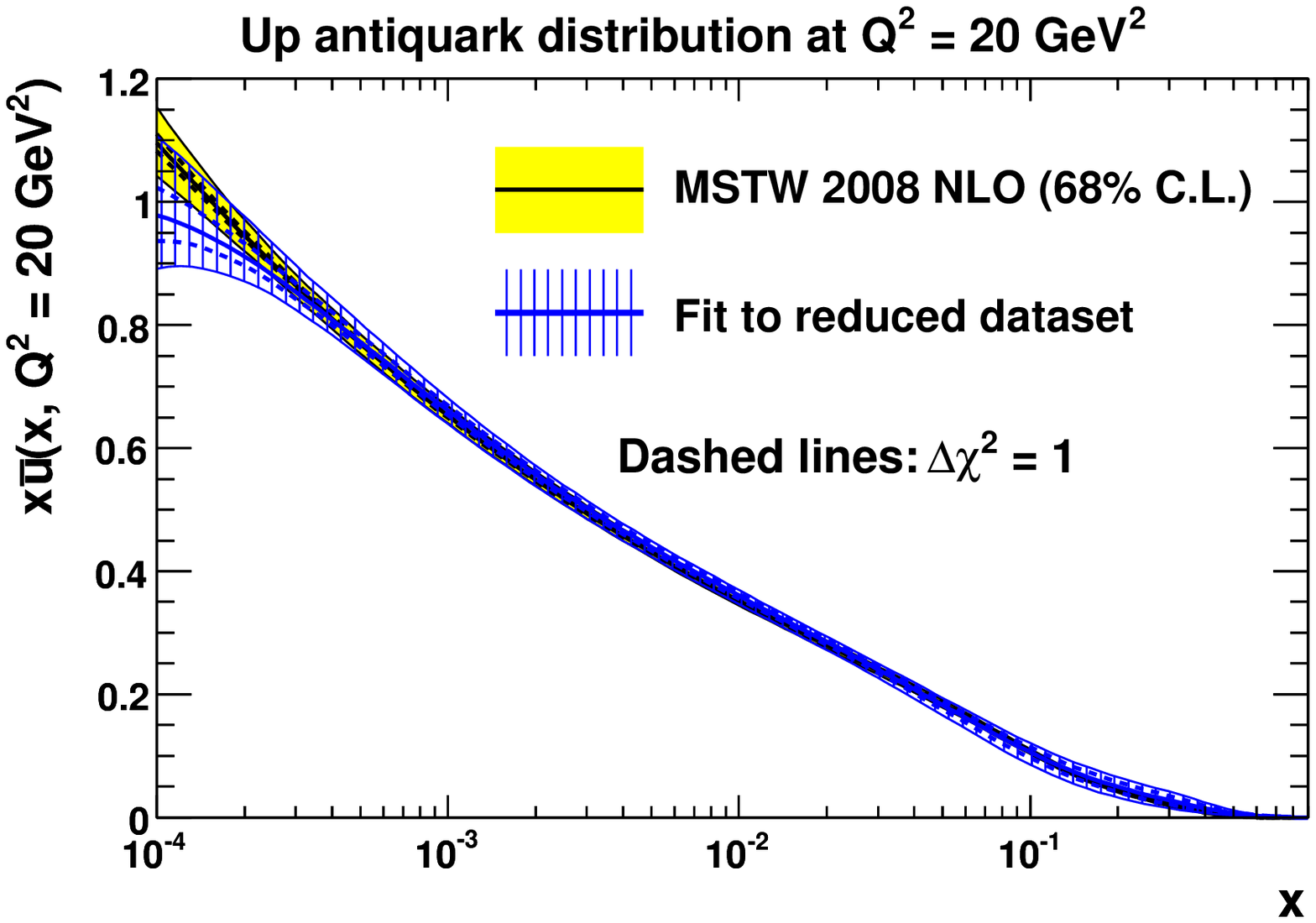}%
  \includegraphics[width=0.5\textwidth]{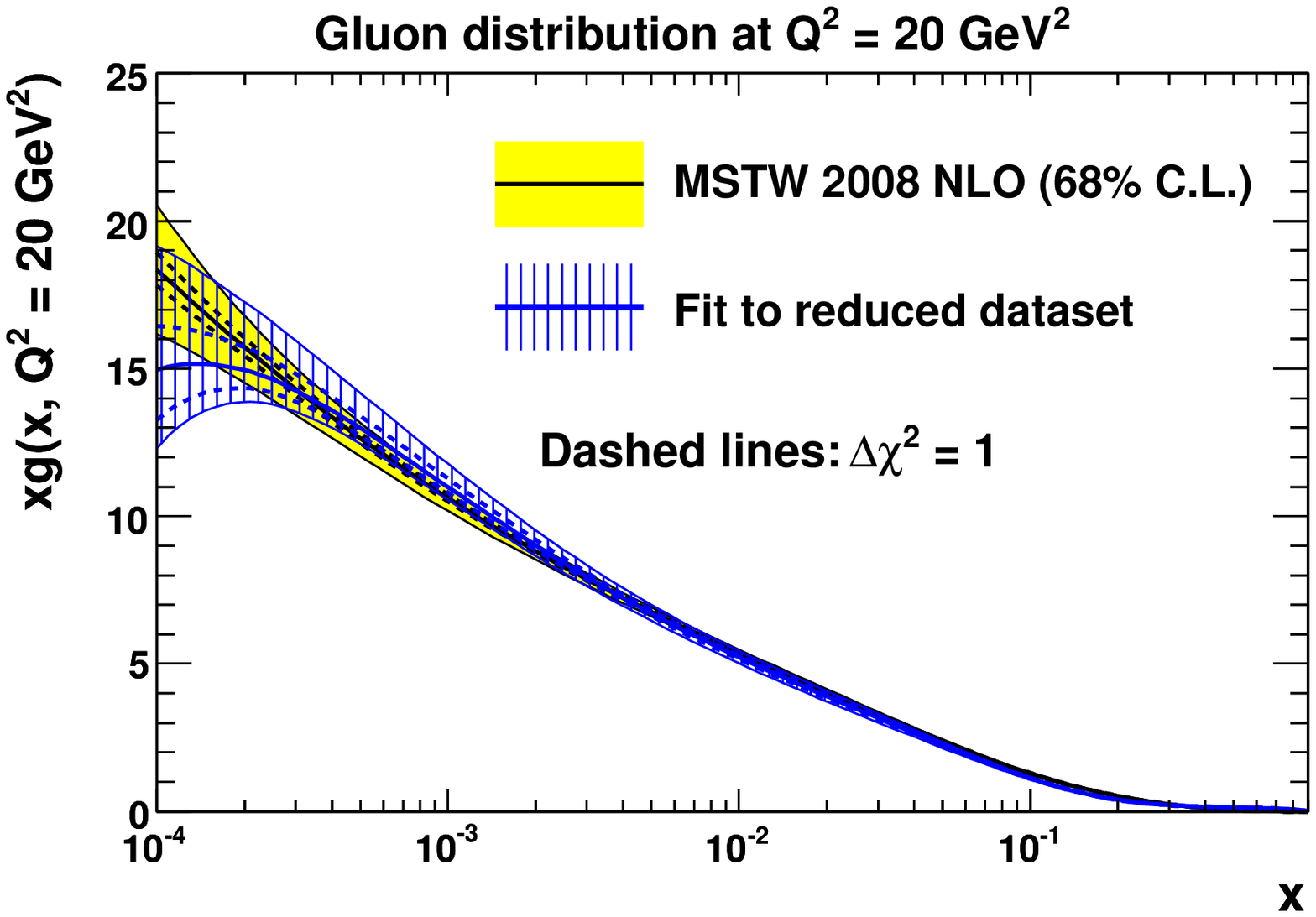}\\
  \includegraphics[width=0.5\textwidth]{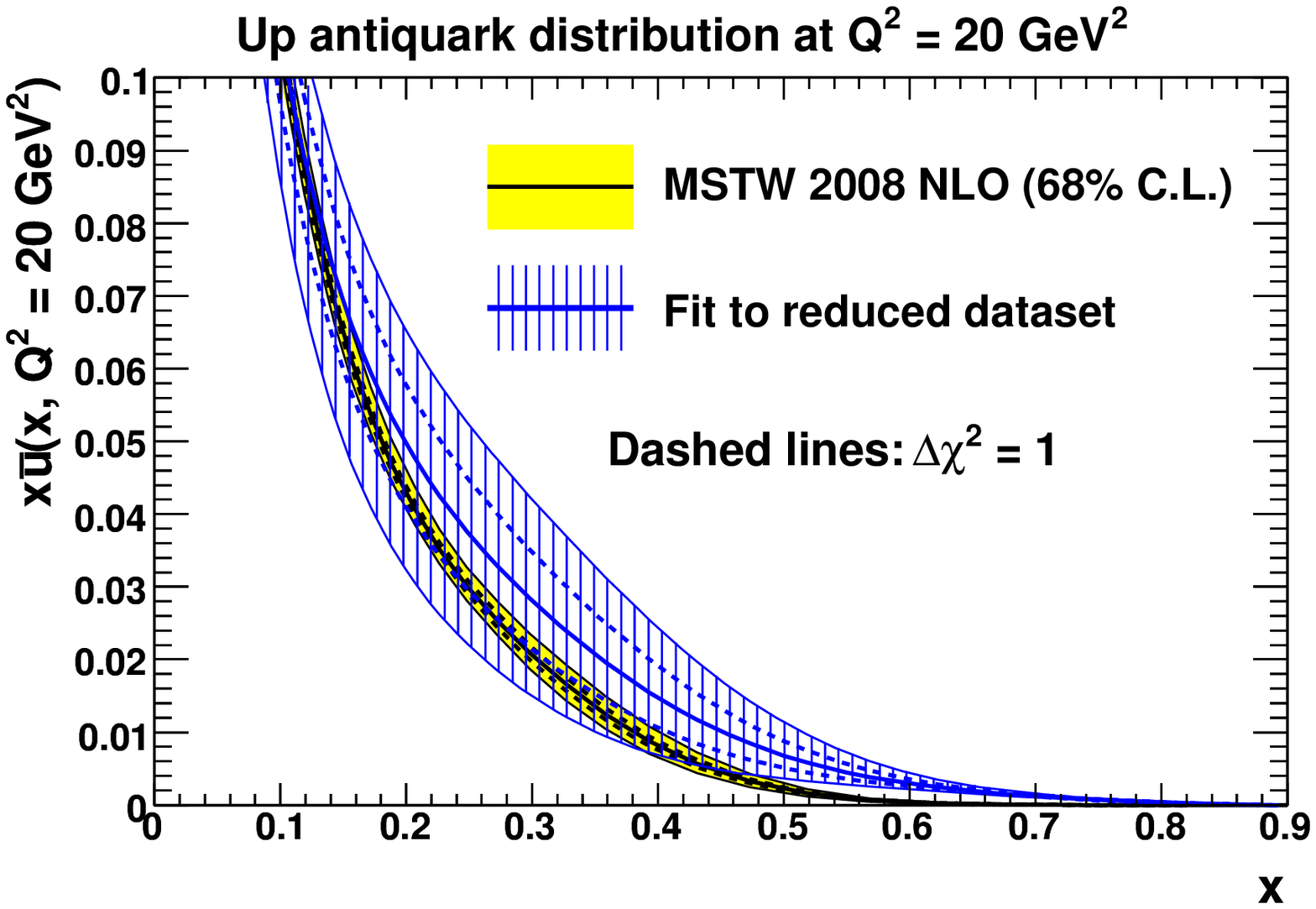}%
  \includegraphics[width=0.5\textwidth]{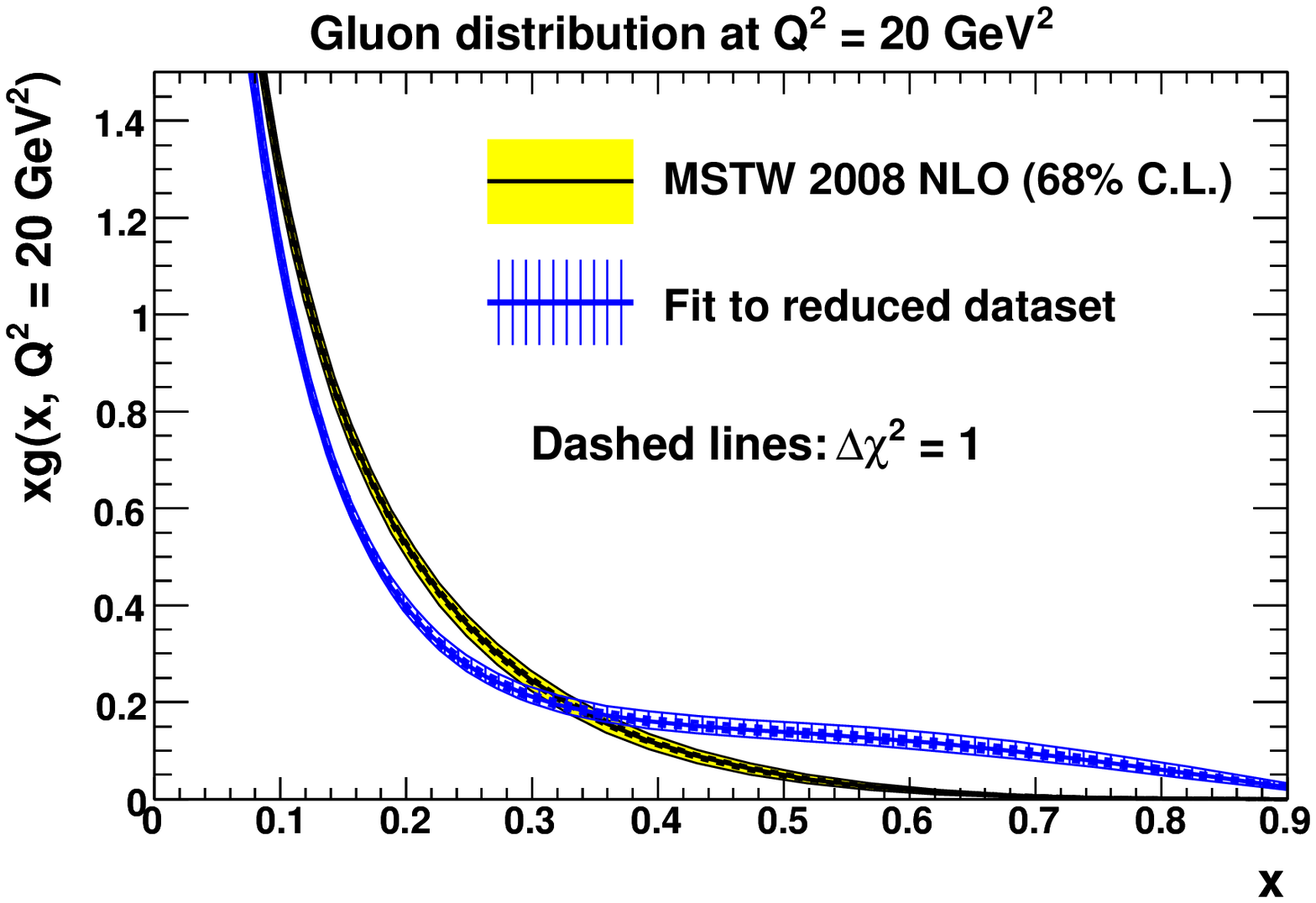}
  \caption{Comparison of (a) the $\bar{u}$ distribution and (b) the gluon distribution from the global and reduced NLO fits.}
  \label{fig:reducedseaplots}
\end{figure}
In Fig.~\ref{fig:reducedseaplots}(a) we see the curves for the $\bar{u}$ distribution, where again there is good compatibility, though the lack of Drell--Yan and neutrino DIS data in the benchmark fit leads to a huge uncertainty at high $x$.  At $x>0.5$ there is some discrepancy.  This is related to the situation for the gluon, seen in Fig.~\ref{fig:reducedseaplots}(b).  At small $x$ the two gluons agree well, again with that from the reduced fit having larger uncertainty.  At high $x$ there is some disagreement.  Using $\alpha_S(M_Z^2)\simeq 0.120$ the extremely good fit to high-$x$ BCDMS data, seen in Table \ref{tab:reducedchisq}, requires a large high-$x$ gluon in order to obtain the sufficiently flat shape in $Q^2$ preferred by the data.  This produces a pull in clear contradiction to other data, and which is overwhelmed in a global fit, but survives in the reduced fit.

Hence, we can see that the ``benchmark'' type of fit to a reduced number of data sets proposed in Ref.~\cite{Dittmar:2005ed} is a useful means of comparing different fitting approaches, as demonstrated in Refs.~\cite{Dittmar:2005ed,Dittmar:2009ii}.  However, the central values of the PDFs obtained and the uncertainties obtained using the Hessian method with the textbook $\Delta\chi^2=1$ should not be taken seriously --- hence our decision not to make these sets publicly available.  As shown above, comparison of the benchmark PDFs with those obtained from more complete fits can be useful in suggesting the reliability of more elaborate methods of determining the PDF uncertainty, giving good support for the ``dynamic tolerance'' approach introduced in this paper.  Benchmark fits can also be useful for highlighting data sets which are inconsistent with the rest of the data in a global fit.  Indeed, it is shown that there is clear tension between the BCDMS $F_2^p$ data and the rest of the data, which is a concern, although the degree of compatibility becomes better when the cut of $Q^2\geq 9$ GeV$^2$ is lowered to our standard choice of $Q^2\geq 2$ GeV$^2$.

%% file: inclusivestructurefunctions.tex
\section{Description of inclusive structure function data} \label{sec:inclusivedata}

\subsection{HERA structure functions} \label{sec:HERAinclusive}

\begin{figure}
  \centering
  \includegraphics[width=0.7\textwidth,clip]{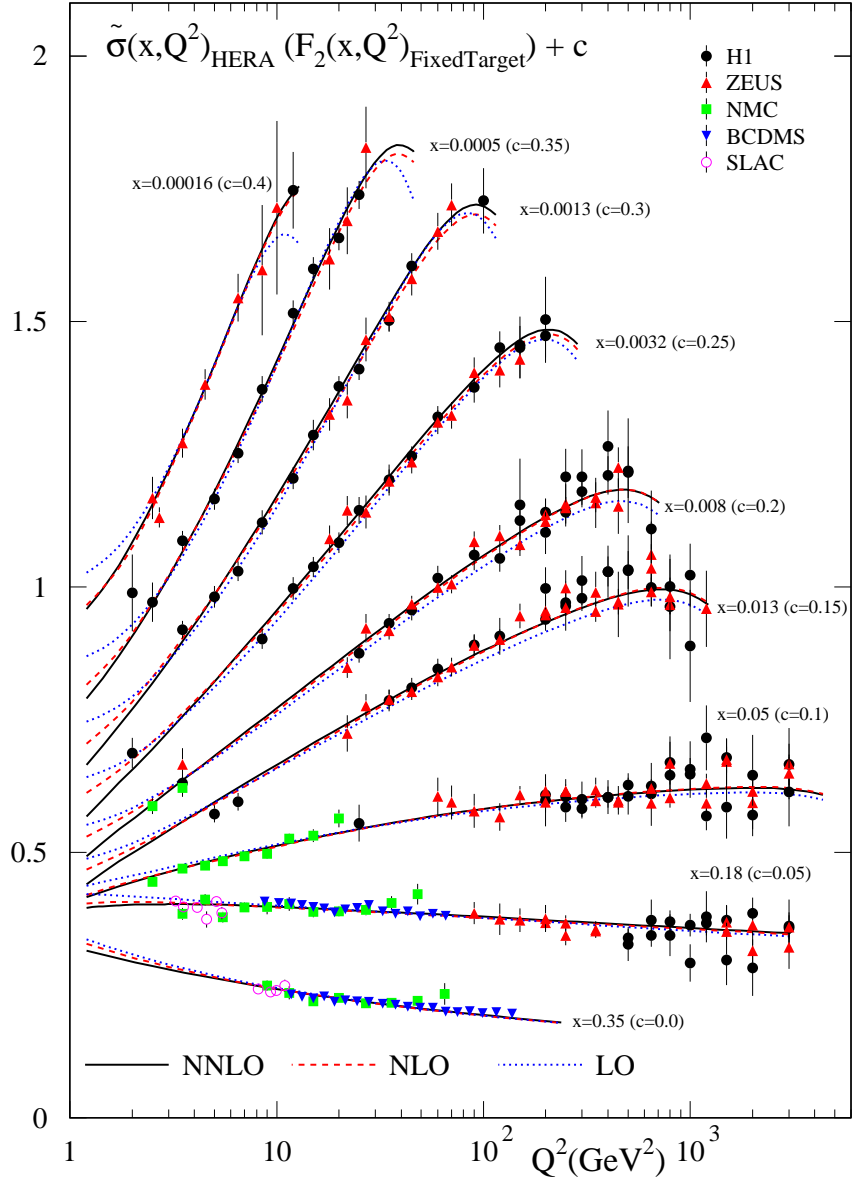}
  \caption{The reduced cross section $\tilde{\sigma}(x,Q^2)$ for a selection of HERA data and the structure function $F_2(x,Q^2)$ for a selection of fixed-target data.  The curves represent $\tilde{\sigma}(x,Q^2)$, evaluated at the HERA centre-of-mass energy, and so for the high $x$ and low $Q^2$ values corresponding to fixed-target data, $y$ is negligible and the curves effectively represent the required $F_2(x,Q^2)$.  The normalisations of data sets are those in the best fit at NNLO, found in Table \ref{tab:normalisations}.  These are very similar to those at NLO, but at LO the HERA data would be $0.5$--$1\%$ lower in general.}
  \label{fig:totf2}
\end{figure}
We show the fit to a selection of the HERA neutral-current data for the reduced cross section,
\begin{equation}
  \tilde{\sigma}(x,Q^2) = F_2(x,Q^2) - \frac{y^2}{1+(1-y^2)}\,F_L(x,Q^2),
\end{equation}
in Fig.~\ref{fig:totf2}.  This illustrates the general features, and the quality of the fit can be judged from the $\chi^2$ values in Table \ref{tab:chisquared} in Section \ref{sec:globalfit}.  At each order the quality of the fit is very similar to previous MRST fits.  A number of general features may be observed by examination of Fig.~\ref{fig:totf2}.  It is clear that the LO fit is not able to generate sufficient evolution to give a genuinely good fit, being flatter at both low and high $Q^2$ in each $x$ bin.  This is due to the presence of large positive contributions to the splitting function $P_{qg}$ beyond LO, and even the very large value of $\alpha_S$ in the LO fit cannot fully compensate.  We also note that the turnover of $\tilde{\sigma}(x,Q^2)$ at high $Q^2$ (equivalent to high $y$) in each $x$ bin is largest at LO, particularly at small $x$.  As we will see later in Section \ref{sec:longitudinal} this is because $F_L(x,Q^2)$ is very large at LO.  The NLO and NNLO fits are very similar to each other.  The main difference is that the NNLO curves are rather steeper at low $x$ and $Q^2$, being driven by an extra $\ln(1/x)$ factor in the NNLO $P_{qg}$.  Although it is difficult to see from the plot this does actually lead to a worse fit at NNLO for the H1 97 minimum bias $e^+p$ data~\cite{Adloff:2000qk}, see Table \ref{tab:chisquared}, the data preferring a slightly flatter shape.  Overall, the quality of the fit is very good.  The only flaw is a slight trend for the evolution to not persist quite strongly enough at high $Q^2$ for $x\sim 0.001$--$0.01$, although there is quite a large spread between the H1 and ZEUS data at high $Q^2$ in these $x$ bins.  This might possibly be a sign of further required corrections to the theory of some sort, and indeed an improvement due to small-$x$ resummation was shown to be possible in Ref.~\cite{White:2006yh}.

\begin{figure}
  \centering
  \includegraphics[width=0.8\textwidth,clip]{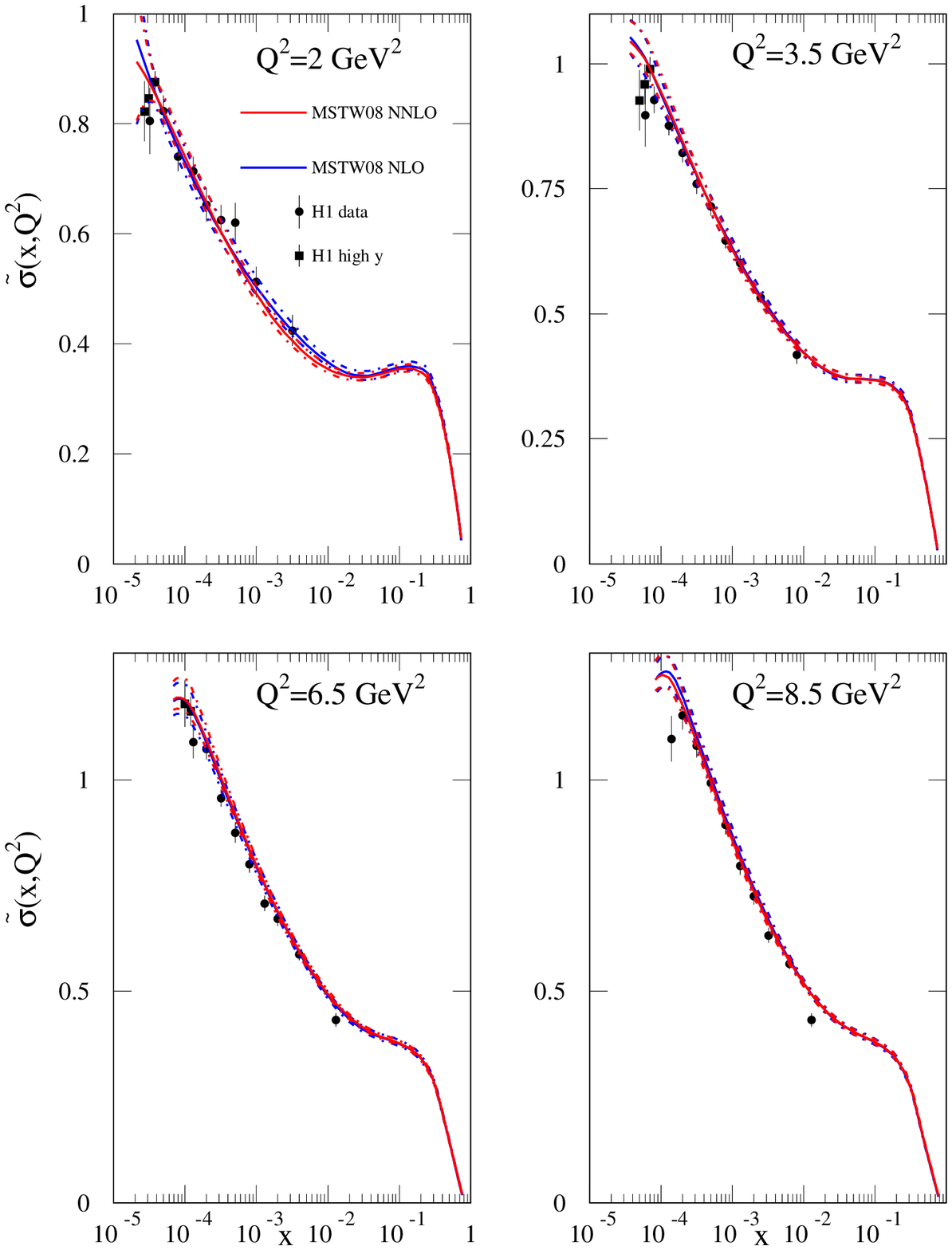}
  \caption{Predictions for the reduced cross section compared to HERA data in bins of $Q^2$.  The dashed lines show the one-sigma PDF uncertainty bands.}
  \label{fig:sigredmstw}
\end{figure}
The only HERA structure function data completely new to our analysis are the H1 MB 99 $e^+p$ data~\cite{Lobodzinska:2003yd} --- a handful of points which probe the highest $y$ values and hence display sensitivity to $F_L(x,Q^2)$.  In Fig.~\ref{fig:sigredmstw} we show our results in bins of $Q^2$ to best compare to these specific data.  The dashed lines show the one-sigma PDF uncertainty bands.  We noted in Ref.~\cite{Martin:2006qv} that while the turnover at small $x$ (high $y$) did not occur with our central NLO fit, it did occur with the central NNLO fit due to the increase of $F_L(x,Q^2)$ from the NNLO coefficient function. It is striking that the central value of the NNLO result no longer fully turns over at low $Q^2$ and small $x$.  However, from Table \ref{tab:chisquared} one can see that the fit to the H1 MB 99 $e^+p$ data is still a little better at NNLO, but also that it is not very bad at LO or NLO. The data, while being indicative, are not in practice very constraining. It is also clear from the plot that the full turnover is allowed within the uncertainties of the PDFs.  We will discuss more direct results about, and relating to, $F_L(x,Q^2)$ in Section \ref{sec:longitudinal}.

Charged-current data from $e^+p$ scattering~\cite{Adloff:2003uh,Chekanov:2003vw} are also included in the fit.  They are fit perfectly well, as seen in Table \ref{tab:chisquared}, but at present do not provide a strong constraint on the fit.  Much more precise data are expected from HERA II analyses.

\subsection{Fixed-target neutral-current structure functions} \label{sec:fixedtargetinclusive}

The fit to a selection of fixed-target data is also shown in Fig.~\ref{fig:totf2}, where in this case the data and the curve (to an extremely good approximation) represent the structure function $F_2(x,Q^2)$.  These data have appeared in numerous previous MRST analyses, and at NLO and NNLO the fit is of the same quality as previously.  These data are one of the main constraints on the value of $\alpha_S(M_Z^2)$, though it is correlated to the high-$x$ gluon.  Indeed, the value of $\alpha_S(M_Z^2)$ at NNLO is slightly smaller than in our most recent analyses~\cite{Martin:2007bv}, and so the high-$x$ gluon is correspondingly smaller.  Perhaps the most striking new feature is that at LO the value of $\alpha_S$ has increased sufficiently that the fit quality is much closer to that at NLO and NNLO, the size of the coupling largely making up for the increased evolution at higher orders in the high-$x$ regime.  Indeed, at $x=0.35$ we see that at very low $Q^2$ (below our $W^2$ cut on data) the LO curve in Fig.~\ref{fig:totf2} is falling most quickly --- a feature not seen previously.

\subsection{Neutrino structure functions} \label{sec:neutrinoinclusive}

There have been significant changes in both the data to which we fit, and the details of the procedure used, for neutrino structure functions.  The NuTeV data~\cite{Tzanov:2005kr} on $F_2\equiv F_2^{\nu N} = F_2^{\bar{\nu} N}$ and $xF_3\equiv (xF_3^{\nu N}+xF_3^{\bar{\nu}N})/2$, where $N=(n+p)/2$, replace the previous CCFR data~\cite{Yang:2000ju}.  These two experiments both use an iron target, corrected to an isoscalar target, and cover much the same kinematic range, but the NuTeV data are more precise.  The two sets agree fairly well, except in the very high-$x$ (i.e.~$x>0.5$) region, where the main source of the discrepancy is the calibration of the magnetic field map of the muon spectrometer.  In practice we find the high-$x$ NuTeV data very difficult to fit.  At high-$x$ the predictions are mainly determined by the valence up quark distribution, which is very well constrained by the neutral-current structure function data.  Given the degree of experimental uncertainty in this region, we do not fit to neutrino data with $x>0.5$.  The main information from these data comes from $x\lesssim 0.3$, where we are sensitive to the details of the different valence and sea quark contributions, which are weighted differently for neutrino structure function data compared to neutral-current data.  We also include the recent CHORUS $F_2$ and $xF_3$ data~\cite{Onengut:2005kv} which cover a similar range in $x$ but at slightly lower $Q^2$ on average, and are taken using a lead target.  These are completely compatible with the NuTeV data for $x<0.5$, but seem to lie a little lower at the highest common $x$ value of $0.65$.  We apply the same cut to the CHORUS data of $x>0.5$ for consistency.  As mentioned in Section \ref{sec:globalfit} we apply a cut of $W^2\ge25$ GeV$^2$ on the $xF_3$ data to guard against potentially larger (than $F_2$) higher-twist contributions.

\begin{figure}
  \centering
  \includegraphics[width=0.7\textwidth,clip]{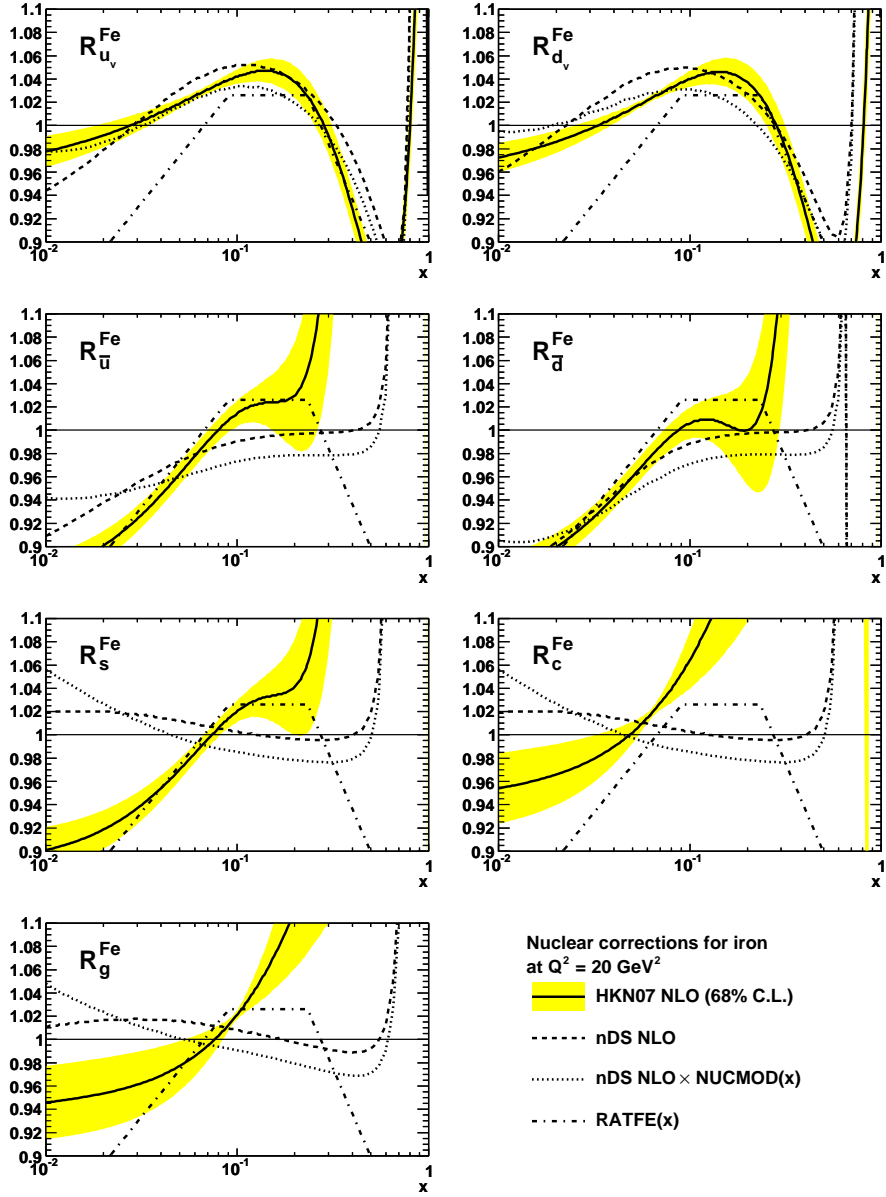}
  \caption{Nuclear corrections for iron ($A=56$, $Z=26$), at $Q^2 = 20$ GeV$^2$.  The solid lines, and the associated one-sigma uncertainty bands, are from an analysis by Hirai, Kumano and Nagai (HKN07)~\cite{Hirai:2007sx}.  The dashed lines are from an analysis by de Florian and Sassot (nDS)~\cite{deFlorian:2003qf}, and the dotted lines are multiplied by the correction factor, \eqref{eq:nucmod}, determined from our NLO global fit.  We take the nDS nuclear corrections for charm (and bottom) quarks to be the same as those for strange quarks.  The flavour- and $Q^2$-independent ``RATFE(x)'' corrections~\cite{Martin:1998sq}, shown by the dot-dashed lines, were applied in previous MRST fits.}  
  \label{fig:nuccorr}
\end{figure}
In the analysis of these data, we have adopted a more sophisticated treatment of nuclear corrections compared to previous fits.  We apply the nuclear corrections $R_f$, defined as $f^A(x,Q^2)=R_f(x,Q^2,A) f(x,Q^2)$, separately for each parton flavour $f$ using the results of a NLO fit by de Florian and Sassot (nDS)~\cite{deFlorian:2003qf}, using the GRV98~\cite{Gluck:1998xa} free proton PDFs as input; see Fig.~\ref{fig:nuccorr}.  Here, $f^A$ are defined to be the parton distributions of a \emph{proton} bound in a nucleus of mass number $A$.  In the figure we compare to a NLO analysis by Hirai, Kumano and Nagai (HKN07)~\cite{Hirai:2007sx} using the MRST98~\cite{Martin:1998sq} free proton PDFs as input, and to the simple ``RATFE'' flavour- and $Q^2$-independent  corrections~\cite{Martin:1998sq}, applied in previous MRST fits.  Since the nDS analysis~\cite{deFlorian:2003qf} neglects heavy flavours, we take the nuclear corrections for charm (and bottom) quarks to be the same as those for strange quarks.  The nDS and HKN07 nuclear corrections are similar for the up and down quarks, which are most important for the neutrino structure functions, but differ significantly for the less important gluon and other quark flavours.\footnote{The EPS09~\cite{Eskola:2009uj} corrections are in good agreement with those from nDS~\cite{deFlorian:2003qf} for valence and sea quarks.}  An alternative approach consists of extracting nuclear PDFs from data on a single nuclear target (iron)~\cite{Schienbein:2007fs}.

The nDS nuclear corrections do not have an associated uncertainty.  We assume that the uncertainty is likely to be of order a few percent, and account for it by multiplying the nuclear corrections $R_f(x,Q^2,A)$ by a flavour-independent modification function of the form:
\begin{equation} \label{eq:nucmod}
  {\rm NUCMOD}(x) = \begin{cases}(1+0.03\,r_1)[1+0.015\,r_2\,\ln^2(x_m/x)] 
&:\quad x < x_m\\ (1+0.03\,r_1)[1+0.015\,r_3\,\ln^2(x/x_m)] &:\quad x \ge 
x_m\end{cases},
\end{equation}
where $x_m = \exp(-2.5)$ is chosen to be roughly in the middle (logarithmically) of the $x$ range spanned by the neutrino data.  The parameters $r_1$, $r_2$ and $r_3$ are allowed to go free when we obtain our fits.  This function allows, for example, a little more variation than the uncertainties in the correction factors quoted in HKN07~\cite{Hirai:2007sx}.  The extent to which the central values of the nDS nuclear corrections are modified by \eqref{eq:nucmod} is shown in Fig.~\ref{fig:nuccorr}: 2--3\% is typical, and we regard this as perfectly acceptable.

\begin{figure}
  \centering
  \includegraphics[width=0.8\textwidth]{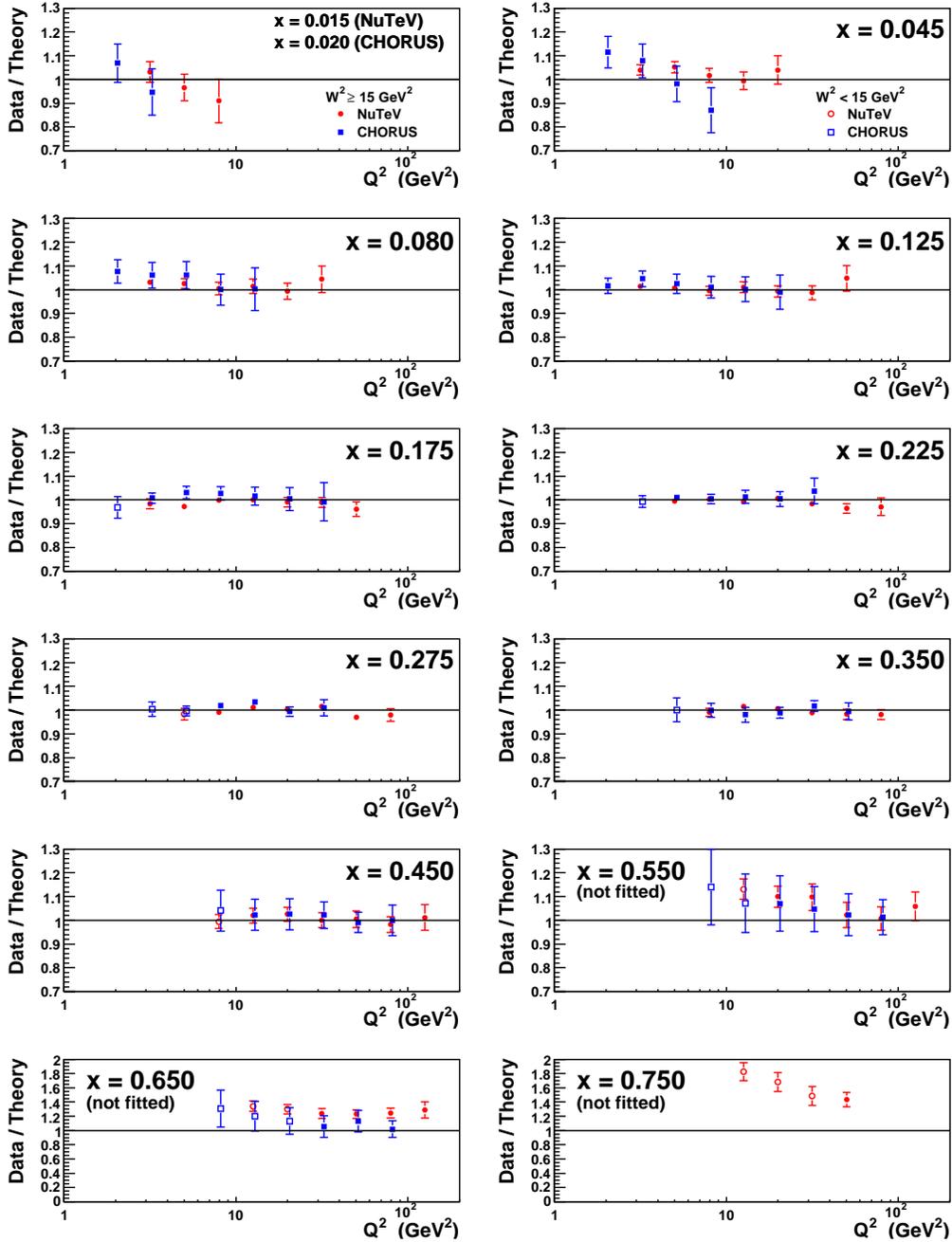}
  \caption{The quality of the NNLO fit to $F_2\equiv F_2^{\nu N} = F_2^{\bar{\nu} N}$ data from NuTeV~\cite{Tzanov:2005kr} and CHORUS~\cite{Onengut:2005kv}.  The data which fail to satisfy the $W^2$ cut and the $x<0.5$ cut are also shown.}
  \label{fig:neutrinoF2nnlo}
\end{figure}
The quality of the NNLO fit to $F_2$ data is shown in Fig.~\ref{fig:neutrinoF2nnlo}, where the data which fail to satisfy the $W^2$ cut and the $x<0.5$ cut are also shown.  The fit quality is perfectly good, as seen in Table \ref{tab:chisquared}, and the NuTeV and CHORUS data agree very well in the region of overlap, the latter lying a little lower in $Q^2$.  For $x>0.5$ there is a slight tendency for CHORUS data to lie above the prediction, which is dominated by the $u_v$ quark determined by neutral-current DIS data.  The NuTeV data, however, lie consistently well above the prediction.

\begin{figure}
  \centering
  \includegraphics[width=0.8\textwidth]{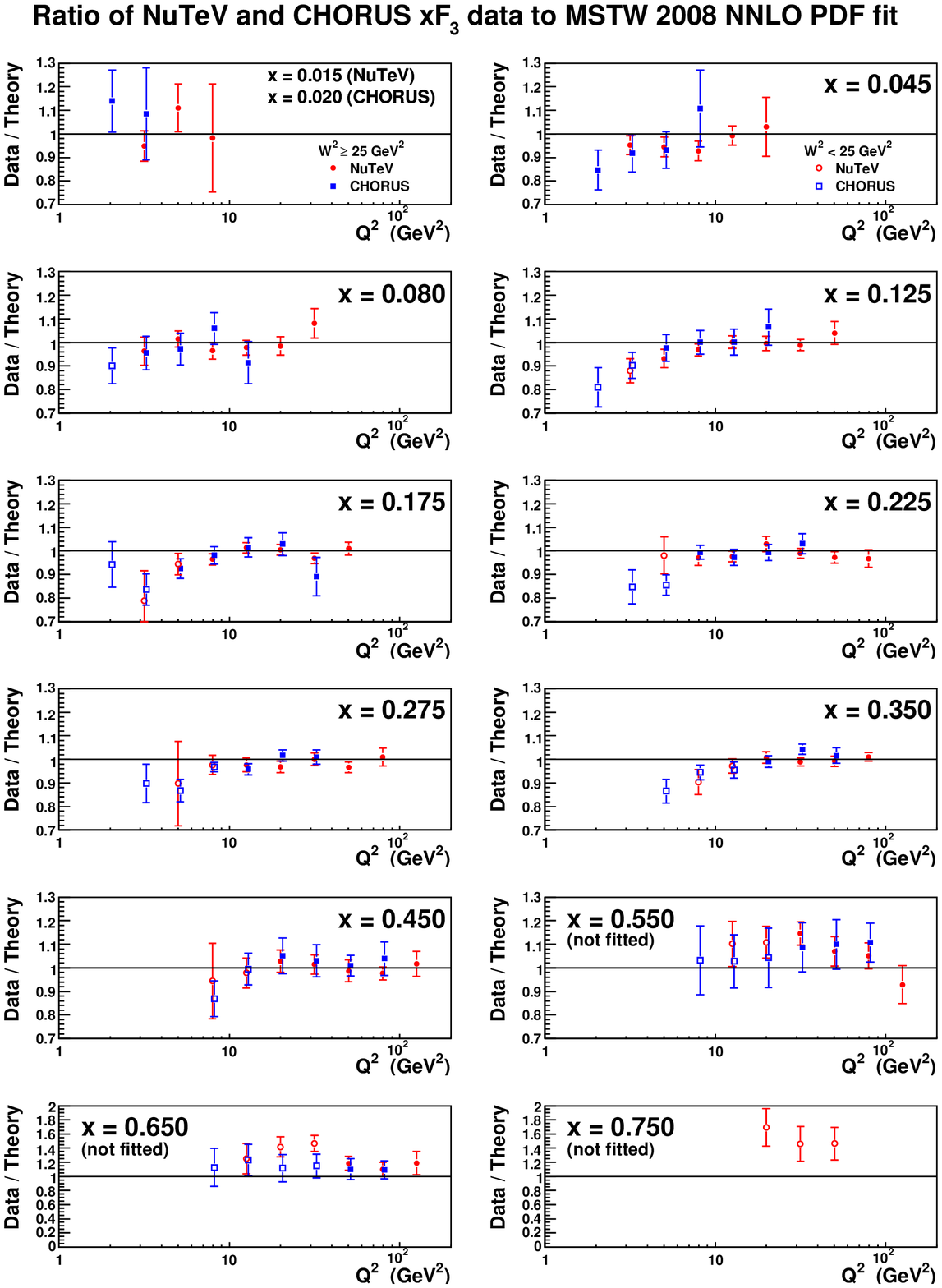}
  \caption{The quality of the NNLO fit to $xF_3\equiv(xF_3^{\nu N}+xF_3^{\bar{\nu}N})/2$ data from NuTeV~\cite{Tzanov:2005kr} and CHORUS~\cite{Onengut:2005kv}.  The data which fail to satisfy the $W^2$ cut and the $x<0.5$ cut are also shown.}
  \label{fig:neutrinoxF3nnlo}
\end{figure}
\begin{figure}
  \centering
  \includegraphics[width=0.8\textwidth]{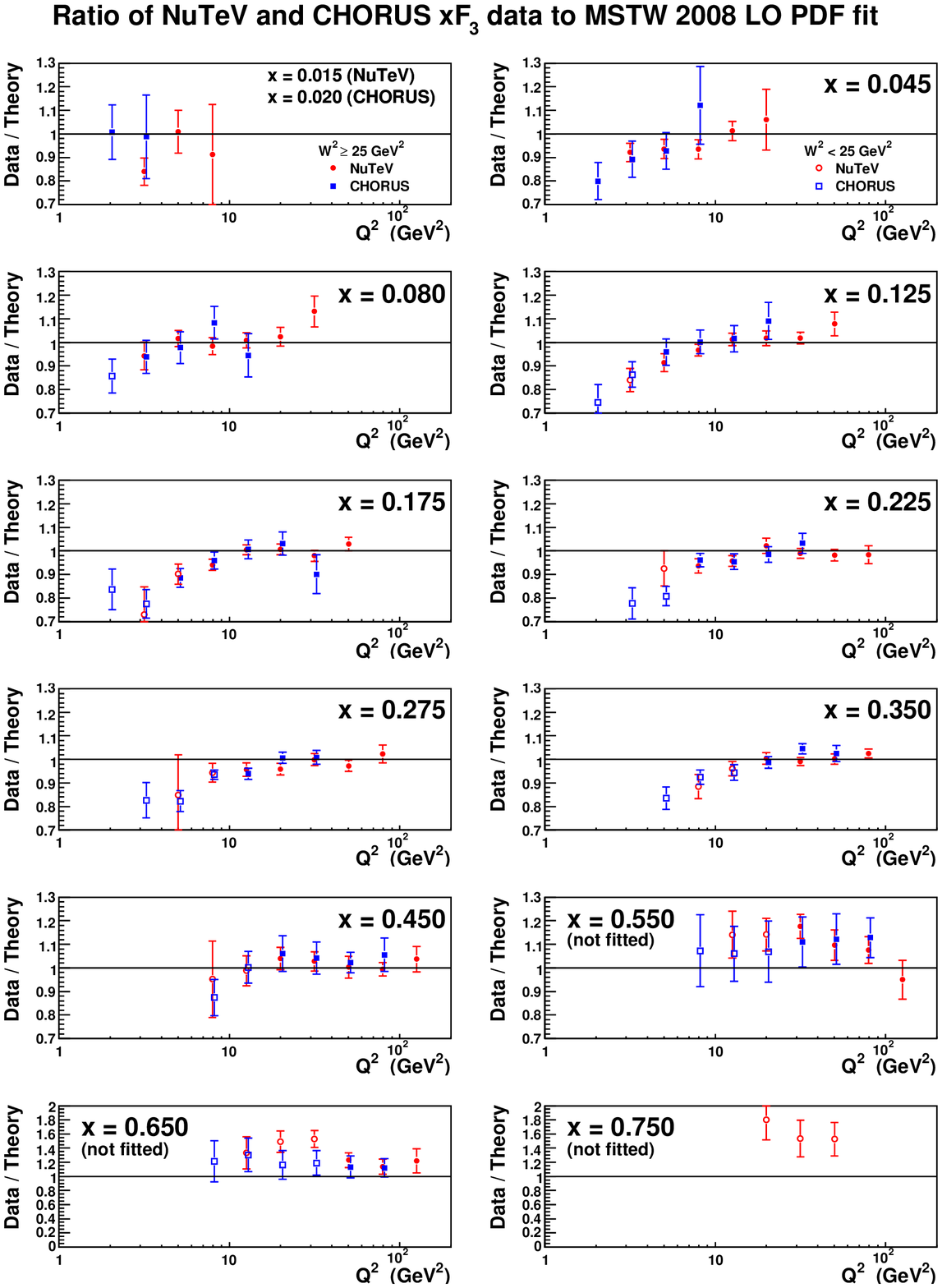}
  \caption{The quality of the LO fit to $xF_3\equiv(xF_3^{\nu N}+xF_3^{\bar{\nu}N})/2$ data from NuTeV~\cite{Tzanov:2005kr} and CHORUS~\cite{Onengut:2005kv}.  The data which fail to satisfy the $W^2$ cut and the $x<0.5$ cut are also shown.}
  \label{fig:neutrinoxF3lo}
\end{figure}
The quality of the NNLO and LO fits to $xF_3$ data is shown in Figs.~\ref{fig:neutrinoxF3nnlo} and \ref{fig:neutrinoxF3lo}, respectively.  One obvious result of the LO fit is that the $Q^2$ slope of the theory is consistently too flat, except for $x>0.275$, and the fit is correspondingly poor.  This failure is largely cured in the NNLO fit, though perhaps not completely at the lowest $x$ values.  Indeed, it is clear from the $\chi^2$ values in Table \ref{tab:chisquared} that the fit to $xF_3$ data, for both experiments, improves significantly from LO to NLO, and then even more at NNLO.  Since
\begin{equation}
  xF_3\equiv \frac{1}{2}\left(xF_3^{\nu N}+xF_3^{\bar{\nu} N}\right) = C_{3} \otimes \sum_i(q_i -\bar q_i),
\end{equation}
this is a particularly clean test of perturbative QCD (and $\alpha_S$) with no complications from mixing in parton evolution --- once the single parton combination is determined at one scale we have a complete prediction (in terms of $\alpha_S$) at higher $Q^2$.  In this case the improvements due to the change in shape with $Q^2$ of the structure function at increasing order are clearly seen.  As seen from Table~\ref{tab:chisquared}, the $\chi^2$ values for the 45 NuTeV $xF_3$ points improve from $62\to 40\to 34$ in going from LO$\to$NLO$\to$NNLO, while the $\chi^2$ values for the 33 CHORUS $xF_3$ points improve from $44\to 31\to 26$.  It is not certain that the improvement has converged at NNLO, and there is perhaps a little scope for higher-order corrections to improve matters further.  This is quite possibly the case for many quantities, but the mixing between parton flavours obscures the interpretation for most.  It is also interesting that these data favour fairly high values of $\alpha_S(M_Z^2)$, contrary to the claim often made about the results obtained from fitting structure function data alone~\cite{Alekhin:2002fv,Alekhin:2005gq,Blumlein:2006be}.  The trend of the $xF_3$ data at $x>0.5$ to lie above the theory predictions is the same as for $F_2$.  In Figs.~\ref{fig:neutrinoxF3nnlo} and \ref{fig:neutrinoxF3lo} we also show the data below our higher (than for $F_2$) $W^2\ge25$ GeV$^2$ cut.  We see that for $x<0.35$ these data do indeed lie systematically below the NNLO result, implying a large negative higher-twist contribution, which may also very slightly influence data above our cut.  We leave a systematic analysis of such higher-twist contributions to a future study.

%% file: dimuondata.tex
\section{Neutrino dimuon production: constraints on strangeness} \label{sec:dimuon}

\begin{figure}[ht]
  \centering
  \includegraphics[width=0.5\textwidth,clip]{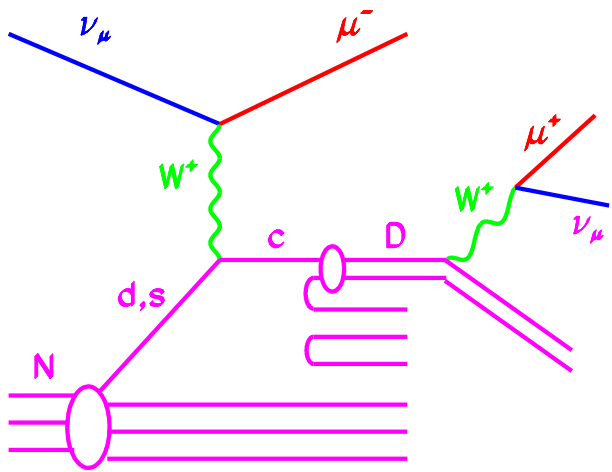}
  \caption{Dimuon production in neutrino--nucleon DIS from charm production via scattering of the $W^+$ boson off a strange or down quark from a nucleon in the heavy nuclear target $N$.  Diagram taken from Ref.~\cite{Goncharov:2001qe}.}
  \label{fig:dimuondiag}
\end{figure}

Our new analysis allows much more flexibility for the strange quark and antistrange quark distributions.  This is both for the total and the two separate contributions.  Protons have no valence strange quarks, that is,
\begin{equation} \label{eq:sval}
  \int_0^1\!{\rm d}x\;\left[s(x,Q^2)-\bar{s}(x,Q^2)\right] = 0.
\end{equation}
But this does \emph{not} necessarily mean that $s(x,Q^2) = \bar{s}(x,Q^2)$ for all $x$.  For example, in a meson cloud model, the proton can fluctuate into a $\Lambda(uds)$ and a $K$ or $K^*(u\bar{s})$, where the $\bar{s}$ in the $K$ or $K^*$ carries a larger momentum fraction of its parent hadron than the $s$ quark in the $\Lambda$.  Nevertheless, the assumption in earlier PDF fits has been that the input distributions satisfy
\begin{equation}
  s = \bar{s} = \frac{\kappa}{2} (\bar{u} + \bar{d}),
\end{equation}
with a constant $\kappa\approx 0.4$--$0.5$, justified by a simplified analysis of dimuon data by the CCFR experiment~\cite{Bazarko:1994tt}.  Updated CCFR and NuTeV dimuon cross sections are now available, therefore we include these data in the global fit to constrain $s$ and $\bar{s}$ separately; see, for example, the process $\nu_{\mu}N \to \mu^+\mu^-X$ proceeding via the subprocess $W^+s\to c$ shown in Fig.~\ref{fig:dimuondiag}.  Previous studies have been made in Refs.~\cite{Mason:2006qa,Olness:2003wz,Lai:2007dq,Alekhin:2008mb}.

\subsection{Description of dimuon cross sections}

The CCFR and NuTeV dimuon cross sections, $\nu_{\mu}N \to \mu^+\mu^-X$, can be related to the underlying process, $\nu_{\mu}N \to \mu^-cX$, using
\begin{equation}
  \frac{{\rm d}^2\sigma}{{\rm d}x\,{\rm d}y}(\nu_\mu N \to \mu^+\mu^- X) = 
B_c\,\mathcal{A}\,\frac{{\rm d}^2\sigma}{{\rm d}x\,{\rm d}y}(\nu_\mu N \to \mu^- c\,X),
\end{equation}
and similarly for the antineutrino data.  Here, $B_c=0.099\pm0.012$ is the charm semileptonic branching ratio~\cite{Mason:2006qa}.  The kinematic acceptance correction $\mathcal{A}$ accounts for the 5 GeV cut on the energy of the charm decay muon and was calculated~\cite{Mason:2006qa} for each bin using the \textsc{disco} Monte Carlo simulation~\cite{Kretzer:2001tc} assuming a charm mass of 1.4 GeV and Collins--Spiller~\cite{Collins:1984ms} fragmentation parameter $\epsilon=0.6$.  Allowing for the CKM mixing, the cross section has the $\mathcal{O}(\alpha_S^0)$ form:
\begin{equation}
  \frac{{\rm d}\sigma}{{\rm d}x\,{\rm d}y}(\nu_\mu N \to \mu^- c\,X) \propto |V_{cs}|^2
\xi s^A(\xi,Q^2) + |V_{cd}|^2 \xi\left[\frac{Z}{A} d^A(\xi,Q^2) 
+ \left(1-\frac{Z}{A}\right) u^A(\xi,Q^2)\right],
\end{equation}
where the rescaling variable $\xi\equiv x(1+m_c^2/Q^2)$, and where $Z = 23.403$ and $A = 49.618$ for the CCFR/NuTeV target~\cite{Mason:2006qa}.  Thus, the $\nu_{\mu}$ and $\bar{\nu}_\mu$ dimuon cross sections are direct measures of $s$ and $\bar{s}$, respectively.  We have assumed isospin symmetry, $d^n = u^p\equiv u$, and, as usual, have taken the renormalisation and factorisation scales to be $\mu_R=\mu_F=Q$.  Finally, just as for the inclusive neutrino data, we apply the nuclear corrections of de Florian and Sassot (nDS)~\cite{deFlorian:2003qf} for each parton flavour, and hence relate the nuclear PDFs, $f^A(x,Q^2)$, to the proton PDFs, $f(x,Q^2)$; see Section \ref{sec:neutrinoinclusive}.  As evident from the difference between the nDS~\cite{deFlorian:2003qf} and HKN07~\cite{Hirai:2007sx} parameterisations for nuclear corrections to strange quarks shown in Fig.~\ref{fig:nuccorr}, the uncertainty is likely to be a few percent.  However, this uncertainty is smaller than the experimental uncertainty on $s+\bar{s}$ at the relatively low scales where the data are probed.  Additionally, there is no evidence for any systematic problem with shape in the fit quality.

It is convenient to parameterise $s \pm \bar{s}$, rather than $s$ and $\bar{s}$ separately, as explained in Section \ref{sec:erroranalysis}.  Recall that at the input scale $Q^2_0=1$ GeV$^2$ we take
\begin{align}
  x(s+\bar{s}) & = A_{+}\,x^{\delta_S}\,(1-x)^{\eta_{+}} (1 + \epsilon_S\,\sqrt{x} + \gamma_S\,x),\\
  x(s-\bar{s}) & = A_{-}\,x^{\delta_{-}} (1-x)^{\eta_{-}} (1-x/x_0), \label{eq:ssm}
\end{align}
where the parameters $\delta_S,~ \epsilon_S$ and $\gamma_S$ are in common with the parameterisation of the total light quark sea, $S$, see \eqref{eq:S}.  The final factor in \eqref{eq:ssm} is to ensure that the number sum rule \eqref{eq:sval} is satisfied, and implies that $(s-\bar{s})$ changes sign at $x=x_0$.

\begin{figure}
  \centering
  \includegraphics[width=0.8\textwidth,clip]{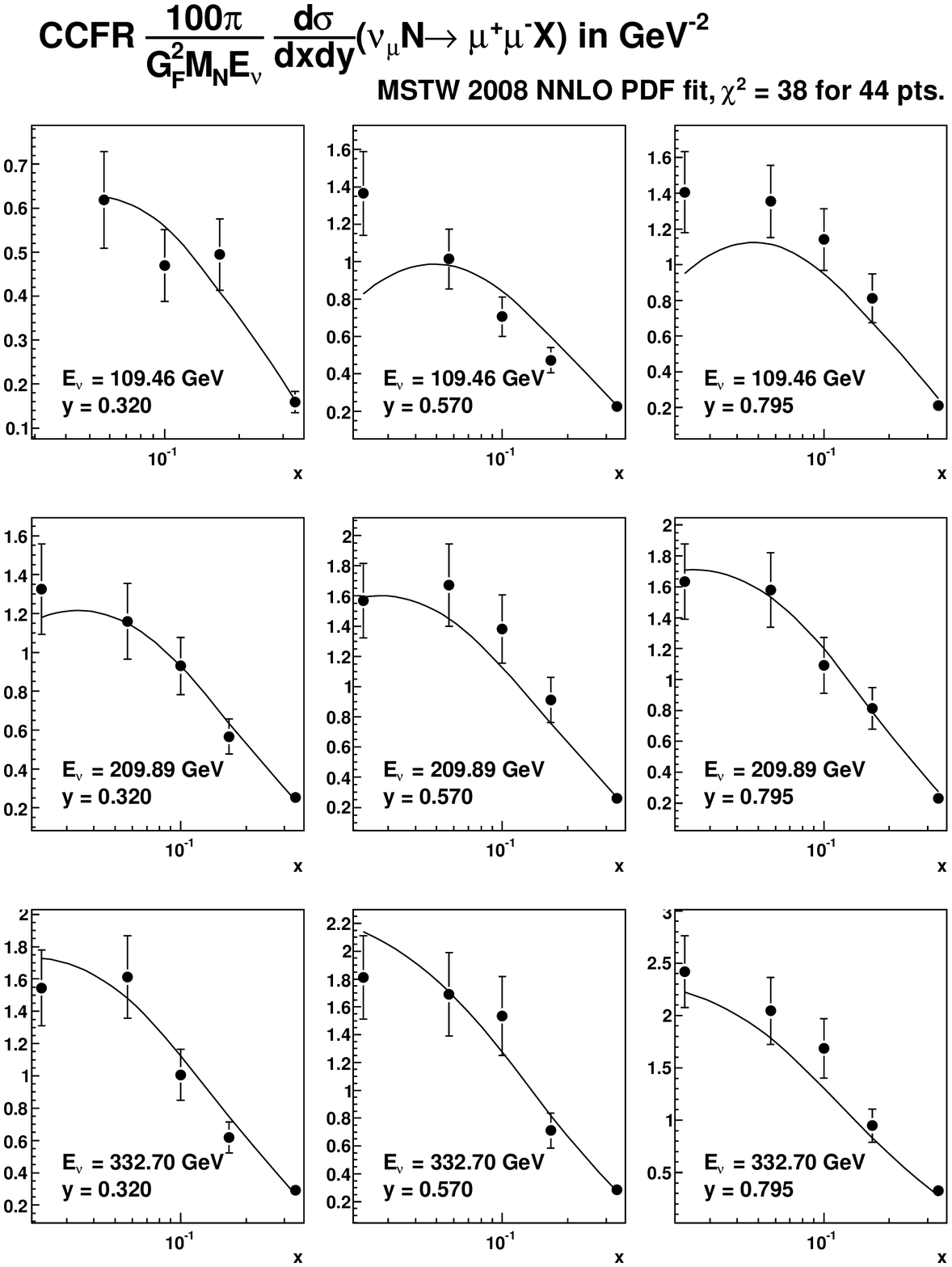}
  \caption{The quality of the NNLO fit to the CCFR neutrino-initiated dimuon production.}
  \label{fig:ccfrdimuonnu}
\end{figure}
\begin{figure}
  \centering
  \includegraphics[width=0.8\textwidth,clip]{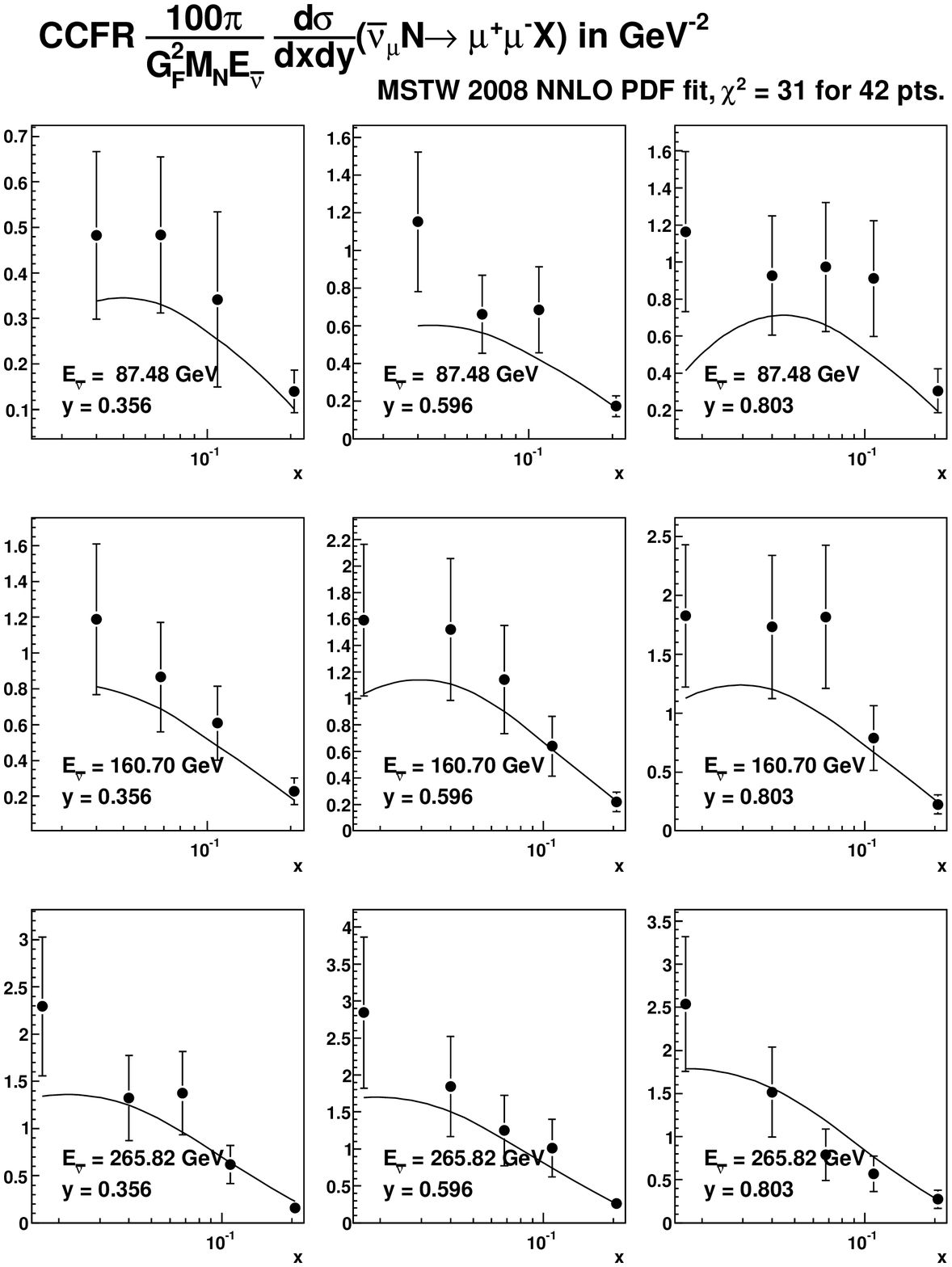}
  \caption{The quality of the NNLO fit to the CCFR antineutrino-initiated dimuon production.}
  \label{fig:ccfrdimuonantinu}
\end{figure}
\begin{figure}
  \centering
  \includegraphics[width=0.8\textwidth,clip]{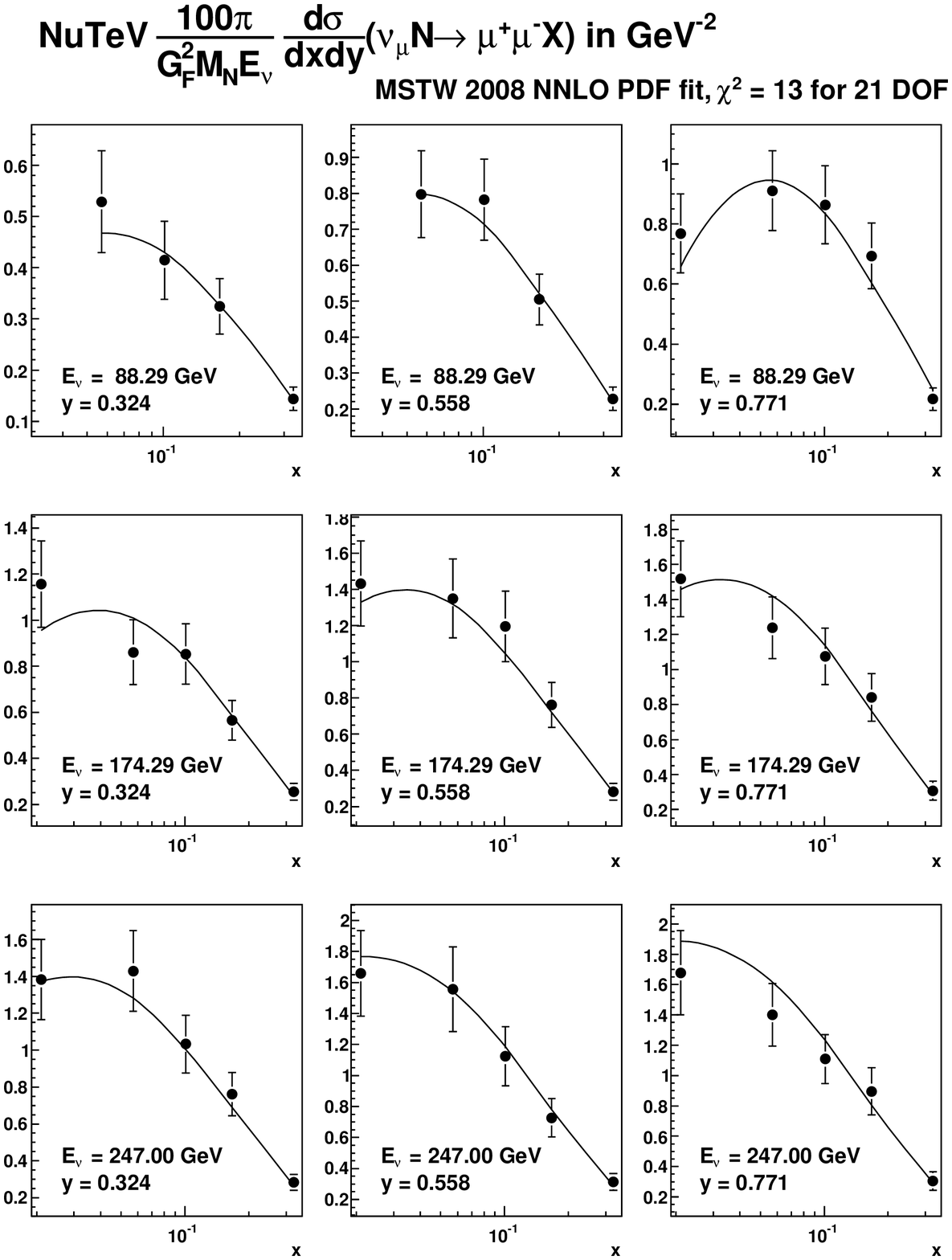}
  \caption{The quality of the NNLO fit to the NuTeV neutrino-initiated dimuon production.}
  \label{fig:nutevdimuonnu}
\end{figure}
\begin{figure}
  \centering
  \includegraphics[width=0.8\textwidth,clip]{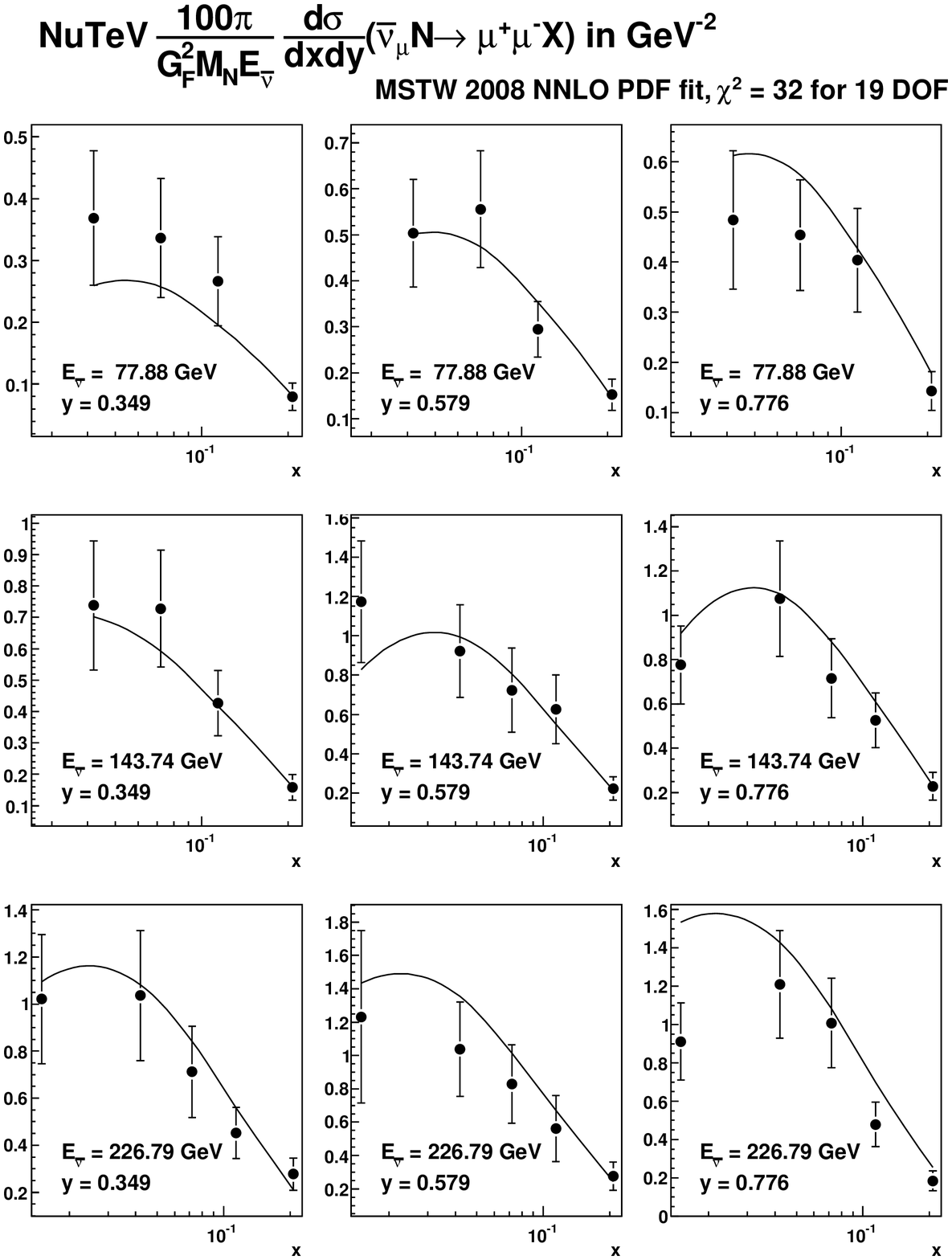}
  \caption{The quality of the NNLO fit to the NuTeV antineutrino-initiated dimuon production.}
  \label{fig:nutevdimuonantinu}
\end{figure}
The description of the CCFR and NuTeV neutrino and antineutrino dimuon data given by the NNLO fit is shown in Figs.~\ref{fig:ccfrdimuonnu}, \ref{fig:ccfrdimuonantinu}, \ref{fig:nutevdimuonnu} and \ref{fig:nutevdimuonantinu}.  Clearly the quality of the fit is very good. There is perhaps a slight tendency for the CCFR antineutrino data to lie a little above the curves and the NuTeV antineutrino data to lie a little below the curves, suggesting that the more recent NuTeV data favour a lower $\bar{s}(x,Q^2)$ distribution in the region of the data and are mainly responsible for the asymmetry.  The results are similar at LO, NLO and NNLO, as suggested by the quality of the fits shown in Table \ref{tab:chisquared}.  Note that the branching ratio $B_c=0.099$ is fixed in the fits.  If allowed to go free there is a distinct tendency for it to choose a high value.  This is the same sort of effect as data set normalisations sometimes floating to one end of the allowed uncertainty.  The combination $4/9(u+\bar{u})+1/9(d+\bar{d}+s+\bar{s})$ is very well constrained by structure function data.  The fit to this can be obtained with the minimum momentum of partons if there is a large proportion of up and antiup quarks, so in the absence of higher order corrections the fit tends to choose this scenario if given the choice, and hence push the strange proportion down.  Indeed it was recently noted~\cite{Rojo:2008ke} that if the strange distribution is left completely free in a fit to structure function data alone it settles on a marginally negative value (we have also observed this phenomenon), albeit with very large uncertainty.  We remove this tendency from the fit by fixing the branching ratio at $B_c=0.099$ due to complete correlation with the $(s+\bar{s})$ normalisation.

\subsection{Impact on strange and antistrange distributions}

\begin{figure}
  (a)\hspace{0.5\textwidth}(b)\\
  \includegraphics[width=0.5\textwidth]{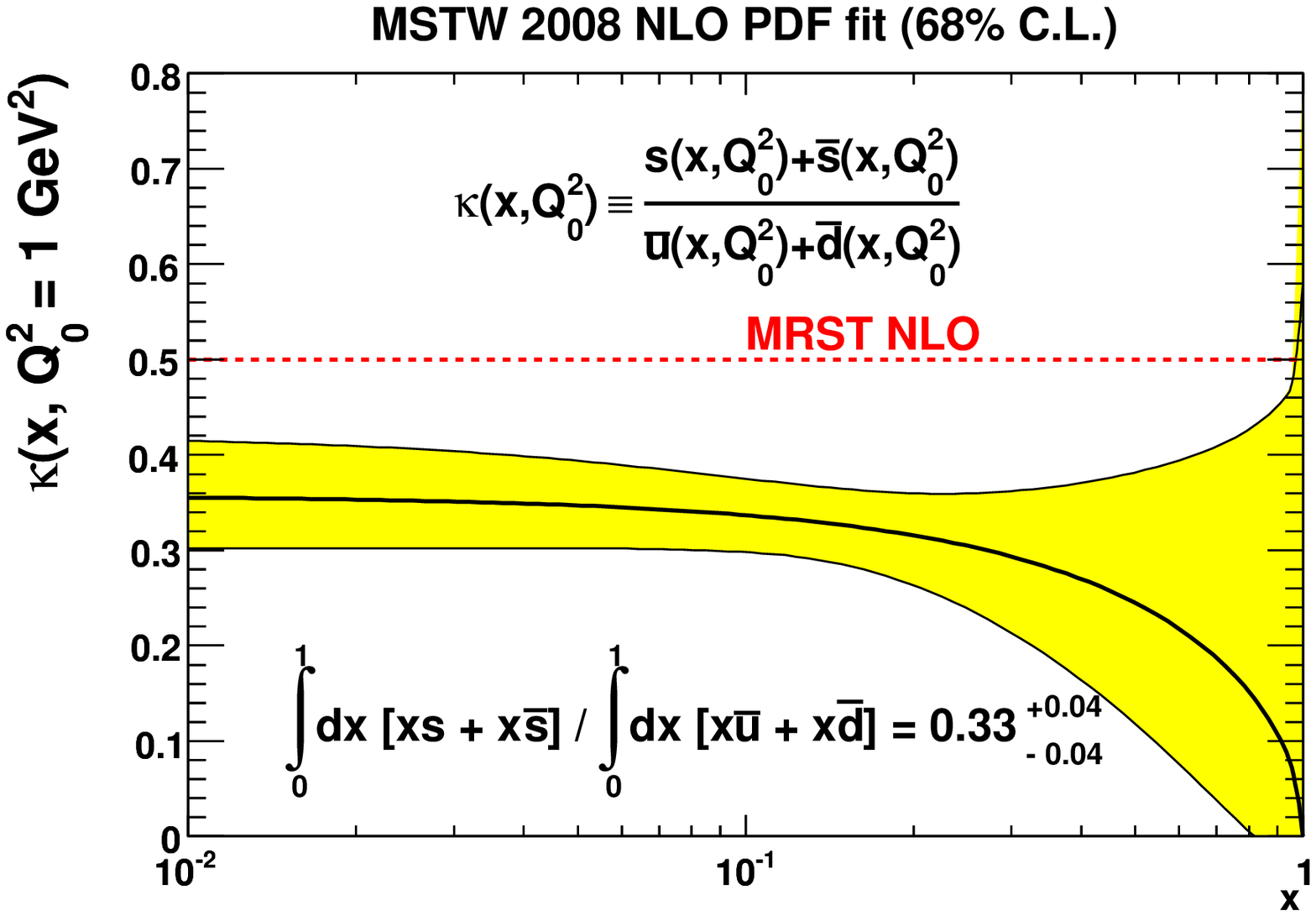}%
  \includegraphics[width=0.5\textwidth]{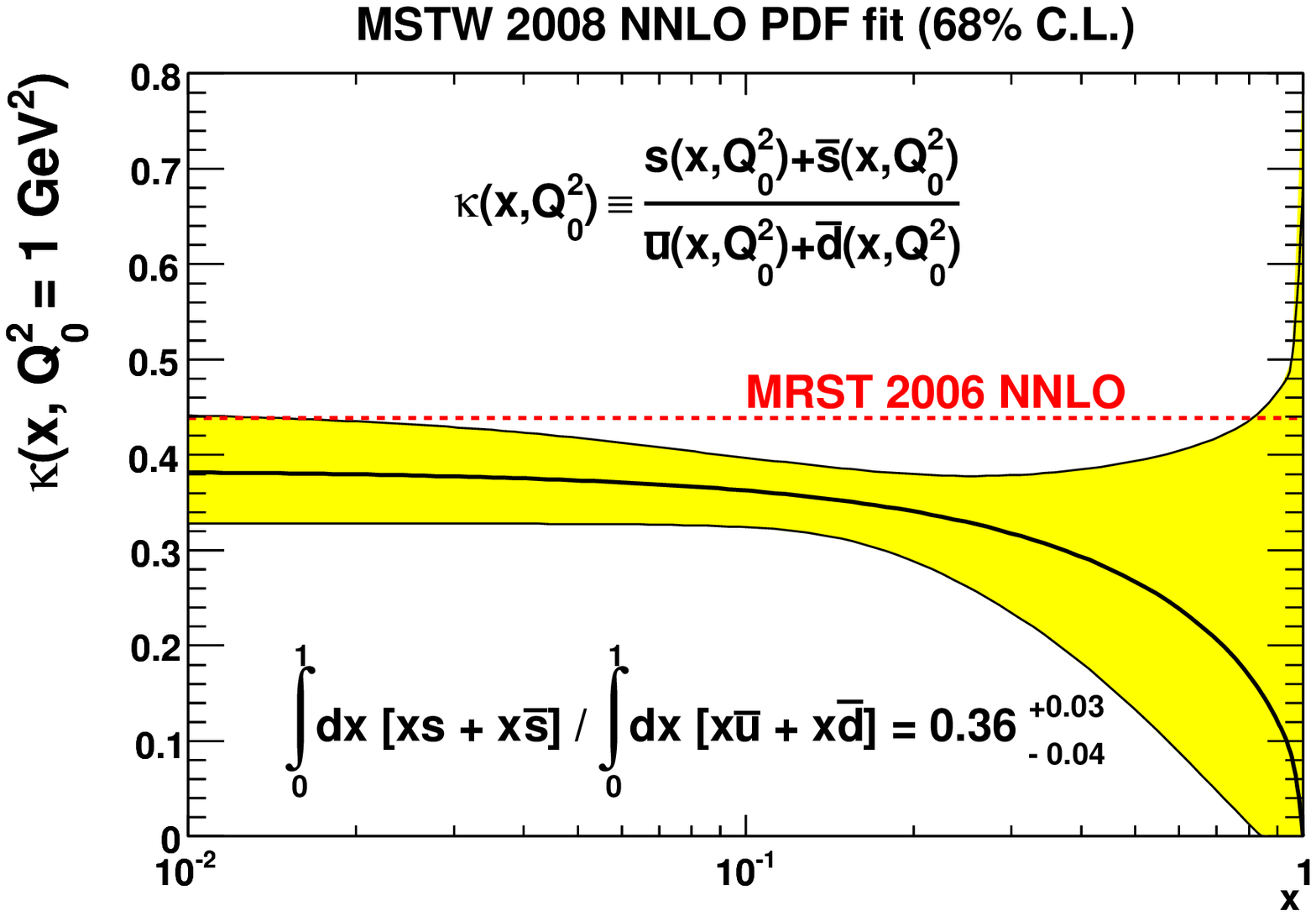}
  \caption{Ratio of the strange to the non-strange distributions at the input scale $Q_0^2 = 1$ GeV$^2$ and the associated one-sigma uncertainty band for both (a) NLO and (b) NNLO distributions.}
  \label{fig:nnlosplus}
\end{figure}
\begin{figure}
  (a)\hspace{0.5\textwidth}(b)\\
  \includegraphics[width=0.5\textwidth]{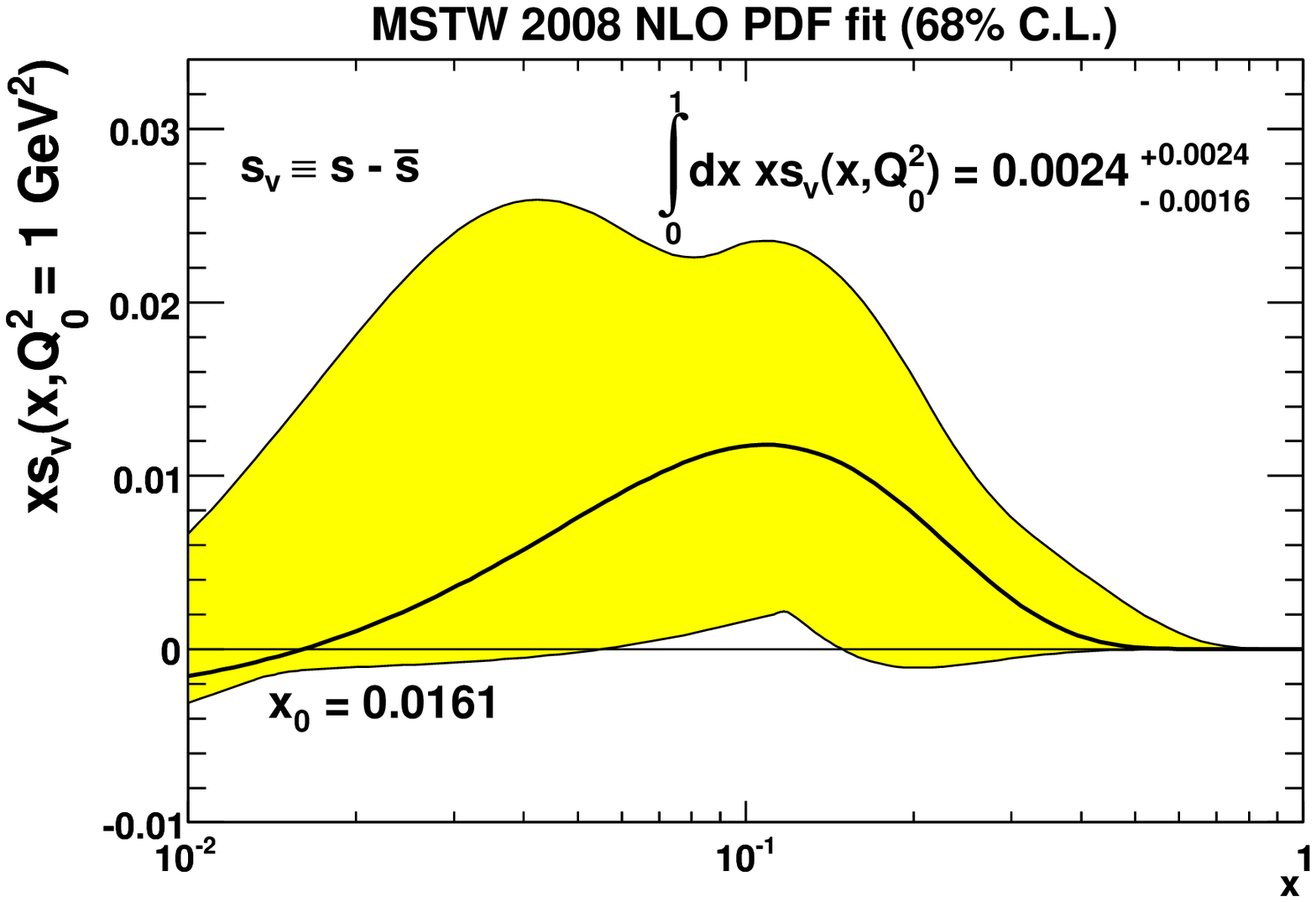}%
  \includegraphics[width=0.5\textwidth]{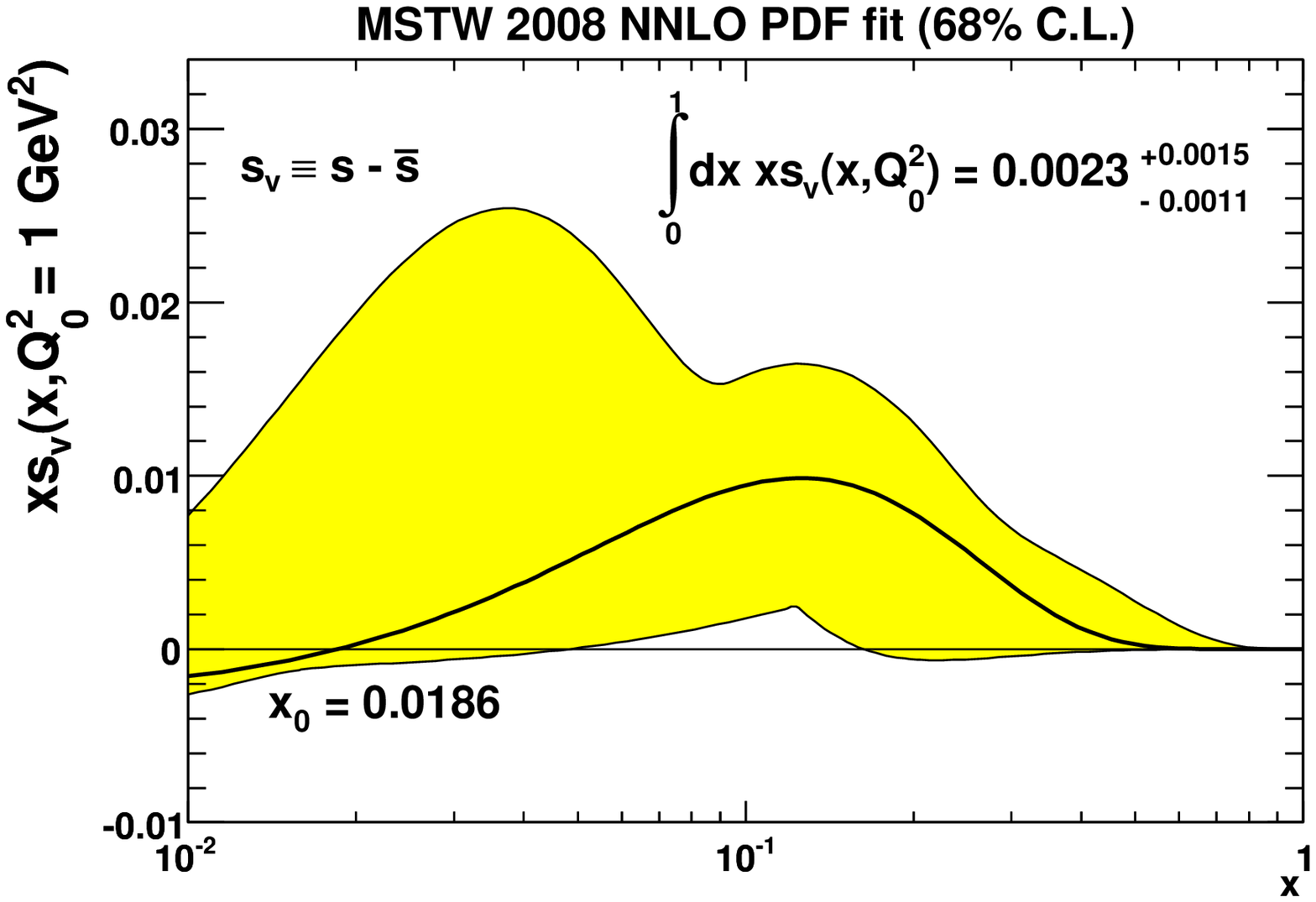}
  \caption{The strange asymmetry, $xs-x\bar{s}$, at the input scale $Q_0^2 = 1$ GeV$^2$ and the associated one-sigma uncertainty band for both (a) NLO and (b) NNLO distributions.}
  \label{fig:nnlosminus}
\end{figure}
In Fig.~\ref{fig:nnlosplus} we show the ratio of the strange sea to the non-strange sea,
\begin{equation}
  \kappa(x,Q^2) \equiv \frac{s(x,Q^2)+\bar{s}(x,Q^2)}{\bar{u}(x,Q^2)+\bar{d}(x,Q^2)},
\end{equation}
at the input scale $Q_0^2 = 1$ GeV$^2$ at both NLO and NNLO.  At small $x$ this suppression factor is just a little below the previous default values of 0.5 at NLO and 0.44 at NNLO. However, there is a strong suggestion of additional suppression of the strange sea at large $x$ relative to the non-strange sea.  This is as one might expect if the suppression is due to the mass of the strange quark since at high $x$ one is nearer to the production threshold. As seen in Section \ref{sec:inputparamunc}, and illustrated in Fig.~\ref{fig:massratio}(a), the suppression of the strange distribution is consistent with the expectation from considering the strange quark to have a mass $m_s^2 \sim 0.1$ GeV$^2$.  In Fig.~\ref{fig:nnlosminus} we also show the difference between $s$ and $\bar{s}$ at the input scale.  Note that $s$ and $\bar{s}$ are constrained by the dimuon data for $0.01\lesssim x \lesssim 0.2$, the region where the central value of the asymmetry is non-zero.  However, note that the asymmetry is non-zero just outside the one-sigma uncertainty band, and it is certainly consistent with zero within the 90\% C.L.~uncertainty band.  As with the total strange distribution the plots for the NLO and NNLO analyses are very similar. Indeed, this is quantified in Table~\ref{tab:asymmetry} which shows the integrated strangeness momentum asymmetry,
\begin{equation}
  \int_0^1\!{\rm d}x\;x\left[s(x,Q^2)-\bar{s}(x,Q^2)\right],
\end{equation}
and the relative strength of the strange quark sea
\begin{equation}
  \frac{\int_0^1\!{\rm d}x\;x\left[s(x,Q^2)+\bar{s}(x,Q^2)\right]}{\int_0^1\!{\rm d}x\;x\left[\bar{u}(x,Q^2)+\bar{d}(x,Q^2)\right]},
\end{equation}
for $Q^2 = 1$, $10$, $100$ GeV$^2$, for the LO, NLO and NNLO parton sets.  Notice that in both cases there is good perturbative convergence, though the LO strangeness fraction is a little low compared to the higher orders. This was not guaranteed to be the case, and relies on a careful definition of the GM-VFNS for charged-current structure functions, as outlined in Section \ref{sec:heavyflavour}.

\begin{table}
  \centering
  \begin{tabular}{c|ccc|ccc}
    \hline \hline
    & \multicolumn{3}{c|}{$\int_0^1\!{\rm d}x\;x(s-\bar{s})$} & \multicolumn{3}{c}{$\int_0^1\!{\rm d}x\;x(s+\bar{s}) / \int_0^1\!{\rm d}x\;x(\bar{u}+\bar{d})$} \\ \hline
    $Q^2$/GeV$^2$ & LO & NLO & NNLO & LO & NLO & NNLO \\ \hline
    1 & $0.0026^{+0.0022}_{-0.0017}$ & $0.0024^{+0.0024}_{-0.0016}$ & $0.0023^{+0.0015}_{-0.0011}$ & $0.24^{+0.03}_{-0.03}$ & $0.33^{+0.04}_{-0.04}$ & $0.36^{+0.03}_{-0.04}$ \\
    10 & $0.0019^{+0.0016}_{-0.0012}$ & $0.0018^{+0.0018}_{-0.0012}$ & $0.0016^{+0.0011}_{-0.0009}$ & $0.45^{+0.02}_{-0.03}$ & $0.53^{+0.02}_{-0.03}$ & $0.54^{+0.02}_{-0.03}$ \\
    100 & $0.0016^{+0.0014}_{-0.0010}$ & $0.0015^{+0.0015}_{-0.0010}$ & $0.0013^{+0.0010}_{-0.0007}$ & $0.55^{+0.02}_{-0.02}$ & $0.62^{+0.02}_{-0.02}$ & $0.63^{+0.02}_{-0.02}$ \\
    \hline \hline
  \end{tabular}
  \caption{Integrated strangeness momentum asymmetry and integrated strangeness fraction for $Q^2 = 1$, $10$, $100$ GeV$^2$.  Note that $\int_0^1\!{\rm d}x\;x(s-\bar{s})\approx 0.007$ (for typical $Q^2\sim 20$ GeV$^2$) is required to bring the NuTeV $\sin^2\theta_W$ to the world average value.  The one-sigma PDF uncertainties are given.}
  \label{tab:asymmetry}
\end{table}

The uncertainty on the $s$ and $\bar{s}$ distributions has increased significantly due to this new treatment, since previously it was simply the same (proportionally) as for the much more accurately determined $\bar u$ and $\bar d$ distributions.  This increase is illustrated clearly in Fig.~\ref{fig:massratio}(b), and is one of the main reasons for the increase in uncertainty for the total sea seen in Fig.~\ref{fig:seauncertainty}.  The increased strange uncertainty also leads to a slight increase in the uncertainty for the $\bar{u}$ and $\bar{d}$ distributions.  From constraints on the different weighting in the sum of sea quarks, mainly on the charge-weighted combination in neutral-current DIS, there is an anti-correlation between strange and $\bar{u},\bar{d}$.  However, this is a fairly small effect, and is difficult to disentangle from the consequences of other changes in our procedure (e.g.~accounting for data set normalisation uncertainties).  However, as well as an understanding of the uncertainties we now have much more confidence in our central values for the strange and antistrange distributions.  The precise values are important for current experiments since the strange quark and antiquark distributions make significant contributions to $W^+$ and $W^-$ production at hadron colliders, particularly at the LHC, through $c\bar{s}$ and $\bar{c}s$ partonic fusion.  The $W^{\pm}$ asymmetry is driven mainly by the difference between $u\bar{d}$ and $\bar{u}d$ fusion, but also has some sensitivity to the strange quark asymmetry.  It is worth noting that the associated production processes $pp \to W^- +c+X$ and $pp \to W^+ +\bar{c}+X$ are a promising way to measure the strange quark distributions, where the LO partonic subprocesses are $gs \to W^-c$ or $g\bar{s} \to W^+\bar{c}$~\cite{Lai:2007dq}.  Indeed, the $Wc$ production process has recently been measured at the Tevatron~\cite{Aaltonen:2007dm}.

\subsection{Implications for the NuTeV \texorpdfstring{$\sin^2\theta_W$}{sin2(thetaW)} anomaly} \label{sec:ssbarimplications}

The determination of the momentum asymmetry $\int_0^1\!{\rm d}x\;x(s-\bar{s})$ of the $s$ and $\bar{s}$ distributions has important implications~\cite{Davidson:2001ji} for the anomaly in the measurement of $\sin^2\theta_W$ reported by NuTeV~\cite{Zeller:2001hh}.  The NuTeV extraction of $\sin^2\theta_W$ from
\begin{equation}
  R^- \equiv \frac{\sigma(\nu_\mu N\to\nu_\mu X) - \sigma(\bar{\nu}_\mu N\to\bar{\nu}_\mu X)}{\sigma(\nu_\mu N\to\mu^- X) - \sigma(\bar{\nu}_\mu N\to\mu^+ X)} \approx \frac{1}{2} - \sin^2\theta_W
\end{equation}
is about $\sim$ three-sigma above the global average, and was at first thought to hint at new physics.  However, there are two effects sensitive to parton distributions which should first be included: isospin violation ($u^p\ne d^n$, $d^p\ne u^n$) and the strange sea asymmetry ($s\ne \bar{s}$).  Isospin violation gives
\begin{equation}
  R^- \approx \frac{1}{2} - \sin^2\theta_W \;+\;(1-\frac{7}{3}\sin^2\theta_W)\frac{\int_0^1\!{\rm d}x\;x\left[(u_v^p-d_v^n)-(d_v^p-u_v^n)\right]}{2\int_0^1\!{\rm d}x\;x(u_v+d_v)}.
\end{equation}
In fact, isospin violation is automatically generated by QED corrections to parton evolution, and was found by MRST~\cite{Martin:2004dh} to remove slightly less than half of the total discrepancy in the NuTeV value of $\sin^2\theta_W$ if evolution of the photon distribution (in the proton) from current quark mass scales was applied, while for evolution from constituent quark masses a reduction of about a quarter was obtained.  The second effect is the possible existence of a momentum asymmetry $\int_0^1\!{\rm d}x\;x(s-\bar{s})$ of the strange sea.  It gives
\begin{equation}
  R^- \approx \frac{1}{2} - \sin^2\theta_W \;-\; (1-\frac{7}{3}\sin^2\theta_W)\frac{\int_0^1\!{\rm d}x\;x(s-\bar{s})}{\int_0^1\!{\rm d}x\;x(u_v+d_v)}.
\end{equation}
A positive value of $\int_0^1\!{\rm d}x\;x(s-\bar{s})\sim 0.007$~\cite{Zeller:2002du,McFarland:2003jw} would bring the NuTeV $\sin^2\theta_W$ to the world average~\cite{Amsler:2008zz}.  Indeed, we see that the analysis of the dimuon data gives a preference for a positive strange momentum asymmetry, which goes some way to further reduce the NuTeV $\sin^2\theta_W$ anomaly.  A reduction in the $\sin^2\theta_W$ anomaly of about one-sigma is likely, but a reduction of up to two-sigma (or down to no effect) is within the range of the PDF uncertainties for the strange momentum asymmetry shown in Table \ref{tab:asymmetry}.

%% file: heavyflavourdata.tex
\section{Description of \texorpdfstring{$F_2^{c\bar{c}}$}{F2(charm)} and \texorpdfstring{$F_2^{b\bar{b}}$}{F2(beauty)} data} \label{sec:heavyflavourdata}

\subsection{Comparison to HERA data}

\begin{figure}
  \centering
  \includegraphics[width=0.8\textwidth]{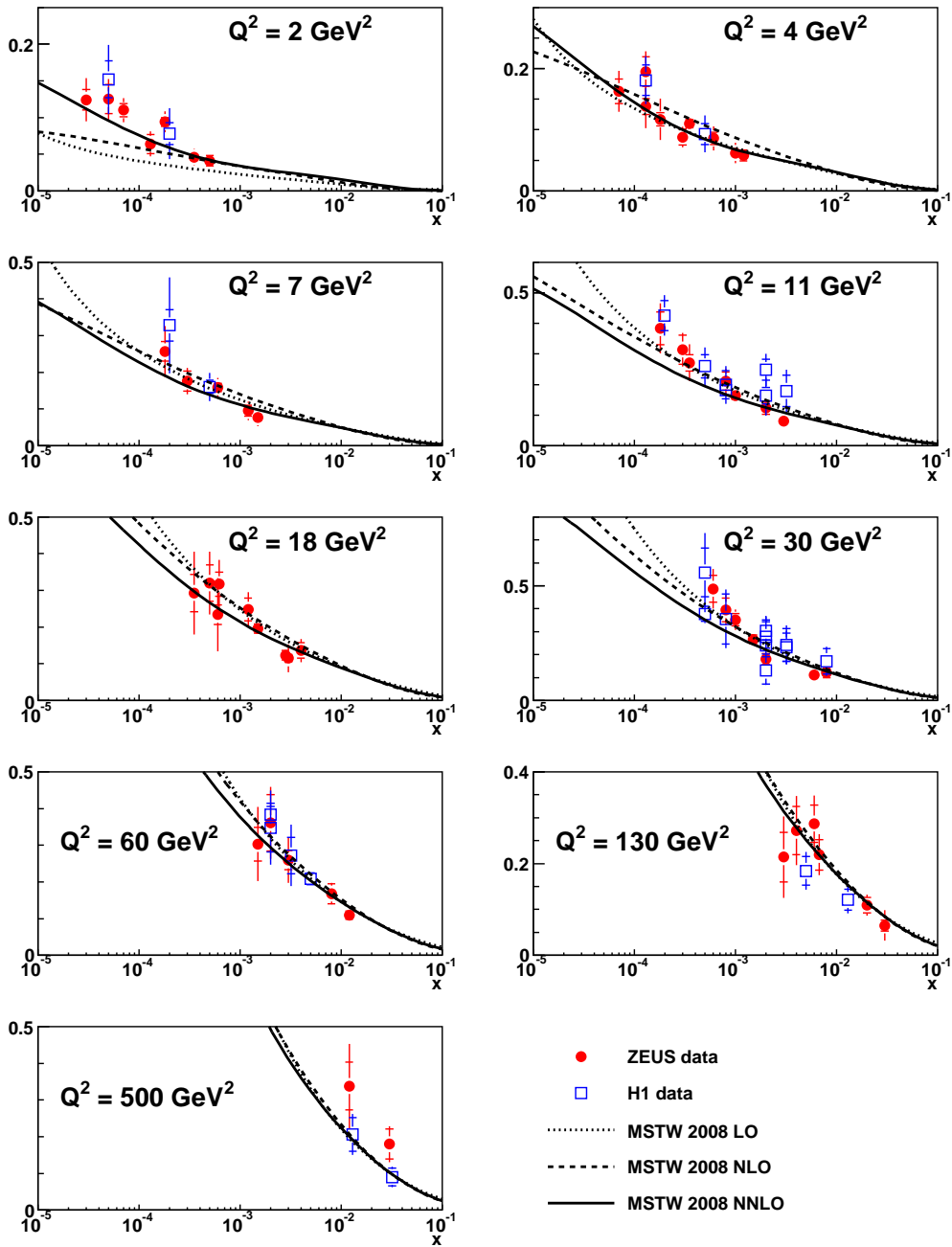}
  \caption{The charm structure function, $F_2^{c\bar{c}}(x,Q^2)$, compared to data from H1~\cite{Adloff:1996xq,Adloff:2001zj,Aktas:2005iw,Aktas:2004az} and ZEUS~\cite{Breitweg:1999ad,Chekanov:2003rb,Chekanov:2007ch}.  The $Q^2$ bins used correspond to the data points in Ref.~\cite{Chekanov:2003rb}; the other data points have been shifted to these $Q^2$ values using the NNLO predictions.}
  \label{fig:f2charm}
\end{figure}
The new $F_2^{c\bar{c}}$ data included in the MSTW fit~\cite{Adloff:1996xq,Adloff:2001zj,Aktas:2005iw,Aktas:2004az,Breitweg:1999ad,Chekanov:2003rb,Chekanov:2007ch} extend the data included in earlier MRST fits~\cite{Adloff:1996xq,Breitweg:1999ad}.  In Fig.~\ref{fig:f2charm} we show the comparisons of our results at LO, NLO and NNLO to the charm structure function $F_2^{c\bar{c}}(x,Q^2)$ for all available H1~\cite{Adloff:1996xq,Adloff:2001zj,Aktas:2005iw,Aktas:2004az} and ZEUS~\cite{Breitweg:1999ad,Chekanov:2003rb,Chekanov:2007ch} data.  As one can see from the figure, and from the values of $\chi^2$ in Table \ref{tab:chisquared}, the quality of the fit is good. At LO the fit undershoots badly for $Q^2=2$ GeV$^2$, the absence of the $1/x$ term in the NLO coefficient function being most important in this region.  However, the quick evolution at LO due to the large gluon and very large $\alpha_S$ quickly compensates and $F_2^{c\bar c}(x,Q^2)$ evolves most quickly at LO.  The NNLO fit quality is best. This is clearly due to the improvement compared to NLO at the lowest $x$ for $Q^2=2$ GeV$^2$. At higher $Q^2$ the NNLO $F_2^{c\bar{c}}(x,Q^2)$ evolves most slowly, a feature that was highlighted as a consequence of heavy flavour at NNLO in Ref.~\cite{Thorne:2006qt}.  This is most easily understood in the GM-VFNS, where the positive corrections to the coefficient functions at low $Q^2$, and the negative contributions to the heavy flavour distributions from matching conditions influencing high $Q^2$, both have the effect of flattening the slope.  However, it has recently also been noticed in an approximate FFNS at NNLO (inclusion of threshold corrections and factorisation/renormalisation scale dependent terms)~\cite{Alekhin:2008hc}.  This slower evolution at NNLO is neither clearly preferred nor rejected by the data shown in Fig.~\ref{fig:f2charm}, with some bins favouring NLO and some NNLO.  Future more precise data may be more discriminating.

As outlined in Section \ref{sec:heavyflavour}, the heavy flavour treatment acquires some model dependence at NNLO, in the detailed form of the $\mathcal{O}(\alpha_S^3)$ coefficient functions, which are approximated using threshold and small-$x$ limits.  However, for charm production this term only appears with all scales fixed at $Q^2=m_c^2$ in the fit to data (only depending on $Q^2$ for $Q^2< m_c^2$), and is a small constant correction, which becomes increasingly less important at high $Q^2$.  The model dependence is mainly due to the treatment of the small-$x$ limit.  Reasonable variations of the parameters controlling this leads to variations in $F_2^{c\bar c}(x,Q^2)$ of approximately $\pm$0.03 for $x=10^{-4}$ or $\pm$0.06 for $x=10^{-5}$, and rather less than this for higher $x$, the variation becoming smaller than $\pm0.01$ at $x=0.001$, and quickly dying away above this.  This is enough to move the fit quality  for $F_2^{c\bar c}(x,Q^2)$ data close to that at NLO in one direction, or to move the fit quality in the $Q^2=2$ GeV$^2$ bin close to the best possible in the other direction.  Because this is a correction to $F_2(x,Q^2)$ which is constant in $Q^2$, this uncertainty has very little impact on the gluon distribution.  The only real uncertainty it leads to is in the input for the sea distribution, which would change to compensate the $Q^2$-constant change in $F_2(x,Q^2)$.  The possible change is up to $4$--$5\%$ for $x\lesssim 10^{-4}$, becoming much less at higher $x$.  This is a lot smaller than the input uncertainty shown in Fig.~\ref{fig:seauncertainty}(a).  The relative uncertainty from this source would quickly become much smaller at high $Q^2$, and would be a negligible influence on the uncertainty at $Q^2 = 100$ GeV$^2$ shown in Fig.~\ref{fig:seauncertainty}.

\begin{figure}
  \centering
  \includegraphics[width=0.8\textwidth]{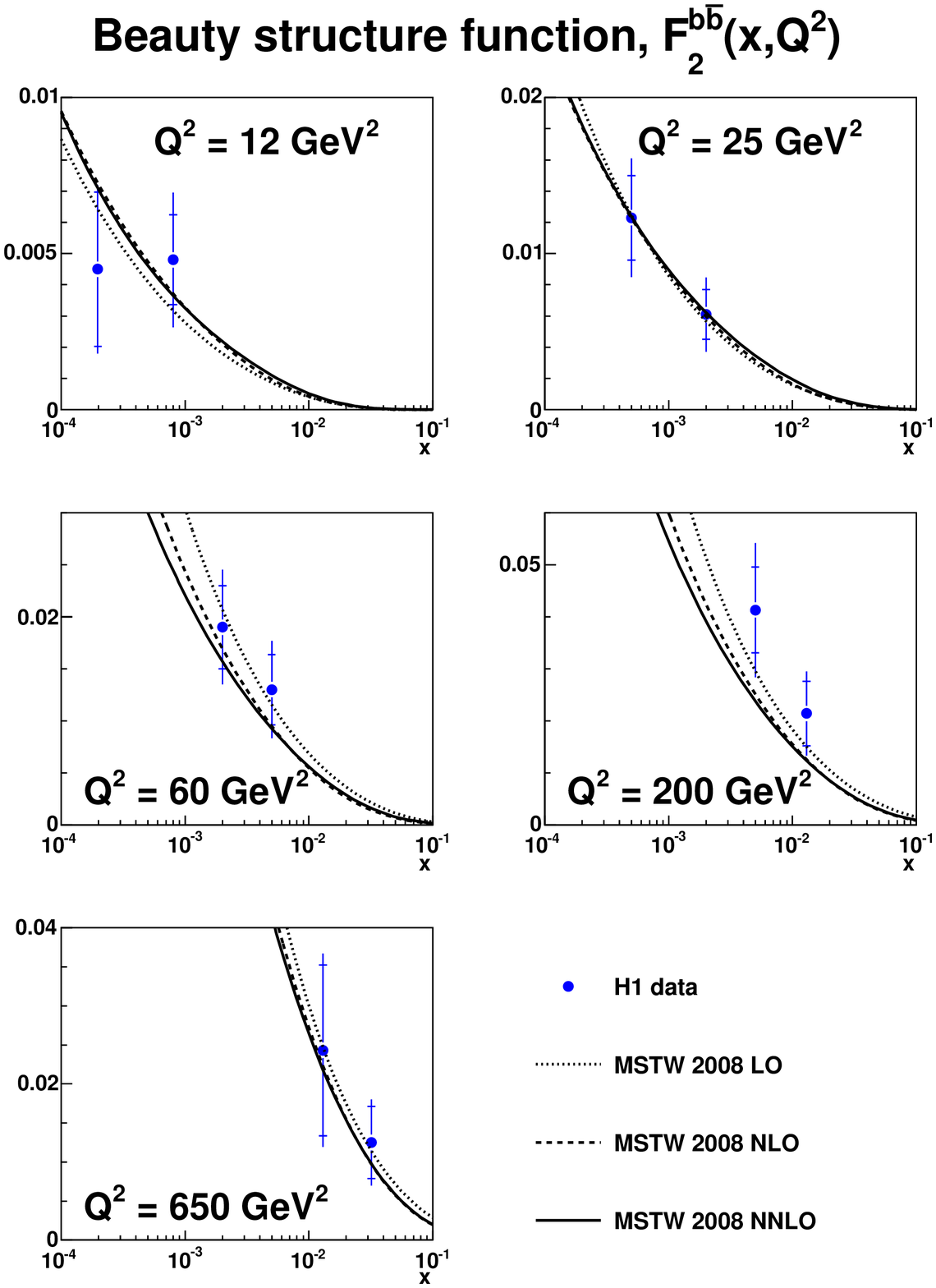}
  \caption{Predictions for the beauty structure function, $F_2^{b\bar{b}}(x,Q^2)$, compared to data from H1~\cite{Aktas:2005iw,Aktas:2004az}.}
  \label{fig:f2beauty}
\end{figure}
In Fig.~\ref{fig:f2beauty} we show the predictions for the beauty structure function $F_2^{b\bar{b}}(x,Q^2)$ compared to the published H1 data~\cite{Aktas:2005iw,Aktas:2004az}.  Clearly at all orders the comparison is good.  It is also clear that these data currently provide no useful constraint on the gluon distribution.  Since much of the data is in the vicinity of $Q^2=m_b^2$, or not much higher, more precise data will be sensitive to the details of threshold contributions and in particular to the value of $m_b$ rather than to the gluon.

\subsection{Constraints on intrinsic charm}

As explained in Section \ref{sec:heavyflavour}, our variable flavour number scheme allows for the possibility of including intrinsic charm with no theoretical problems.  However, we simply choose not to include it in the default fit since we have no reason to believe it is anything other than a very small component of the total charm distribution, and none of the data in our global fit provide any constraint.  The nonperturbatively generated contribution to the charm distribution is in general of ${\cal O}(\Lambda_{QCD}^2/m_c^2)$, but may be enhanced at large $x$~\cite{Brodsky:1980pb}, whereas all data on $F_2^{c\bar c}(x,Q^2)$ from HERA are at small $x$.

\begin{figure}
  \centering
  \includegraphics[width=0.8\textwidth]{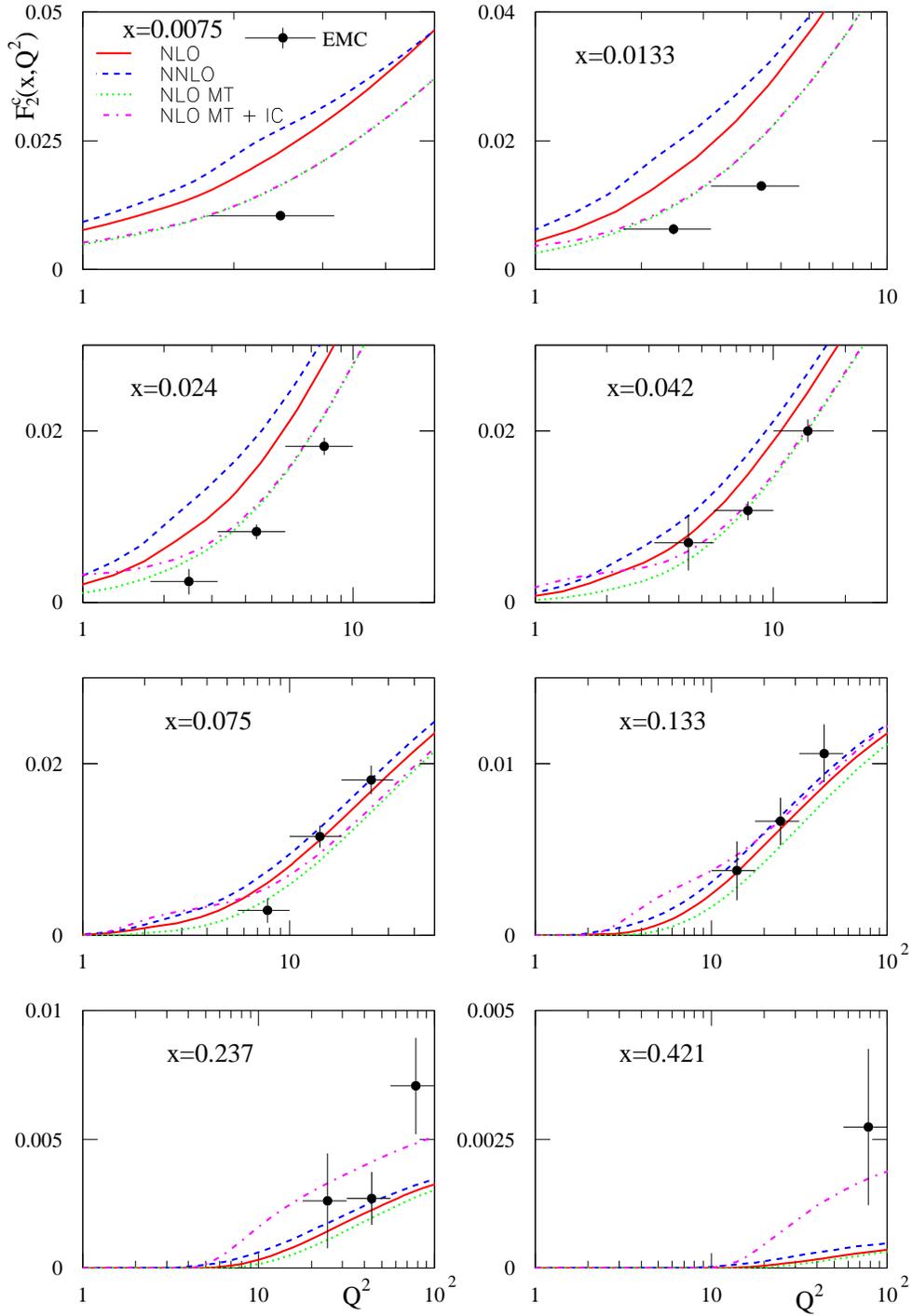}
  \caption{The comparison of the EMC charm data~\cite{Aubert:1982tt} to our predictions at NLO and NNLO.  MT stands for the modified threshold approach and IC stands for inclusion of intrinsic charm.}
  \label{fig:charmemc}
\end{figure}
In this section we use the only data which currently do show any sensitivity, the  EMC data on $F_2^{c\bar c}(x,Q^2)$~\cite{Aubert:1982tt}.  In Fig.~\ref{fig:charmemc} we show the comparison of the EMC charm data to our predictions at NLO and NNLO.  In general the comparison to data is quite poor, tending to overshoot at the lower values of $x$ and $Q^2$.  This is worse at NNLO than at NLO because the NNLO threshold corrections cause a significant increase.  Hence, at fixed NLO and NNLO with $m_c=1.4$ GeV there is some room for intrinsic charm at $x >0.2$, but much more tendency to overshoot data at smaller $x$. This latter problem must be addressed before drawing conclusions about the intrinsic charm contribution.  When considering this comparison it is pertinent to remember that when calculating heavy flavour structure functions we use $W^2=4m_c^2$ as the threshold, whereas in reality we need to produce mesons, i.e.~$W^2$ is really required to be a little greater.  In practice this is a higher-twist effect, but the EMC data are not so far from threshold, and are much more sensitive to these details than HERA data.  Hence, we consider a comparison where we replace $m_c^2$ by $m_c^2(1+\Lambda^2/m_c^2)$ in the threshold dependent parts of coefficient functions, where $\Lambda^2$ is a binding energy which we take to be $\Lambda = 0.2$ GeV.  We make no change in the off-threshold parts of coefficient functions, nor in PDF definitions.  The result is labelled as NLO MT in Fig.~\ref{fig:charmemc}.  It clearly matches lower $Q^2$ EMC data better, and the prediction is slightly smaller at higher $x$ (even with this modification NNLO predictions would generally be too large).  The curve labelled NLO MT + IC includes a $0.3\%$ integrated number density contribution of intrinsic charm (and the same for anticharm) using the parameterisation in Ref.~\cite{Brodsky:1980pb} with the upper limit on $x$ modified to the correct threshold value.  We note that this is only $\approx 1/10$ of the upper limit obtained by CTEQ~\cite{Pumplin:2007wg}.  This seems to be about the maximum that can be included even with nonperturbative suppression of standard fixed-order contributions, and even then only at NLO.  A much larger contribution would overshoot most data at $x>0.1$.\footnote{Note that $0.3\%$ was presented as the most likely value in 1983 using very old PDFs and only the LO perturbative contribution~\cite{Hoffmann:1983ah}.}  Hence, if the EMC data are to be believed, there is no room for a very sizeable intrinsic charm contribution, although there is a suggestion of some deficit at the highest $x$ values.

%% file: lowenergydrellyan.tex
\section{Description of low-energy Drell--Yan data} \label{sec:lowmassDY}
As in recent analyses~\cite{Martin:2004ir,Martin:2007bv} we fit to the E866/NuSea Drell--Yan dilepton production data in $pp$ collisions~\cite{Webb:2003bj}, now corrected for electromagnetic radiative corrections; moreover we do not include the $pd$ collision data as we observe significant systematic differences between our predictions and these results.  The Drell--Yan process is known up to NNLO for the rapidity ($y$) distribution~\cite{Anastasiou:2003ds}.  However, the data are binned in the Feynman-$x$ variable, $x_F=x_1-x_2$.  Previous MRST fits used LO kinematics ($p_T=0$) when transforming between the two variables.  However, the mean $p_T$ of each data bin is available, and hence we now use the exact kinematics:
\begin{equation}
  y = \ln\left(H+\sqrt{H^2+1}\right),\quad\text{where }H \equiv \frac{x_F\,\sqrt{s}}{2\sqrt{M^2+p_T^2}}.
\end{equation}
At LO a good fit to the data is impossible while simultaneously fitting to structure function data which probe the same range of $x$ and hard scale.  This is because there is a large difference in the size of the corrections to the respective coefficient functions when going from the spacelike to the timelike regime, i.e.~there is a factor of $1 + (\alpha_S(M)/\pi) C_F\pi^2/2$ for Drell--Yan production at NLO.  Since this is missing at LO, the normalisations of the two calculated quantities cannot be made consistent with the structure function data and Drell--Yan data at the same time.  Previous MRST fits used a constant factor of 1.3 to enhance the LO Drell--Yan cross section, but we have improved this and now use a $K$-factor consisting of the $\pi^2$-enhanced term.  At NLO and NNLO we isolate the $\alpha_S$ dependence from the $K$-factor by writing:
\begin{align}
  K^{\rm LO}(M) &= 1 + \left(\frac{\alpha_S(M)}{\pi}\right)\,\frac{C_F\pi^2}{2}, \label{eq:KDYLO}\\
  K^{\rm NLO}(M,y) &= 1 + \left(\frac{\alpha_S(M)}{\pi}\right)\,C(M,y),\label{eq:KDYNLO}\\
  K^{\rm NNLO}(M,y) &= 1 + \left(\frac{\alpha_S(M)}{\pi}\right)\,D(M,y) + \left(\frac{\alpha_S(M)}{\pi}\right)^2\,E(M,y).\label{eq:KDYNNLO}
\end{align}
We calculate the $C$-, $D$- and $E$-factors with the \textsc{dyrap} program~\cite{Anastasiou:2003ds} using NLO or NNLO PDFs at the respective order.  This method, first used in Ref.~\cite{Martin:2007bv}, is a considerable improvement on previous MRST fits where only the overall $K(M,y)$ was calculated for a fixed value of $\alpha_S$.  (Additionally, NNLO fits prior to Ref.~\cite{Martin:2007bv} used only NLO $K$-factors.)

\begin{figure}
  \centering
  \includegraphics[width=0.8\textwidth,clip]{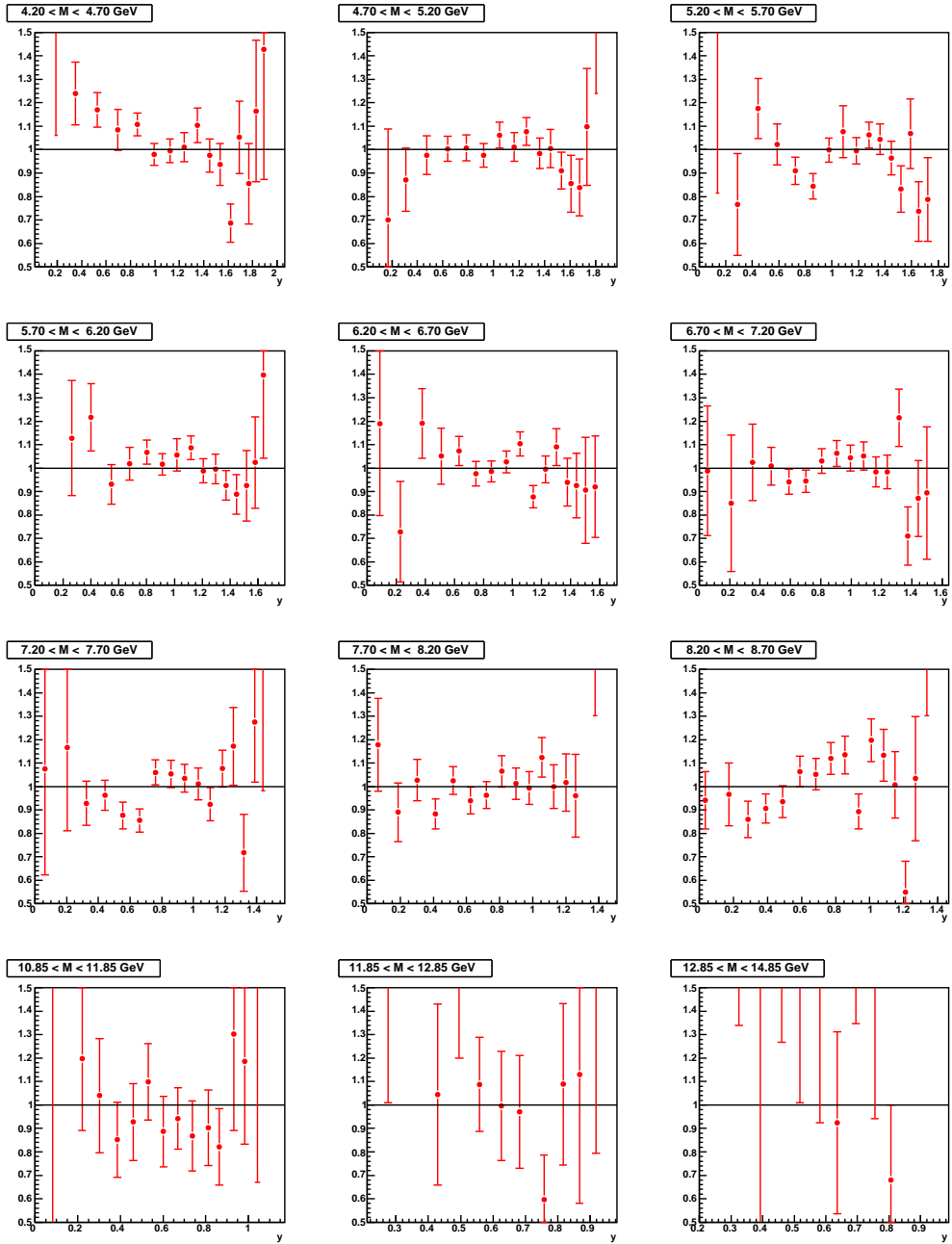}
  \caption{The ratio of E866/NuSea $pp$ Drell--Yan data points to the predictions of the NNLO PDF fit as a function of rapidity for different mass bins.}
  \label{fig:drellyyan}
\end{figure}
The quality of the fit to the E866/NuSea data is illustrated in Fig.~\ref{fig:drellyyan} at NNLO.  As seen in Table \ref{tab:chisquared}, the fit quality is very similar at LO and NLO and the figures of data against theory would be similar to that at NNLO.  As one can see, there is no systematic failure in the comparison between data and theory --- the relatively large $\chi^2$ value is due to a large scatter in the data points.  The only exception is that in the lowest mass bin the data lie somewhat above the theory at low rapidities, although low $M$ is where additional theoretical corrections are most likely to play some role and also where there may be complications in the mass reconstruction from the proximity to the $J/\psi$ and $\Upsilon$ resonances.  Additionally, unlike alternative studies~\cite{Alekhin:2006zm}, we do not see any significant inconsistency between the E866/NuSea $pp$ data and the rest of the global fit.  Our $\chi^2$ for the remaining data only decreases by a few units when these Drell--Yan data are removed, i.e.~by much less than the tolerance on any of our eigenvectors.  We do note, however, that the positive corrections of order $10\%$ to the cross section at NNLO require the data to move up to a normalisation of $1.087$, i.e.~slightly outside the quoted uncertainty of $6.5\%$.

As in previous analyses we include the E866/NuSea $pd/pp$ ratio data~\cite{Towell:2001nh}, which constrains the $\bar{d}-\bar{u}$ difference.  The experimental extraction of the $pd/pp$ ratio by E866/NuSea used a larger data set than the individual cross section measurements, although there was some overlap.  Moreover, the analysis was of a different form using different cuts and averages over much larger data subsets, and therefore had little sensitivity to those lower statistics regions where the discrepancies with the $pd$ data occur, in particular at the highest $x_F$ values.  We have improved the theoretical treatment of the $pd/pp$ ratio compared to previous analyses by calculating the separate $K$-factors for $pp$ and $pd$ collisions.  This has an effect of at most $2\%$, and does aid the quality of the global fit very slightly.  There is no significant change in our results for the difference $\bar{d}-\bar{u}$, though the uncertainty is reduced a little compared to previous analyses, as seen in Fig.~\ref{fig:Deltauncertainty}.

%% file: wandzattevatron.tex
\section{\texorpdfstring{$W$}{W} and \texorpdfstring{$Z$}{Z} boson production at the Tevatron} \label{sec:wztevatron}

$W$ and $Z$ boson production at high-energy hadron colliders has traditionally provided important constraints on parton distributions.  While low-energy Drell--Yan production directly probes the quark sea, $W$ and $Z$ production (via $q\bar{q}\to V$, with $V=W^{\pm},Z$) at proton--antiproton colliders are mainly sensitive to the valence $u$ and $d$ quarks, in different combinations to structure function measurements in deep-inelastic scattering.

Because of the overall experimental luminosity uncertainty (of order $\pm 6\% $ at the Tevatron), the \emph{total} $W$ and $Z$ cross section measurements do not provide competitive constraints on the PDFs, although they do provide an important cross-check on the parton distributions and on the hard-scattering framework (see Section~\ref{sec:totalpredictions} below).  The shape of the $W$ and $Z$ rapidity distributions, however, is sensitive to the shape of the parton distributions, and \emph{does} provide a powerful constraint.  We have for many years used CDF data on the $W\to \ell\nu$ charge asymmetry~\cite{Abe:1998rv} in our global fits, and in the following subsection we study the impact of recent CDF and D{\O} measurements in the updated MSTW fit.  There are also nowadays very precise Tevatron data on the $Z$ rapidity distribution, which provides complementary information to the $W\to\ell\nu$ charge asymmetry measurement.  We include these data in the global fit for the first time, and study their consequences.

As for the Drell--Yan process, the $W$ and $Z$ rapidity distributions have been calculated to NNLO~\cite{Anastasiou:2003ds}, and so we are able to perform consistent fits at LO, NLO and NNLO.

\subsection{Description of Tevatron \texorpdfstring{$W$}{W} asymmetry data}

The $W$ charge asymmetry at the Tevatron is defined by
\begin{equation}
  A_W(y_W) = \frac{{\rm d}\sigma(W^+)/{\rm d} y_W - {\rm d}\sigma(W^-)/{\rm d} y_W}{{\rm d}\sigma(W^+)/{\rm d} y_W + {\rm d}\sigma(W^-)/{\rm d} y_W}\approx \frac{u(x_1)d(x_2)-d(x_1)u(x_2)}{u(x_1)d(x_2)+ d(x_1)u(x_2)},
\end{equation}
where $x_{1,2}=(M_W/\sqrt{s})\exp(\pm y_W)$.  This description in terms of the PDFs is valid at leading order when sea-quark contributions are neglected. In practice, it is usually the lepton charge asymmetry which is measured, defined in a similar way as
\begin{equation} \label{eq:leptonasymm}
  A(\eta_\ell) = \frac{{\rm d}\sigma(\ell^+)/{\rm d}\eta_{\ell}-{\rm d}\sigma(\ell^-)/{\rm d}\eta_{\ell}}{{\rm d}\sigma(\ell^+)/{\rm d}\eta_{\ell}+{\rm d}\sigma(\ell^-)/{\rm d}\eta_{\ell}},
\end{equation}
where $\eta_{\ell}$ is the pseudorapidity of the charged lepton.  Defining the emission angle of the charged lepton relative to the proton beam in the $W$ rest frame by $\cos^2\theta^* = 1 - 4E_T^2/M_W^2$ leads to
\begin{equation}
  y_\ell = y_{W} +  \frac{1}{2}\ln\left(\frac{1+\cos\theta^*}{1-\cos\theta^*}\right).
\end{equation}
Care must be taken when using a valence-quark-only approximation to the lepton asymmetry, since near the edges of phase space, i.e.~$\cos\theta^* \approx \pm 1$, sea-quark contributions can become important.  In fact, neglecting overall factors, the numerator of \eqref{eq:leptonasymm} can be approximated by\footnote{Contributions involving $s$, $c$ and $b$ quarks are not displayed, as these are very small at Tevatron energies.}
\begin{equation} \label{eq:leptonasymmapprox}
  u(x_1)d(x_2)(1-\cos\theta^*)^2 + \bar{d}(x_1)\bar{u}(x_2)(1+\cos\theta^*)^2 - d(x_1)u(x_2)(1+\cos\theta^*)^2 - \bar{u}(x_1)\bar{d}(x_2)(1-\cos\theta^*)^2,
\end{equation}
where for $\cos\theta^* \approx 1$ the leading ($\bar d\bar u$) sea--sea contribution for $W^+$ production is enhanced relative to the valence--valence contribution by the large $(1+\cos\theta^*)^2$ term arising from the $V-A$ decay to leptons.  Hence if the lepton charge asymmetry is measured in different bins of lepton $E_T$, the sensitivity to antiquark PDFs is greatest at lower $E_T$.  For example, if $E_T=25$ GeV then $\cos\theta^* =\pm 0.78$ and, taking the positive value, then from \eqref{eq:leptonasymmapprox} the antiquark contribution will be enhanced relative to the quark contribution by a factor $(1+\cos\theta^*)^2/(1-\cos\theta^*)^2=68$.  In this case, in the forward direction, leptons from $W^+$ produced from quarks will be mainly from bosons with rapidity $y_W = y_\ell + 1.05$ (taking $\cos\theta^* = -0.78$), while leptons from $W^+$ produced from antiquarks will be mainly from bosons with rapidity $y_W = y_\ell - 1.05$ (taking $\cos\theta^* = +0.78$).  At high $\eta_\ell\simeq y_\ell$ (not too close to the kinematic limit) this can lead to the latter being a significant proportion, or even dominating.  Hence, in addition to measuring the overall shapes of the quark distributions, the lepton charge asymmetry also probes the separation into valence and sea quarks.  In practice, this is particularly so for the less well-constrained down quark.

In the global fit, we of course include all quark and gluon contributions to the appropriate order in perturbation theory. We calculate $K$-factors for the differential cross sections ${\rm d}\sigma(\ell^\pm)/{\rm d}\eta_{\ell}$ using the \textsc{fewz} code~\cite{Melnikov:2006kv} with LO, NLO or NNLO PDFs at the respective order.\footnote{We find that the \textsc{vegas} Monte Carlo integration in \textsc{fewz} does not converge at NNLO for lepton production for bin widths of $\Delta\eta_\ell=0.2$ to sufficient accuracy of $\mathcal{O}(1\%)$ within a reasonable run time of $\mathcal{O}(\rm months)$.  This problem has been noticed by other groups (see, for example, Refs.~\cite{Adam:2008ge,Adam:2008pc}).  Since the effect of NNLO corrections is expected to cancel almost entirely when calculating the asymmetry $A(\eta_\ell)$, we instead use NLO $K$-factors calculated with NNLO PDFs. We have confirmed that the NNLO corrections to the $W$ asymmetry (as a function of the $W$ rapidity) are extremely small; see Fig.~11 of Ref.~\cite{Anastasiou:2003ds}.}  We also include the effect of a finite $W$ width in these $K$-factors.\footnote{The future use of \textsc{applgrid}~\cite{Carli:2005ji} may eliminate the need to resort to calculating LO $W$ and $Z$ cross sections multiplied by fixed $K$-factors during PDF fits, although the latter method should still be a very good approximation.}

\begin{figure}
  \centering
  \includegraphics[width=0.8\textwidth,clip]{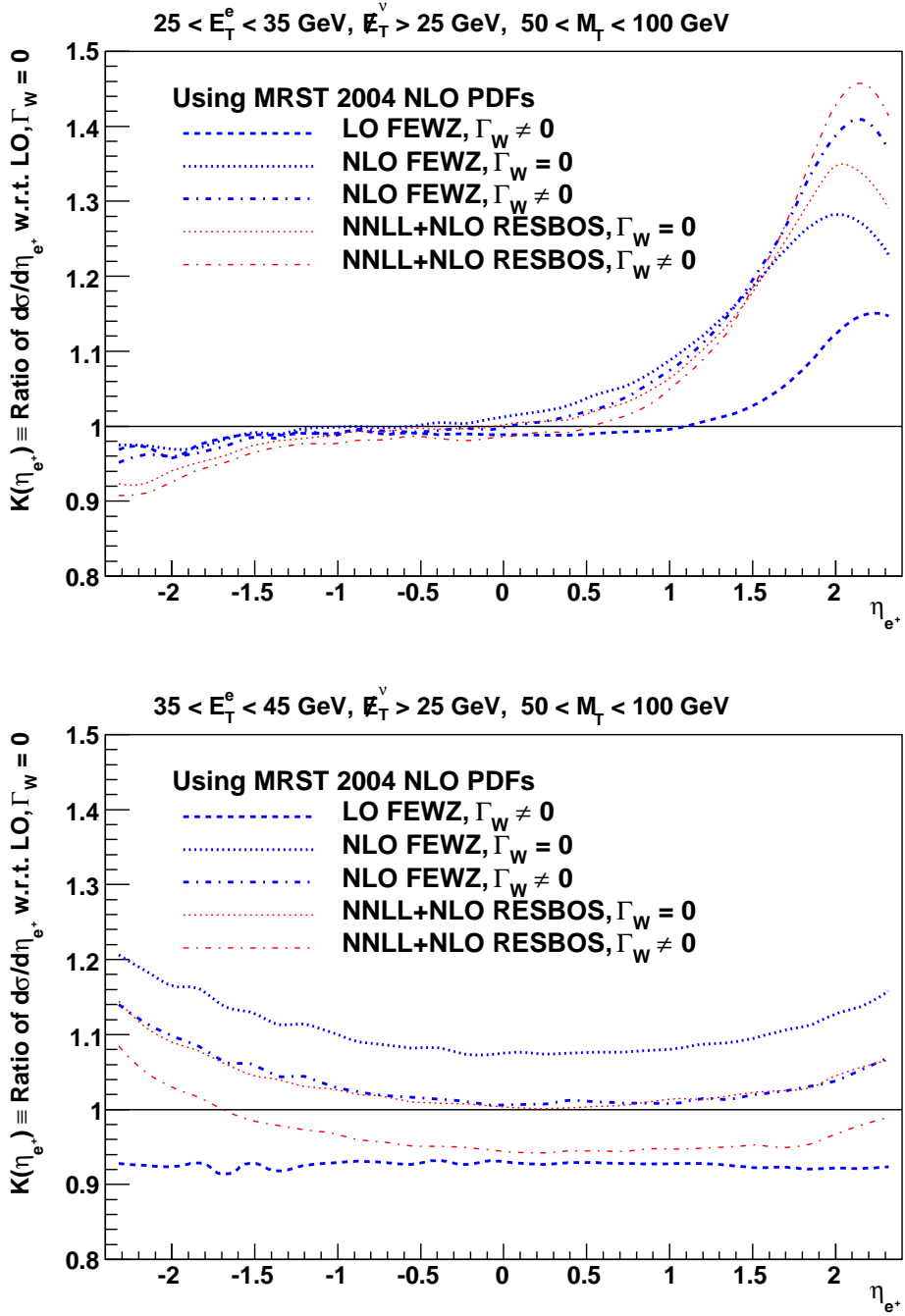}
  \caption{${\rm d}\sigma/{\rm d}\eta_{e^+}$ $K$-factors for CDF kinematic cuts~\cite{Acosta:2005ud} for various methods of calculation.}
  \label{fig:cdfkfactors04}
\end{figure}
In Fig.~\ref{fig:cdfkfactors04} we show the effect of a finite $W$ width, NLO corrections and NNLL $p_T^W$-resummation on the lepton pseudorapidity distribution when using a common set of parton distributions (MRST 2004 NLO).
\begin{figure}
  \centering
  \includegraphics[width=0.8\textwidth,clip]{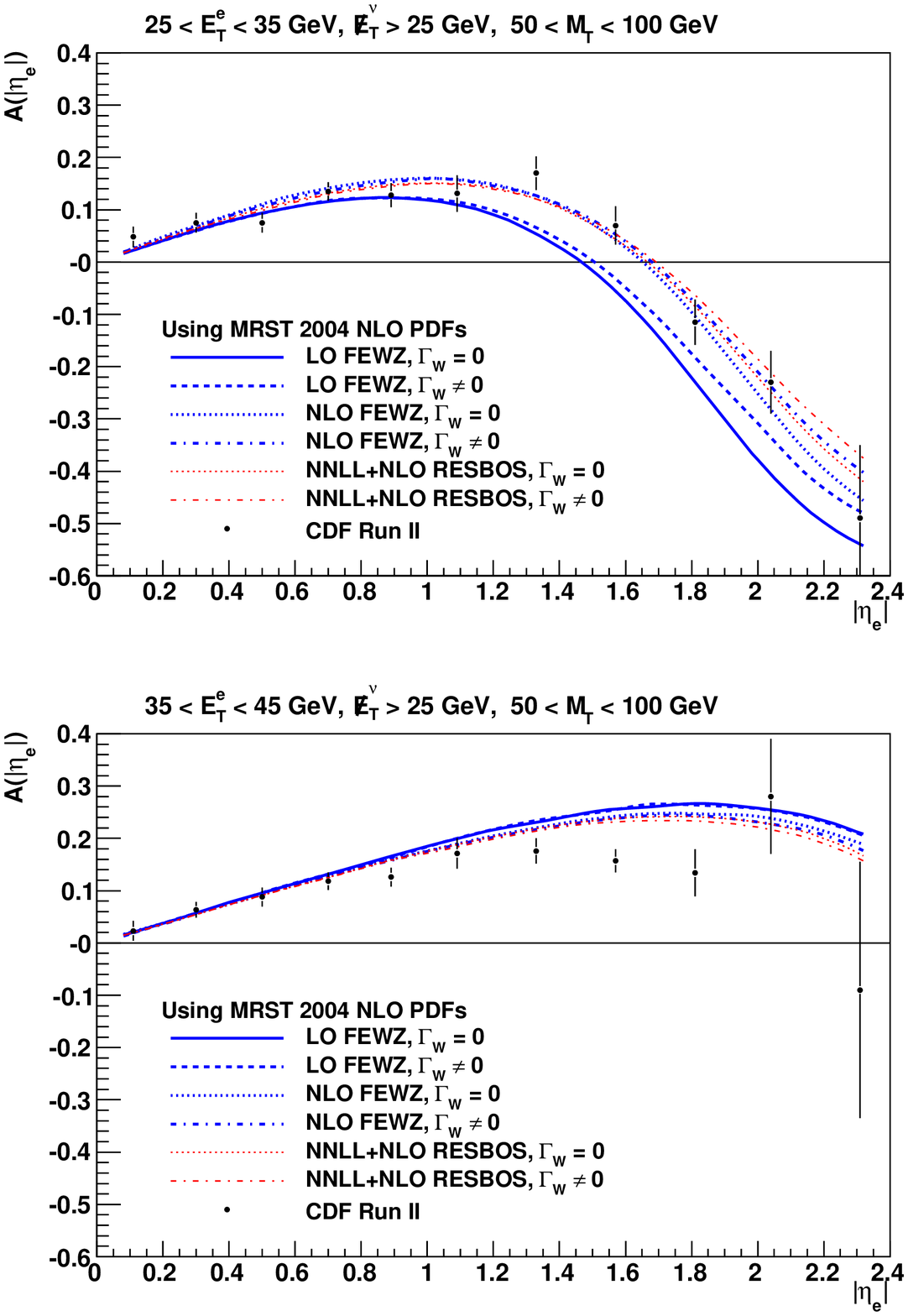}
  \caption{$W\to e\nu$ charge asymmetry from CDF~\cite{Acosta:2005ud} for various methods of calculation.}
  \label{fig:cdfasymmetry04}
\end{figure}
Then in Fig.~\ref{fig:cdfasymmetry04} we show the corresponding predictions for the lepton charge asymmetry.  In the lower $E_T^e$ bin, finite $W$ width and higher-order effects are important only in the forward direction, meaning that these features have a significant effect on the asymmetry shown in Fig.~\ref{fig:cdfasymmetry04}.  (Recall that CP invariance requires that ${\rm d}\sigma(\ell^+)/{\rm d}\eta_{\ell} = {\rm d}\sigma(\ell^-)/{\rm d}(-\eta_{\ell})$.)  In the upper $E_T^e$ bin, these features are almost symmetric about $\eta_{e}=0$, meaning that the effect on the asymmetry is small.  From Fig.~\ref{fig:cdfasymmetry04} we see that at NLO the effect on the asymmetry of the finite $W$ width is smaller than the experimental uncertainties, but is not insignificant, and hence we include this.  The further effect of resummations is generally smaller still, so given that our fit is meant to correspond to PDFs at a fixed perturbative order we do not include the effect of resummations in our $K$-factors.

\begin{figure}
  \centering
  \includegraphics[width=0.8\textwidth,clip]{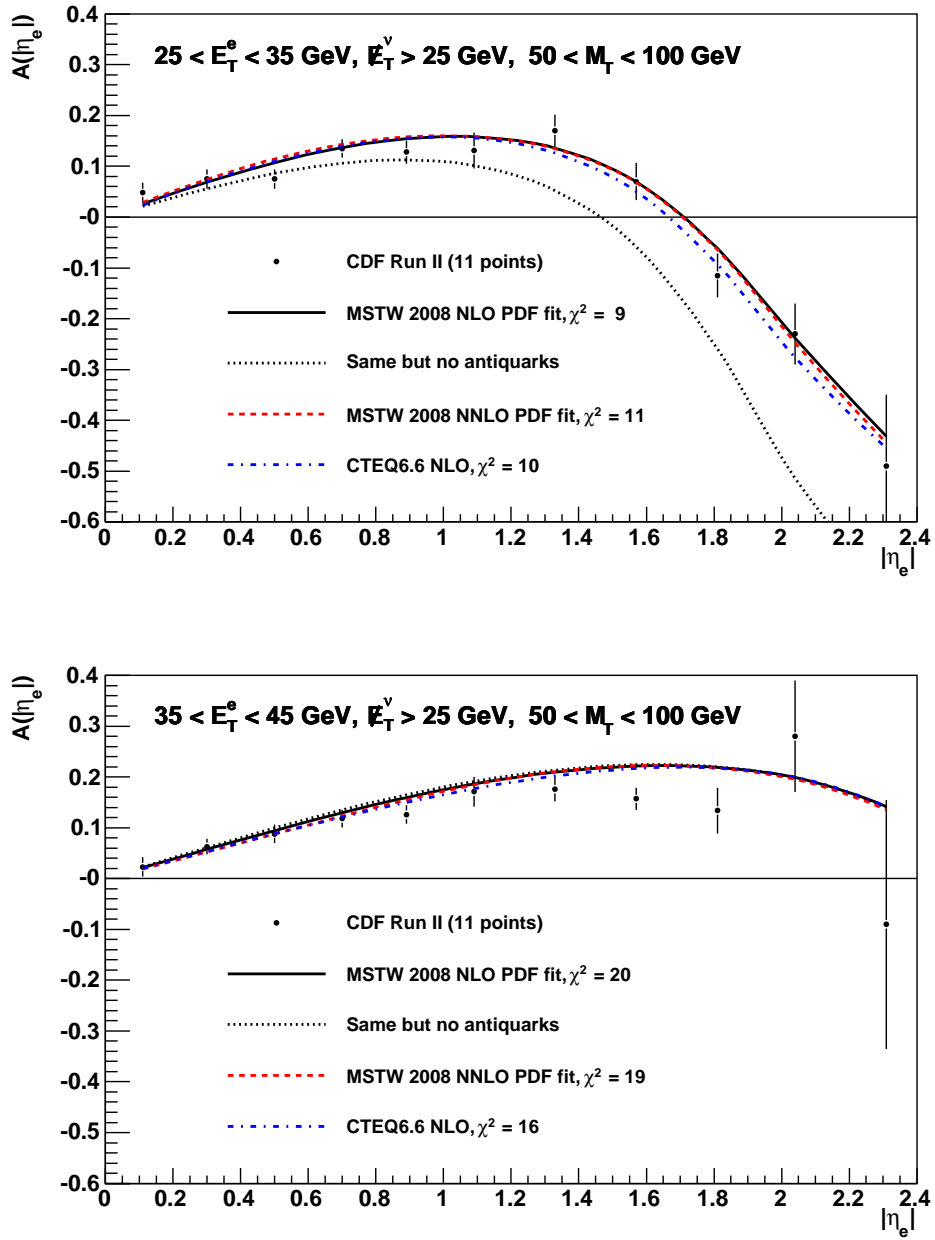}
  \caption{A comparison of theory and data for $W\to e\nu$ charge asymmetry from CDF~\cite{Acosta:2005ud}.}
  \label{fig:cdfasymmetry}
\end{figure}
Figure~\ref{fig:cdfasymmetry} shows the fit to the most recent CDF $W\to e \nu$ lepton charge asymmetry data~\cite{Acosta:2005ud}.  The data are presented in two bins of lepton $E_T$.  In each case the data are reasonably well fitted at both NLO and NNLO.  Also shown is the NLO fit with the antiquark contributions removed.  The difference is much more significant in the lower $E_T$ bin, as expected.  In the higher $E_T$ bin there appears to be some evidence for a systematic discrepancy between the fit and the data, particularly at high $\vert\eta_e\vert$ (these data seem to be described slightly better by CTEQ6.6 PDFs).  This could in principle be removed by increasing the $d$ quark distribution in the $x \sim 0.1$ region, thereby enhancing the negative $d(x_1)$ contribution in the numerator in \eqref{eq:leptonasymmapprox}, but this would produce a tension with DIS structure function data.  The overall impact on the $d$-quark distribution will be discussed further below.

\begin{figure}
  \centering
  \includegraphics[width=0.8\textwidth,clip]{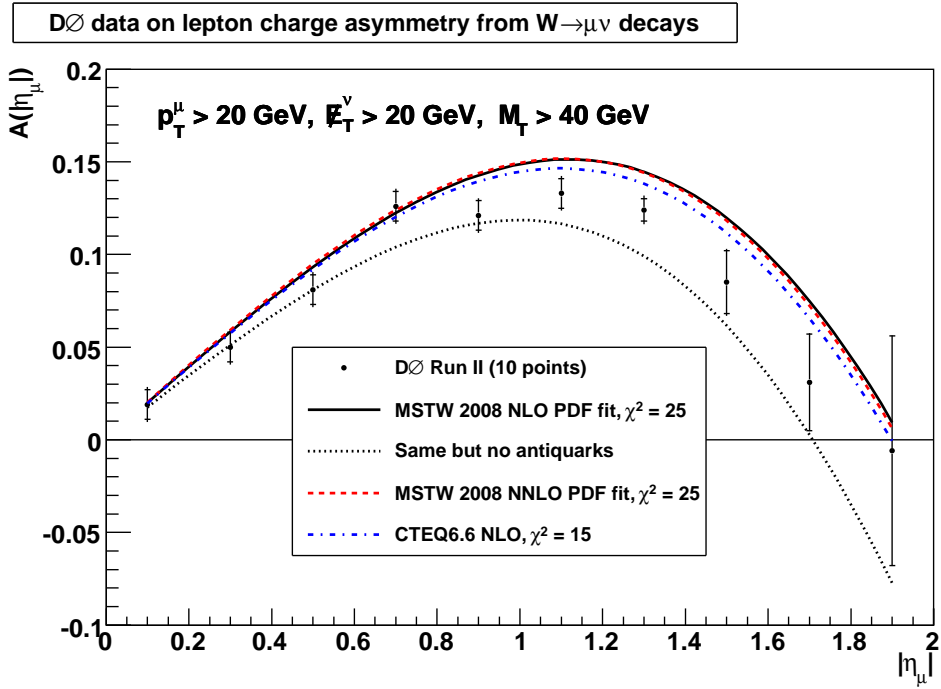}
  \caption{A comparison of theory and data for $W\to \mu\nu$ charge asymmetry from D{\O}~\cite{Abazov:2007pm}.}
  \label{fig:d0asymmetry}
\end{figure}
The corresponding fits to the D{\O} $W\to \mu \nu$ lepton charge asymmetry data from Ref.~\cite{Abazov:2007pm} are shown in Fig.~\ref{fig:d0asymmetry}.  The data are presented in a single lepton $p_T^\mu > 20$~GeV bin, and the antiquark contributions are therefore sizeable.  Some systematic discrepancy at high $\vert\eta_\mu\vert $ is again apparent.  Note that there are more recent D{\O} electron data~\cite{Abazov:2008qv} than the muon data shown in Fig.~\ref{fig:d0asymmetry}.  However when we include these D{\O} electron data in the global analysis we are unable to obtain a good quality fit, with significant tension between the new D{\O} electron asymmetry data and the DIS structure function data (both proton and deuterium) and with low-mass Drell--Yan data.  Furthermore, we notice considerable tension between the D{\O} electron asymmetry data~\cite{Abazov:2008qv} and the preliminary CDF $W$ asymmetry data~\cite{Han:2008zzc}.\footnote{The CDF $W$ asymmetry data have since been published~\cite{Aaltonen:2009ta}.}  Pending further investigation of these issues, we have decided not to include either the more recent D{\O} data~\cite{Abazov:2008qv} or the CDF $W$ asymmetry data~\cite{Han:2008zzc,Aaltonen:2009ta} in the current fit.

\subsection{Description of Tevatron \texorpdfstring{$Z$}{Z} rapidity distribution data}

The integrated luminosity for Run II at the Tevatron has allowed a high-precision measurement of the $Z$ rapidity distribution.  Analogously to the low-energy Drell--Yan distribution ${\rm d}^2\sigma/{\rm d} M{\rm d} y$ discussed in Section \ref{sec:lowmassDY}, this provides a constraint on the (mainly quark) PDFs over a broad range of $x$ at $Q^2 \sim M_Z^2$.  The mix of quark distributions probed is approximately
\begin{equation}
  0.29\,u(x_1)\,u(x_2) + 0.37\,d(x_1)\,d(x_2),
\end{equation}
in contrast to the ($4 u\bar{u} + d\bar{d}$) combination probed in $pp$ Drell--Yan production.  The $Z$ rapidity distribution is therefore particularly sensitive to the $d$ distribution, providing complementary information to the $W$ charge asymmetry discussed above.

The theoretical calculations are performed using the same form of $K$-factors as for the low-mass Drell--Yan data \eqref{eq:KDYLO}, \eqref{eq:KDYNLO} and \eqref{eq:KDYNNLO}, with the NLO and NNLO coefficients calculated exactly using the \textsc{vrap} program~\cite{Anastasiou:2003ds}.

\begin{figure}
  \centering
  \includegraphics[width=0.8\textwidth,clip]{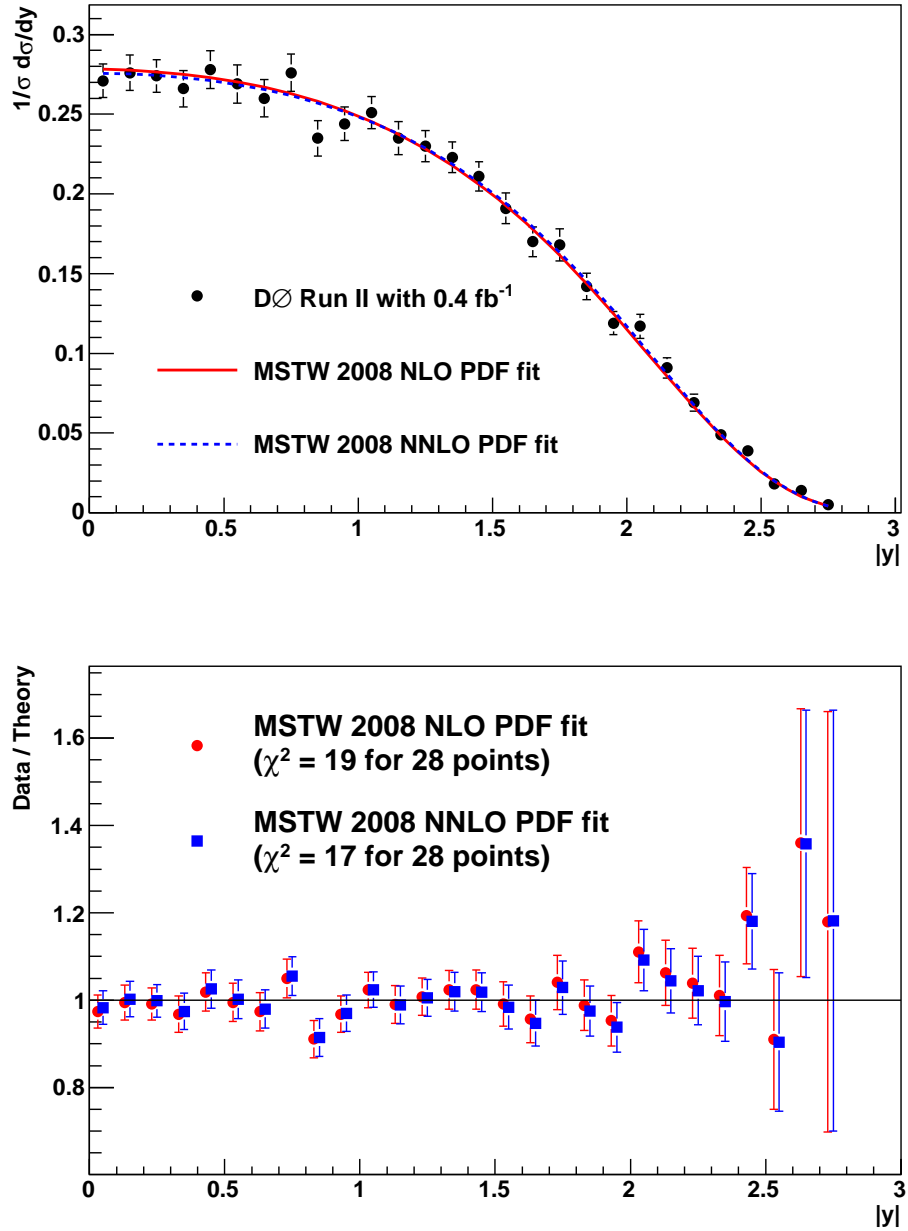}
  \caption{The comparison of data and theory for the $Z/\gamma^*$ rapidity shape distribution from D{\O}~\cite{Abazov:2007jy}.}
  \label{fig:d0zrap}
\end{figure}
In Fig.~\ref{fig:d0zrap} we show the results of the NLO and NNLO fits compared to the $Z/\gamma^*$ rapidity shape distribution measured in the mass range $71 < M_{ee} < 111$ GeV by D{\O} using 0.4~fb$^{-1}$ of Run II data~\cite{Abazov:2007jy} .
\begin{figure}
  \centering
  \includegraphics[width=0.8\textwidth,clip]{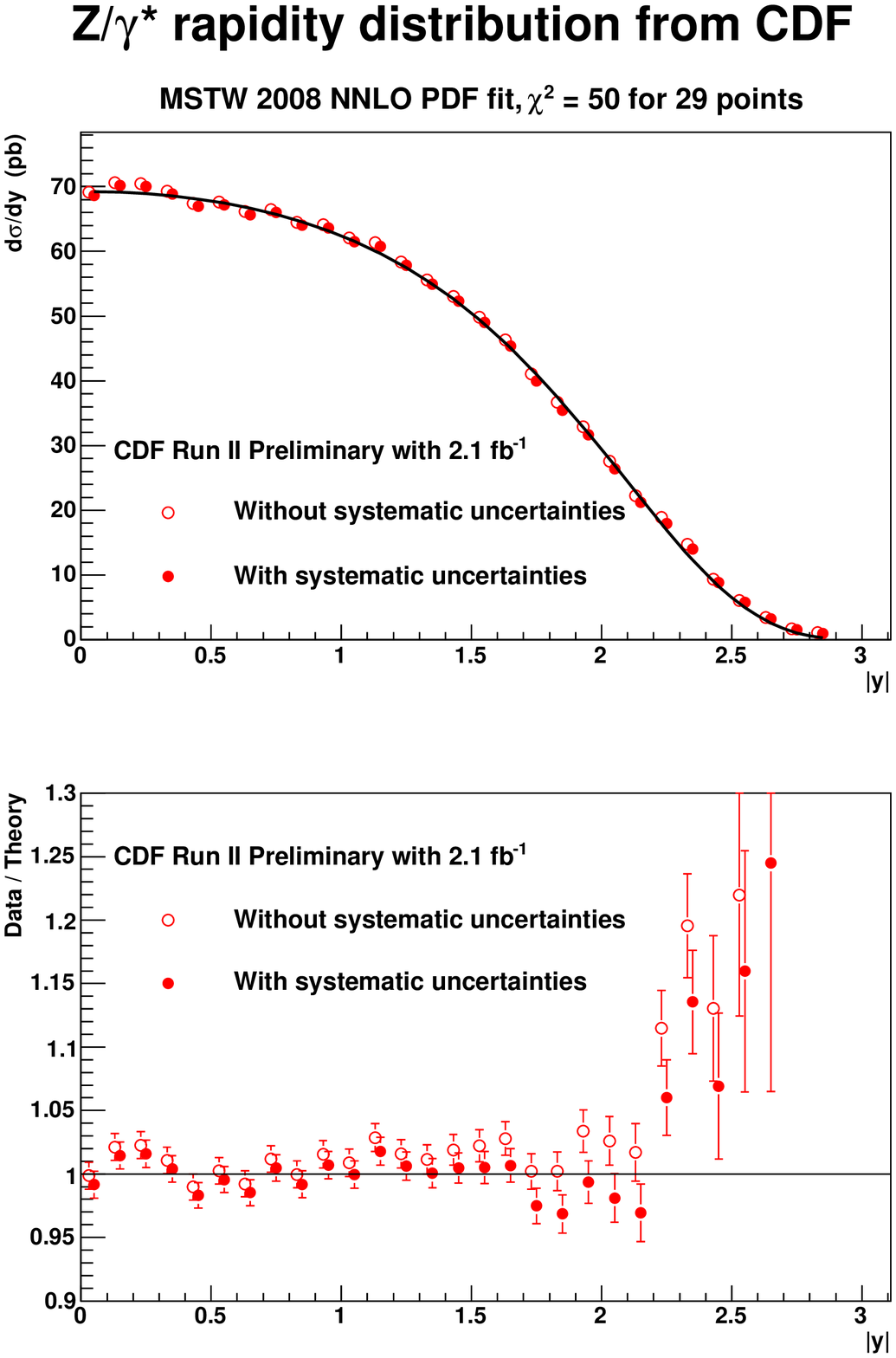}
  \caption{The comparison of data and theory for the $Z/\gamma^*$ rapidity distribution from CDF~\cite{Han:2008}.}
  \label{fig:cdfzrap}
\end{figure}
In Fig.~\ref{fig:cdfzrap} we show the results of the NNLO fit compared to the preliminary $Z/\gamma^*$ rapidity distribution measured in the mass range $66 < M_{ee} < 116$ GeV by CDF using 2.1~fb$^{-1}$ of Run II data~\cite{Han:2008}.  The data are well fitted, particularly when the systematic uncertainties are taken into account.  There is evidence of a slight excess of data over theory at high rapidity ($y>2.2$ $\Rightarrow$ $x_1>0.4$ and $x_2<0.005$).

During the fit the $Z$ rapidity distribution was calculated at LO in the narrow-width approximation ($M_{ee} = M_Z$) neglecting the $\gamma^*$ contribution, then applying the appropriate $K$-factors to account for higher-order QCD corrections.  To be more precise, the invariant mass range $66 < M_{ee} < 116$ GeV should be explicitly integrated over in the theory prediction allowing for a finite $Z$ width and including the $\gamma^*$ contribution.  This more precise treatment has the effect of lowering the theoretical prediction by less than 1\% over most of the rapidity range, rising above unity at $|y|\approx 2$, then rising more rapidly as the kinematic limit is approached at still larger rapidities.  Refitting the PDFs with this additional factor included gives an increase in the fitted CDF normalisation by 1\% and a better description of the CDF $Z/\gamma^*$ rapidity data at large rapidity, with smaller shifts needed in the correlated systematic uncertainties.  For the NLO fit, the $\chi^2$ for the 29 CDF data points improves from 49 to 42, while the $\chi^2$ for the 28 D{\O} data points improves from 19 to 17.  However, the refitted PDFs are almost unchanged, with the differences very small compared to the quoted uncertainties.  Since the CDF data are still unpublished and may change before publication, we will update our fits in the near future using the final $Z/\gamma^*$ data and accounting for the precise $M_{ee}$ range and $\gamma^*$ contribution in the theoretical calculation.

\subsection{Impact on the down valence quark distribution}

\begin{figure}
  \centering
  \includegraphics[width=0.8\textwidth,clip]{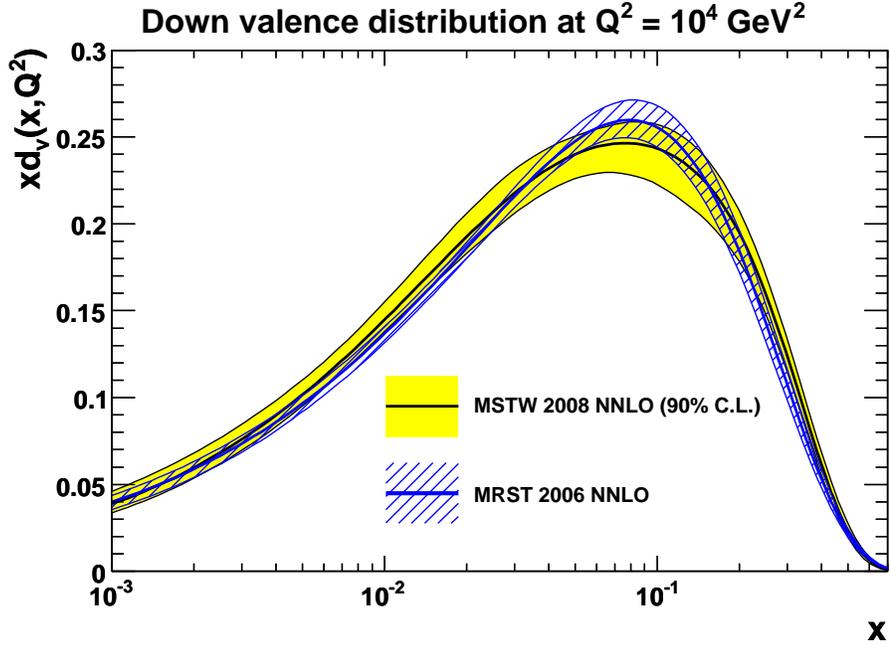}
  \caption{The impact of new data on the down valence quark distribution at NNLO.}
  \label{fig:downvalence}
\end{figure}

The $W$ and $Z$ data at the Tevatron are mainly sensitive to the up and down quark distributions of the proton.  However, the up quark distribution is already well constrained by proton structure function data where it appears charge-weighted.  Therefore, in practice, the Tevatron $W$ and $Z$ data mainly constrain the down quark distribution.  In Fig.~\ref{fig:downvalence} we compare the down valence distribution at NNLO with that from the MRST 2006 NNLO set.  Note that we now make a better choice of parameters in the input $d_v$ distribution \eqref{eq:dv} to allow for more flexibility for the eigenvector PDF sets, which generally gives larger uncertainties than previously, despite the presence of more constraining data.  Clearly there is a significant change in the shape, though the new and previous versions generally agree within the $90\%$ C.L.~limits.  This change is mainly due to the inclusion of new Tevatron data, though there is also some influence from the new neutrino structure function data.  The most significant change is the increase in $d_v(x,Q^2)$ for $x\sim 0.3$, seen more clearly in Fig.~\ref{fig:comparisonNNLO}(b), which is influenced by the lepton asymmetry data. From the number sum rule this has to be compensated for by a decrease elsewhere, which is mainly for $x \sim 0.07$. 

%% file: jetdata.tex
\section{Description of jet data} \label{sec:jetdata}
Knowledge of the high-$x$ gluon distribution is important for calculations of both signal and background in new physics searches at the LHC.  The main direct constraint comes from Tevatron data on inclusive jet production.  This is well illustrated by looking at the analyses of the Run I Tevatron jet data.

\subsection{Description of Run I Tevatron jet data}
Historically, early CDF Run I inclusive jet data at central rapidity~\cite{Abe:1996wy} indicated an excess in the high-$E_T$ jet spectrum using NLO QCD with the available parton distributions of the time.  This discrepancy was initially taken as a possible indication of quark compositeness.  It was found to be ``impossible''~\cite{Glover:1996ae} to describe simultaneously both the CDF jet data and the DIS structure function data with a common set of parton distributions having a physically motivated input form.  However, it was found that the data could be accommodated in a global fit with an input gluon parameterisation with a somewhat contrived shoulder at high $x$~\cite{Lai:1996mg}.  Subsequently, D{\O} Run I jet data became available, which measured the $E_T$ distribution at a range of rapidities~\cite{Abbott:2000ew}.  NLO global analyses were able to fit all the jet data with physically reasonable forms, though there was some tension between the description of the high-$E_T$ CDF and D{\O} data.  This ``saga'' indicated the need to provide PDFs with associated uncertainties so that it can easily be seen in which regions of $x$ the PDFs are relatively constrained or unconstrained.  (These issues are further discussed in, for example, Refs.~\cite{Stump:2003yu,Mangano:2008ag}.)

The Tevatron Run I jet data have been routinely included in the default global analyses since CTEQ4~\cite{Lai:1996mg} and MRST 2001~\cite{Martin:2001es}.  The MRST 2004 analysis~\cite{Martin:2004ir} showed that a physical parameterisation of the input gluon distribution could be obtained by taking the input form \eqref{eq:inputxg} in the DIS scheme then transforming to the conventional $\overline{\rm MS}$ scheme.  The resultant $\overline{\rm MS}$ parameterisation was found to automatically give a shoulder-like form at large $x$, which produced a better description of the Tevatron Run I inclusive jet data~\cite{Abbott:2000ew,Affolder:2001fa} than the previous MRST 2001 PDFs.

At NLO, the cross section for inclusive jet production at the Tevatron can be written as
\begin{equation} \label{eq:jetcross}
  \sigma_{p\bar{p}} = \alpha_S^2(\mu_R^2) \sum_{a,b=q,g}\left[\hat{\sigma}_{ab}^{\rm LO}+\alpha_S(\mu_R^2)\hat{\sigma}_{ab}^{\rm NLO}\right]\otimes f_{a/p}(x_a,\mu_F^2) \otimes f_{b/\bar{p}}(x_b,\mu_F^2).
\end{equation}
Usually, multidimensional (Monte Carlo) phase space integration is used to evaluate \eqref{eq:jetcross} with cuts on, for example, the transverse energy $E_T$ and pseudorapidity $\eta$ of the produced jets.  However, this can take hours or days of CPU time, so it is impractical to include jet data in a PDF fit in this way.  The old (approximate) solution used by MRST was to calculate ``$K$-factors'', defined as $\sigma_{p\bar{p}}^{\rm NLO}/\sigma_{p\bar{p}}^{\rm LO}$, for a given set of PDFs.  Then six pseudogluon ``data'' points at $\mu_F^2=2000$ GeV$^2$ (and a value of $\Lambda_{\rm QCD}$ corresponding to $\alpha_S(M_Z^2) \approx 0.115$) were inferred that would correspond to the best $\sigma_{p\bar{p}}$ needed to describe the data.  These six pseudogluon points (with a free normalisation) and the $\Lambda_{\rm QCD}$ point were included in the MRST fit rather than the original jet data points.  The comparison to the actual jet data points, including the inclusion of the correlated systematic uncertainties, was only made \emph{after} the fit, though it was checked that the $\chi^2$ changes induced in the pseudodata were quantitatively similar to those in the final comparison to the jet data.  The CTEQ group \emph{do} include the jet data points explicitly in their fit, but the jet cross section is still calculated at LO and a NLO $K$-factor applied.

A much improved solution, which we adopt here, is to interpolate the PDFs and $\alpha_S$ so that they can be factorised from $\hat{\sigma}_{ab}$ in \eqref{eq:jetcross}.  Then the convolution integrals of \eqref{eq:jetcross} are replaced with multiplication.  The phase space integration then only needs to be done once for a given set of kinematic cuts, with the result stored in a grid for later use during a PDF fit.  This technique has previously been used by H1~\cite{Adloff:2000tq} to determine the gluon density from DIS jet data, and by ZEUS~\cite{Chekanov:2005nn} in a PDF fit including jet data from DIS and photoproduction.  Grids corresponding to the kinematic cuts used in published Tevatron and HERA jet data are provided by the \textsc{fastnlo} project~\cite{Kluge:2006xs} which uses the \textsc{nlojet++} calculations~\cite{Nagy:2001fj,Nagy:2003tz}.  (Another project with similar aims, which has also been applied to calculate $W$ and $Z$ production using \textsc{mcfm}~\cite{Campbell:2000bg}, is the \textsc{applgrid} project~\cite{Carli:2005ji}.)

\subsection{Description of Run II Tevatron jet data} \label{sec:run2jets}
Recently, Tevatron Run II inclusive jet data from both CDF~\cite{Abulencia:2007ez,Aaltonen:2008eq} and D{\O}~\cite{Abazov:2008hu} have become available with much higher statistics and much better control over the corrections necessary to go from the ``raw'' data to measurements at the parton ``jet'' level.  For example, D{\O} now claim a relative uncertainty on the jet $p_T$ of less than $2\%$.  Our present global analysis uses all the available Run II jet data, including the complete information on correlated systematic uncertainties in the $\chi^2$ definition \eqref{eq:chisqcorr}.  Due to the increased reliability, and statistics, of the Run II jet data, we now omit the Run I data from the global fit.

Another change is that we now use a renormalisation and factorisation scale $\mu_R=\mu_F=p_T$, rather than $\mu_R=\mu_F=p_T/2$.\footnote{Note that the Run II data are binned in terms of transverse momentum $p_T$ and rapidity $y$, while the Run I data were binned in terms of transverse energy $E_T$ and pseudorapidity $\eta$, although $E_T=p_T$ and $\eta=y$ for the massless jets used in Run I, but not for the massive jets used in Run II.}  The latter scale appears to be very convergent for central rapidities, but becomes unstable for larger $y$.  For example, the NLO contribution is some 60--70$\%$ of that at LO for $y \sim 2.5$.  In comparison, the scale choice $\mu_R=\mu_F=p_T$ leads to NLO corrections of $\approx 20$--$30\%$ for all $y$ (except at the highest $y$ and $p_T$, where they are smaller) and hence, while not as convergent near $y=0$, the scale choice $\mu_R=\mu_F=p_T$ is reasonably stable for all rapidities of the data.\footnote{We thank M.~Wobisch for bringing this to our attention.}

\begin{figure}
  \centering
  \includegraphics[width=\textwidth,clip]{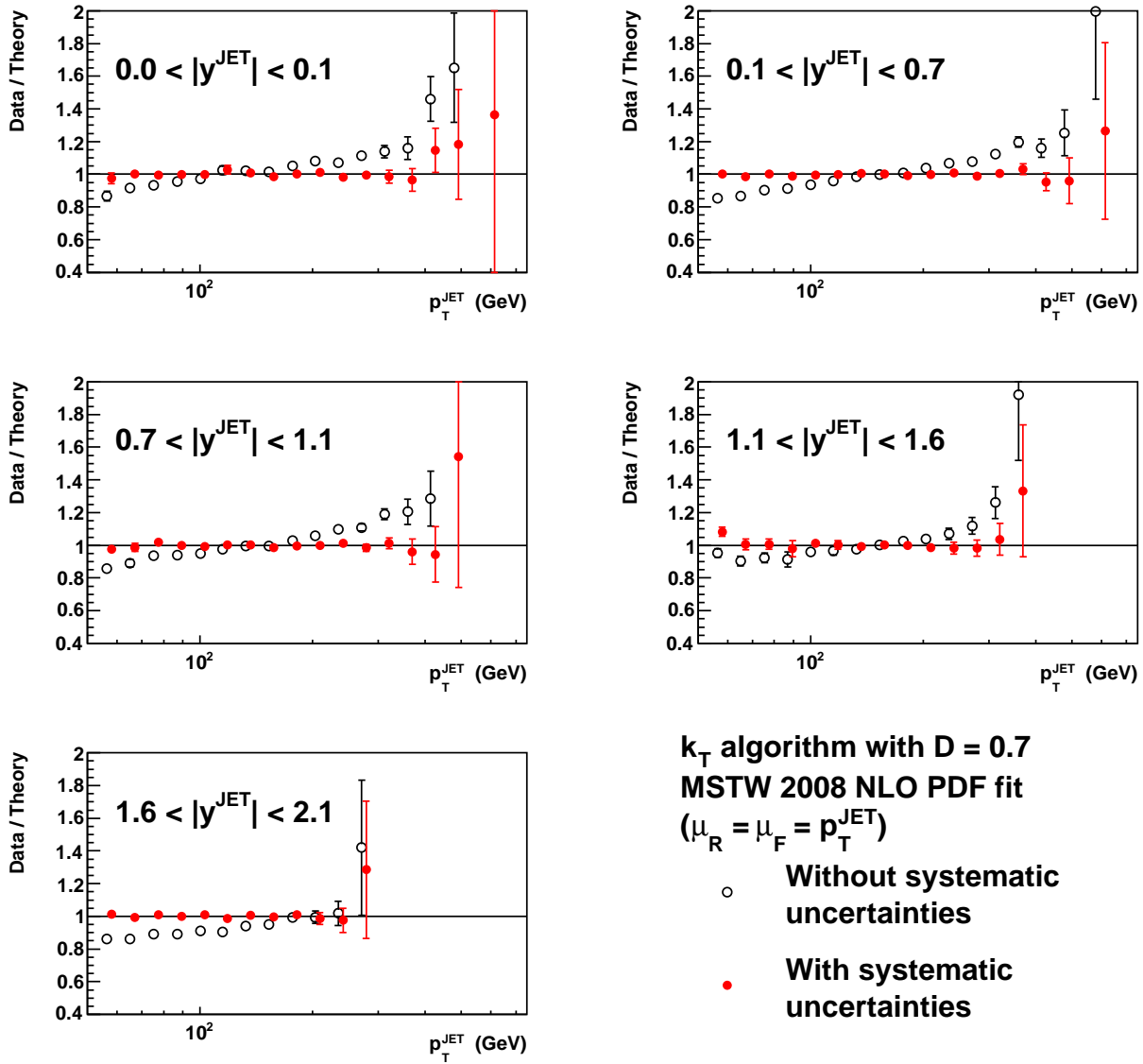}
  \caption{Data/theory ratio for the CDF Run II data obtained using the $k_T$ jet clustering algorithm~\cite{Abulencia:2007ez} and the MSTW 2008 NLO fit.}
  \label{fig:cdfjet2}
\end{figure}
In Fig.~\ref{fig:cdfjet2} we show the comparison of the NLO fit with CDF Run II inclusive jet data using the $k_T$ jet algorithm~\cite{Abulencia:2007ez}.  There is also an analysis of the same data using the cone-based Midpoint algorithm~\cite{Aaltonen:2008eq}, but the $k_T$ algorithm data~\cite{Abulencia:2007ez} were published and implemented in our analysis first, therefore only the $k_T$ algorithm data are included in the final global fit.  Furthermore, there is a theoretical preference for using the $k_T$ jet algorithm rather than a seeded cone algorithm, since the latter is not formally infrared safe, which in practice means that the perturbative calculation is only accurate to a given order of the expansion (NLO for the Midpoint algorithm).
\begin{figure}
  \centering
  \includegraphics[width=\textwidth,clip]{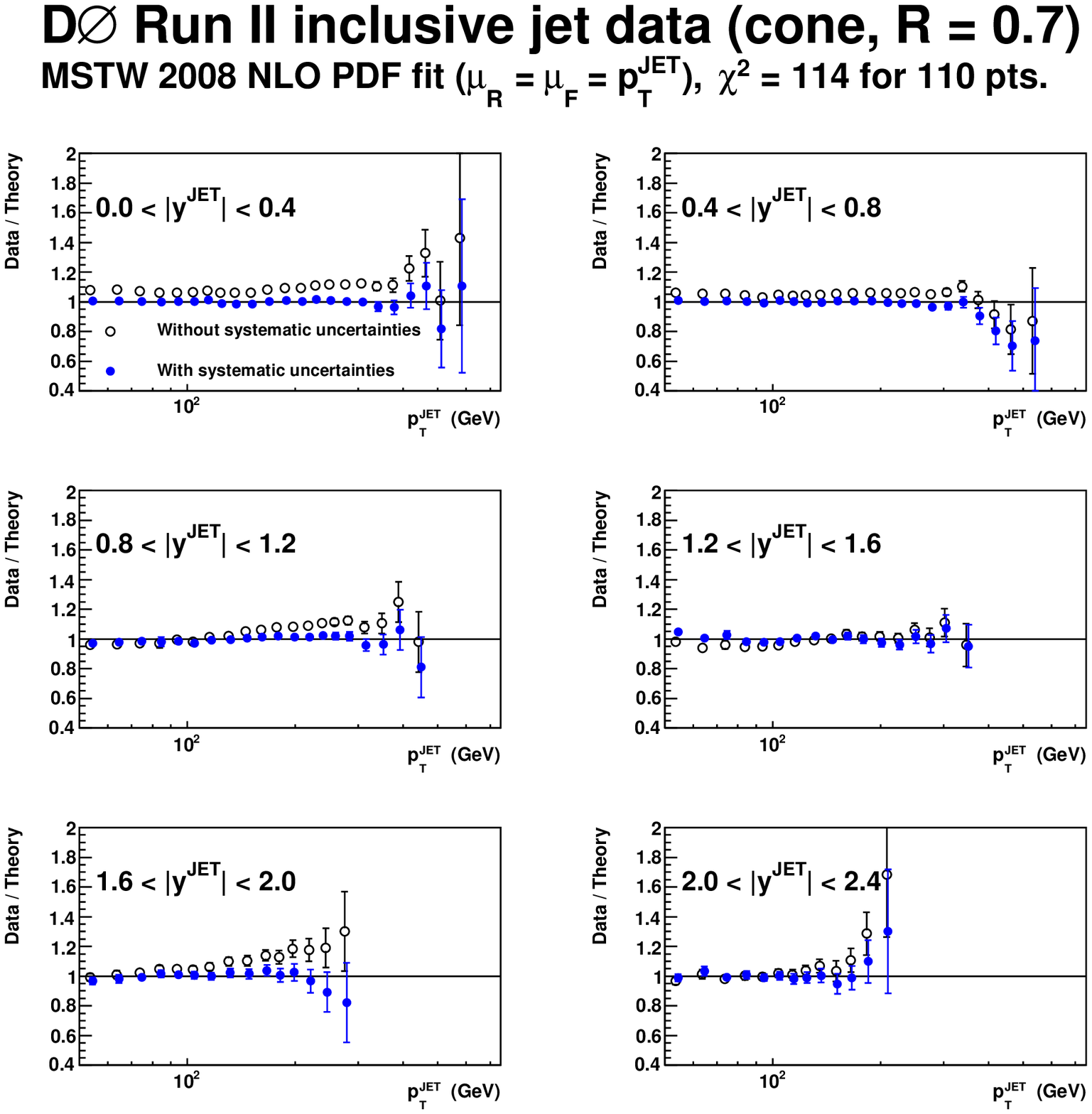}
  \caption{Data/theory ratio for the D{\O} Run II data~\cite{Abazov:2008hu} and the MSTW 2008 NLO fit.}
  \label{fig:d0jet2}
\end{figure}
In Fig.~\ref{fig:d0jet2} we show the comparison of the NLO fit to D{\O} Run II inclusive jet data using a cone jet algorithm~\cite{Abazov:2008hu}.  We achieve an excellent description of both data sets shown in Figs.~\ref{fig:cdfjet2} and \ref{fig:d0jet2}.  Note that the data points are allowed to shift by the correlated systematic errors according to the $\chi^2$ definition \eqref{eq:chisqcorr}.  Recall from Table~\ref{tab:normalisations} that the CDF Run II inclusive jet data are forced to share the same normalisation as the CDF Run II $Z$ rapidity distribution~\cite{Han:2008} due to the common luminosity uncertainty.  In practice, the shared normalisation of 0.96 at NLO (or 0.99 at NNLO) is determined by the much more constraining $Z$ rapidity distribution.

\begin{figure}
  \centering
  \includegraphics[width=\textwidth,clip]{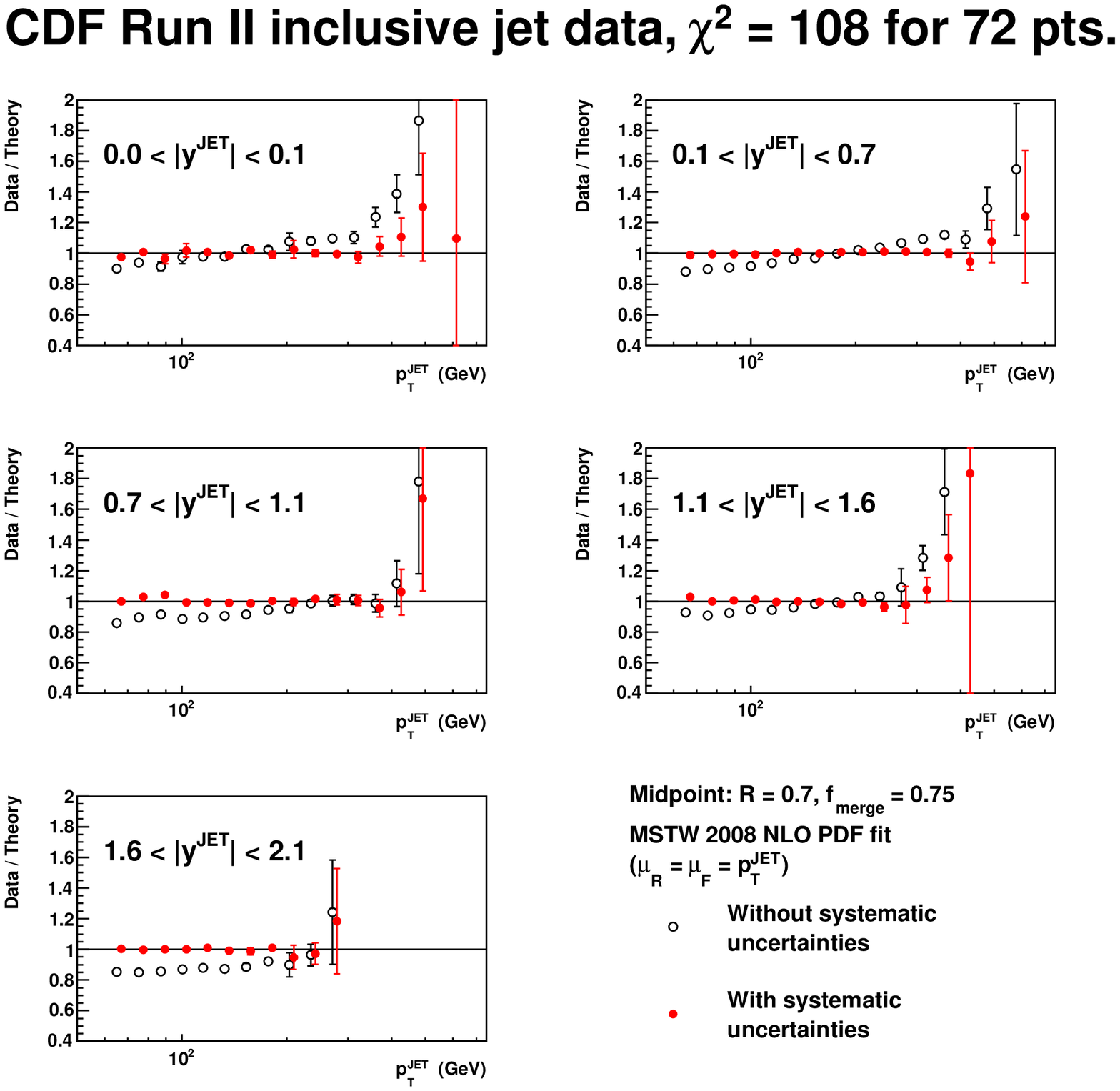}
  \caption{Data/theory ratio for the CDF Run II data obtained using the cone-based Midpoint jet clustering algorithm~\cite{Aaltonen:2008eq} and the MSTW 2008 NLO fit.  These data are not included in the fit.}
  \label{fig:cdfjet2mid}
\end{figure}
In Fig.~\ref{fig:cdfjet2mid} we compare the predictions using the MSTW 2008 NLO PDFs to the CDF Run II data obtained using the Midpoint cone algorithm~\cite{Aaltonen:2008eq}, which were not included in the fit.\footnote{We include the complete 24 (excluding luminosity) sources of correlated systematic uncertainty~\cite{Aaltonen:2008eq}.  Note that the data tables originally published in Ref.~\cite{Aaltonen:2008eq} have received minor corrections, and the corrected data are shown in Fig.~\ref{fig:cdfjet2mid}.}  The agreement is generally good, but with the second and third point for $0.7<|y^{\rm JET}|<1.1$, and the first point for $1.1<|y^{\rm JET}|<1.6$, each contributing about 11--13 to the total $\chi^2 = 108$ for 72 points.\footnote{We note that the rapidity region $0.7<|y^{\rm JET}|<1.1$ corresponds to a crack between the central and plug calorimeters~\cite{Aaltonen:2008eq}.}  These three points appear simply to be outliers.  Replacing the CDF Run II data measured using the $k_T$ jet algorithm by these cone algorithm data and refitting improves the $\chi^2$ by 6 and the $\chi^2_{\rm global}$ by 4.  The $\chi^2$ of the D{\O} Run II cone algorithm data is unchanged on refitting.  We conclude that there is no significant tension between the three Tevatron Run II data sets on inclusive jet production~\cite{Abulencia:2007ez,Abazov:2008hu,Aaltonen:2008eq} obtained using either the cone~\cite{Abazov:2008hu,Aaltonen:2008eq} or $k_T$~\cite{Abulencia:2007ez} jet clustering algorithms.\footnote{If, instead of tying the normalisation to the CDF Run II $Z$ rapidity distribution~\cite{Han:2008}, we allow a separate normalisation for the CDF Run II inclusive jet data, the preferred normalisation at NLO for the Midpoint algorithm data is 1.06, with an improvement in $\chi^2$ to 91 for 72 points.  By contrast, the $k_T$ algorithm data prefer a normalisation of 0.98 with almost no improvement in $\chi^2$ compared to the default normalisation of 0.96.}

\subsection{Impact on the gluon distribution}
The impact of the Tevatron Run II jet data on the high-$x$ behaviour of the gluon distribution is interesting.  These data prefer a smaller high-$x$ gluon distribution than the Run I data.  Indeed, we find that it is no longer necessary to take the input gluon parameterisation \eqref{eq:inputxg} in the DIS scheme and then transform to the $\overline{\rm MS}$ scheme, as was done in the MRST 2004~\cite{Martin:2004ir} and MRST 2006~\cite{Martin:2007bv} analyses in order to give an enhanced gluon at high $x$ to better fit the Tevatron Run I data.  Instead, the input gluon form \eqref{eq:inputxg} is now taken directly in the $\overline{\rm MS}$ scheme.

\begin{figure}
  (a)\\
  \begin{minipage}{\textwidth}
    \centering
    \includegraphics[width=0.8\textwidth,clip]{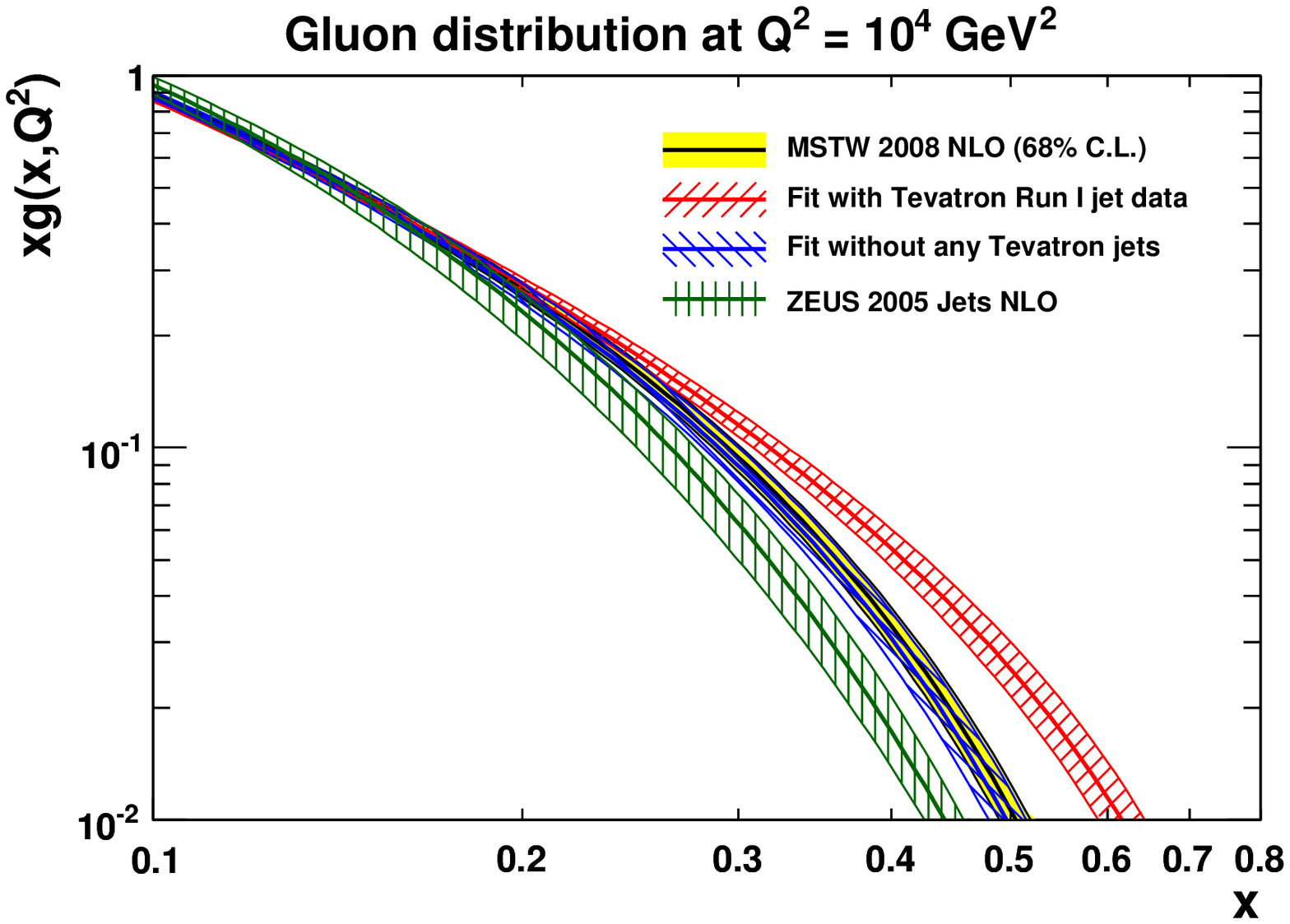}
  \end{minipage}
  (b)\\
  \begin{minipage}{\textwidth}
    \centering
    \includegraphics[width=0.8\textwidth,clip]{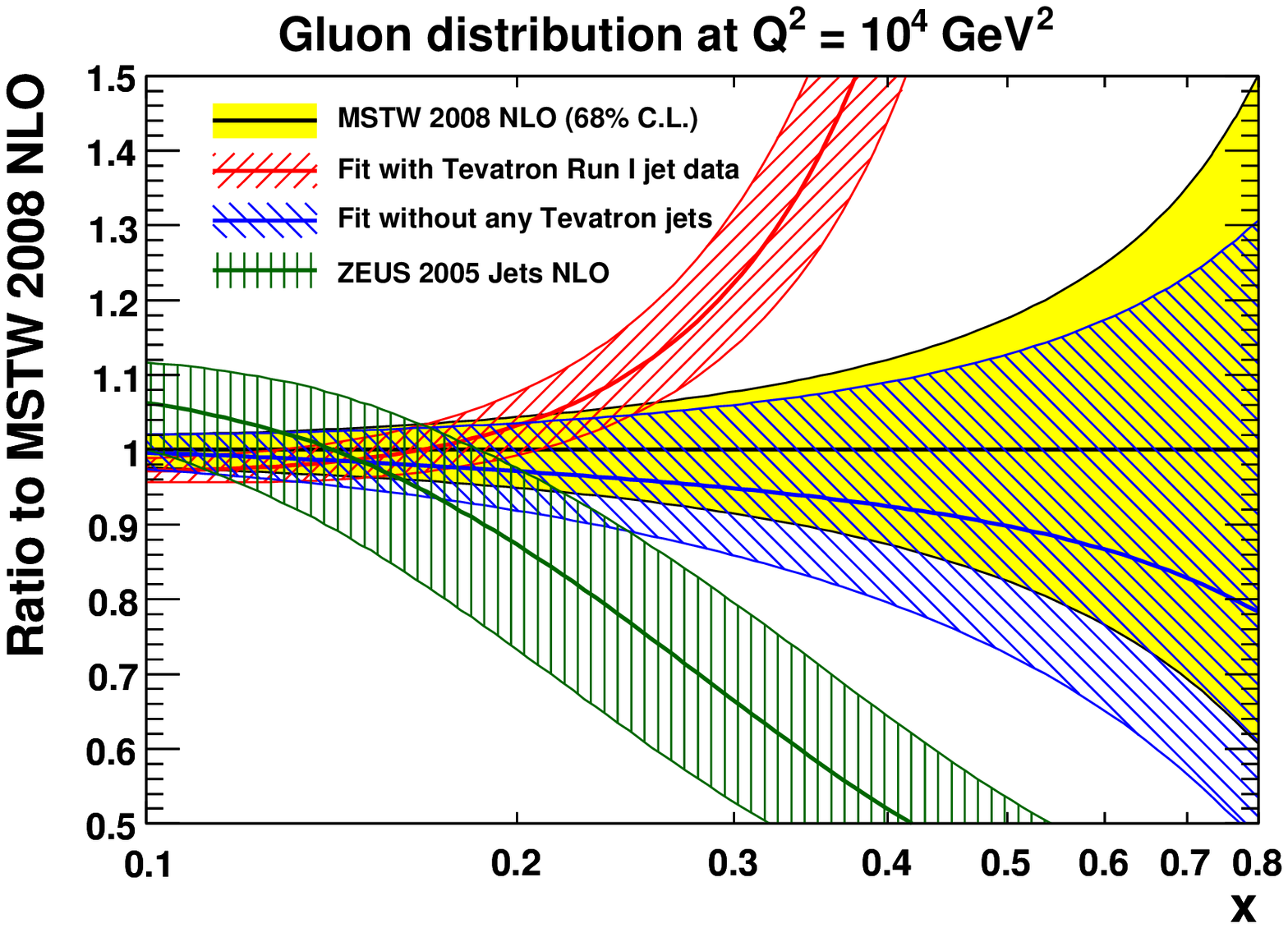}
  \end{minipage}
  \caption{The impact of the Tevatron jet data on the high-$x$ gluon distribution at NLO.}
  \label{fig:gluonhighx}
\end{figure}
In Fig.~\ref{fig:gluonhighx} we show the high-$x$ gluon distribution from the MSTW 2008 NLO analysis compared to that obtained from a fit with the Tevatron Run II jet data replaced by Tevatron Run I jet data, and from a fit without any Tevatron jet data.  The fit to Tevatron Run I data uses \textsc{fastnlo}~\cite{Kluge:2006xs} and includes hadronisation corrections that were previously neglected in all global fits, but can reach the level of a $\mathcal{O}(10\%)$ correction in the lower $E_T$ bins.  We manage to fit the Tevatron Run I jet data well using the input gluon parameterisation \eqref{eq:inputxg} in the $\overline{\rm MS}$ scheme, without needing to perform the DIS$\to\overline{\rm MS}$ transformation used in the MRST 2004~\cite{Martin:2004ir} and MRST 2006~\cite{Martin:2007bv} analyses.  The 68\% C.L.~uncertainty bands from the fits to Run I and Run II data clearly disagree; the 90\% C.L.~uncertainty bands also disagree for $x\gtrsim 0.4$ (not shown here).

We also compare in Fig.~\ref{fig:gluonhighx} to the high-$x$ gluon distribution of the ZEUS 2005 Jets analysis~\cite{Chekanov:2005nn}, which is even smaller than our fit without the Tevatron jet data.  This is likely to at least partially be due to the omission of fixed-target data in the ZEUS 2005 Jets analysis.  In a global analysis, where $\alpha_S(M_Z^2)$ takes a value of $0.117$--$0.120$, the fixed-target data constrain the high-$x$ gluon distribution to be moderately large, though in general the constraint is correlated with the value of $\alpha_S(M_Z^2)$.  The ZEUS 2005 Jets analysis also has a more restrictive input gluon parameterisation than our distribution.  Note that although the ZEUS fit~\cite{Chekanov:2005nn} included additional data on dijets in photoproduction, which are not included in our fit, these data were found only to reduce the uncertainties without changing the central values, so do not provide an explanation for the discrepancy.

\begin{table}
  \centering
  \begin{tabular}{cc|cc|c|c}
    \hline\hline
    CDFI~\cite{Affolder:2001fa} & D{\O}I~\cite{Abbott:2000ew} & CDFII($k_T$)~\cite{Abulencia:2007ez} & D{\O}II~\cite{Abazov:2008hu}& $\Delta\chi^2_{\text{non-jet}}$ & $\alpha_S(M_Z^2)$ \\
    (33 pts.) & (90 pts.) & (76 pts.) & (110 pts.) & (2513 pts.) & \\ \hline
    53 & 119 & 64 & 117 & 0 & 0.1197 \\
    {\bf\large 51} & {\bf\large 48} & 132 & 180 & 9 & 0.1214 \\
    56 & 110 & {\bf\large 56} & {\bf\large 114} & 2 & 0.1202 \\
    {\bf\large 53} & {\bf\large 85} & {\bf\large 68} & {\bf\large 117} & 1 & 0.1204 \\
    \hline\hline
  \end{tabular}
  \caption{$\chi^2$ values for Tevatron data on inclusive jet production from various global fits.  The large bold values indicate the data sets explicitly included in each fit.  The ``MSTW 2008'' fit includes only the Run II data.  We also indicate the increase in the $\chi^2$ for all data sets other than Tevatron inclusive jets when these data are added to the fit.  Finally, the value of $\alpha_S(M_Z^2)$ is given for each fit.  All fits are carried out at NLO with a scale choice of $\mu_R=\mu_F=p_T$.}
  \label{tab:chi2jetskt}
\end{table}
In Table~\ref{tab:chi2jetskt} we show the $\chi^2$ values for the Tevatron inclusive jet production data from various global fits.  The four lines in the table are fits (i) without Tevatron jet data, (ii) with only Run I data, (iii) with only Run II data, and (iv) with both Run I and Run II data included.  The CDF Run I data~\cite{Affolder:2001fa} exist only for a central rapidity bin, and hence are less sensitive to the gluon distribution than the other data sets in Table~\ref{tab:chi2jetskt}.  The $\chi^2$ per data point when fitting only Run I data is significantly larger than 1 for CDF, suggesting unusually large fluctuations, and significantly smaller than 1 for D{\O}, suggesting an overestimation of errors.  The $\chi^2$ per data point when fitting only Run II data is much closer to 1 for both CDF and D{\O}, suggesting that the Run II data are more reliable than the Run I data.  The tension between the Run I and Run II data is evident from Table~\ref{tab:chi2jetskt}, i.e.~the fit to only Run I data gives a poor description of Run II data, and vice versa.  The global $\chi^2$ for the fit without Tevatron jet data is 2371 for 2513 points.  When including the Run I data it is 9 worse for the same data points, but when including the Run II data it is only 2 worse.  The $\alpha_S$ values are also slightly different: the fit without Tevatron jets has $\alpha_S(M_Z^2) = 0.1197$ compared to the fit with Run I data which has $\alpha_S(M_Z^2) = 0.1214$ and the standard fit with Run II data which has $\alpha_S(M_Z^2) = 0.1202$.  The Run II jet data are therefore slightly more consistent with the other data sets included in the global fit than the Run I jet data, and lead to a significantly lower value of $\alpha_S$, which influences the gluon distribution.

\begin{figure}
  (a)\\
  \begin{minipage}{\textwidth}
    \centering
    \includegraphics[width=0.8\textwidth,clip]{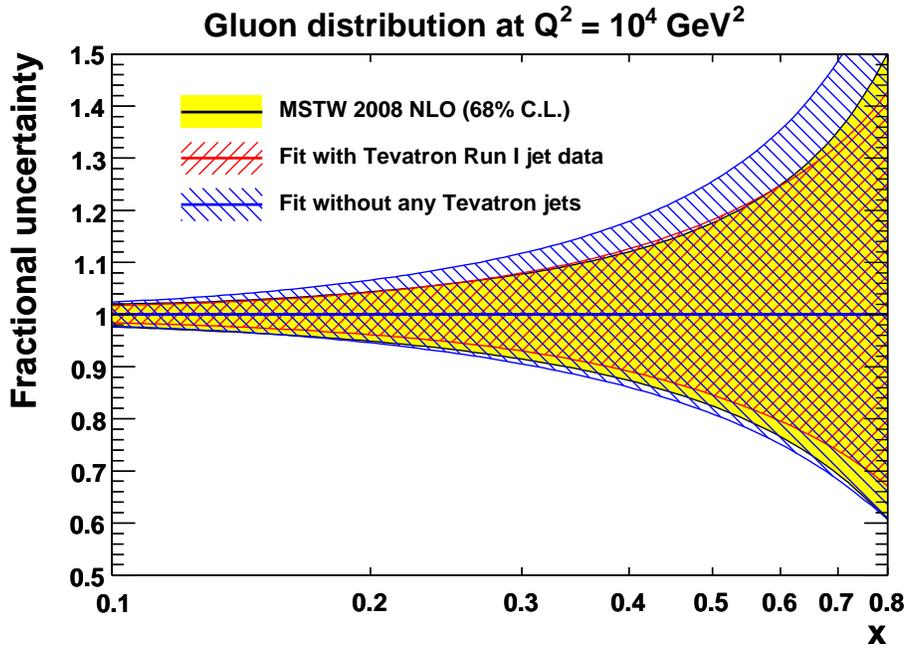}
  \end{minipage}
  (b)\\
  \begin{minipage}{\textwidth}
    \centering
    \includegraphics[width=0.8\textwidth,clip]{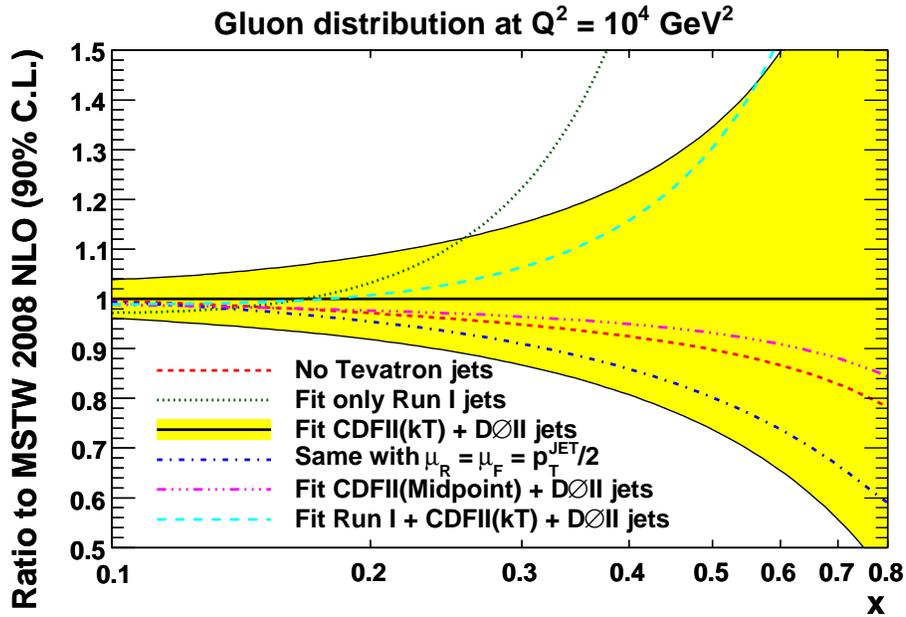}
  \end{minipage}
  \caption{(a) The impact of the Tevatron jet data on the fractional uncertainty of the high-$x$ gluon distribution at NLO.  (b) The impact on the high-$x$ gluon distribution of replacing CDF Run II jet data obtained with the $k_T$ jet algorithm~\cite{Abulencia:2007ez} by data obtained with the Midpoint cone algorithm~\cite{Aaltonen:2008eq}.  Also shown is the effect of taking an alternative scale choice.}
  \label{fig:fracgluonhighx}
\end{figure}
In Fig.~\ref{fig:fracgluonhighx}(a) we show that the inclusion of Tevatron jet data does reduce the fractional uncertainty on the high-$x$ gluon, albeit not dramatically.  This rather disappointing result is due to the large number of sizeable correlated systematic uncertainties on the jet data, which are free to adjust to best fit the data according to the $\chi^2$ definition \eqref{eq:chisqcorr}.  There is therefore a considerable trade-off between these systematic shifts and the input gluon parameters.

The high-$x$ gluon distribution from the fit using CDF Run II data obtained with the Midpoint cone algorithm~\cite{Aaltonen:2008eq} instead of the $k_T$ algorithm~\cite{Abulencia:2007ez} lies slightly below the central value of the MSTW 2008 NLO fit, but is well within the uncertainty band, lying slightly above the gluon from the fit without Tevatron jet data; see Fig.~\ref{fig:fracgluonhighx}(b).  (The value of $\alpha_S(M_Z^2) = 0.1196$.)  This fact confirms the statement made in Section \ref{sec:run2jets} that the two CDF Run II data sets are consistent.  In Fig.~\ref{fig:fracgluonhighx}(b) we also show the effect of taking a scale choice of $\mu_R=\mu_F=p_T/2$ rather than our default choice of $\mu_R=\mu_F=p_T$ for Tevatron jets.  The lower scale choice results in a smaller high-$x$ gluon distribution, which is within the 90\% C.L.~uncertainty band (but outside the 68\% C.L.~uncertainty band, not shown here).  The value of $\alpha_S(M_Z^2) = 0.1192$ is also smaller with the lower scale choice.  The quality of the description to either of the CDF Run II data sets is similar with either scale choice, but the $\chi^2$ of the D{\O} Run II data is $\approx$10 worse with the lower scale choice.  Therefore, overall, the Tevatron Run II jet data prefer our default scale choice of $\mu_R=\mu_F=p_T$.  By contrast, when fitting only to Tevatron Run I jet data, the $\chi^2$ of the D{\O} Run I data is $\approx$10 \emph{better} with $\mu_R=\mu_F=p_T/2$, i.e.~the default scale choice used in previous MRST and CTEQ analyses.  Finally, in Fig.~\ref{fig:fracgluonhighx}(b) we also show the high-$x$ gluon distribution obtained from a simultaneous fit to Run I and Run II jet data, i.e.~the last row in Table~\ref{tab:chi2jetskt}.  Unsurprisingly, it lies in-between the gluon distributions obtained from the two separate fits to either Run I or Run II jet data alone.  Note, however, that all gluon distributions shown in Fig.~\ref{fig:fracgluonhighx}(b) are obtained from fits with slightly different values of $\alpha_S$, and hence are not really directly comparable.

\begin{figure}
  (a)\\
  \begin{minipage}{\textwidth}
    \centering
    \includegraphics[width=0.8\textwidth,clip]{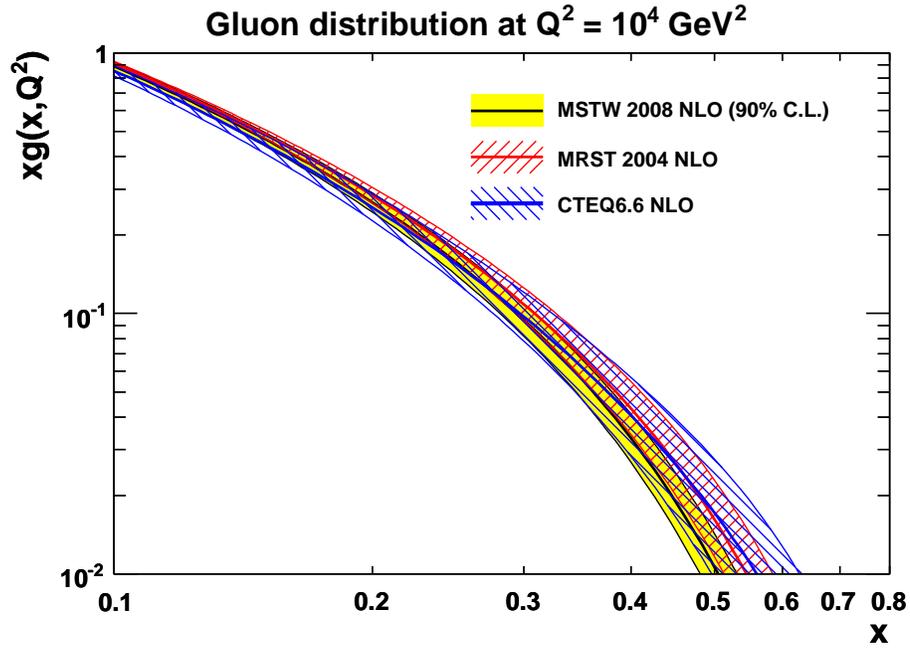}
  \end{minipage}
  (b)\\
  \begin{minipage}{\textwidth}
    \centering
    \includegraphics[width=0.8\textwidth,clip]{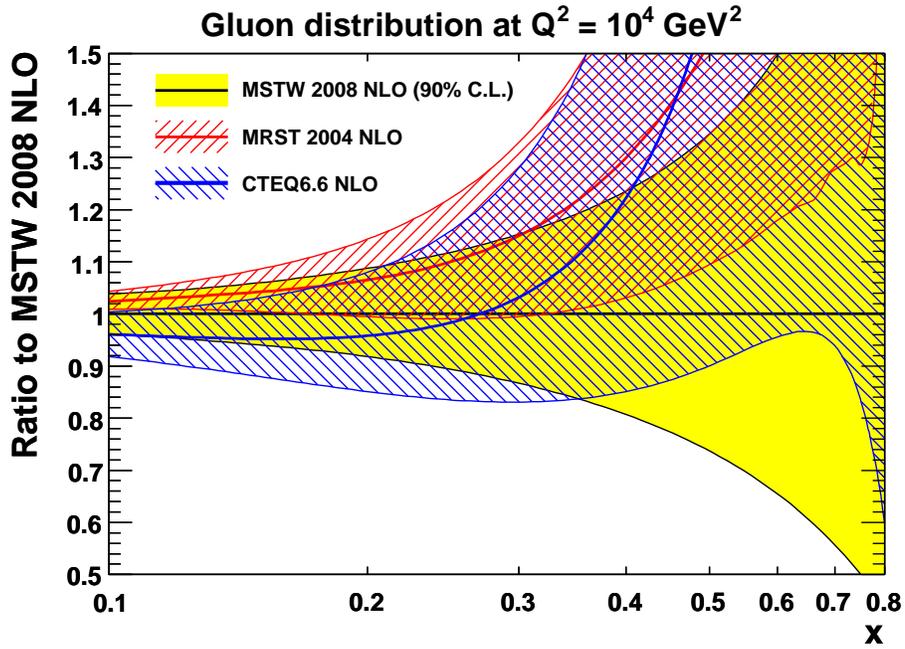}
  \end{minipage}
  \caption{Comparison of the current high-$x$ gluon distribution at NLO with that from the previous MRST~\cite{Martin:2004ir} and CTEQ~\cite{Nadolsky:2008zw} analyses.}
  \label{fig:gluonhighx1}
\end{figure}
A comparison with earlier gluon distributions is shown in Fig.~\ref{fig:gluonhighx1}.  The plots show that the inclusion of the Run I Tevatron jet data in the MRST 2004~\cite{Martin:2004ir} and CTEQ6.6~\cite{Nadolsky:2008zw} analyses led to a significantly harder gluon than the present analysis (MSTW 2008) which is based on the Run II Tevatron jet data.  (For the MRST 2004 set, where eigenvector PDF sets are not available, we assume that the fractional PDF uncertainties are the same as for the MRST 2001 set.)  In fact, with the Run II data, we are almost back to the ``soft'' MRST 2001 gluon, which did not describe the Tevatron Run I jet data nearly as well as the MRST 2004 analysis.

As explained in Section \ref{sec:choicedata}, we include the Tevatron Run II jet data in the NNLO analysis on the basis that the NNLO theoretical correction is very likely to be small and a smooth function of $p_T$, and to be less than the size of the correlated systematic uncertainties on the data.  As can be implied from Figs.~\ref{fig:gluonhighx} and \ref{fig:fracgluonhighx}, the inclusion of these data does not influence the central value of the high-$x$ gluon distribution at all significantly, but does help constrain the uncertainty.

\subsection{Jet production at HERA} \label{sec:HERAjets}

\begin{figure}
  \centering
  \includegraphics[width=0.8\textwidth,clip]{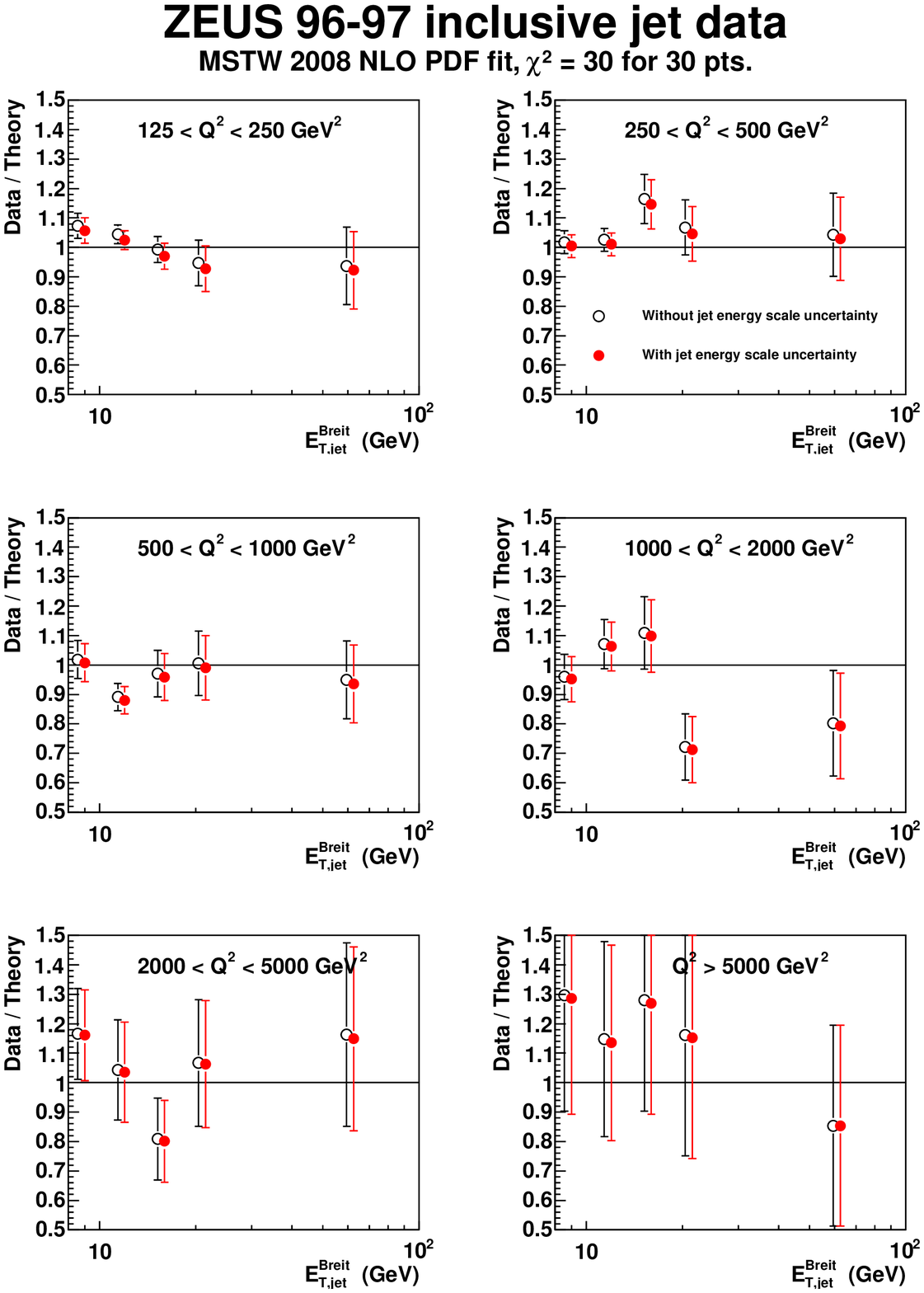}
  \caption{Data/theory ratio for the ZEUS 96--97 inclusive jet data~\cite{Chekanov:2002be} and the MSTW 2008 NLO fit.}
  \label{fig:zeusjet}
\end{figure}
\begin{figure}
  \centering
  \includegraphics[width=0.8\textwidth,clip]{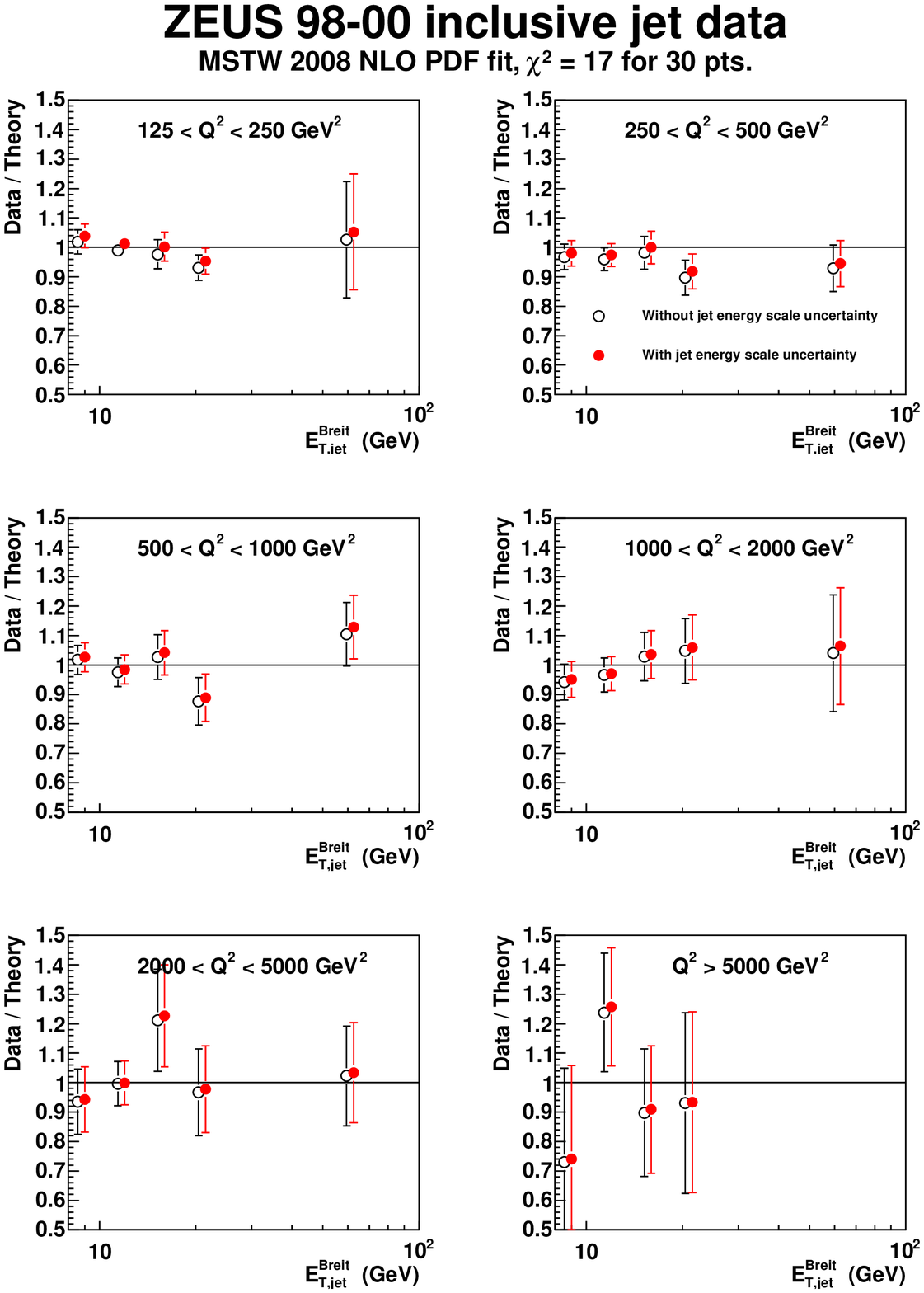}
  \caption{Data/theory ratio for the ZEUS 98--00 inclusive jet data~\cite{Chekanov:2006xr} and the MSTW 2008 NLO fit.}
  \label{fig:zeusjet2}
\end{figure}
\begin{figure}
  \centering
  \includegraphics[width=0.8\textwidth,clip]{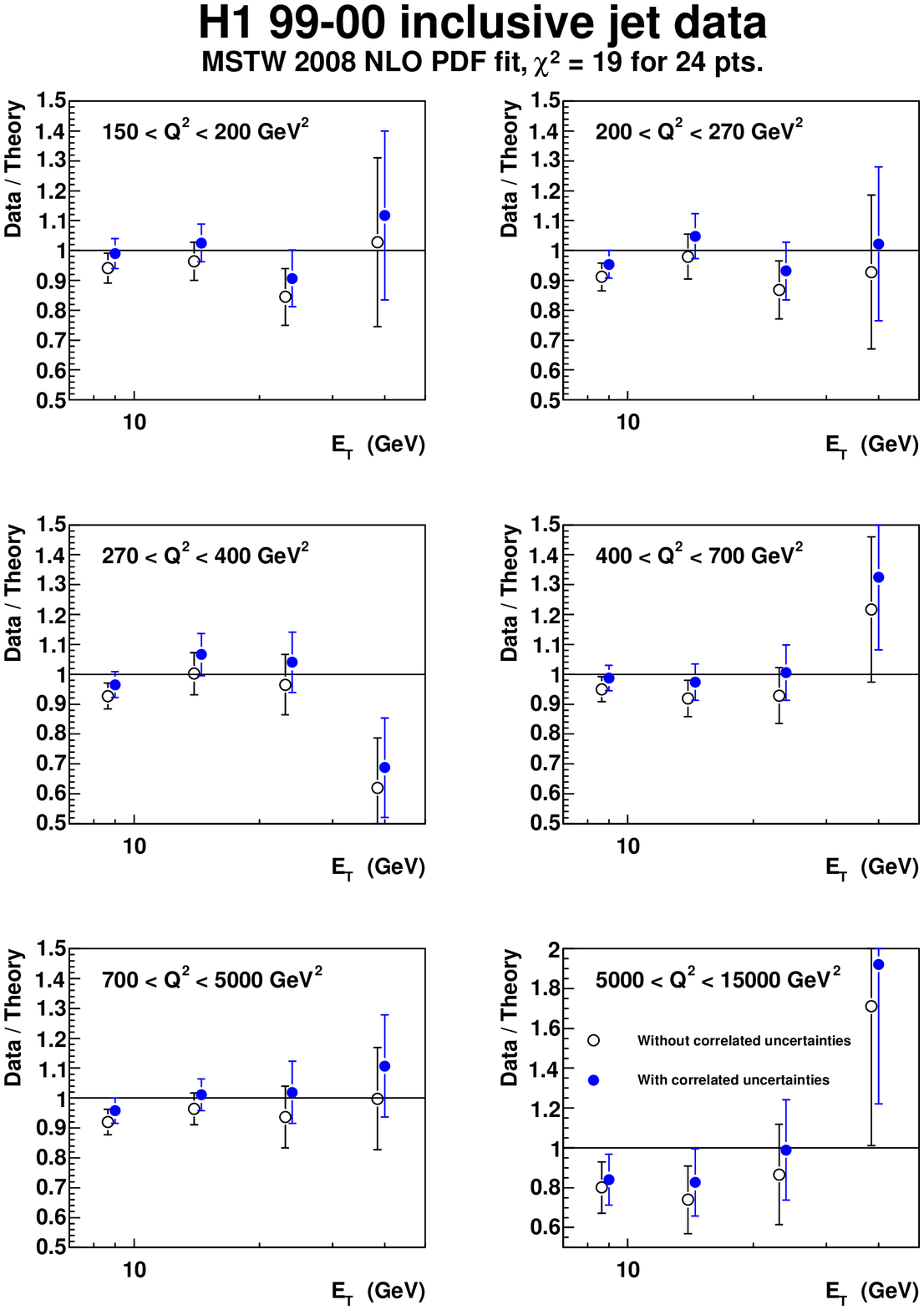}
  \caption{Data/theory ratio for the H1 99--00 inclusive jet data~\cite{Aktas:2007pb} and the MSTW 2008 NLO fit.}
  \label{fig:h1jet}
\end{figure}
In Figs.~\ref{fig:zeusjet} and \ref{fig:zeusjet2} we show the comparison of the NLO fit to ZEUS data~\cite{Chekanov:2002be,Chekanov:2006xr} on inclusive jet production, and in Fig.~\ref{fig:h1jet} we show the comparison of the NLO fit to H1 data~\cite{Aktas:2007pb}.  Jet data have been included previously by the ZEUS Collaboration in their own PDF analysis~\cite{Chekanov:2005nn}.  H1 have similarly included inclusive jet data in their determination of diffractive PDFs~\cite{Aktas:2007bv}.  However, we find that the HERA jet data have little influence on either the central values or the PDF uncertainties: the constraint they provide with the current accuracy on the data being mimicked by other data sets in a fully global fit.  We do not include the dijet photoproduction data as this involves further systematic uncertainties due to the choice of photon PDFs.

We omit the HERA jet data from the NNLO fit since it is quite probable that the unknown NNLO corrections are fairly large.  However, the description of these data is excellent if we use NLO hard cross sections with NNLO parton distributions, i.e.~comparable to the fit we obtain at NLO.  If the NNLO correction is equivalent to a fairly smooth (and not too large) $K$-factor, the good fit could likely be maintained by the interplay of data relative to theory using the correlated systematic uncertainties.

%% file: longitudinal.tex
\section{Low-\texorpdfstring{$x$}{x} gluon distribution and \texorpdfstring{$F_L$}{FL}} \label{sec:longitudinal}

In this section we outline our results for the small-$x$ gluon distribution and the implications of a measurement of $F_L(x,Q^2)$.  This is, in principle, a very important quantity to compare to predictions since it gives an independent test of the gluon distribution at low $x$ to accompany the indirect determination from $\partial F_2(x,Q^2)/ \partial \ln Q^2$, which beyond LO results in a perhaps surprisingly small, or even negative, gluon distribution.  However, it is also a direct test of the success of extensions of our theory beyond fixed-order calculations.  Until recently we have been limited to consistency checks on the relationship between $F_2(x,Q^2)$ and $F_L(x,Q^2)$ at high $y$, where both contribute to the total cross section $\tilde \sigma(x,Q^2)=F_2(x,Q^2)-y^2/(1+(1-y)^2)F_L(x,Q^2)$.  Extracting $F_L$ {\em data} from this requires an extrapolation in $y$ making some theory assumptions.  It is more useful to fit directly to  $\tilde \sigma(x,Q^2)$ where there is a turnover at the highest $y$.  This was discussed in Section \ref{sec:HERAinclusive}, where we saw that although the full turnover in $\tilde \sigma(x,Q^2)$ implied by the data was not seen at NLO, or even in full for the central fit at NNLO, the quality of fit to the data was acceptable given the large uncertainties.  Clearly the precision of such studies is limited, and can be affected by systematics, e.g.~the photoproduction background uncertainty.  However, using the final low-energy run at HERA, a direct measurement of $F_L(x,Q^2)$ at HERA has been possible, and some results have already been presented~\cite{Aaron:2008tx,Chekelian:2008br}.

\begin{figure}
  \centering
  \includegraphics[width=0.8\textwidth,clip]{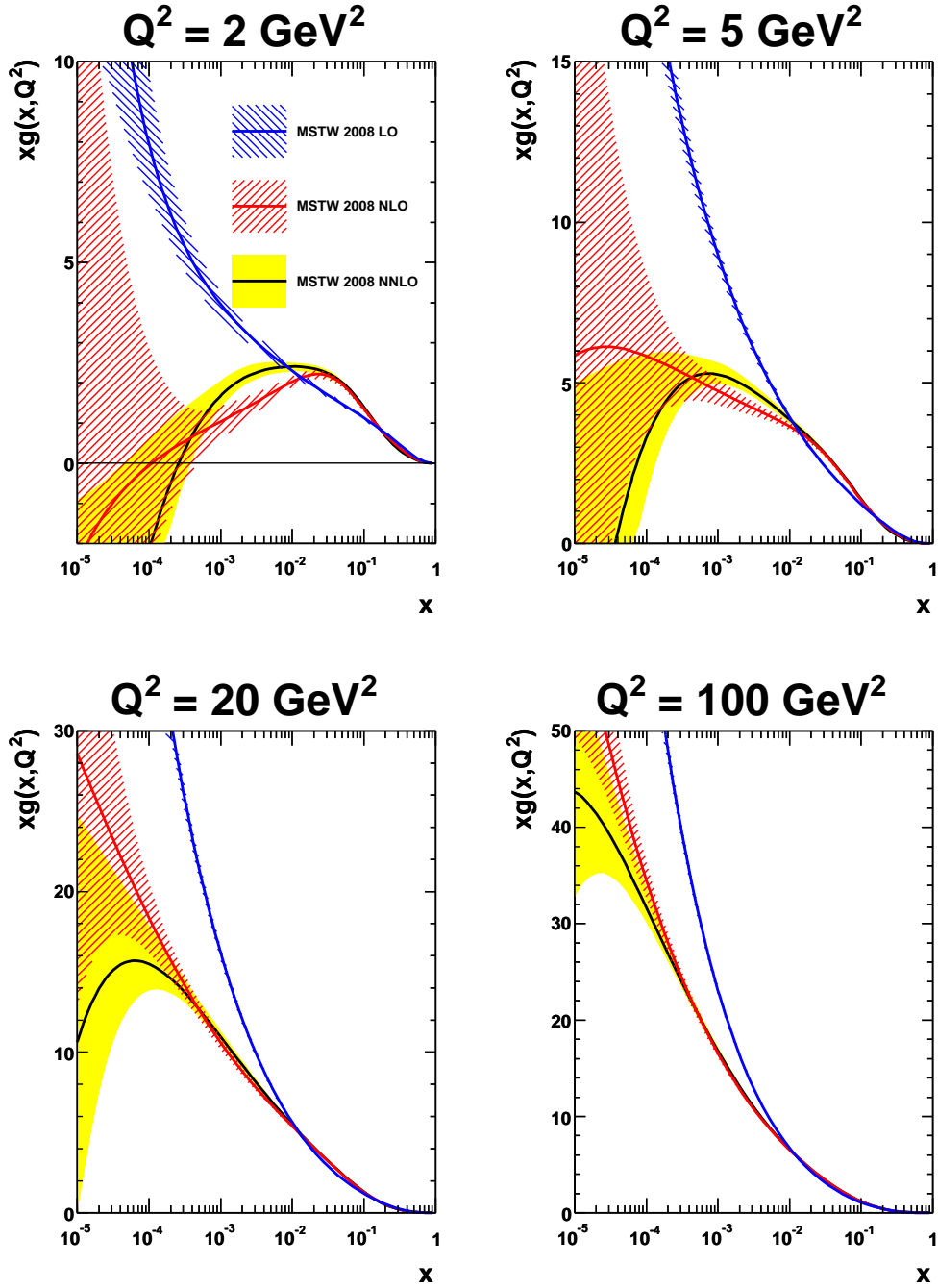}
  \caption{The gluon distribution at LO, NLO and NNLO including the one-sigma PDF uncertainty bands.}
  \label{fig:flxg}
\end{figure}
The prediction for $F_L(x,Q^2)$ is mainly determined by the form of the gluon distribution extracted from the fit.  This is shown in Fig.~\ref{fig:flxg}.  There is poor stability as one changes the perturbative order of the calculation, particularly at small $x$ and $Q^2$.  At small $x$ there is a large order-by-order change in the splitting functions, particularly $P_{qg}$, and this leads to a large variation in the gluon.  At NLO a small-$x$ divergence appears for the first time, making the evolution of the quarks much quicker, and hence the NLO gluon is dramatically smaller at small $x$.  Most extractions find a NLO gluon which is small and perhaps valence-like at $Q^2 \lesssim 2$ GeV$^2$ and very small $x$.  However, the gluon parameterisation we have used since Ref.~\cite{Martin:2001es} allows the possibility for the gluon distribution to become negative, and for the best fit it chooses to do so for $x \lesssim 10^{-4}$ at $Q^2=2$ GeV$^2$.  It is frequently claimed that higher twists, in particular gluon recombination effects, will be a cure for this, but simply applying these corrections, on top of NLO leading-twist QCD, results in little change~\cite{Martin:2003sk,Watt:2005iu}.  At NNLO there is an additional $\ln(1/x)/x$ divergence, and at the smallest $x$ the NNLO gluon becomes smaller to compensate, or in practice more negative for very small values of $x$.  However, it is important to note that the gluon beyond leading order does not have a simple \emph{probabilistic} interpretation.  This is particularly the case at high orders at small $x$, where perturbative corrections to cross sections can be large, and even more so in unphysically motivated schemes such as the $\overline{\rm MS}$ factorisation scheme.

\begin{figure}
  \centering
  \includegraphics[width=0.8\textwidth,clip]{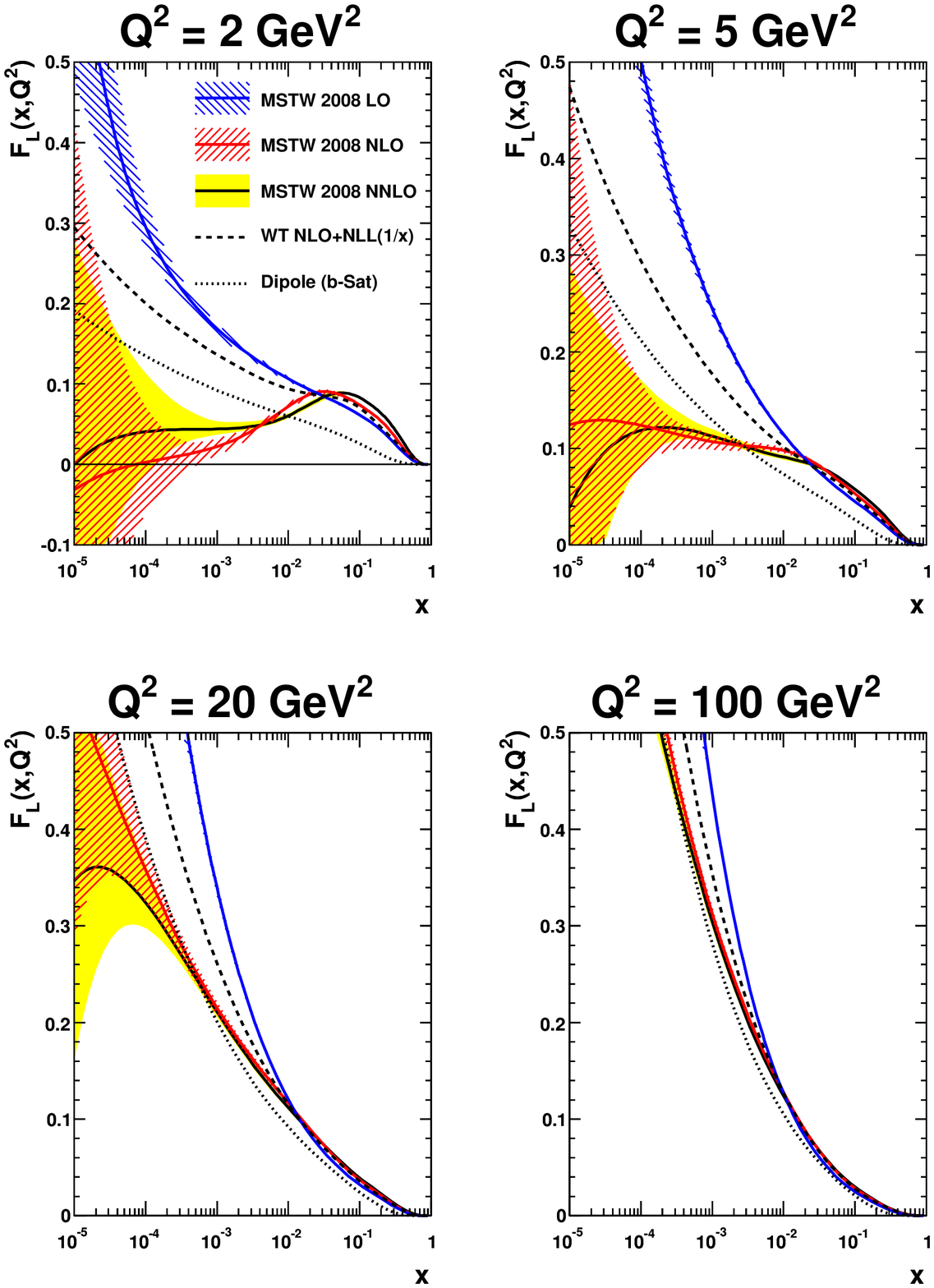}
  \caption{Predictions for the longitudinal proton structure function, $F_L(x,Q^2)$, at LO, NLO and NNLO including the one-sigma PDF uncertainty bands.  Also shown are resummed NLL BFKL predictions~\cite{White:2006yh} and dipole model predictions~\cite{Kowalski:2006hc}.}
  \label{fig:fl}
\end{figure}
The fact that the gluon is not directly physical is well-illustrated by the NNLO ${\cal O}(\alpha_s^3)$ longitudinal coefficient function $C^{(3)}_{L,g}(x)$~\cite{Moch:2004xu,Vermaseren:2005qc}.  This has a large positive contribution at small $x$, and this counters the decrease in the small-$x$ gluon.  The predictions for $F_L(x,Q^2)$ at LO, NLO and NNLO are shown in Fig.~\ref{fig:fl}.  The $F_L(x,Q^2)$ prediction is more stable in going from NLO to NNLO than the gluon distribution at small $x$.  Despite the behaviour of the gluon distribution, the central value of $F_L$ is positive at NNLO for all $x > 10^{-5}$ at $Q^2=2$ GeV$^2$, and is greater than at NLO at the smallest $x$.  The uncertainty becomes enormous as $x$ decreases below $10^{-4}$, but at the lowest $Q^2$ the NLO and NNLO predictions are discrepant in some regions.  The LO prediction is far larger than either, reflecting the huge correction in  the small-$x$ gluon going to NLO.
\begin{figure}
  \centering
  \includegraphics[width=0.8\textwidth,clip]{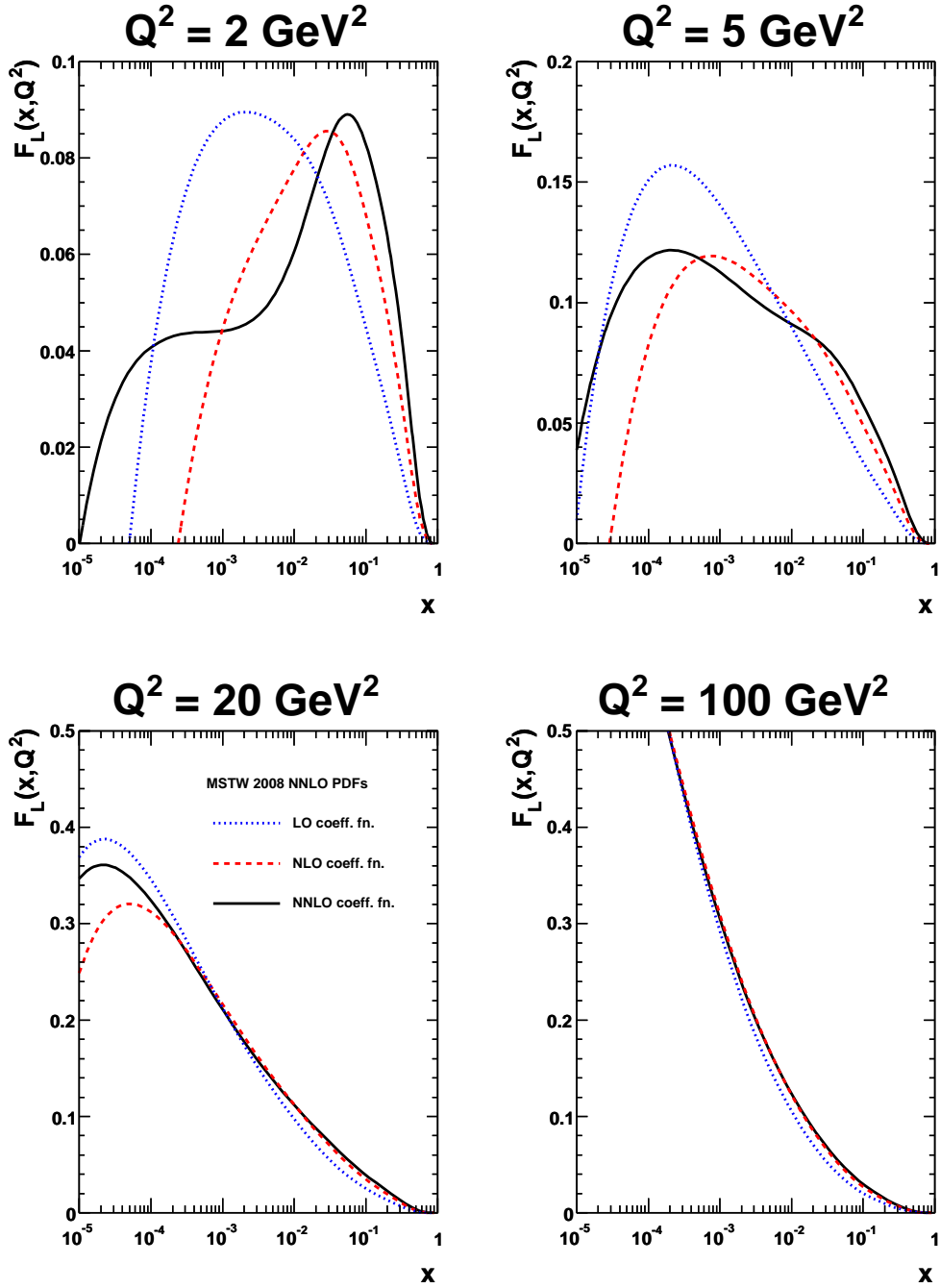}
  \caption{Predictions for the longitudinal proton structure function, $F_L(x,Q^2)$, using NNLO PDFs and LO, NLO and NNLO coefficient functions.}
  \label{fig:flnnlo}
\end{figure}
In order to more clearly separate the relative effect of different gluon distributions, in Fig.~\ref{fig:flnnlo} we show the predictions for $F_L$ with LO, NLO and NNLO coefficient functions but using common NNLO PDFs.  There is a positive correction at high $x$ at each order. NLO then results in a negative correction at lower $x$, whereas NNLO gives an additional negative correction at intermediate $x$ before becoming positive again at sufficiently small $x$.  However, the value of $x$ at which these transitions occur is very sensitive to $Q^2$.  As the gluon becomes steeper the regions of the gluon distribution probed in the convolution become more local to the value of $x$, and small-$x$ divergences in the coefficient functions become less important, along with the relative importance of higher orders decreasing as the coupling becomes weaker.  Hence, the ultimate positive effect of the $\ln(1/x)/x$ term in the NNLO coefficient function is only seen to increase the NNLO result above LO for $x>10^{-5}$ in the two lowest $Q^2$ bins.

As the relative lack of stability in fixed-order predictions implies, there are various potentially large corrections beyond fixed-order perturbation theory.  It is possible that there is a large higher-twist contribution from renormalons in the quark sector~\cite{Stein:1996wk}, and the implications are discussed in Refs.~\cite{Martin:2006qv,Thorne:2008aj}.  Perhaps more significantly, since the small-$x$ NNLO correction is itself rather large, even higher orders might be important.  There are leading $\ln(1/x)$ terms of the form:
\begin{equation}
  xP_{gg}(x) \sim \alpha_S^n \ln^{n-1}(1/x),\quad
  xP_{qg}(x) \sim \alpha_S^n \ln^{n-2}(1/x)\quad\text{and}\quad
  xC_{L,g}(x) \sim \alpha_S^n \ln^{n-2}(1/x).
\end{equation}

A fit which performs a double resummation of leading and next-to-leading $\ln(1/x)$ and running coupling contributions leads to a better fit to small-$x$ data than a conventional perturbative fit~\cite{White:2006yh}.  The gluon distribution from this resummed fit, defined in a more physical scheme, is larger at small $x$ and $Q^2$ than NLO or NNLO, and indeed is always positive for $Q^2\geq 1$ GeV$^2$.  This is reflected also in the prediction for $F_L(x,Q^2)$.  Similar approaches~\cite{Altarelli:2008aj,Ciafaloni:2007gf} all lead to rather comparable results for the calculated splitting functions, but only in Ref.~\cite{White:2006yh} have detailed phenomenological studies taken place.  A partially overlapping set of additional corrections are considered in the dipole picture.  As with small-$x$ resummations, this approach can be cast in the language of $f(x,k^2)$, the unintegrated gluon distribution, which is directly related to the dipole--proton cross section.  The structure functions are obtained by convoluting this dipole cross section with the wave functions for the photon to fluctuate into a quark--antiquark pair.  Hence, this picture includes some of the resummation effects, but also higher-twist contributions, and is designed to approach $Q^2=0$ smoothly.  However, it misses quark and higher-$x$ contributions.  In this framework, higher-twist corrections are not small in either $F_L$ or $F_T$ separately, but largely cancel in $F_2=F_L+F_T$~\cite{Bartels:2000hv,Dittmar:2009ii}.  Overall the $F_L(x,Q^2)$ predicted in the dipole model approach is steeper at small $x$ than fixed-order predictions, and is automatically stable at lowest $Q^2$.  The general features are rather insensitive to whether saturation effects are included in the dipole cross section.  Resummed NLL BFKL predictions~\cite{White:2006yh} and dipole model predictions~\cite{Kowalski:2006hc} are additionally shown in Fig.~\ref{fig:fl} illustrating these features.

The first published direct measurement of $F_L(x,Q^2)$ can be found in Ref.~\cite{Aaron:2008tx} for $12\le Q^2\le 90$ GeV$^2$, based on data taken in the last few months of HERA running, when the proton beam energy was lowered from the nominal value of 920 GeV to values of 460 GeV and 575 GeV.
\begin{figure}[t]
  \centering
  \includegraphics[width=0.8\textwidth,clip]{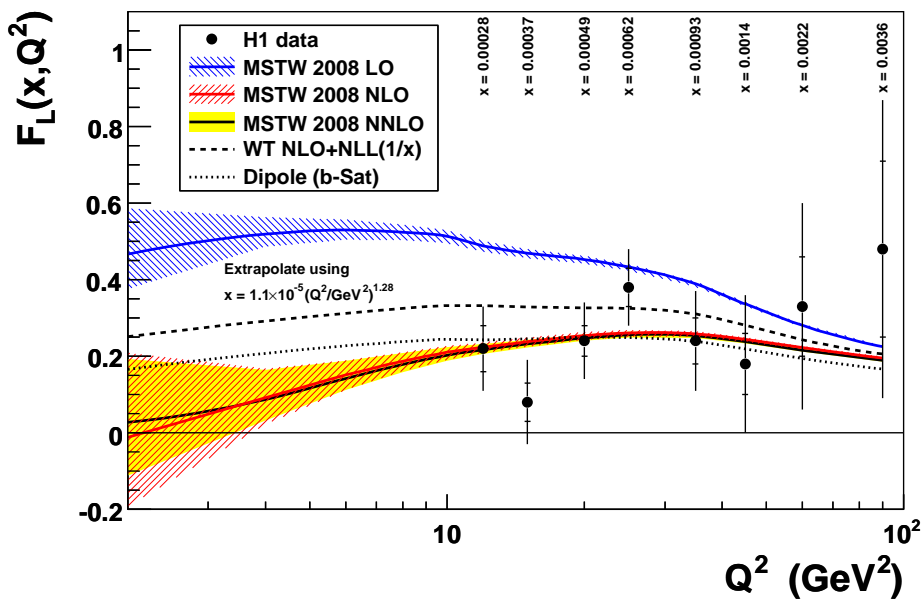}
  \caption{The longitudinal proton structure function, $F_L(x,Q^2)$, measured by H1~\cite{Aaron:2008tx} compared to predictions at LO, NLO and NNLO including the one-sigma PDF uncertainty bands.  Also shown are resummed NLL BFKL predictions~\cite{White:2006yh} and dipole model predictions~\cite{Kowalski:2006hc}.  (The dependence on the form of the dipole cross section is mild~\cite{Watt:2007nr}.)}
  \label{fig:h1fl}
\end{figure}
In Fig.~\ref{fig:h1fl} we show these data compared to our predictions calculated at LO, NLO and NNLO using the appropriate parton distributions at each order.  The uncertainty bands obtained from the 40 alternative eigenvector PDF sets are shown at each order.  In the region of the existing H1 data the NLO and NNLO predictions agree well with the data, while the LO prediction overshoots the data at the lower $Q^2$ values.  We also show in Fig.~\ref{fig:h1fl} the predictions at lower $Q^2\lesssim 10$ GeV$^2$.  We extrapolate the values of $x$ accordingly, by fitting the $x$ values of the H1 data as a power law function of $Q^2$, obtaining $x = (1.09\times10^{-5})(Q^2/Q_0^2)^{1.28}$ with $Q_0^2 = 1$ GeV$^2$.  The NLO and NNLO predictions are very similar along this line, but as seen in Fig.~\ref{fig:fl} this is partially accidental. Note that the NNLO uncertainty band at low $x$ and $Q^2$ is smaller than the NLO uncertainty band due to the lesser sensitivity to small-$x$ PDFs in the convolution integrals at NNLO compared to NLO, because the more divergent coefficient function samples the PDFs further from $x$.  Again we also show resummed NLL BFKL predictions~\cite{White:2006yh} and dipole model predictions~\cite{Kowalski:2006hc}, and both agree well with the published data.  For $Q^2\in[5,10]$ GeV$^2$ the uncertainty in NLO/NNLO predictions for $F_L(x,Q^2)$ due to the gluon uncertainty increases to more than $20\%$.  A good measurement of $F_L(x,Q^2)$ here would automatically improve the gluon determination.  Resummations and dipole models suggest a higher low-$Q^2$ $F_L(x,Q^2)$ by an absolute value of up to $0.15$ --- well outside the fixed-order uncertainties.  So, moreover, a good measurement of $F_L(x,Q^2)$ in this region would start to discriminate between theories.

%% file: comparison.tex
\section{Comparison with other PDF sets} \label{sec:comparison}
In this section we make a comparison with various other PDF sets obtained from analyses of a variety of data sets by different groups.  Some of the most important features pertaining to previous MRST sets have already been mentioned, but here we will present a general overview of the similarities to, and differences from, other sets, and suggest reasons to explain these differences.

\begin{figure}
  (a)\hspace{0.5\textwidth}(b)\\
  \includegraphics[width=0.5\textwidth]{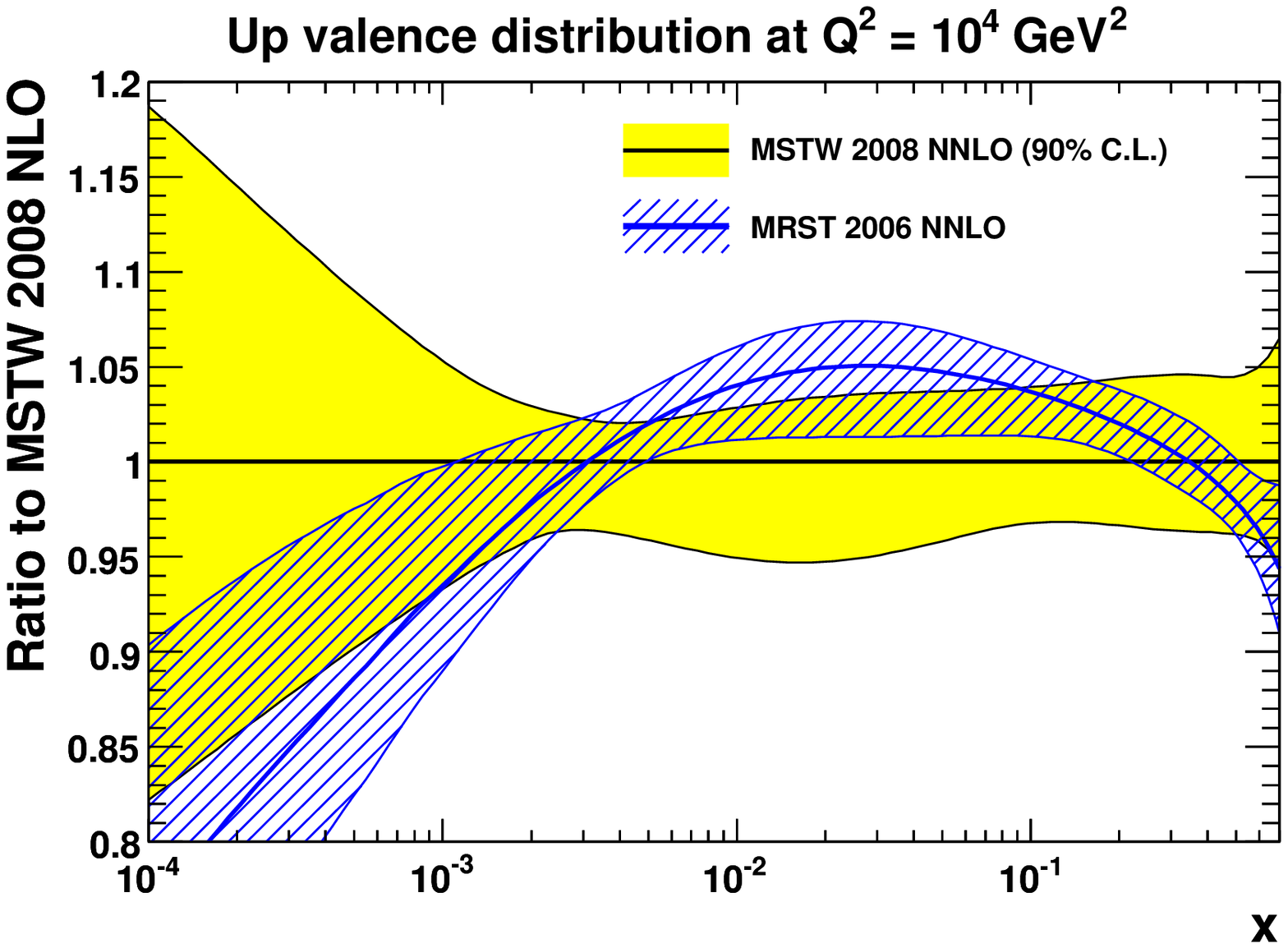}%
  \includegraphics[width=0.5\textwidth]{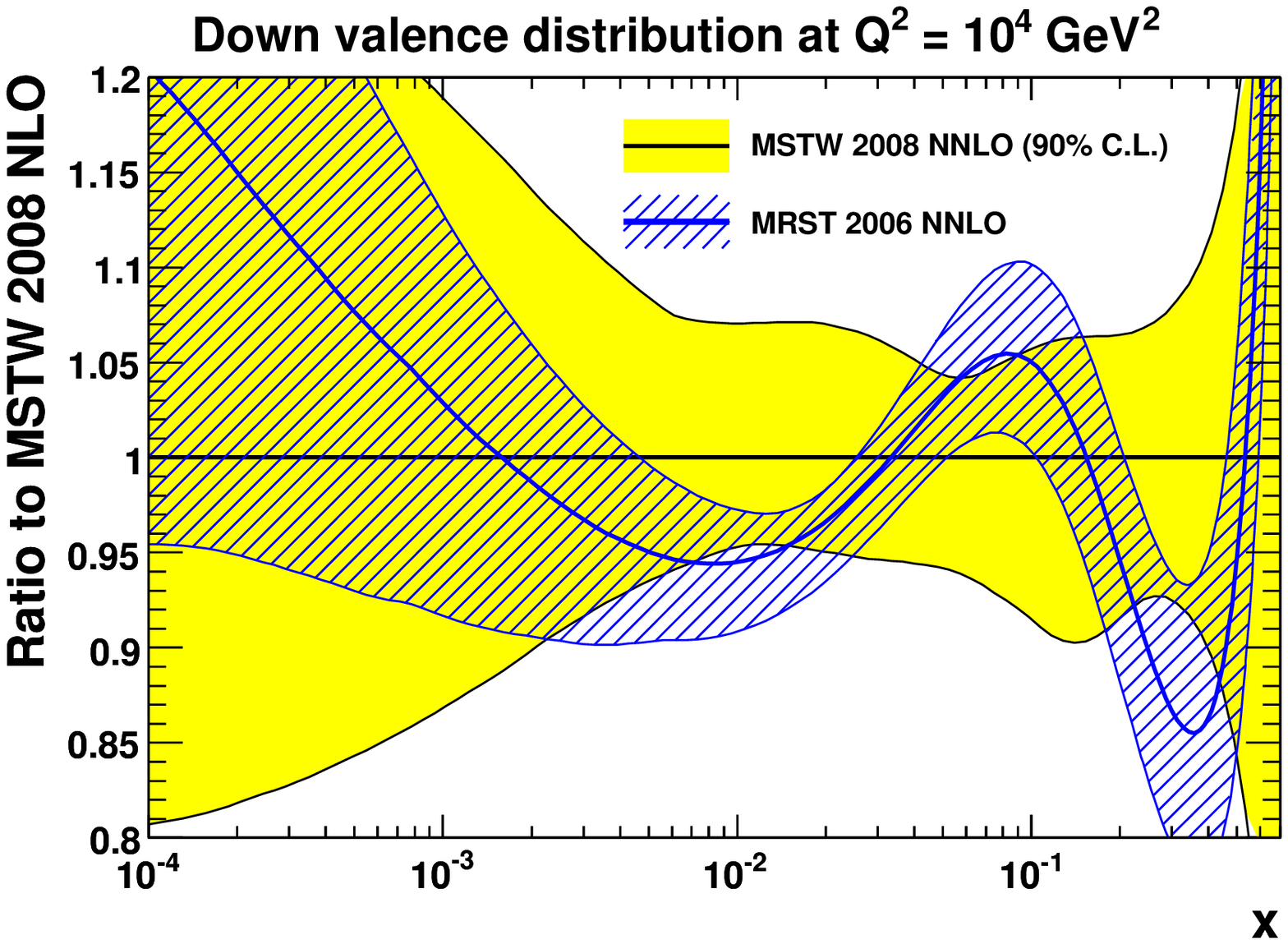}\\
  (c)\hspace{0.5\textwidth}(d)\\
  \includegraphics[width=0.5\textwidth]{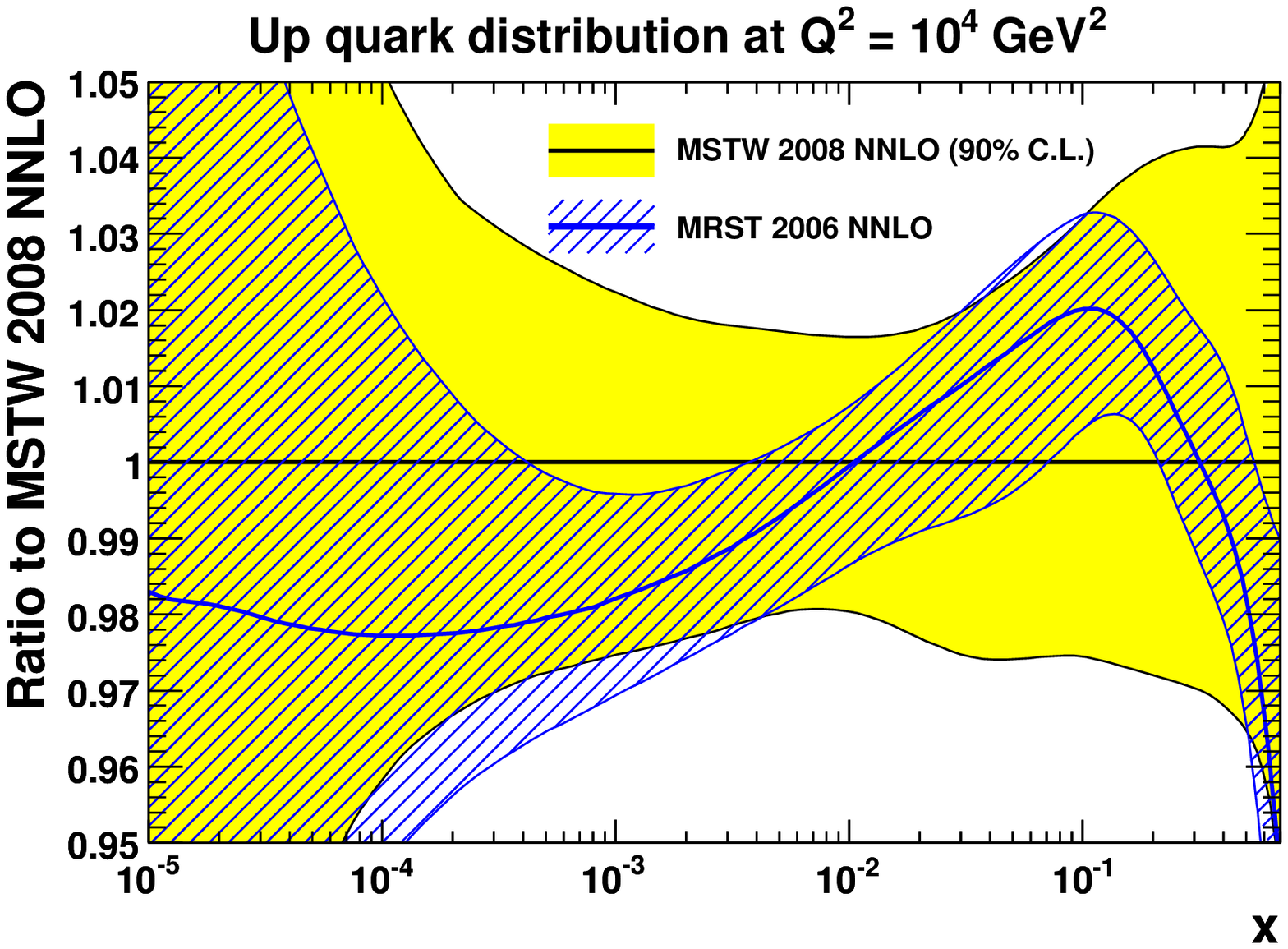}%
  \includegraphics[width=0.5\textwidth]{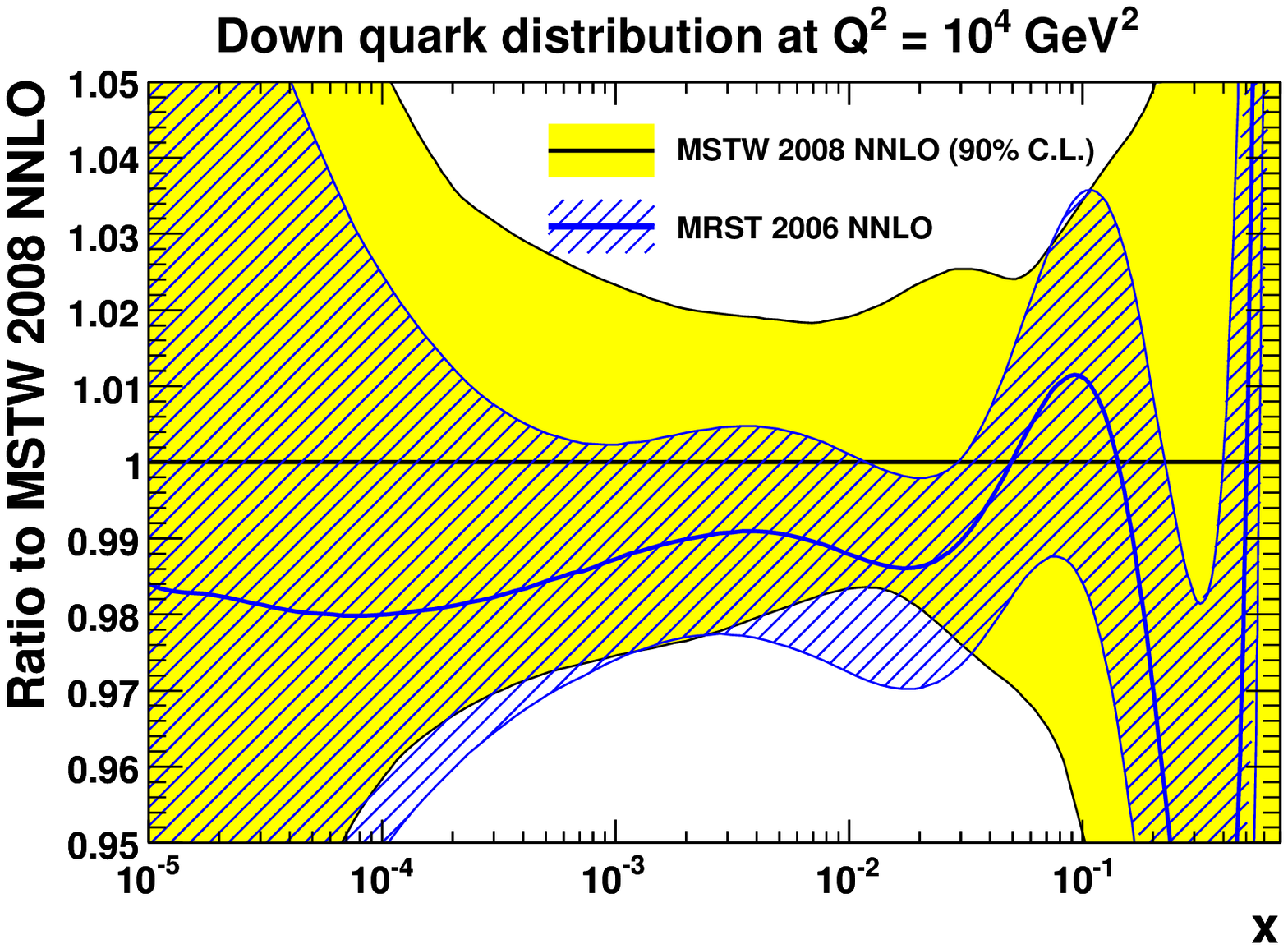}\\
  (e)\hspace{0.5\textwidth}(f)\\
  \includegraphics[width=0.5\textwidth]{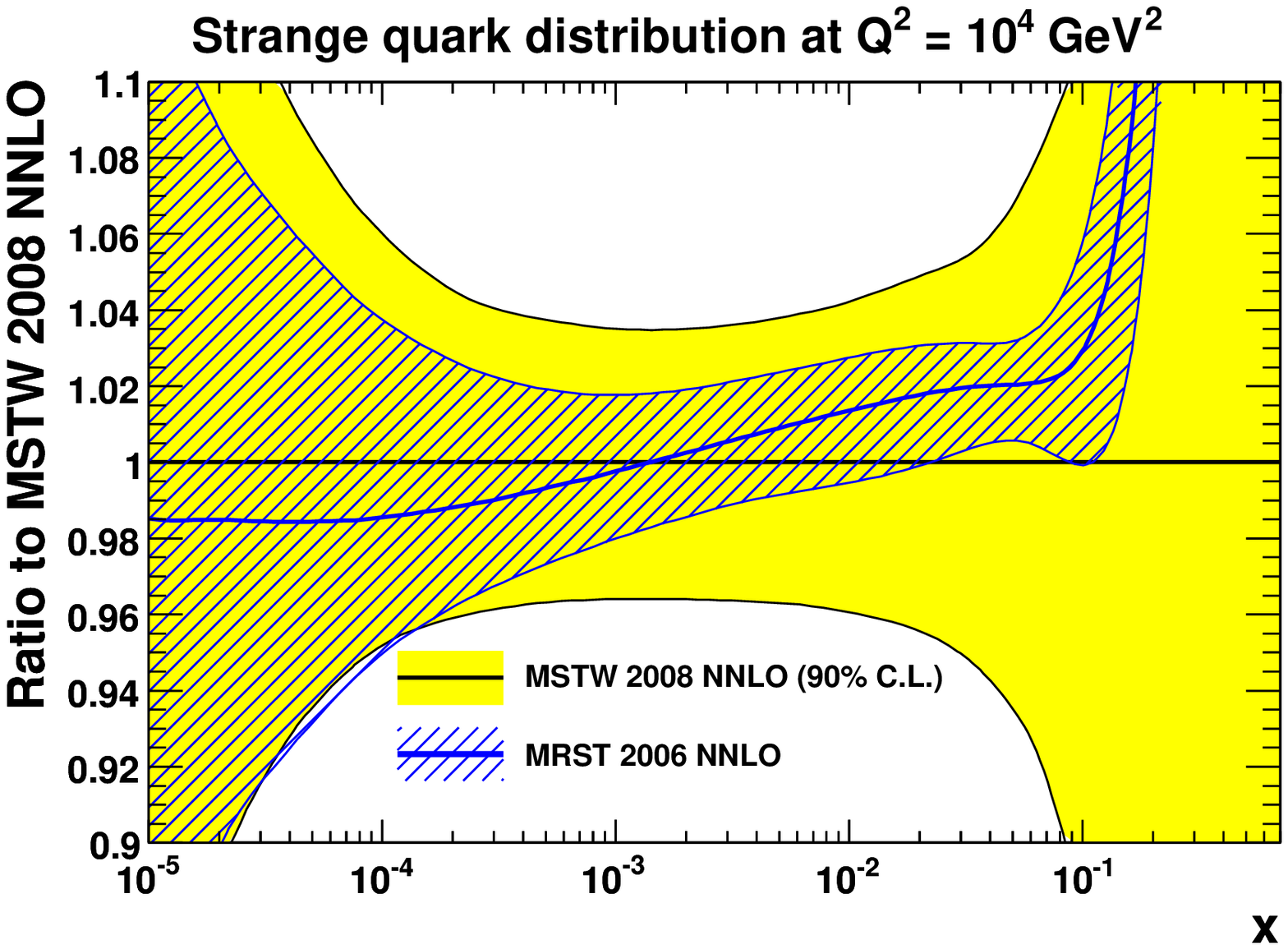}%
  \includegraphics[width=0.5\textwidth]{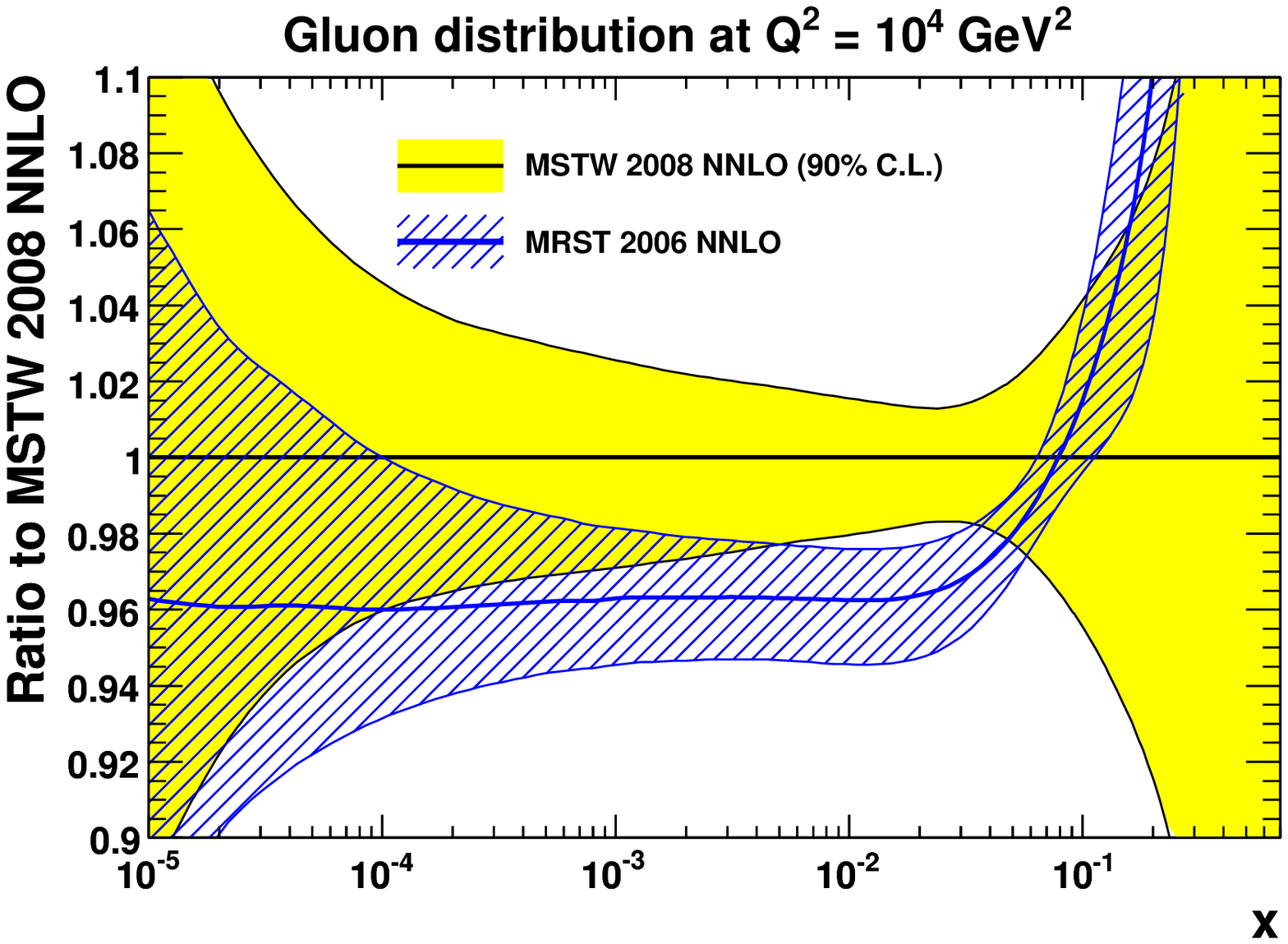}
  \caption{MSTW 2008 NNLO PDFs compared to MRST 2006 NNLO PDFs.}
  \label{fig:comparisonNNLO}
\end{figure}
We begin with a historical perspective.  In Fig.~\ref{fig:comparisonNNLO} we show the current NNLO PDFs compared to those from MRST 2006~\cite{Martin:2007bv}.  The comparison would in general terms be rather similar if comparing our current NLO set to the (unpublished) 2006 NLO set, but because the change in $\alpha_S(M_Z^2)$ has been slightly larger at NNLO some differences in PDFs are a little more apparent in the NNLO comparison.  Comparing with older NLO sets would mean a change in the definition of the GM-VFNS, and the consequences of this have already been illustrated in Fig.~\ref{fig:06to04}.  Beginning from the top of Fig.~\ref{fig:comparisonNNLO}, we see that the shape of both valence distributions has changed significantly, though there is still compatibility with MRST 2006 within the 90\% C.L.~uncertainty bands.  This is partially due to the smaller value of $\alpha_S(M_Z^2)$ --- the coefficient functions give a positive effect at very high $x$ and negative for $x\sim 0.1$, and this is smaller in 2008 than 2006, so the quarks compensate.  However, the details, and the large changes in $d_v$ are due to the new Tevatron Run II $W$ and $Z$ data and the NuTeV and CHORUS neutrino DIS data.  For $x<0.01$ the form of both valence quark distributions is determined mainly by the requirement to satisfy the number sum rules.  The gluon distribution has changed significantly: the smallness at high-$x$ being mainly due to the Tevatron Run II jet data, as discussed in detail in Section~\ref{sec:jetdata}.  This is part of the reason for the decrease in $\alpha_S(M_Z^2)$, and the weaker coupling and the momentum sum rule both contribute to the significantly larger gluon distribution at lower $x$ in the 2008 set.  This feeds through, via evolution, to the sea quarks being higher at small $x$, though well within previous uncertainties.  For the gluon and light quarks the uncertainty is generally a little larger than for the MRST 2006 set.  This is due largely to the inclusion of data normalisation uncertainties into the PDF error analysis, but also due to an improvement in the flexibility of the parameterisation for the gluon and down valence distribution, and for light sea quarks the independent determination of the strange distribution also leads to an increase.  The fact that we fit the strange distribution directly results in the very large increase in the uncertainty for $x>0.001$.  But as we assume that the input strange has the same shape as $x\to 0$ as the light quark distributions, the uncertainty becomes the same as for $u$ and $d$ at very small $x$ and high $Q^2$.

\begin{figure}
  (a)\hspace{0.5\textwidth}(b)\\
  \includegraphics[width=0.5\textwidth]{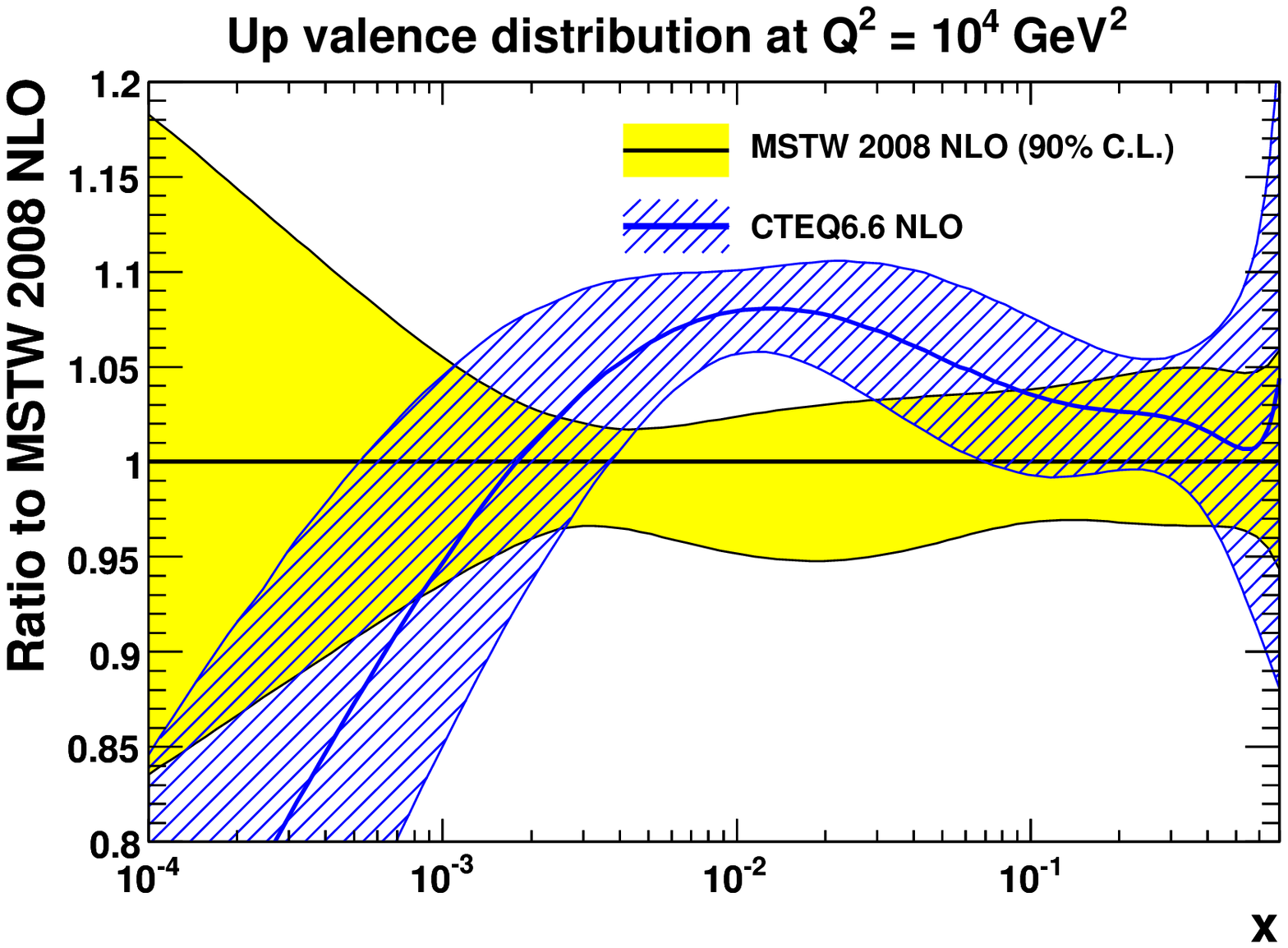}%
  \includegraphics[width=0.5\textwidth]{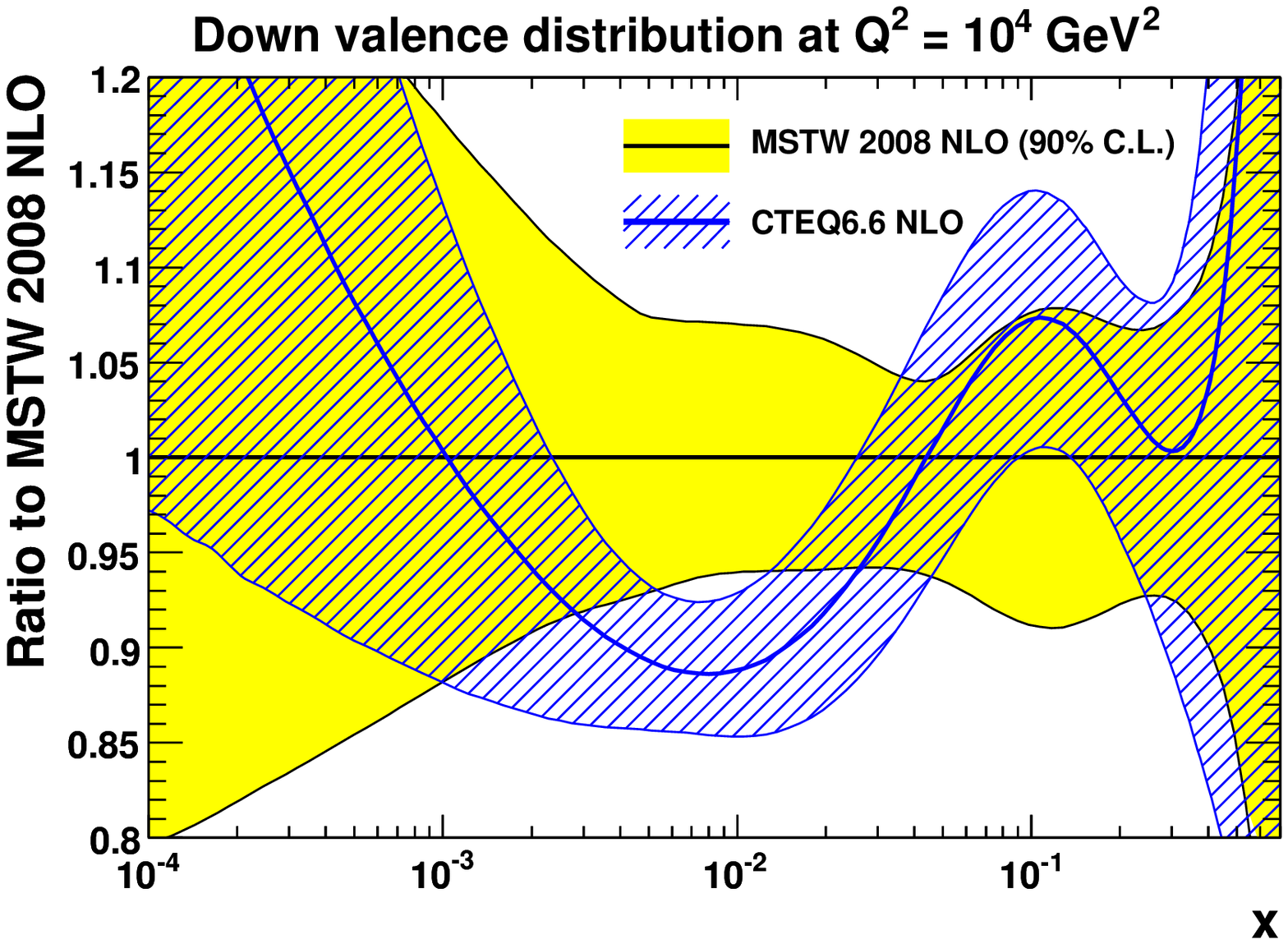}\\
  (c)\hspace{0.5\textwidth}(d)\\
  \includegraphics[width=0.5\textwidth]{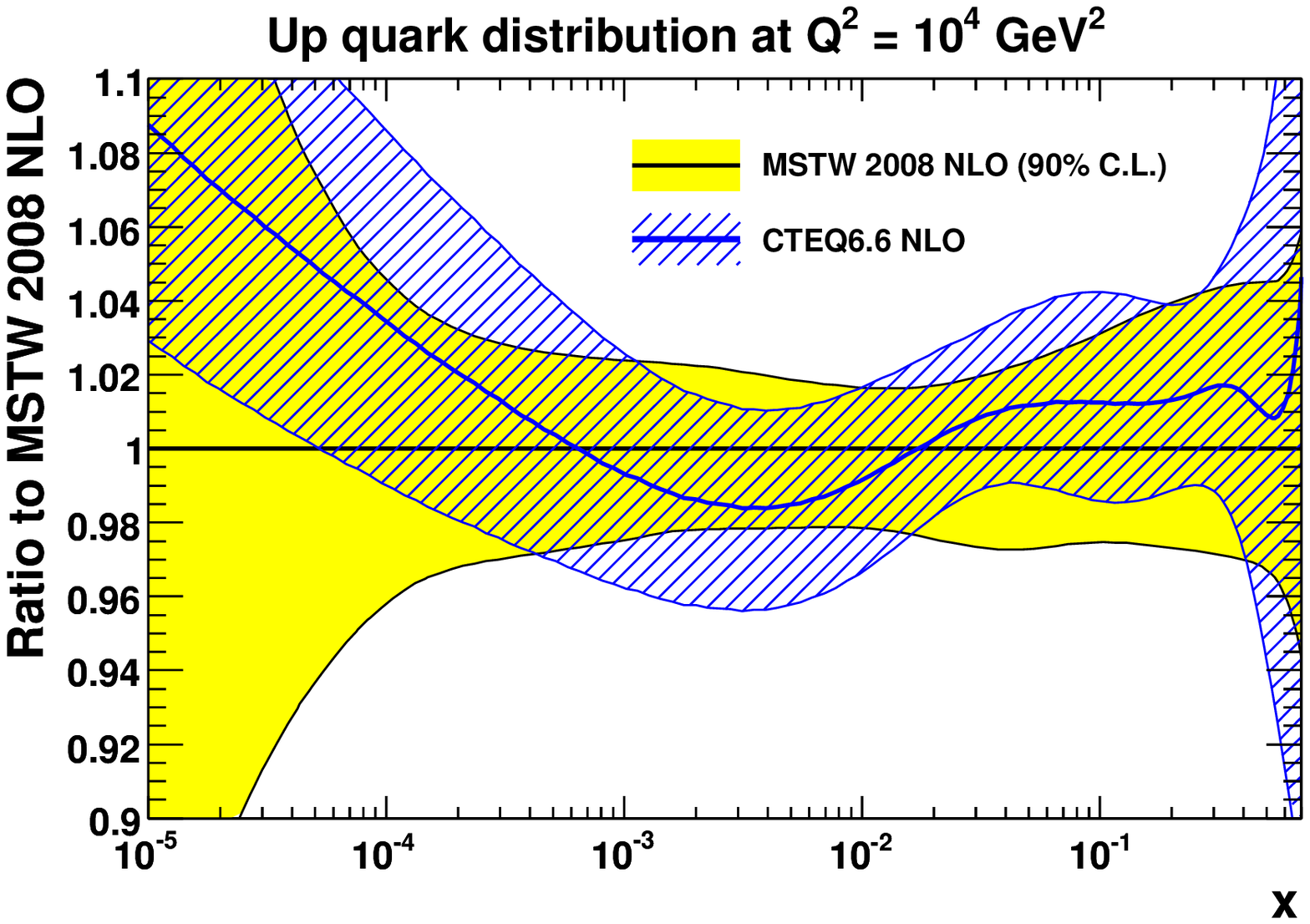}%
  \includegraphics[width=0.5\textwidth]{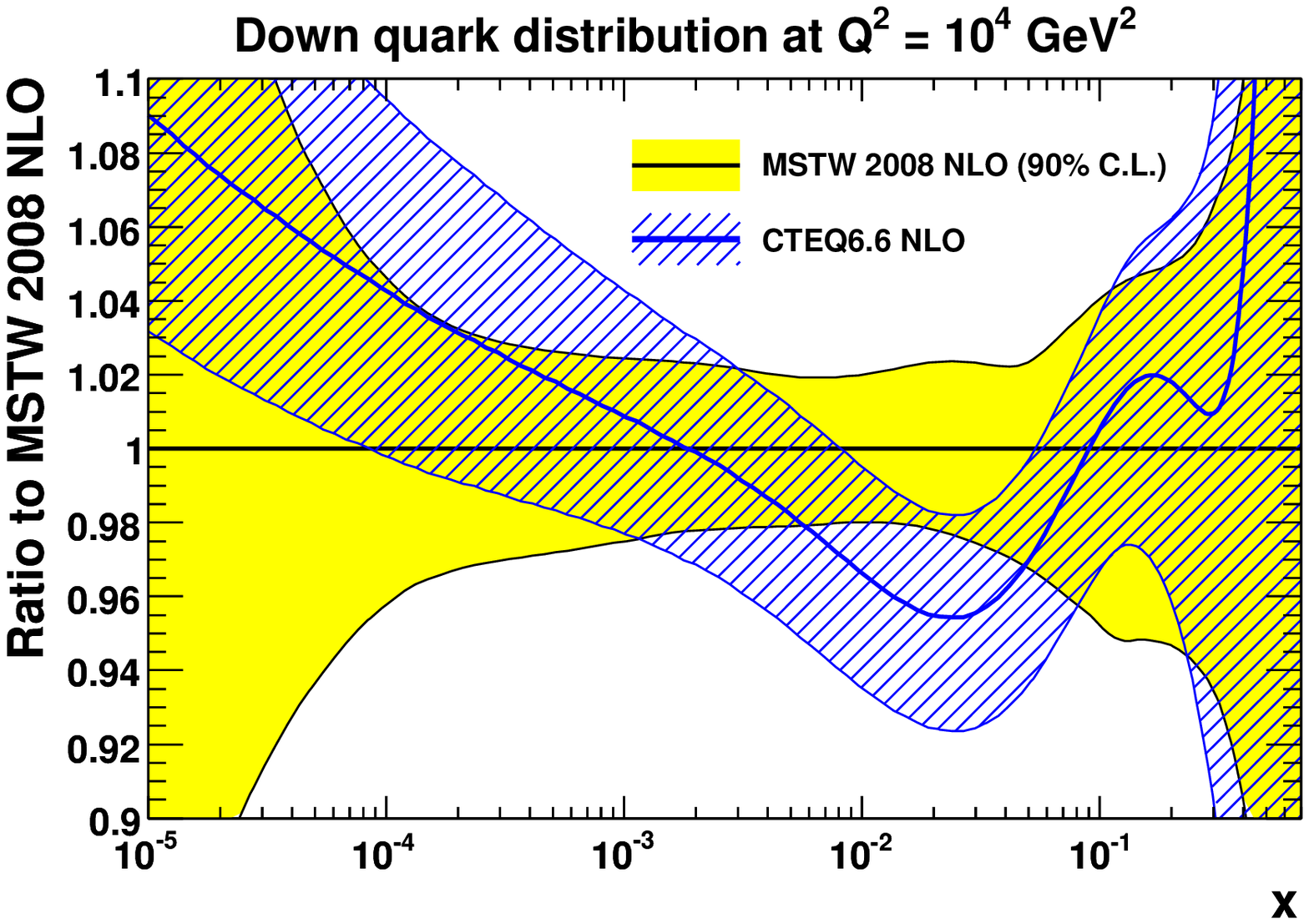}\\
  (e)\hspace{0.5\textwidth}(f)\\
  \includegraphics[width=0.5\textwidth]{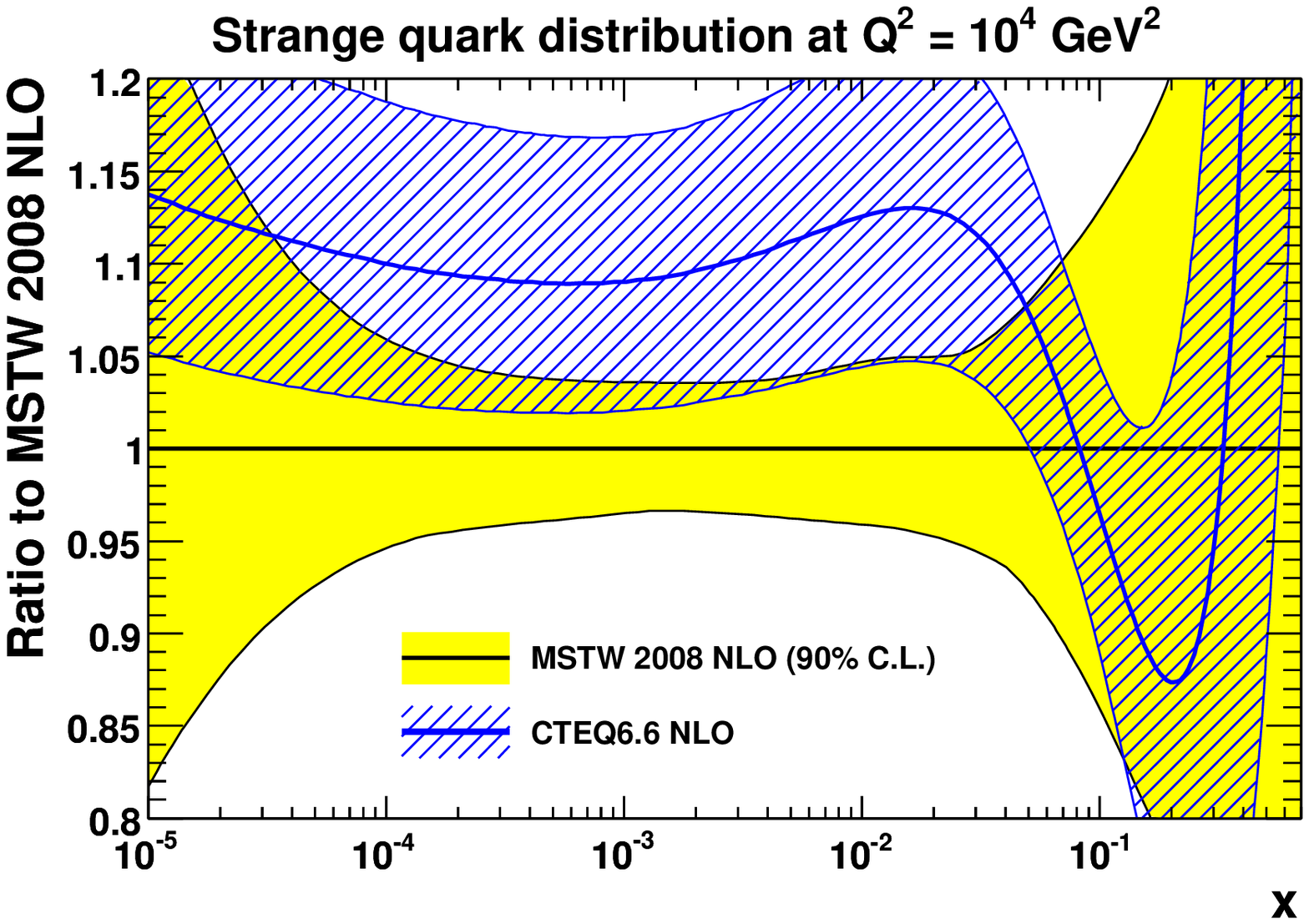}%
  \includegraphics[width=0.5\textwidth]{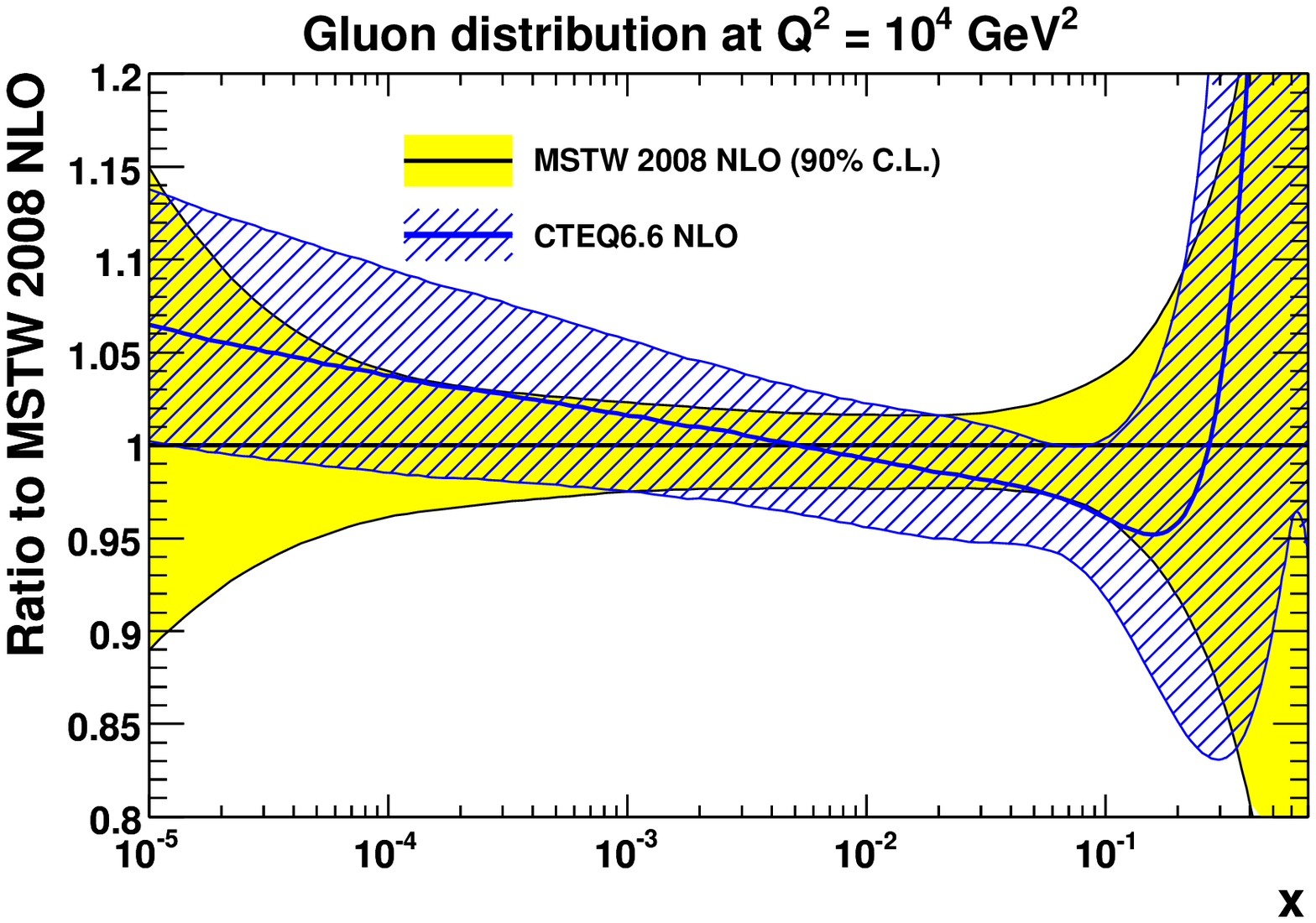}
  \caption{MSTW 2008 NLO PDFs compared to CTEQ6.6 NLO PDFs.}
  \label{fig:comparisonNLO}
\end{figure}
In Fig.~\ref{fig:comparisonNLO} we show the NLO PDFs at $Q^2=10^4$ GeV$^2$ compared to those from CTEQ6.6~\cite{Nadolsky:2008zw}.  The CTEQ6.6 analysis uses a very similar total set of data, but is not as up-to-date, in particular not including any Tevatron Run II data or the neutrino DIS data sets from NuTeV and CHORUS.  Overall, there is good agreement between the two sets, with very few regions where the $90\%$ C.L.~limits do not overlap. It is striking that the difference between the MSTW 2008 and CTEQ6.6 valence quarks is rather similar to that between the MSTW 2008 and MRST 2006 distributions.  This suggests that the difference is largely due to the new NuTeV, CHORUS and Tevatron Run II data included in our analysis.  The strange quark distribution is rather different, though the $90\%$ C.L.~uncertainty bands overlap. This is due to the different assumptions made in parameterising the strange distribution, and was addressed in Section~\ref{sec:inputparamunc}.  As in previous comparisons the gluons have a systematic difference.  Since MSTW use the flexible parameterisation which allows the distribution to be negative at very small $x$ at input (and at $1.69$ GeV$^2$, the CTEQ input scale) while the CTEQ input is valencelike, the latter is bigger at the smallest $x$ values.  The gluon in the CTEQ6.6 set is determined at high $x$ by Run I Tevatron jet data, so is higher in this region, and from the momentum sum rule must be smaller at intermediate $x \sim 0.1$ to compensate.  However, the greater similarity in the heavy flavour treatments now employed perhaps leads to better agreement in the intermediate $x$ range than previously.  The comparison of the $u$ and $d$ distributions for $x<0.1$, i.e.~outside the valence region, very much reflects the gluon difference.  Finally, we note that the MSTW 2008 uncertainty bands are not as small compared to CTEQ bands as in previous comparisons, despite the addition of more data and on average a smaller tolerance for the former.  This is a reflection of the data normalisation uncertainties and the improved flexibility in parameterisations.

\begin{figure}
  (a)\hspace{0.5\textwidth}(b)\\
  \includegraphics[width=0.5\textwidth]{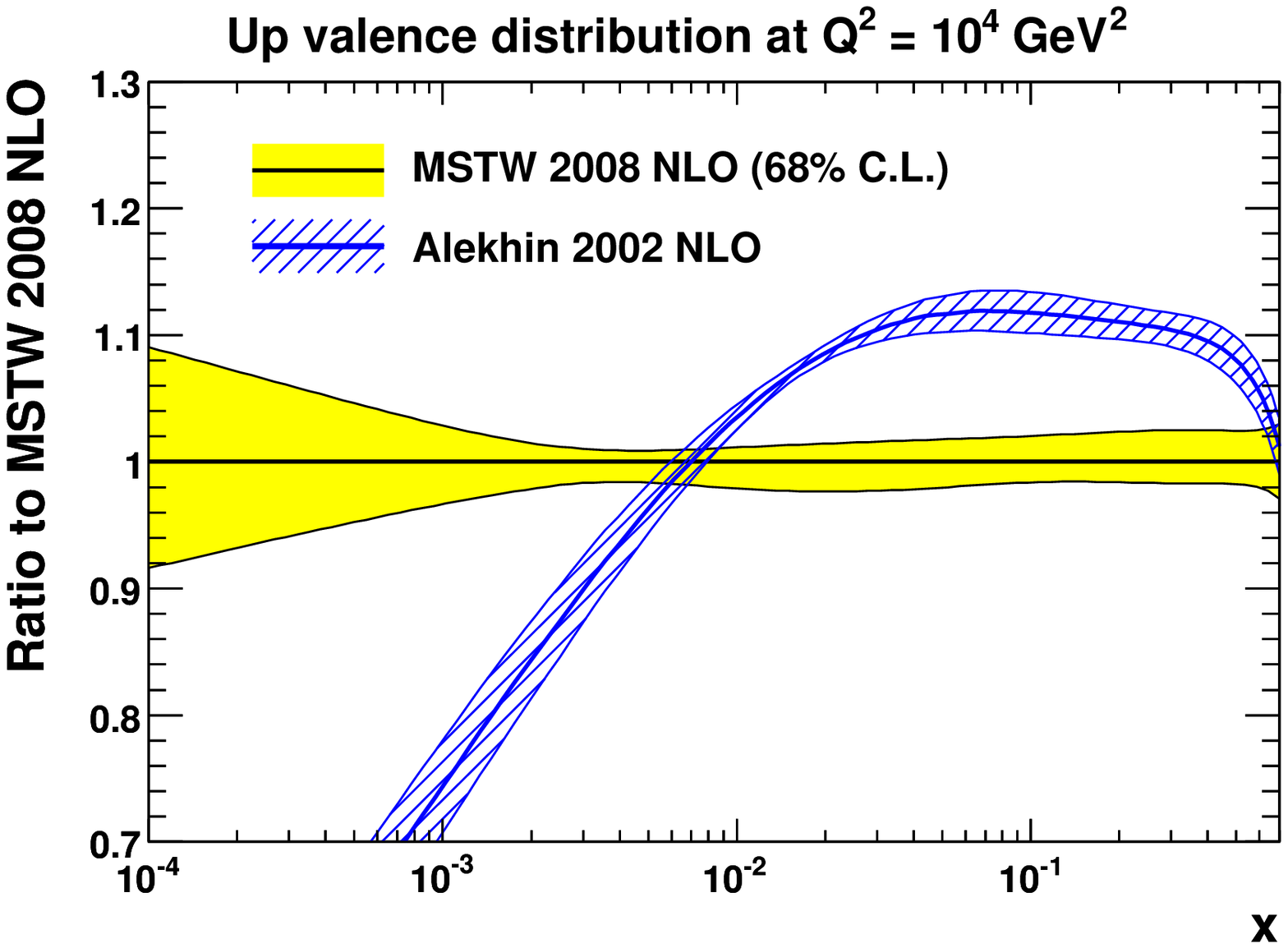}%
  \includegraphics[width=0.5\textwidth]{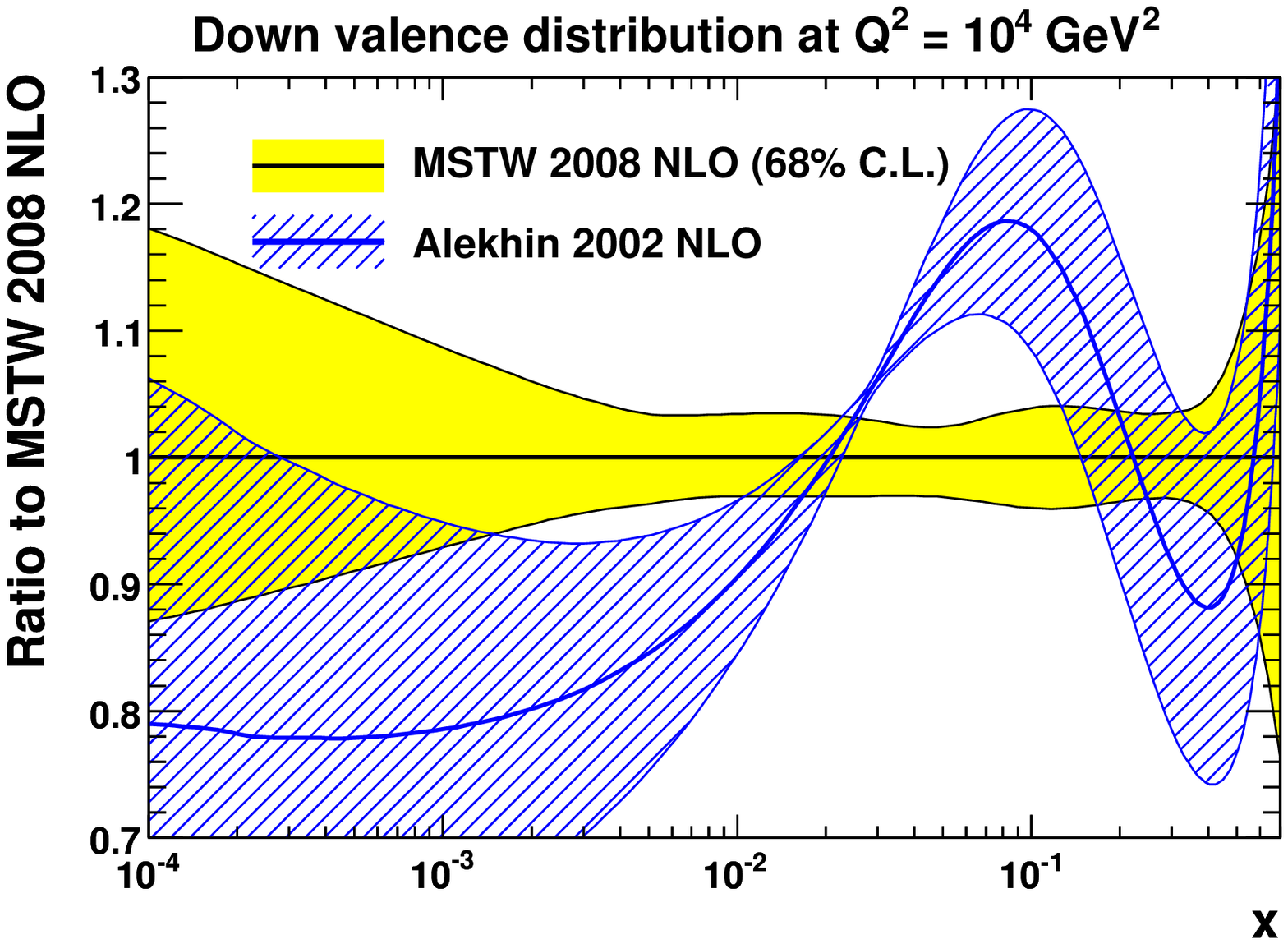}\\
  (c)\hspace{0.5\textwidth}(d)\\
  \includegraphics[width=0.5\textwidth]{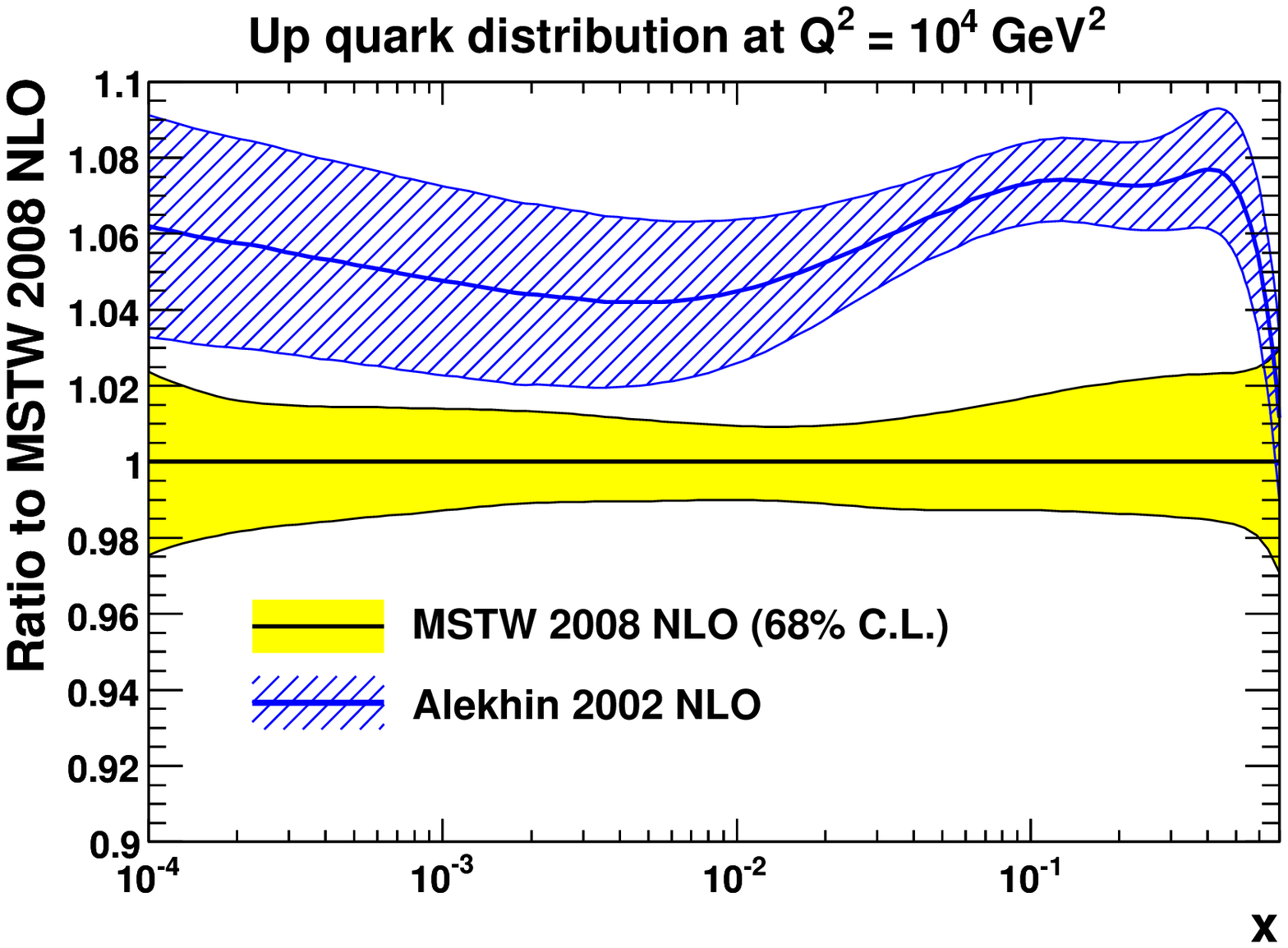}%
  \includegraphics[width=0.5\textwidth]{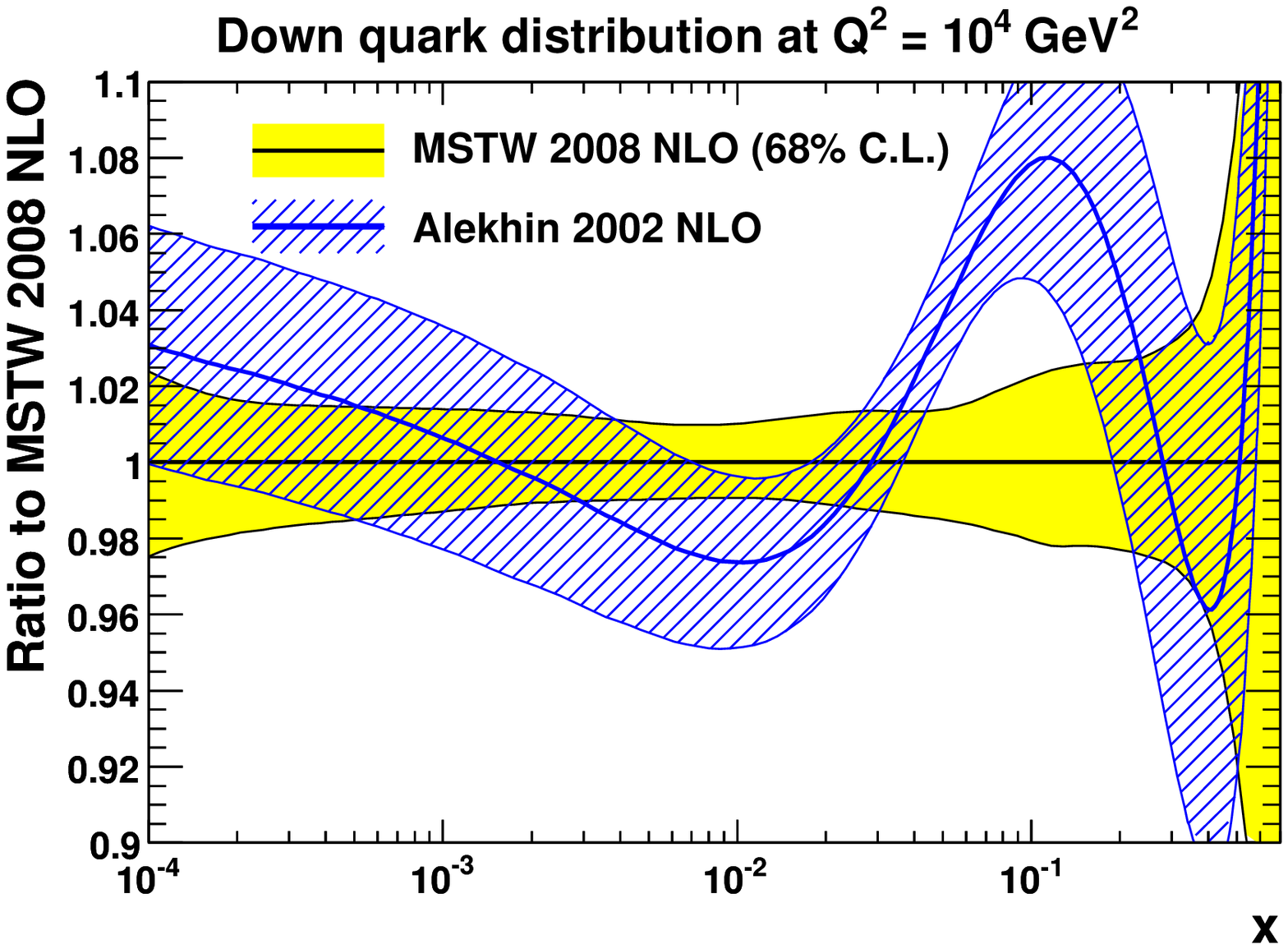}\\
  (e)\hspace{0.5\textwidth}(f)\\
  \includegraphics[width=0.5\textwidth]{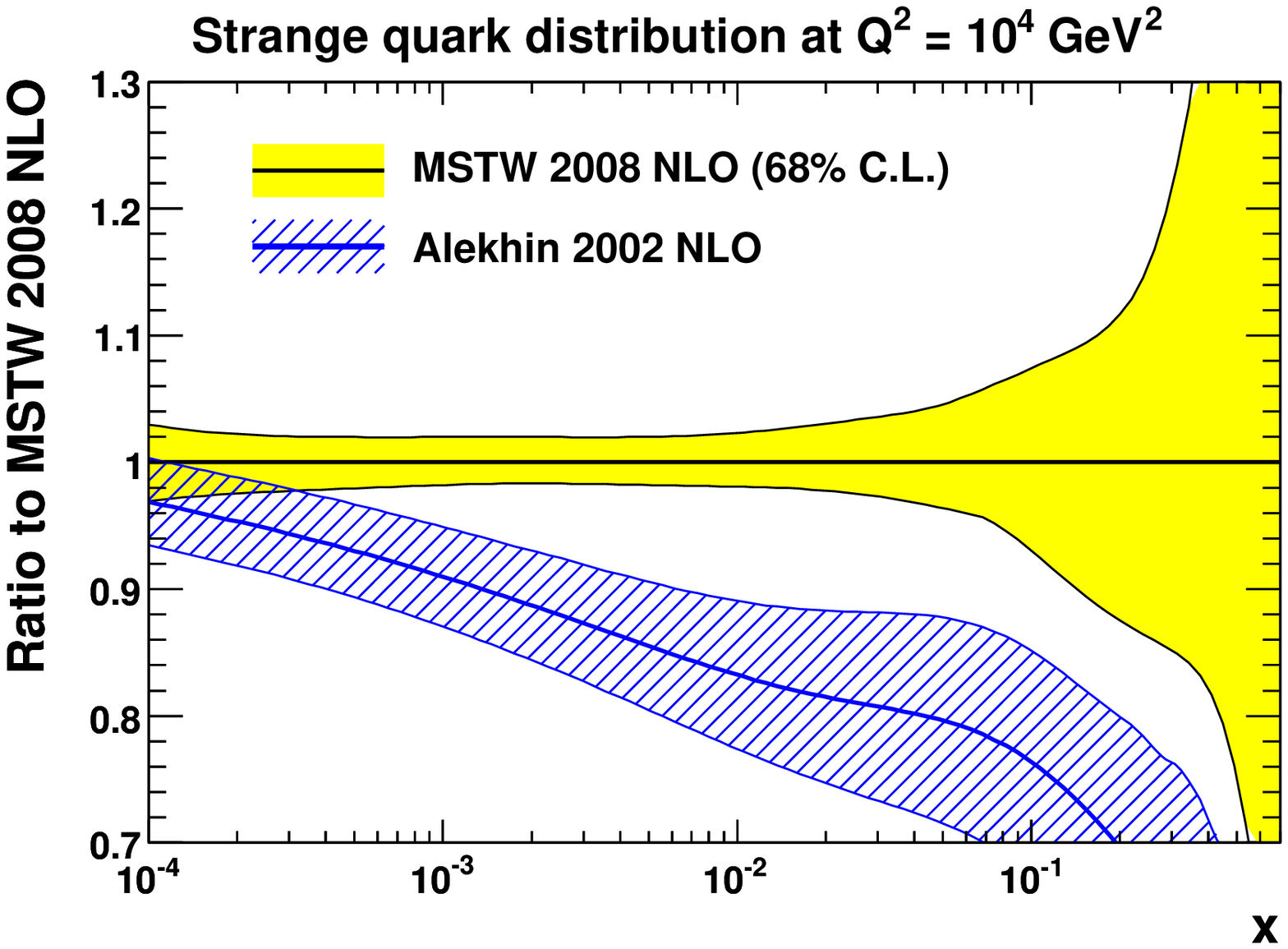}%
  \includegraphics[width=0.5\textwidth]{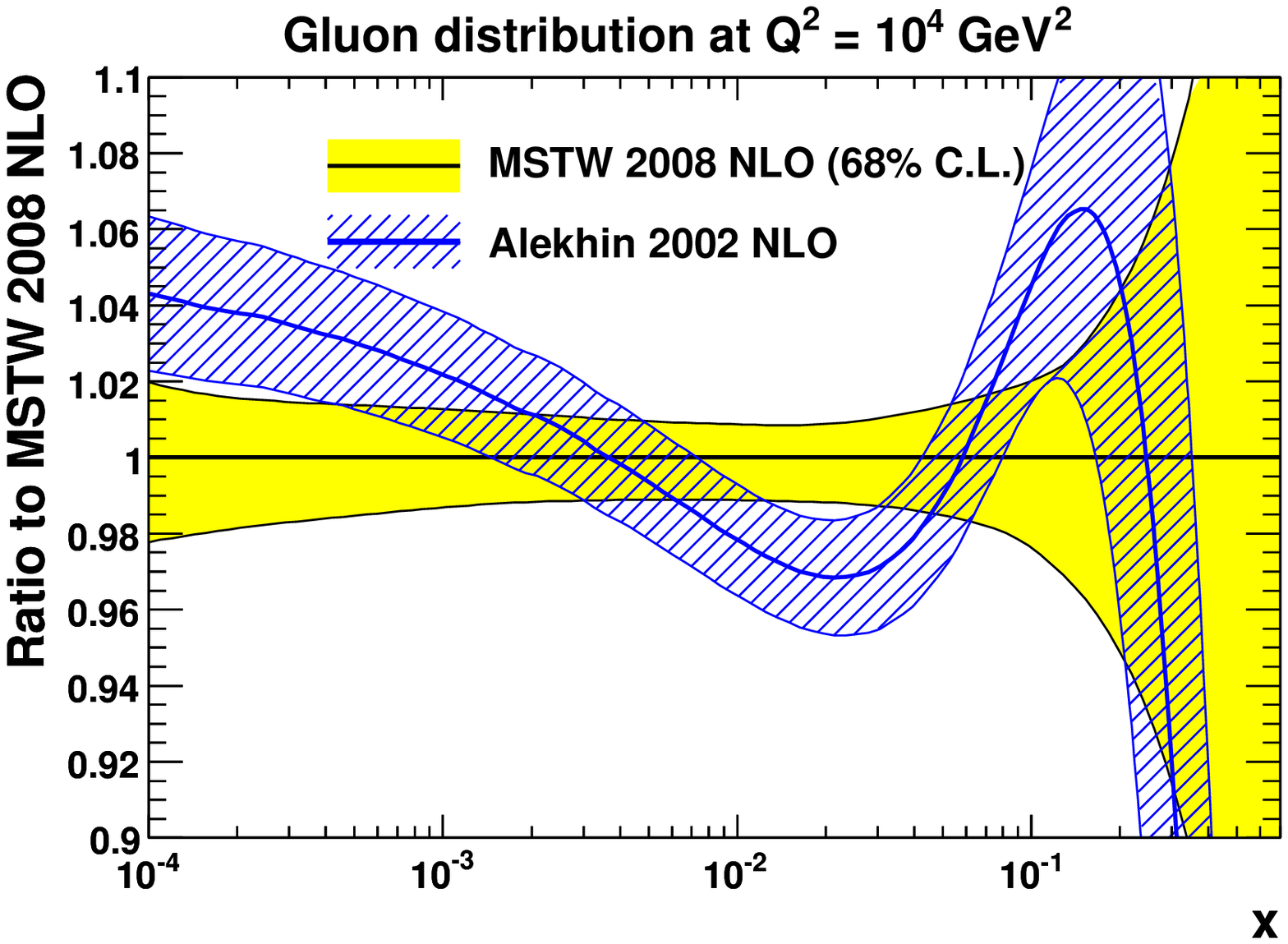}
  \caption{MSTW 2008 NLO PDFs compared to Alekhin NLO PDFs.}
  \label{fig:comparisonAlekhinNLO}
\end{figure}
In Fig.~\ref{fig:comparisonAlekhinNLO} we show the NLO PDFs compared to those from Alekhin~\cite{Alekhin:2002fv}.  As one can see there are many significant differences.  The Alekhin NLO PDFs are obtained from a fit to structure function data only. An update including some Drell--Yan data, and in particular the E866/NuSea $pd/pp$ data, was made in Ref.~\cite{Alekhin:2006zm}, but only NNLO sets were made available.  Since NLO sets are more widely used, and there are some systematic differences in the approaches at NNLO, we instead compare to the NLO set.  There is clearly a significant discrepancy in quark distributions everywhere, despite the fact that largely the same neutral-current DIS data are fit.  This highlights the importance of fitting neutrino DIS data and Drell--Yan and vector boson hadroproduction data to obtain the quark flavour decomposition.  The absence of Tevatron jet data plays a part in the gluon distribution in the Alekhin set being much smaller at $x>0.3$.  The smaller value of $\alpha_S(M_Z^2)=0.117$ also contributes to differences, as does the fact that the PDFs correspond to the ZM-VFNS.  However, these certainly do not explain the major systematic differences.

\begin{figure}
  (a)\hspace{0.5\textwidth}(b)\\
  \includegraphics[width=0.5\textwidth]{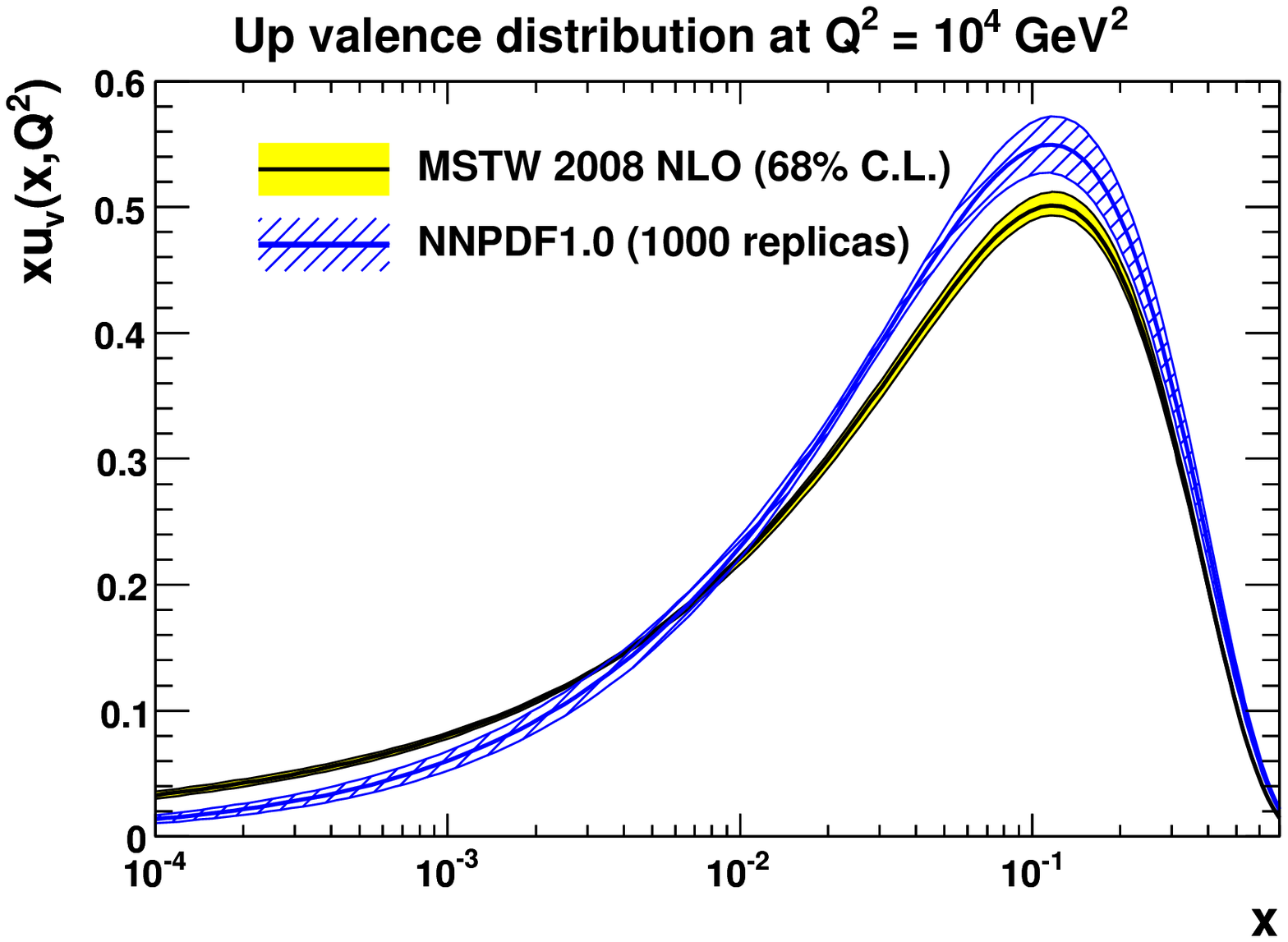}%
  \includegraphics[width=0.5\textwidth]{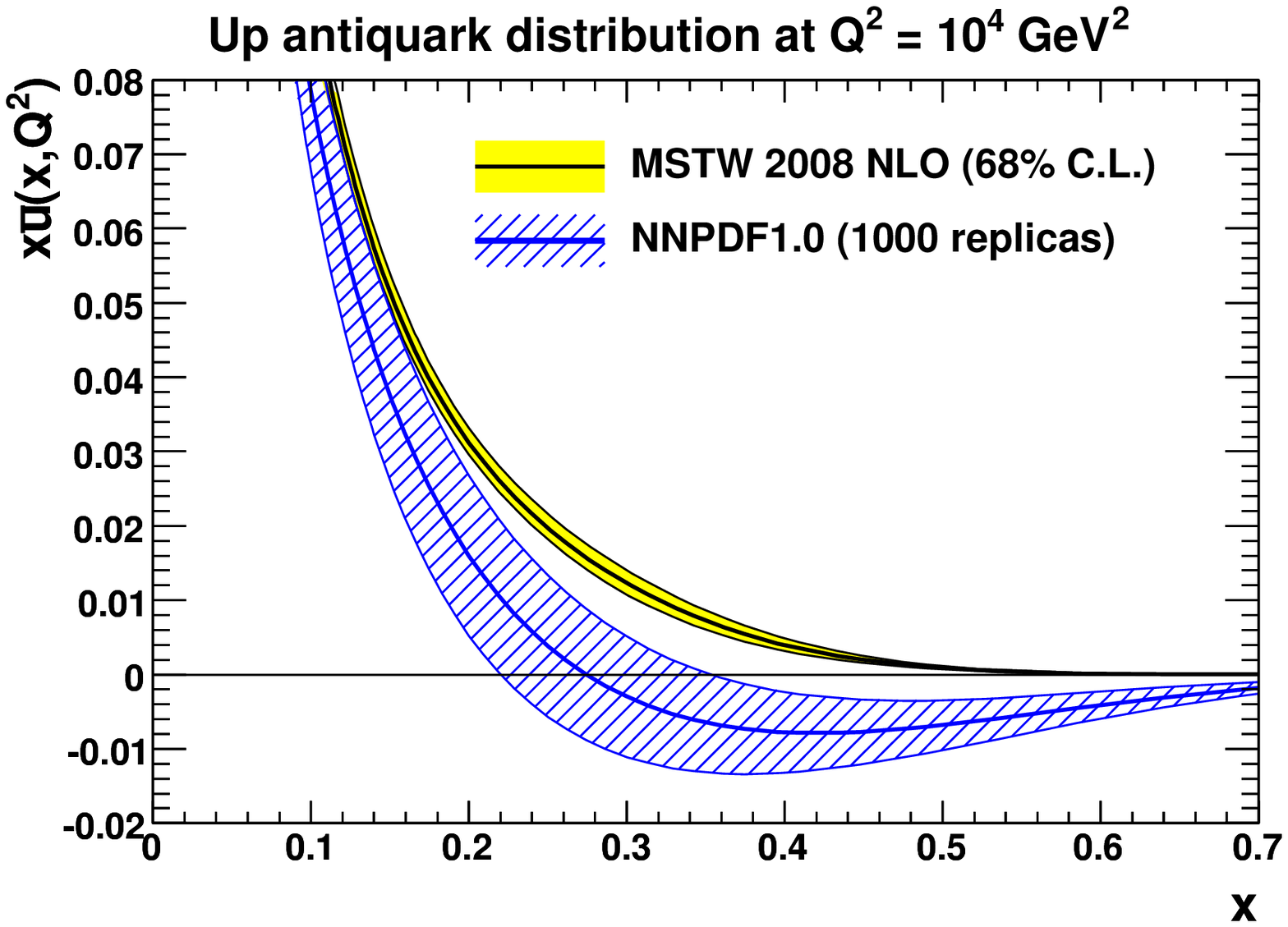}
  \caption{MSTW 2008 NLO (a) up valence quark and (b) high-$x$ $\bar u$ distribution compared to those from NNPDF1.0.}
  \label{fig:comparisonNNPDF1}
\end{figure}
Finally we compare to the very recent NNPDF1.0 set~\cite{Ball:2008by}, which uses proton and deuterium target DIS data, as well as CHORUS data (without nuclear corrections), again in the ZM-VFNS.  This approach relaxes the restriction in uncertainty due to fixed parameterisation limitations, though in practice relies on a very large number of parameters.  It also proceeds by making a Monte Carlo sample of the distribution of experimental data by generating a large number of replicas of data centred on each data point with full inclusion of the information from errors and their correlations.  Each replica is used to generate a PDF set and the mean for a quantity obtained from averaging, and uncertainties from standard deviations.  The data are split into training and validation sets for each replica, and the $\chi^2$ for one is monitored while the other minimised, thus avoiding over-complicating the input PDFs by stopping when the fit to the validation sets stop improving.  As with the Alekhin set there are some significant systematic differences between these and the MSTW 2008 sets.  However, the use of some neutrino DIS data and the fact that a fixed $\alpha_S(M_Z^2)=0.119$ is used in NNPDF1.0, together with a more flexible input parameterisation compared to Alekhin, means that there are less dramatic variations.  Some of the most significant variations are shown in Fig.~\ref{fig:comparisonNNPDF1} where we compare our up valence and up antiquark distribution to those from NNPDF1.0~\cite{Ball:2008by}.  Some significant differences are also observed in the total up quark distribution and the small-$x$ gluon at high $Q^2$.  From Fig.~\ref{fig:comparisonNNPDF1} we see there are sizeable differences in the central values of the up valence distribution, even where the sea is very small, perhaps due to the fact that relative data set normalisations are fitted in our approach but not in the NNPDF1.0 analysis.  The difference in the valence distribution is also undoubtedly due to the lack of data distinguishing between valence and sea quarks.  Indeed, the most striking disagreement is in the up antiquark distribution at very large $x$.  In the global fit this would be constrained to be positive by Drell--Yan data.  Some differences in central values is not a great surprise but it is more surprising that the error propagation used in the NNPDF1.0 analysis does not completely encompass the MSTW 2008 values.  Perhaps this is a reflection of the fact that some data sets used in the fit are incompatible, and discrepancies will sometimes occur, as we found comparing our benchmark fit and standard fit for the high-$x$ gluon in Section~\ref{sec:reduced}.
\begin{figure}
  (a)\hspace{0.5\textwidth}(b)\\
  \includegraphics[width=0.5\textwidth]{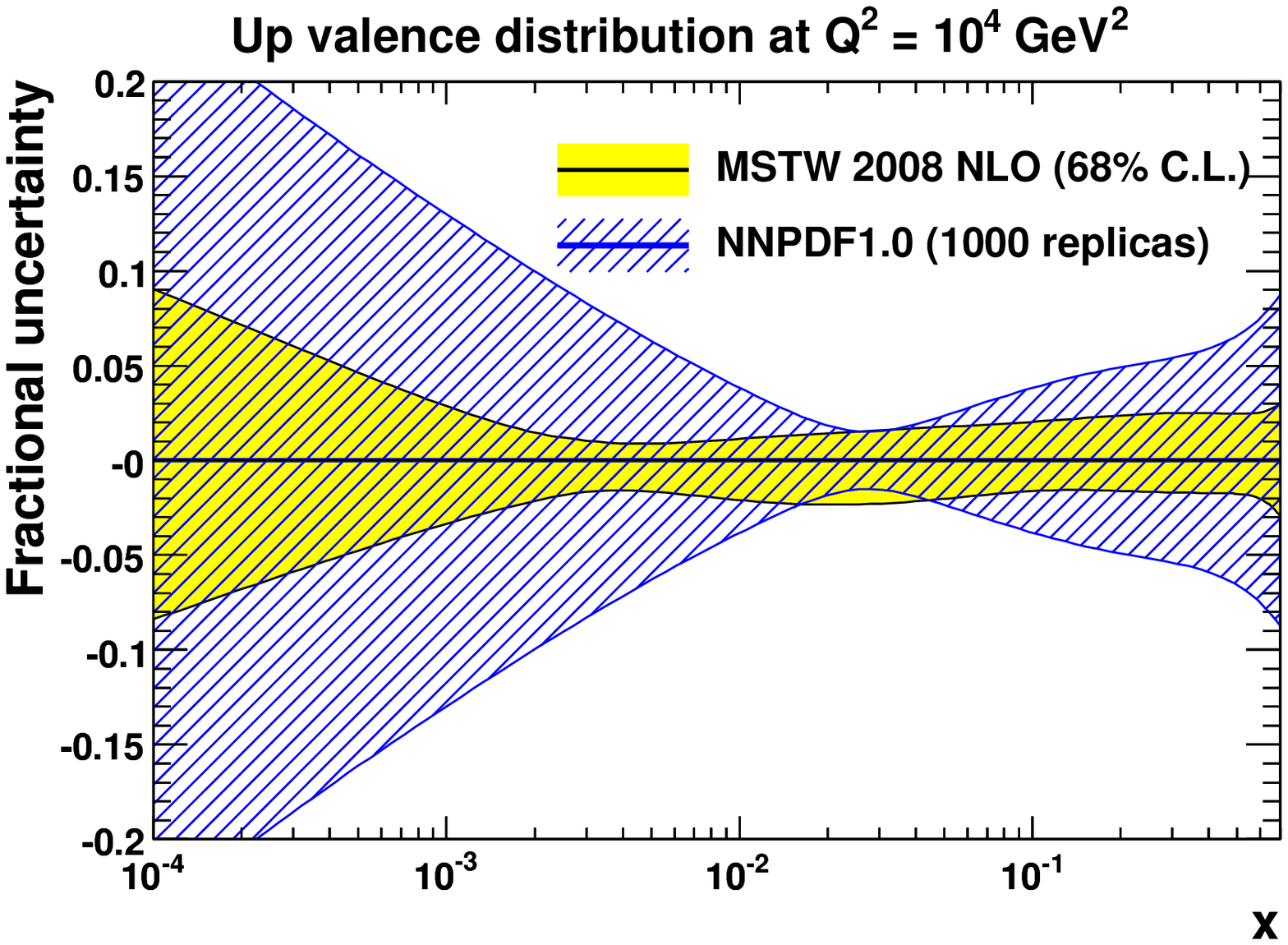}%
  \includegraphics[width=0.5\textwidth]{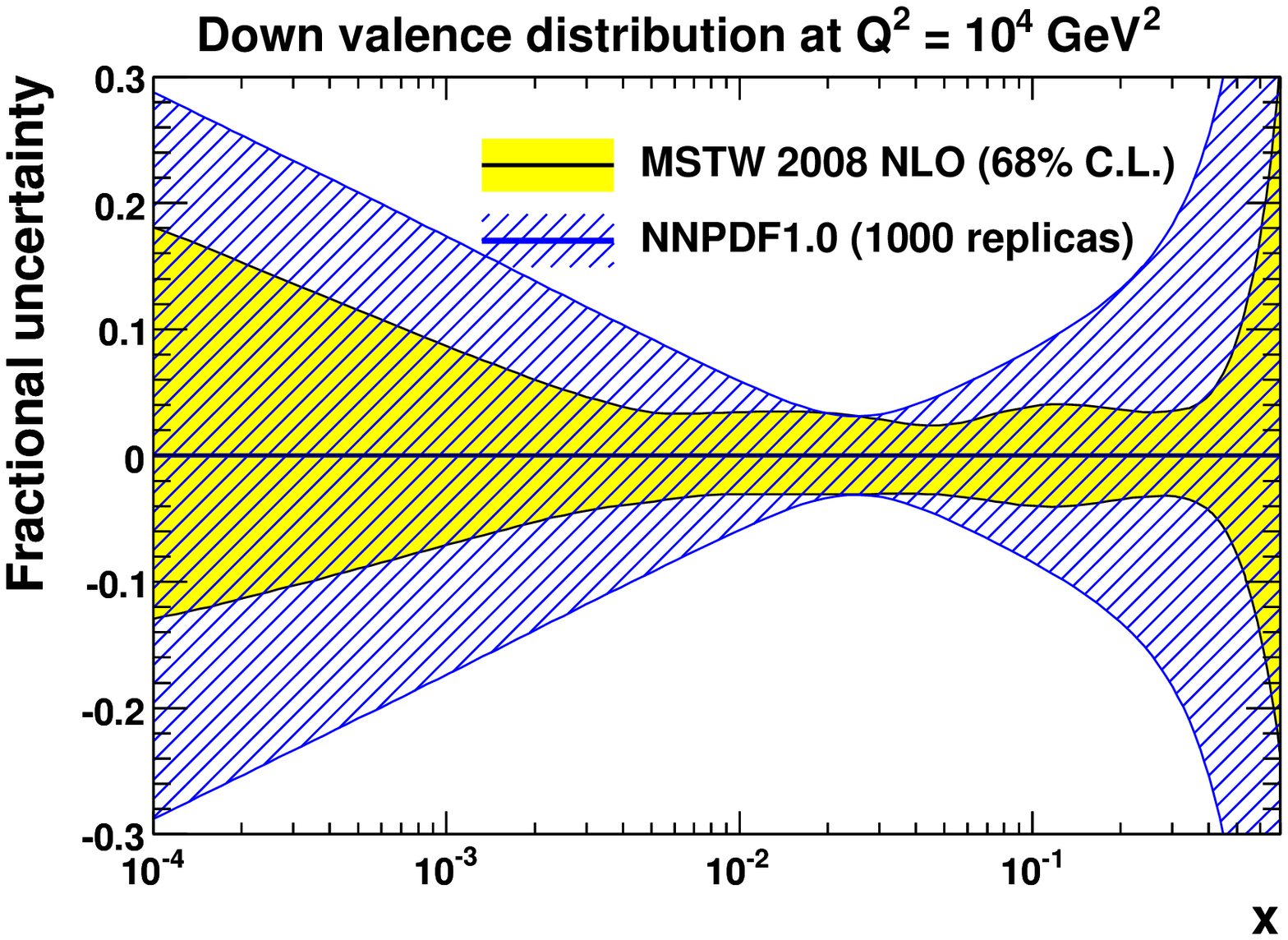}\\
  (c)\hspace{0.5\textwidth}(d)\\
  \includegraphics[width=0.5\textwidth]{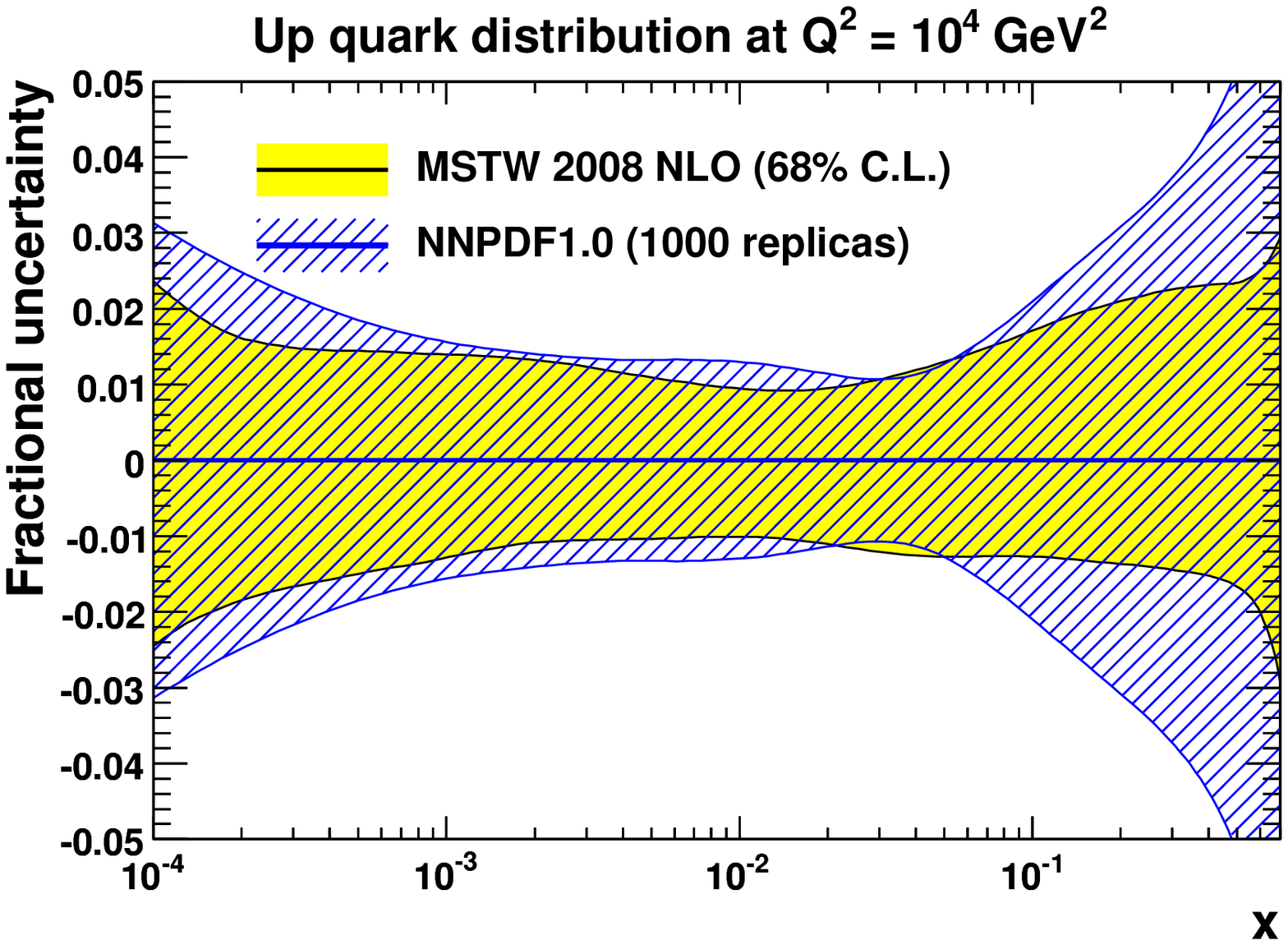}%
  \includegraphics[width=0.5\textwidth]{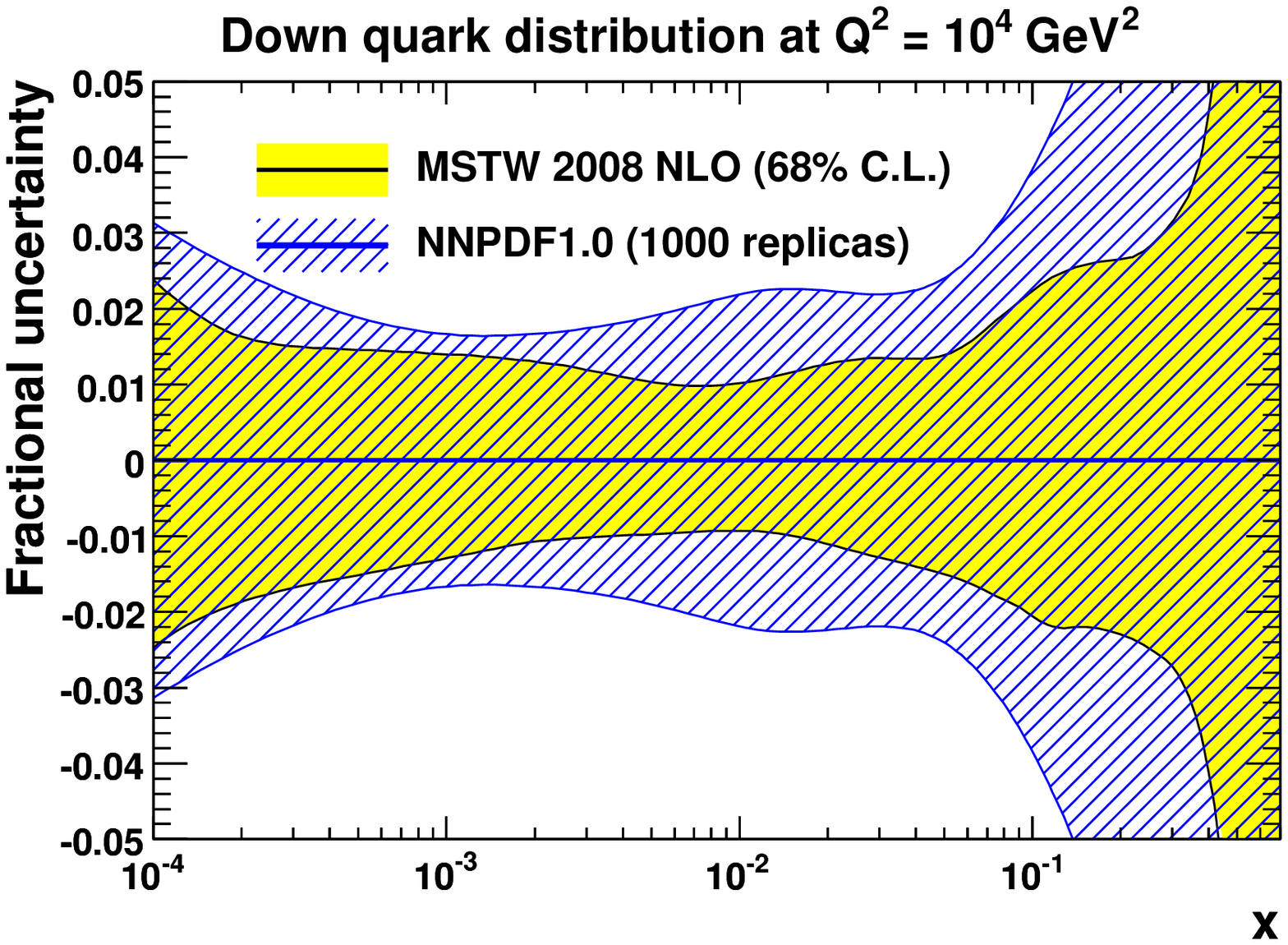}\\
  (e)\hspace{0.5\textwidth}(f)\\
  \includegraphics[width=0.5\textwidth]{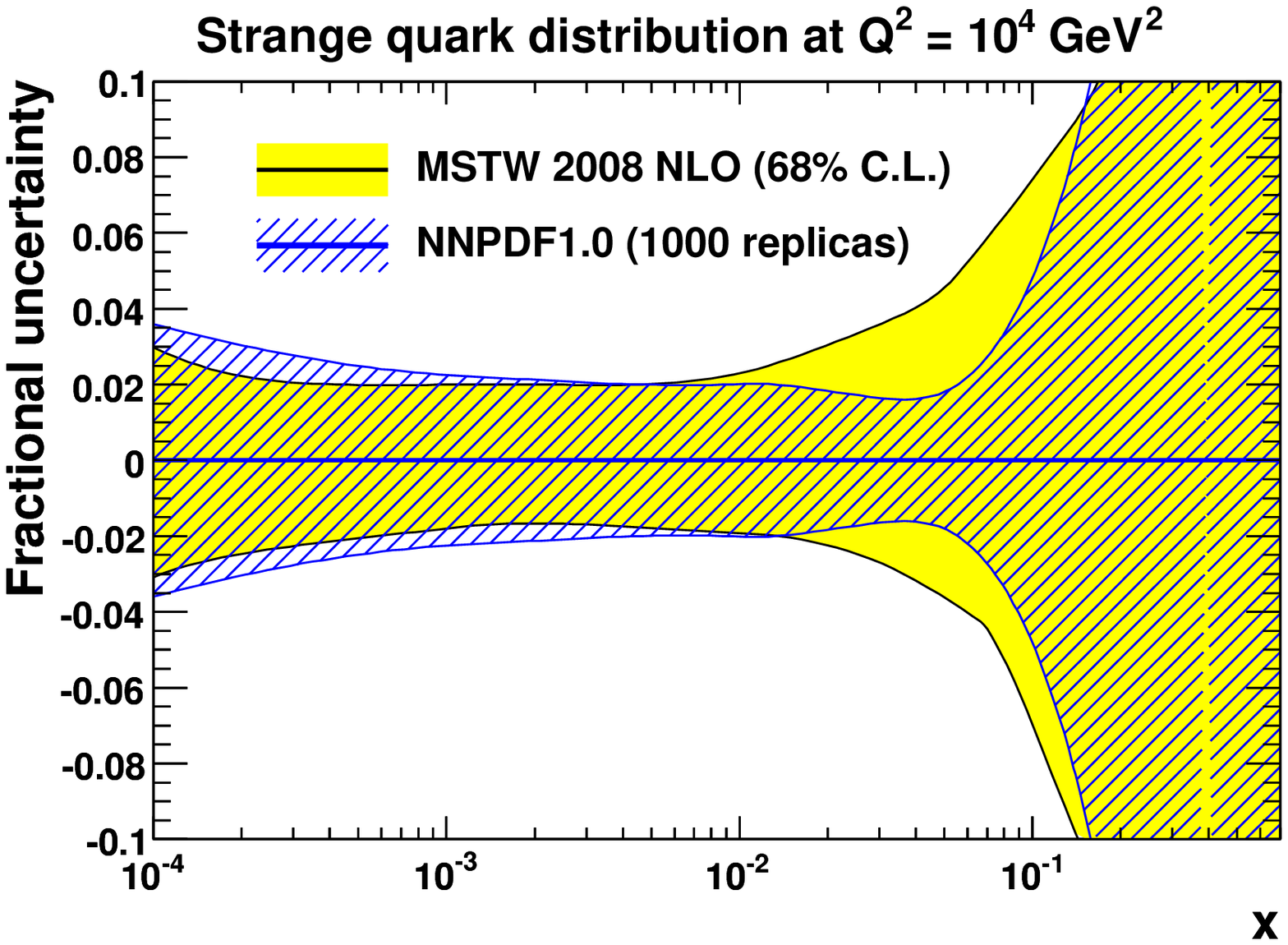}%
  \includegraphics[width=0.5\textwidth]{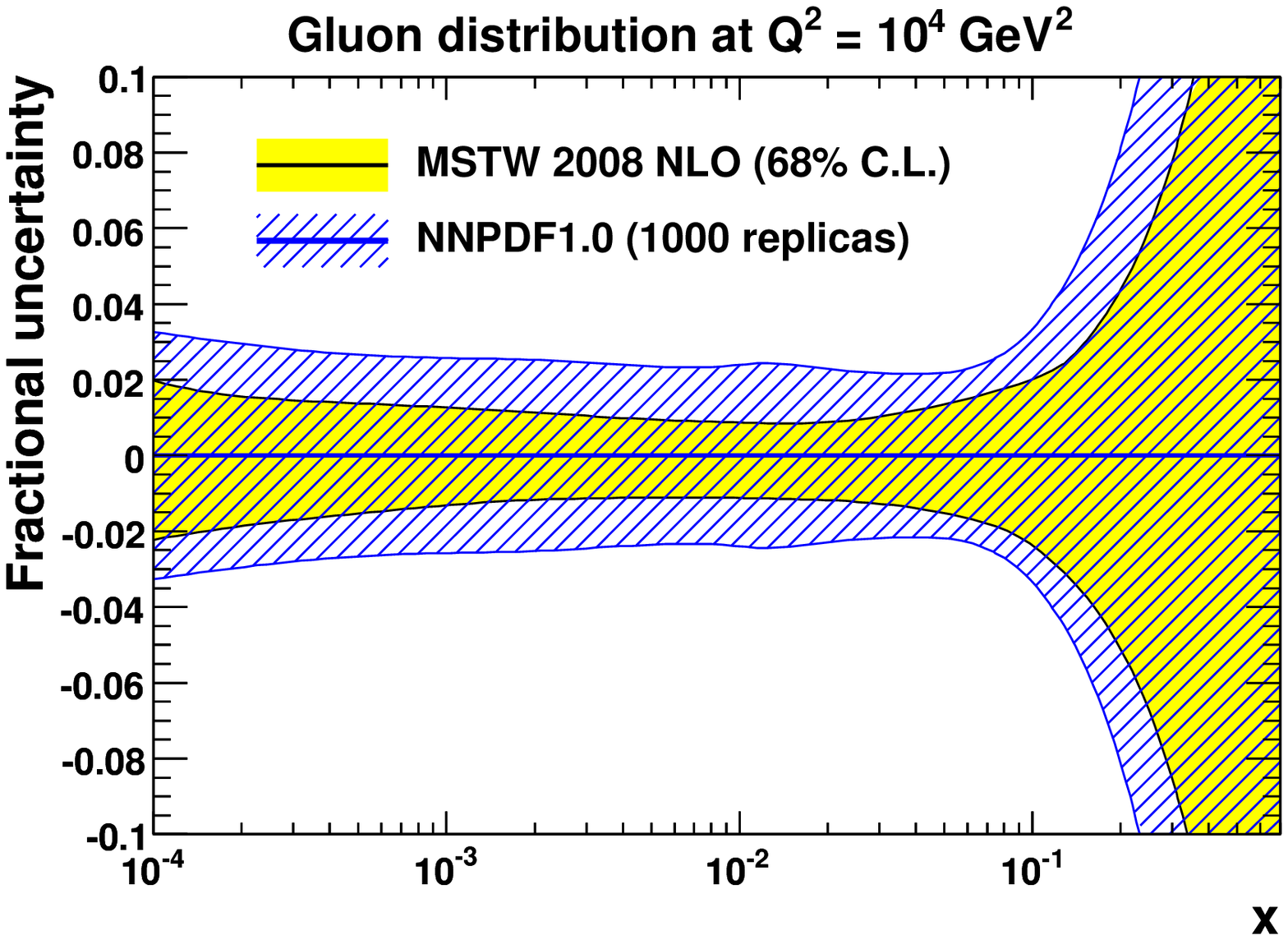}
  \caption{MSTW 2008 NLO PDF uncertainties compared to those from NNPDF1.0.}
  \label{fig:comparisonNNPDF}
\end{figure}
Since it is the uncertainties that are currently most interesting for the NNPDF1.0 set we concentrate our comparison on this.  In Fig.~\ref{fig:comparisonNNPDF} we compare our fractional uncertainties to those from NNPDF1.0~\cite{Ball:2008by}.  For total quark and gluon distributions the NNPDF uncertainties are only a little larger, and it must be remembered that we have many more data sets in the fit.  For the strange quark the fact that we fit directly, whereas in NNPDF1.0 the strange distribution at input is a fixed fraction of the total light sea, actually leads to our uncertainty being larger for $0.01\lesssim x \lesssim 0.2$.  The increased flexibility in the NNPDF parameterisation does lead to a larger uncertainty at high $x$ and small $x$ for valence distributions, though we can argue that we use more constraining data for $0.01\lesssim x \lesssim 0.7$, and have less flexibility from the sum rule outside this range.  The NNPDF1.0 valence quark uncertainty narrows in the region $x\simeq0.025$ at $Q^2=10^4$ GeV$^2$ (and at $x\simeq0.06$ at $Q^2=2$ GeV$^2$) which we do not understand, as this is just below the lower limit of the BCDMS data, which is one of the best constraints.  An update, NNPDF1.1~\cite{Rojo:2008ke}, has recently been reported.  The uncertainties are larger in this analysis due both to a randomisation of the preprocessing exponents and due to extra freedom introduced in the strangeness sector.  Since there are no dimuon data used to constrain the strange PDF, and very little constraint from any other data set, the uncertainty on $s(x,Q^2)+\bar s(x,Q^2)$ in NNPDF1.1 is huge.  This feeds into the other PDFs, with uncertainties increasing by factors of three for many quark combinations, and it is difficult to tell what effect the randomisation alone has.  This illustrates how important the direct constraint on the strange PDF, and some theoretical input in the parameterisation, can be when obtaining PDFs with genuinely representative uncertainties.

%% file: predictions.tex
\section{\texorpdfstring{$W$}{W} and \texorpdfstring{$Z$}{Z} total cross section predictions} \label{sec:totalpredictions}

In this section we present and discuss predictions for the total cross sections for $W$ and $Z$ production multiplied by leptonic branching ratios at the Tevatron and LHC.  To begin with we consider $W \equiv W^+ + W^-$ production, and subsequently consider the separate $W^+$ and $W^-$ total cross sections at the LHC.  (At the Tevatron, the $W^+$ and $W^-$ total cross sections are equal.)

To avoid confusion, we first define exactly what we mean by the total $W$ and $Z$ cross sections.  We assume a narrow-width approximation, in which the $W$ and $Z$ are treated as on-shell particles, of mass $M_W$ and $M_Z$ respectively, in the calculation of the production cross sections.  These cross sections are then multiplied by the appropriate leptonic branching ratios.  We assume that the cross sections measured at the leptonic level over a restricted region of phase space can be corrected back to this ``theoretical'' quantity.  We present predictions at LO, NLO and NNLO in perturbative QCD, using a program based on that of van Neerven {\it et al.}~\cite{Hamberg:1990np} incorporating subsequent corrections from Harlander and Kilgore~\cite{Harlander:2002wh}.  We use LO electroweak perturbation theory, with the $qqW$ and $qqZ$ couplings defined by 
\begin{equation}
  g_W^2 =  G_F M_W^2 / \sqrt{2}, \qquad g_Z^2 = G_F M_Z^2 \sqrt{2}. 
\end{equation}
The electroweak mixing angle $\sin^2\theta_W^{\rm eff}$ determines the relevant strengths of the vector and axial-vector $Z$ couplings to the quarks.  Since our cross section predictions depend to some extent on the electroweak parameters, we use the latest (2008) values from the Particle Data Group (PDG)~\cite{Amsler:2008zz}.  We sum over $u,d,s,c,b$ quarks in the initial state, using the 2008 PDG (unitarity constrained) values of the CKM weak mixing matrix elements $|V_{qq^\prime}|$ in the case of $W^\pm$ production.

The MSTW 2008 predictions for the total $W$ and $Z$ cross sections, including the one-sigma PDF uncertainties, are given for the Tevatron and LHC in Table \ref{tab:wztot}.
\begin{table}
  \centering
  \begin{tabular}{l|c|c|c}
    \hline\hline
    Tevatron, $\sqrt{s} = 1.96$ TeV & $B_{\ell\nu} \cdot \sigma_W$ (nb) & $B_{\ell^+\ell^-}\cdot\sigma_Z$ (nb) & $R_{WZ}$ \\
    \hline
    MSTW 2008 LO   & $1.963^{+0.025}_{-0.028}$ $\left(^{+1.2\%}_{-1.4\%}\right)$ & $0.1788^{+0.0023}_{-0.0025}$ $\left(^{+1.3\%}_{-1.4\%}\right)$ & $10.98^{+0.02}_{-0.03}$ $\left(^{+0.2\%}_{-0.3\%}\right)$ \\
    MSTW 2008 NLO  & $2.659^{+0.057}_{-0.045}$ $\left(^{+2.1\%}_{-1.7\%}\right)$ & $0.2426^{+0.0054}_{-0.0043}$ $\left(^{+2.2\%}_{-1.8\%}\right)$ & $10.96^{+0.03}_{-0.02}$ $\left(^{+0.3\%}_{-0.2\%}\right)$ \\
    MSTW 2008 NNLO & $2.747^{+0.049}_{-0.042}$ $\left(^{+1.8\%}_{-1.5\%}\right)$ & $0.2507^{+0.0048}_{-0.0041}$ $\left(^{+1.9\%}_{-1.6\%}\right)$ & $10.96^{+0.03}_{-0.03}$ $\left(^{+0.2\%}_{-0.2\%}\right)$ \\
    \hline\hline\multicolumn{4}{c}{}\\\hline\hline
    LHC, $\sqrt{s} = 10$ TeV & $B_{\ell\nu} \cdot \sigma_W$ (nb) & $B_{\ell^+\ell^-}\cdot\sigma_Z$ (nb) & $R_{WZ}$ \\
    \hline
    MSTW 2008 LO   & $12.57^{+0.13}_{-0.19}$ $\left(^{+1.1\%}_{-1.5\%}\right)$ & $1.163^{+0.011}_{-0.017}$ $\left(^{+1.0\%}_{-1.5\%}\right)$ & $10.81^{+0.02}_{-0.02}$ $\left(^{+0.2\%}_{-0.2\%}\right)$ \\
    MSTW 2008 NLO  & $14.92^{+0.31}_{-0.24}$ $\left(^{+2.1\%}_{-1.6\%}\right)$ & $1.390^{+0.029}_{-0.022}$ $\left(^{+2.1\%}_{-1.5\%}\right)$ & $10.73^{+0.02}_{-0.02}$ $\left(^{+0.2\%}_{-0.2\%}\right)$ \\
    MSTW 2008 NNLO & $15.35^{+0.26}_{-0.25}$ $\left(^{+1.7\%}_{-1.6\%}\right)$ & $1.429^{+0.024}_{-0.022}$ $\left(^{+1.7\%}_{-1.6\%}\right)$ & $10.74^{+0.02}_{-0.02}$ $\left(^{+0.2\%}_{-0.2\%}\right)$ \\
    \hline\hline\multicolumn{4}{c}{}\\\hline\hline
    LHC, $\sqrt{s} = 14$ TeV & $B_{\ell\nu} \cdot \sigma_W$ (nb) & $B_{\ell^+\ell^-}\cdot\sigma_Z$ (nb) & $R_{WZ}$ \\
    \hline
    MSTW 2008 LO   & $18.51^{+0.22}_{-0.32}$ $\left(^{+1.2\%}_{-1.7\%}\right)$ & $1.736^{+0.019}_{-0.028}$ $\left(^{+1.1\%}_{-1.6\%}\right)$ & $10.66^{+0.02}_{-0.02}$ $\left(^{+0.2\%}_{-0.2\%}\right)$ \\
    MSTW 2008 NLO  & $21.17^{+0.42}_{-0.36}$ $\left(^{+2.0\%}_{-1.7\%}\right)$ & $2.001^{+0.040}_{-0.032}$ $\left(^{+2.0\%}_{-1.6\%}\right)$ & $10.58^{+0.02}_{-0.02}$ $\left(^{+0.2\%}_{-0.2\%}\right)$ \\
    MSTW 2008 NNLO & $21.72^{+0.36}_{-0.36}$ $\left(^{+1.7\%}_{-1.7\%}\right)$ & $2.051^{+0.035}_{-0.033}$ $\left(^{+1.7\%}_{-1.6\%}\right)$ & $10.59^{+0.02}_{-0.03}$ $\left(^{+0.2\%}_{-0.3\%}\right)$ \\
    \hline\hline
  \end{tabular}
  \caption{Predictions for $W\equiv W^++W^-$ and $Z$ total cross sections at the Tevatron and LHC, including the one-sigma PDF uncertainties, and their ratio $R_{WZ}$.  We take $\mu_R=\mu_F=M_{W,Z}$.}
  \label{tab:wztot}
\end{table}
It is seen that the NNLO PDF uncertainties are around 2\% at both the Tevatron and LHC.  The uncertainty coming from neglected higher-order QCD corrections beyond NNLO, estimated by allowing the factorisation and renormalisation scales to vary so that the ratios $\mu_F/M_{W,Z}$, $\mu_R/M_{W,Z}$ and $\mu_F/\mu_R$ all lie in the range between $0.5$ and $2$ (as in, for example, Ref.~\cite{Cacciari:2008zb}), is around $\pm 0.5\%$ at the Tevatron and $\pm 1\%$ at the LHC.  Uncertainties not well accounted for by scale variations, in particular the effect of small-$x$ resummations, variations in GM-VFNS and changes in $\alpha_S(M_Z^2)$ may all lead to further uncertainties.  These are difficult to quantify at NNLO, but at the LHC may be $1$--$2\%$ in each case.  For the Tevatron only the last, i.e.~$\alpha_S(M_Z^2)$, is likely to be significant.
\begin{figure}
  \begin{center}
    \includegraphics[width=0.8\textwidth]{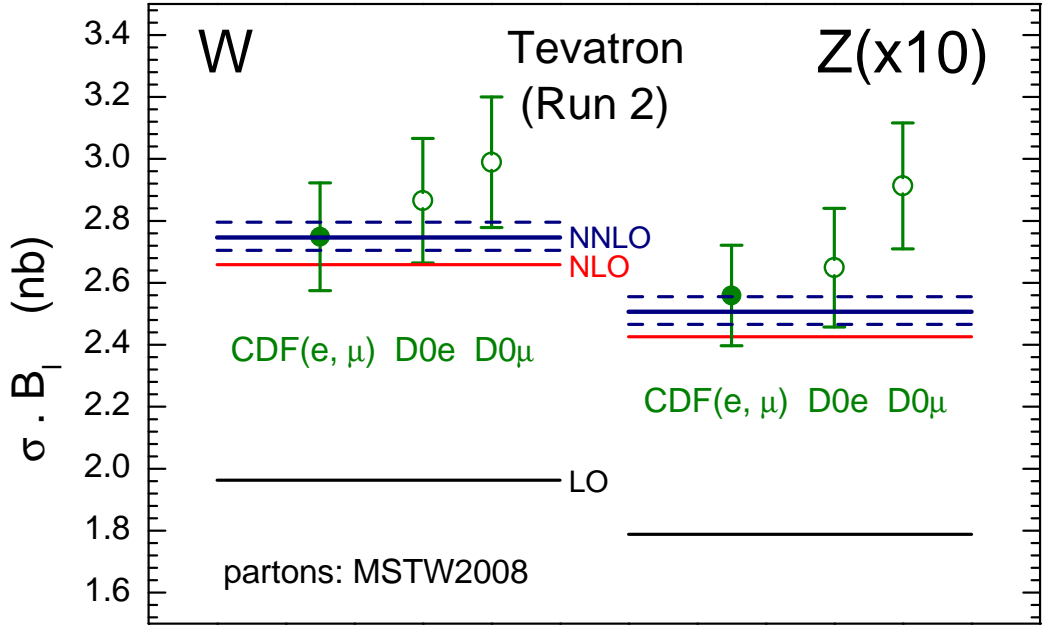}
    \caption{$W\equiv W^++W^-$ and $Z$ total cross sections compared to Tevatron data~\cite{Abulencia:2005ix,D0wztot}, where the dashed lines show the one-sigma PDF uncertainties on the NNLO predictions.  The luminosity uncertainty of $\sim \pm6\%$ is not included on the experimental data points.}
    \label{fig:wztev}
  \end{center}
\end{figure}
In Fig.~\ref{fig:wztev} we compare the MSTW 2008 NLO and NNLO predictions for the $W$ and $Z$ total cross sections to Tevatron data from CDF~\cite{Abulencia:2005ix} and D{\O}~\cite{D0wztot}.  The agreement is good.  The experimental error, which is dominated by the $\sim \pm 6\%$ uncertainty on the machine luminosity measurement, is significantly larger than the uncertainty on the theoretical predictions.

\begin{figure} 
  \begin{center}
    \includegraphics[width=0.8\textwidth]{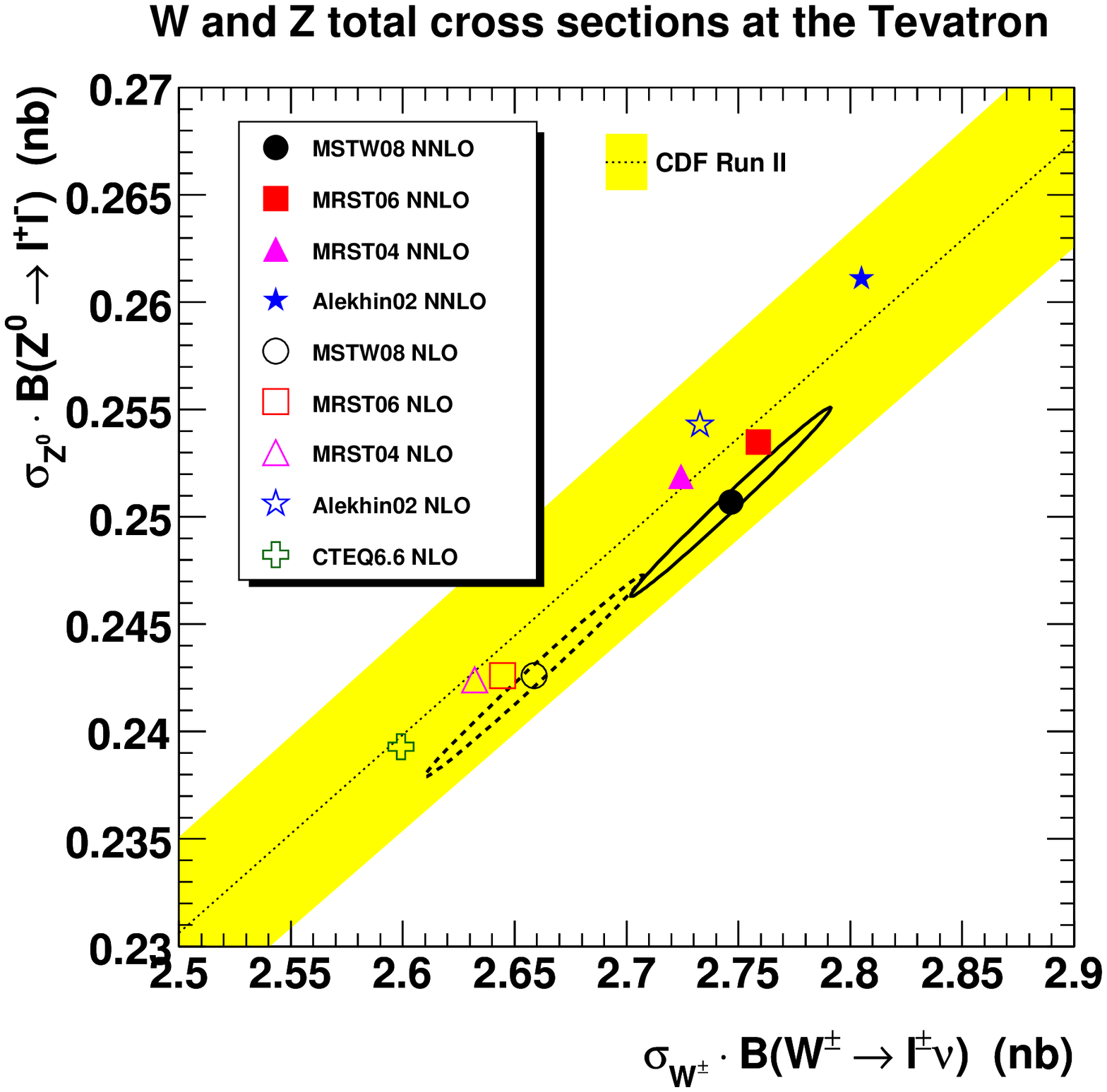}
    \caption{$W\equiv W^++W^-$ and $Z$ total cross sections at the Tevatron ($\sqrt{s} = 1.96$ TeV).  The error ellipses, calculated using the one-sigma error sets, are shown for the MSTW 2008 NLO and NNLO PDFs.  We show the CDF Run II measurement of the ratio $R_{WZ} = 10.84\pm0.15(\rm stat.)\pm0.14(\rm syst.)$~\cite{Abulencia:2005ix} with statistical and systematic uncertainties added in quadrature to get the overall uncertainty band.}
    \label{fig:wztotTev}
  \end{center}
\end{figure}
\begin{figure} 
  \begin{center}
    \includegraphics[width=0.8\textwidth]{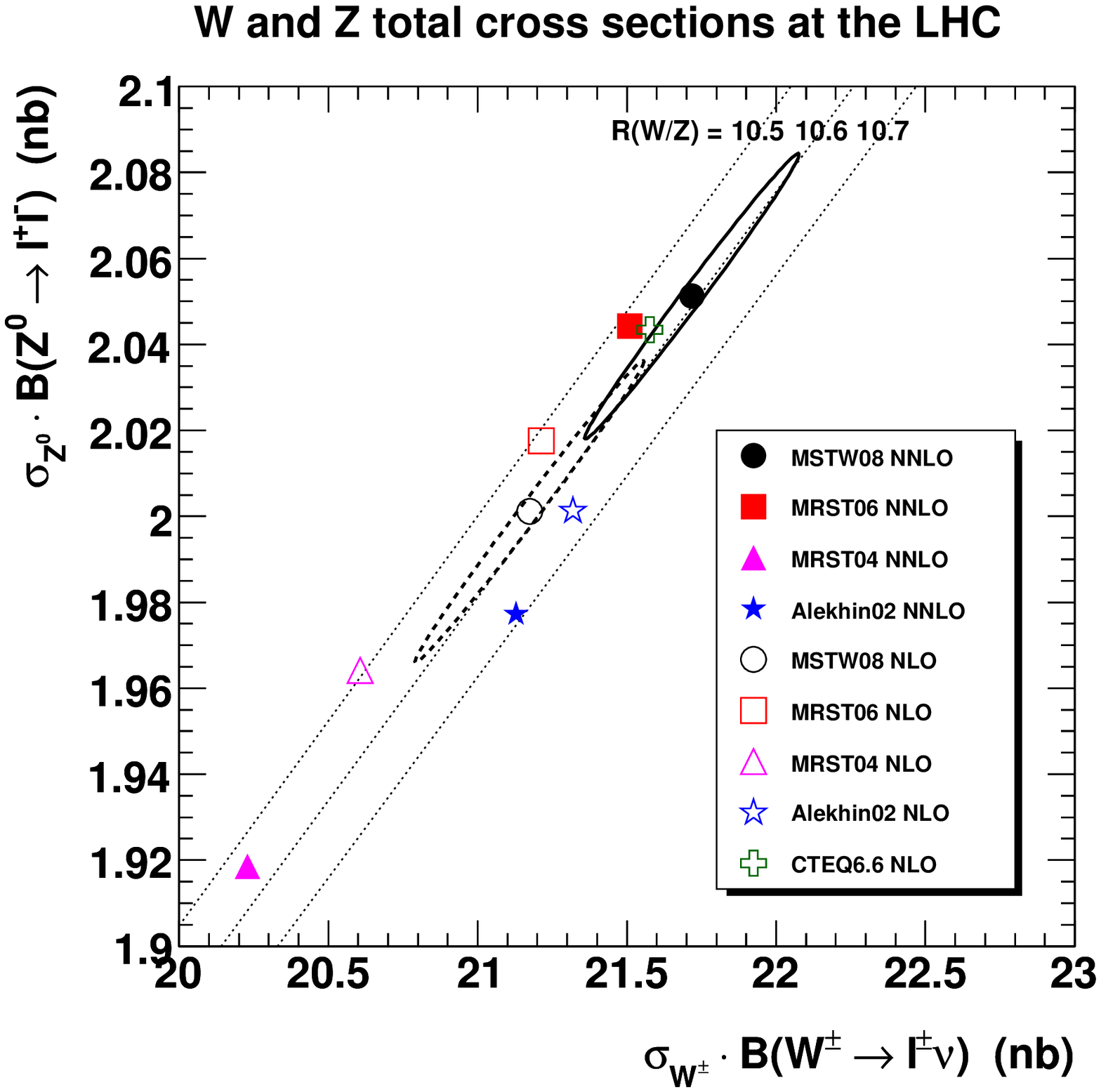}
    \caption{$W\equiv W^++W^-$ and $Z$ total cross sections at the LHC ($\sqrt{s} = 14$ TeV).  The error ellipses, calculated using the one-sigma error sets, are shown for the MSTW 2008 NLO and NNLO PDFs.}
    \label{fig:wztotLHC}
  \end{center}
\end{figure}
For both theory and experiment, the uncertainty in the $W$ and $Z$ cross section {\it ratio} is much smaller than on the individual cross sections.  This is illustrated in Figs.~\ref{fig:wztotTev} and \ref{fig:wztotLHC}, which show two-dimensional plots of the MSTW 2008 NLO and NNLO $W$ cross sections against the corresponding $Z$ cross sections, including the error ellipses calculated using \eqref{eq:ellipseF} and \eqref{eq:ellipseG} with the one-sigma error sets.\footnote{As noted in Ref.~\cite{Nadolsky:2008zw}, the error ellipse calculated with e.g.~one-sigma PDFs corresponds to a confidence level smaller than one-sigma in two dimensions~\cite{Amsler:2008zz,NumericalRecipes}.}  We compare to the central values of cross sections calculated with a variety of other PDF sets.  For the LHC predictions in Fig.~\ref{fig:wztotLHC}, we show lines of constant $R_{WZ} = B_{\ell\nu} \cdot \sigma_W/B_{\ell\bar{\ell}}\cdot\sigma_Z$, and on the corresponding Tevatron plot in Fig.~\ref{fig:wztotTev} we overlay a recent measurement of $R_{WZ} = 10.84\pm0.15(\rm stat.)\pm0.14(\rm syst.)$ from CDF~\cite{Abulencia:2005ix}.

\begin{table}
  \centering
  \begin{tabular}{l|cc|cc}
    \hline\hline
    & \multicolumn{2}{c|}{Tevatron} & \multicolumn{2}{c}{LHC} \\
    Ratio to MSTW 2008 & $\sigma_W$ & $\sigma_Z$ & $\sigma_W$ & $\sigma_Z$ \\
    \hline
    MRST 2006 NNLO              & 1.00 & 1.01 & 0.99 & 1.00 \\ 
    MRST 2004 NNLO              & 0.99 & 1.00 & 0.93 & 0.94 \\ 
    Alekhin02 NNLO              & 1.02 & 1.04 & 0.97 & 0.96 \\ \hline
    MRST 2006 NLO (unpublished) & 0.99 & 1.00 & 1.00 & 1.01 \\
    MRST 2004 NLO               & 0.99 & 1.00 & 0.97 & 0.98 \\
    Alekhin02 NLO               & 1.03 & 1.05 & 1.01 & 1.00 \\
    CTEQ6.6 NLO                 & 0.98 & 0.99 & 1.02 & 1.02 \\
    \hline\hline
  \end{tabular}
  \caption{Ratios of predictions for $W\equiv W^++W^-$ and $Z$ total cross sections at the Tevatron ($\sqrt{s} = 1.96$ TeV) and LHC ($\sqrt{s} = 14$ TeV) calculated using the central values of different PDF sets, with respect to those from the MSTW 2008 sets.}
  \label{tab:wztotpdf}
\end{table}
The ratios of predictions for $W$ and $Z$ cross sections calculated using the central values of different PDF sets, with respect to those from the MSTW 2008 sets, are given in Table \ref{tab:wztotpdf}.  The difference between the predictions of the various PDF sets can be understood by considering the values of the corresponding PDFs in the relevant $x$ and $Q^2$ region. We have already noted that the relevant PDFs in the MRST 2006 and MSTW 2008 sets are very similar, see Fig.~\ref{fig:comparisonNNLO}, and this is reflected in the similarity of the $W$ and $Z$ cross sections.  We have also noted that a more careful treatment of the heavy quark ($c$ and $b$) distributions at NNLO in the MRST 2006 and MSTW 2008 sets gave rise to a significant increase in the light quark distributions at high $Q^2$, and this explains the increase in the $W$ and $Z$ cross sections compared to MRST 2004.\footnote{A similar effect was observed at NLO in the evolution of CTEQ6.1 to CTEQ6.5 and CTEQ6.6~\cite{Tung:2006tb,Nadolsky:2008zw}.}  The CTEQ6.6 NLO predictions are close to those of MSTW 2008 NLO at the Tevatron, but some $2\%$ higher at the LHC. This can be traced back to the more rapid evolution of the CTEQ quarks at small $x$, driven by a (larger) gluon that, unlike MSTW 2008, is constrained to be positive at $Q_0^2$, see Fig.~\ref{fig:comparisonNLO}.  The Alekhin NLO and NNLO predictions for the ratio $R_{WZ}$ at the LHC are larger than the predictions from all other PDF sets, due to the small strange quark distribution in the Alekhin PDF sets; see Section.~\ref{sec:comparison}.

We have so far considered total $W = W^+ + W^-$ production, but of course at the LHC $\sigma_{W^+}\neq\sigma_{W^-}$, and the ratio $R_\pm \equiv  \sigma_{W^+}/\sigma_{W^-}$ is also of interest from the PDF perspective. Assuming a diagonal CKM matrix we have, at leading order,
\begin{equation}
  R_\pm \approx   \frac{u(x_1)\bar d(x_2) + c(x_1) \bar s(x_2)  + (1\leftrightarrow 2)    }{ d(x_1)\bar u(x_2) + s(x_1) \bar c(x_2)  + (1\leftrightarrow 2)     }.
\end{equation}
For central production, which dominates the total cross sections, we have $x_1 \sim x_2 \sim M_W/\sqrt{s}$ and $Q\sim M_W$.  In the limit $\sqrt{s} \to \infty$, $R_\pm \to 1$ since the quark and antiquark distributions become equal in this limit. The fact that $R_\pm \neq 1$ at LHC is a reflection of the residual differences, particularly between $u$ and $\bar u$ and between $d$ and $\bar d$, due to the valence quarks.
\begin{figure}
  \begin{center}
    \includegraphics[width=0.8\textwidth]{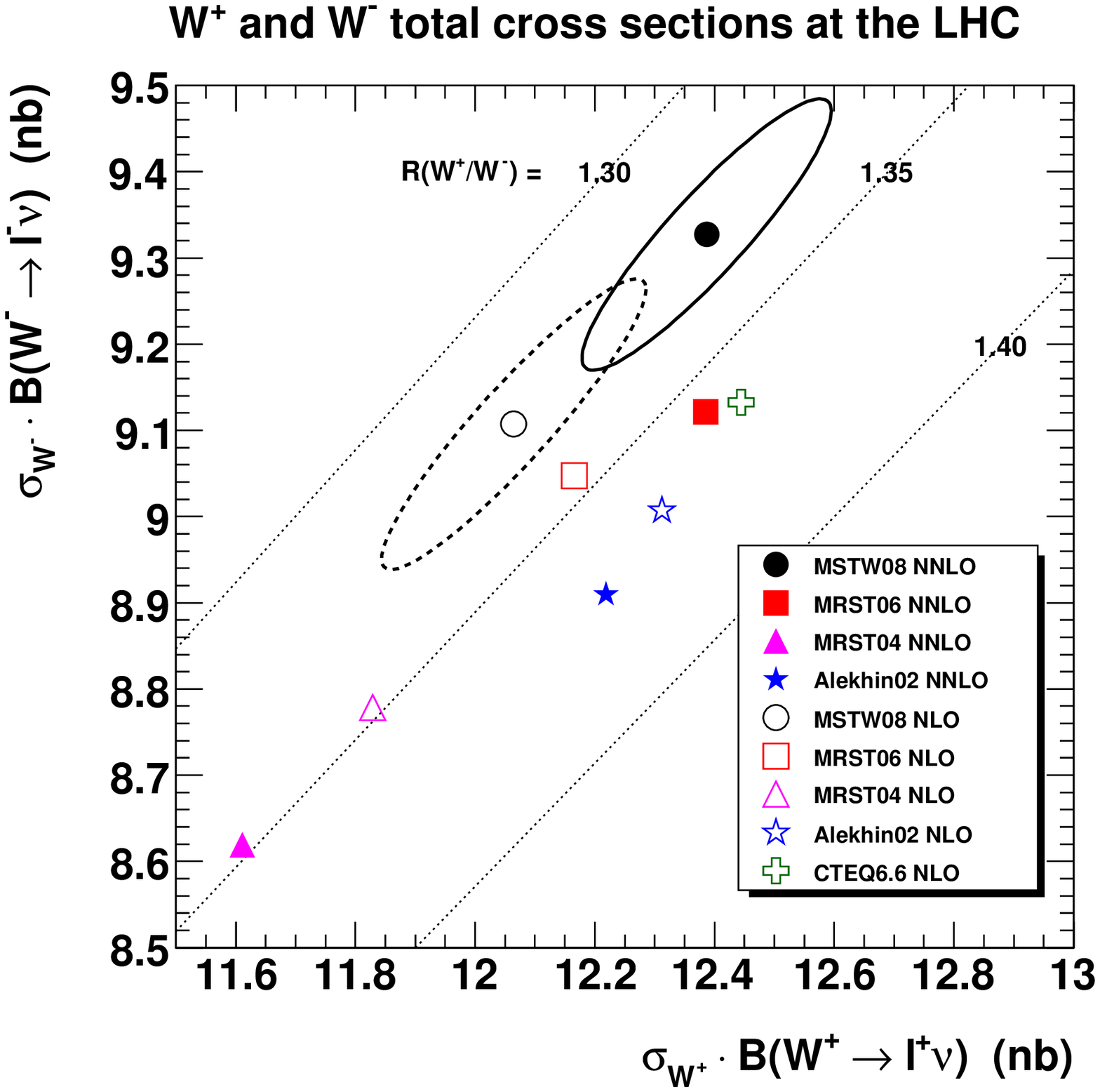}
    \caption{$W^+$ and $W^-$ total cross sections at the LHC ($\sqrt{s} = 14$ TeV).  The error ellipses, calculated using the one-sigma error sets, are shown for the MSTW 2008 NLO and NNLO PDFs.}
    \label{fig:w+w-tot}
  \end{center}
\end{figure}
\begin{table}
  \centering
  \begin{tabular}{l|c|c|c}
    \hline\hline
    LHC, $\sqrt{s} = 10$ TeV & $B_{\ell\nu} \cdot \sigma_{W^+}$ (nb) & $B_{\ell\nu}\cdot\sigma_{W^-}$ (nb) & $R_\pm$ \\
    \hline
    MSTW 2008 LO   & $7.35^{+0.08}_{-0.12}$ $\left(^{+1.1\%}_{-1.6\%}\right)$ & $5.22^{+0.06}_{-0.09}$ $\left(^{+1.1\%}_{-1.7\%}\right)$ & $1.408^{+0.015}_{-0.012}$ $\left(^{+1.0\%}_{-0.8\%}\right)$ \\
    MSTW 2008 NLO  & $8.62^{+0.18}_{-0.14}$ $\left(^{+2.1\%}_{-1.7\%}\right)$ & $6.30^{+0.14}_{-0.11}$ $\left(^{+2.2\%}_{-1.7\%}\right)$ & $1.367^{+0.012}_{-0.010}$ $\left(^{+0.9\%}_{-0.7\%}\right)$ \\
    MSTW 2008 NNLO & $8.88^{+0.15}_{-0.15}$ $\left(^{+1.7\%}_{-1.6\%}\right)$ & $6.47^{+0.11}_{-0.11}$ $\left(^{+1.7\%}_{-1.6\%}\right)$ & $1.373^{+0.012}_{-0.010}$ $\left(^{+0.8\%}_{-0.7\%}\right)$ \\
    \hline\hline\multicolumn{4}{c}{}\\\hline\hline
    LHC, $\sqrt{s} = 14$ TeV & $B_{\ell\nu} \cdot \sigma_{W^+}$ (nb) & $B_{\ell\nu}\cdot\sigma_{W^-}$ (nb) & $R_\pm$ \\
    \hline
    MSTW 2008 LO   & $10.69^{+0.14}_{-0.19}$ $\left(^{+1.3\%}_{-1.8\%}\right)$ & $7.83^{+0.10}_{-0.14}$ $\left(^{+1.2\%}_{-1.8\%}\right)$ & $1.366^{+0.013}_{-0.010}$ $\left(^{+0.9\%}_{-0.8\%}\right)$ \\
    MSTW 2008 NLO  & $12.06^{+0.24}_{-0.21}$ $\left(^{+2.0\%}_{-1.8\%}\right)$ & $9.11^{+0.19}_{-0.16}$ $\left(^{+1.2\%}_{-1.6\%}\right)$ & $1.325^{+0.011}_{-0.009}$ $\left(^{+0.8\%}_{-0.7\%}\right)$ \\
    MSTW 2008 NNLO & $12.39^{+0.22}_{-0.21}$ $\left(^{+1.8\%}_{-1.7\%}\right)$ & $9.33^{+0.16}_{-0.16}$ $\left(^{+1.7\%}_{-1.7\%}\right)$ & $1.328^{+0.011}_{-0.009}$ $\left(^{+0.8\%}_{-0.7\%}\right)$ \\
    \hline\hline
  \end{tabular}
  \caption{Predictions for $W^+$ and $W^-$ total cross sections at the LHC, including the one-sigma PDF uncertainties, and their ratio $R_\pm$.  We take $\mu_R=\mu_F=M_W$.}
  \label{tab:w+w-tot}
\end{table}
Fig.~\ref{fig:w+w-tot} shows various predictions for the $W^+$ and $W^-$ total cross sections at the LHC.  Lines of constant $R_\pm$ are also superimposed.  The MSTW 2008 predictions are listed in Table~\ref{tab:w+w-tot}.

Comparing Fig.~\ref{fig:w+w-tot} with the corresponding Fig.~\ref{fig:wztotLHC} for $W$ and $Z$ production, we see that there is less correlation between $W^+$ and $W^-$ than between $W$ and $Z$.  This is because the combinations of $u$ and $d$ quark and antiquark distributions probed in $W \equiv W^+ + W^-$ and $Z$ production are more similar than in $W^+$ and $W^-$ separately.  It is also interesting that the prediction for $R_\pm$ has decreased significantly in going from MRST 2006 to MSTW 2008, due to a change in the  $u/d$ ratio resulting from the addition of the new Tevatron $W$ and $Z$ data and neutrino DIS data, as discussed in Section~\ref{sec:wztevatron}.

Note that unlike $R_{WZ}$, which additionally depends on electroweak parameter and branching ratio values, the overwhelmingly dominant uncertainties in the theoretical predictions for $R_\pm$ are those due to the PDFs.  The experimental measurement should also be very precise, since it is simply a matter of comparing the number of $\ell^+$ and $\ell^-$ events in a sample of $W\to \ell\nu$ events.  From Fig.~\ref{fig:w+w-tot}, we see that a measurement of $R_\pm$ with an error of less than $1\%$ at the LHC will further constrain the parton distributions, particularly the $u/d$ ratio.

\begin{figure}
  \begin{center}
    \includegraphics[width=0.8\textwidth]{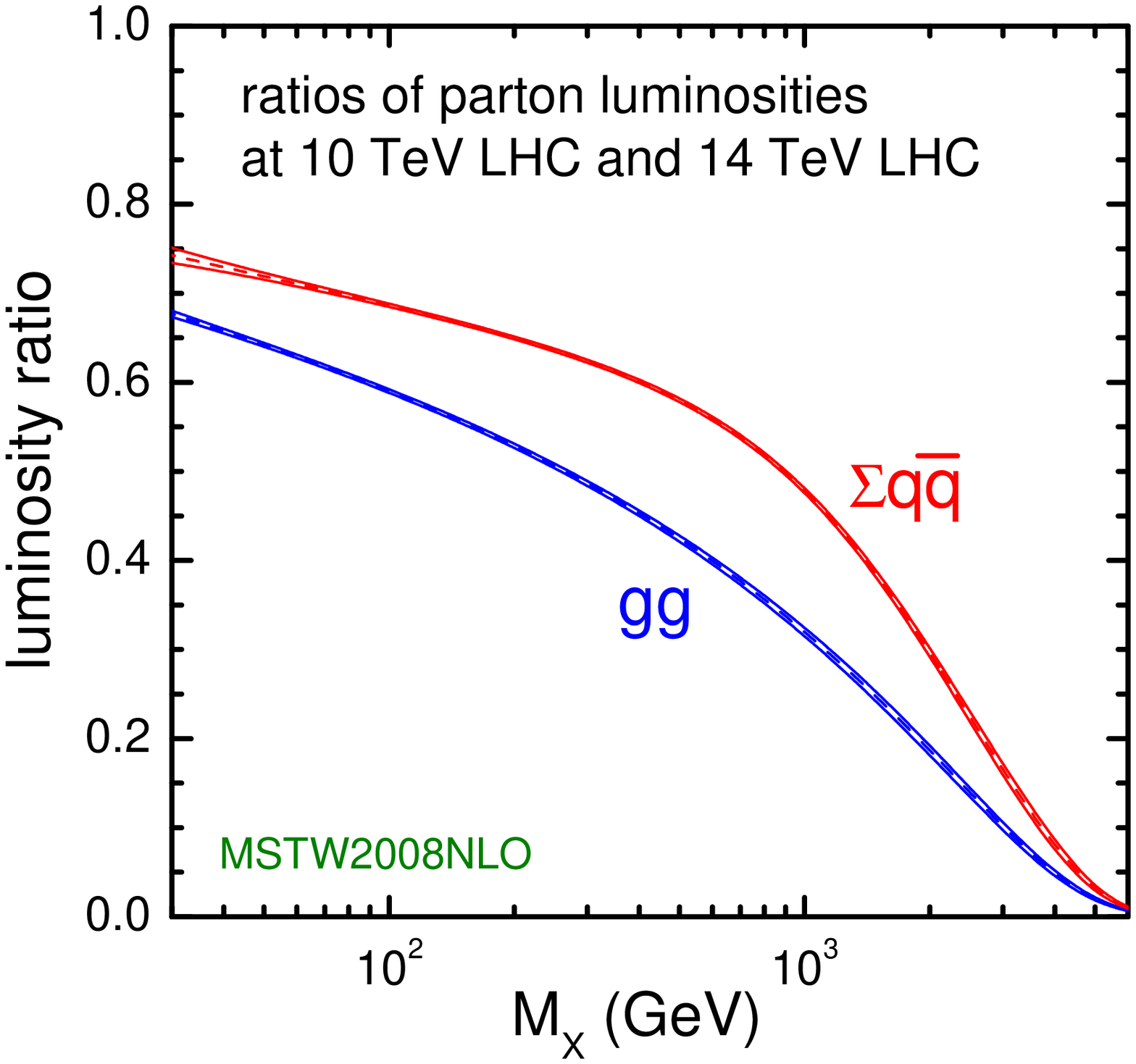}
    \caption{Ratio of parton luminosities at $\sqrt{s} = 10$ TeV compared to $\sqrt{s} = 14$ TeV at the LHC, using the MSTW 2008 NLO PDFs including the one-sigma uncertainty bands.}
    \label{fig:ratiolum}
  \end{center}
\end{figure}
In Fig.~\ref{fig:ratiolum} we show the ratio of the parton luminosities
\begin{equation}
  \frac{\partial {\cal L}_{ab}}{\partial M_X^2} = \frac{1}{s} \int_\tau^1\frac{{\rm d}x}{x}\;f_a(x,M_X^2)
  f_b(\tau/x,M_X^2), \quad \tau = \frac{M_X^2}{s}
\end{equation}
at $\sqrt{s} = 10$ TeV compared to $\sqrt{s} = 14$ TeV at the LHC for $ab=gg, \sum_{q = u,...,b} q\bar q$, using the MSTW 2008 NLO PDFs including the one-sigma uncertainty bands.  The PDF uncertainty on the ratio of the $W$ total cross section at 10 TeV compared to 14 TeV is of order $\pm0.3\%$, and similarly for the ratio of $Z$ total cross sections.  The PDF uncertainty is so small because it is only sensitive to the \emph{slope} (rather than absolute values) of the PDFs in the region between $x = 0.006$ and $x = 0.01$ (at central rapidity).  Other theoretical uncertainties, such as the choice of electroweak parameters, should also cancel out in this ratio.  Experimentally, the accuracy of the measurement of this ratio will depend on whether the \emph{relative} machine luminosity at 10 TeV and 14 TeV can be measured with high precision.

%% file: conclusions.tex
\section{Conclusions} \label{sec:conclusions}

We have presented the new MSTW 2008 parton distribution functions (PDFs) at leading-order (LO), next-to-leading order (NLO) and next-to-next-leading order (NNLO). In each case we include eigenvector sets spanning the uncertainty of the PDFs defined at a fixed value of $\alpha_S(M_Z^2)$.  These sets are a major update to all our previously available sets, and incorporate the maximum amount of information from deep-inelastic scattering (DIS) and other hard-scattering data.  We have also introduced a number of new theoretical refinements to produce what are arguably the most complete and reliable sets available.  The PDFs are specifically designed for use in precision cross section predictions and uncertainties at the LHC.  The grids and interpolation code (in Fortran, C++ and Mathematica formats) are available from \textsc{hepforge}~\cite{mstwpdf}, and are also made available as part of the \textsc{lhapdf} package~\cite{Whalley:2005nh} from version 5.7.0.

The principal new data sets included in the analysis are:
\begin{itemize}
\item neutrino structure function ($F_2$ and $xF_3$) data from NuTeV~\cite{Tzanov:2005kr} and CHORUS~\cite{Onengut:2005kv};
\item neutrino dimuon cross sections from CCFR and NuTeV~\cite{Goncharov:2001qe}, which constrain the strange quarks and antiquarks, allowing these to be fit directly for the first time, instead of using the previous input assumption that $s=\bar{s}=\kappa\,(\bar{u}+\bar{d})/2$ with $\kappa\approx 0.5$;
\item the $Z$ rapidity distribution~\cite{Abazov:2007jy,Han:2008} and the lepton charge asymmetry from $W$ decays~\cite{Acosta:2005ud,Abazov:2007pm} measured at the Tevatron Run II, both of which constrain mainly the down quark distribution;
\item the inclusive jet production data from the Tevatron Run II~\cite{Abazov:2008hu,Abulencia:2007ez}, which prefer a smaller high-$x$ gluon\footnote{The smaller gluon distribution at high $x$ has important consequences.  For example, the predicted Tevatron cross section for production via gluon--gluon fusion of a Standard Model Higgs boson with mass $M_H = 170$ GeV decreases by almost 15\% compared to the result using the previous MRST 2006 NNLO PDFs~\cite{Anastasiou:2008tj,deFlorian:2009hc} (for implications, see Ref.~\cite{Bechtle:2008jh}), although this is also partly due to the smaller value of $\alpha_S$ for MSTW 2008 NNLO.} than the previous Run I data;
\item updated HERA data on $F_2^{c\bar c}$~\cite{Adloff:1996xq,Adloff:2001zj,Aktas:2005iw,Aktas:2004az,Breitweg:1999ad,Chekanov:2003rb,Chekanov:2007ch} and on inclusive jet production in DIS~\cite{Aktas:2007pb,Chekanov:2002be,Chekanov:2006xr}.
\end{itemize}

In addition to including new data, there are a number of important developments in the fitting procedure.
\begin{itemize}
\item Our treatment of heavy flavours, carried out in a general-mass variable flavour number scheme (GM-VFNS), is the most complete available, particularly at NNLO where the improved treatment, first implemented in the MRST 2006 analysis~\cite{Martin:2007bv}, has a significant quantitative effect.  At NLO we estimate the possible theoretical uncertainty on the parton distributions due to variations in the specific choice of GM-VFNS to be around $2\%$, though this should decrease at NNLO.
\item We improve and extend our input parameterisation, not only introducing two new free parameters for each of $s+\bar{s}$ and $s-\bar{s}$, but using one more in the determination of the uncertainty on the gluon distribution, and making a better choice for the down valence quark distribution.
\item We have designed and implemented a new treatment of PDF uncertainties, doing away with the conventional notion of an overall fixed tolerance $\Delta\chi^2_{\rm global}=50$ (for a 90\% confidence-level limit) and replacing it with a new ``dynamical tolerance'' in which the tolerance for each independent eigenvector direction is determined from the condition that all data sets should be described within their 90\% (or 68\%) confidence-level limits.  In the same context, we have also included uncertainties from data set normalisations for the first time.  This can have an important effect on the PDF uncertainties, particularly for light quarks in the range of $x$ and $Q^2$ where they are needed for collider phenomenology.  It also means that, despite the presence of more constraining data in the fit, the PDF uncertainties have got slightly bigger in some cases.
\item We have replaced the unconventional choice of $\alpha_S$ used in the MRST fits, which is not strictly correct at NNLO, with the exact solution of the renormalisation group equation \eqref{eq:rge}, as used in recent public DGLAP evolution codes.  This has enabled our evolution code to be checked against the \textsc{pegasus}~\cite{Vogt:2004ns} and \textsc{hoppet}~\cite{Salam:2008qg} codes for the first time.
\item We use the \textsc{fastnlo} package~\cite{Kluge:2006xs} for NLO (and for Tevatron jets also for approximate NNLO) calculations of jet cross sections directly in the fit.  We have also improved our calculation of $K$-factors for low-mass Drell--Yan lepton pair and $W$ and $Z$ boson production using the \textsc{dyrap}/\textsc{vrap}~\cite{Anastasiou:2003ds} and \textsc{fewz}~\cite{Melnikov:2006kv} codes.
\end{itemize}

We have fitted a wide variety of data ($\sim2700$ data points in all) and overall the quality of the NLO and NNLO fits is similar and is perfectly acceptable, with $\chi^2/N_{\rm pts.} \sim 1$ for almost all data sets fitted.  As has been noticed in previous studies, however, the quality of the LO fit is poor, and there are a number of examples where the data really do require significant NLO corrections.  We therefore do not recommend that the LO set be used, for example, in leading-order parton-shower Monte Carlo event generators.  A modified LO set (see Refs.~\cite{Sherstnev:2007nd,Sherstnev:2008dm}) suitable for this purpose will be produced in a subsequent study.

Each of our best-fit PDF sets comes with a corresponding set of 40 error PDFs (compared to 30 in previous MRST versions), which can be used to estimate the PDF uncertainty on any physical quantity.  It is important to remember that these PDF uncertainties come from the experimental errors on input data only.  They do not include additional theory errors, coming, for example, from missing higher-order terms in the perturbation series. This issue will be addressed in a future study.  While the convergence of the series is an important and outstanding issue for some quantities studied, for example, small $x$ structure functions, it does not appear to be a problem in other cases, for example, $\sigma_{W,Z}$ at the hadron colliders. 

In subsequent studies we will return to some of the more theoretical issues, including the dependence of the fit on the value of $\alpha_S$ and the heavy quark ($c,b$) masses, and the impact on the fits of including ${\cal O}(\alpha)$ QED corrections, as was done in Ref.~\cite{Martin:2004dh}.  We also intend to investigate $W$, $Z$ and Drell--Yan lepton pair production at the LHC in more detail, with a particular focus on extracting information on the parton distributions at very small $x$.

In the meantime, these distributions are obtained from the most up-to-date and complete global fit yet performed, and we believe they are the most reliable sets available as we enter the LHC era.

%% file: acknowledgements.tex
\section*{Acknowledgements}

We thank C.~Gwenlan, K.~Hatakeyama, T.~Kluge, M.~Lancaster, D.~Mason, P.~Nadolsky, F.~Petriello, M.~Wobisch and J.~Zhu for valuable information, and S.~Alekhin, A.~Cooper-Sarkar, S.~Forte, J.~Huston, W.-K.~Tung and A.~Vogt for useful discussions.  R.S.T.~acknowledges the Royal Society for the award of a University Research Fellowship.  G.W.~acknowledges the UK Science and Technology Facilities Council for the award of a Responsive Research Associate position.